\documentclass[useAMS,a4paper,fleqn,usenatbib]{mnras}

\usepackage{mathptmx}

\usepackage[T1]{fontenc}
\usepackage{ae,aecompl}
\usepackage{geometry}

\usepackage{graphicx}	
\usepackage{subfig}
\usepackage{verbatim}
\usepackage{caption}
\usepackage{float}
\usepackage{enumerate}

\newcommand{\kms}{\ensuremath{\textrm{\,km s}^{-1}}}

\newcommand{\chisq}{\ensuremath{\chi^{2}_{red}}}
\newcommand{\chisqlt}{\ensuremath{\chi^{2}_{red} < 1}}

\newcommand{\hb}{\ensuremath{\mbox{H}{\beta}}}
\newcommand{\hg}{\ensuremath{\mbox{H}{\gamma}}}
\newcommand{\hd}{\ensuremath{\mbox{H}{\delta}}}
\newcommand{\ha}{\ensuremath{\mbox{H}{\alpha}}}
\newcommand{\oiii}{[OIII]} 

\newcommand{\ebv}{E(B-V)}
\newcommand{\ca}{Ca {\sc ii} K}
\newcommand{\gmos}{GMOS-S}
\newcommand{\pms}{\ensuremath{\pm}}
\newcommand{\tosp}{\ensuremath{t_{OSP}}}
\newcommand{\tisp}{\ensuremath{t_{ISP}}}
\newcommand{\tysp}{\ensuremath{t_{YSP}}}

\newcommand{\tmax}{\ensuremath{t_{max}}}
\newcommand{\confit}{{\sc confit }}

\title[Young stellar populations in type II quasars]{Young stellar populations in type II quasars: timing the onset of star formation and nuclear activity}

\author[P. S. Bessiere]{
P. S. Bessiere,$^{1,2}$\thanks{E-mail: pbessier@astro.puc.cl}
C.~N.~Tadhunter,$^{2}$
C.~Ramos Almeida$^{3,4}$
M.~Villar Mart\'{i}n$^{5}$
\newauthor
 A.~Cabrera-Lavers$^{3,4,6}$
\\
$^{1}$ Instituto de Astrof\'\i sica, Facultad de F\'\i sica, Pontificia Universidad Cat\'{o}lica de Chile, Casilla 306, Santiago 22, Chile\\
$^{2}$ Department of Physics and Astronomy, University of Sheffield, Sheffield, S3 7RH, UK\\
${^3}$ Instituto de Astrof\'\i sica de Canarias (IAC), C/V\'ia L\'actea, s/n, E-38205, La Laguna, Tenerife, Spain\\
${^4}$ Departamento de Astrof\'\i sica, Universidad de La Laguna, E-38205, La Laguna, Tenerife, Spain\\
${^5}$ Centro de Astrobiolog\'{i}a (INTA-CSIC), Carretera de Ajalvier, km 4, 28850 Torrej\'{o}n de Ardoz, Madrid, Spain\\
${^6}$ Gran Telescopio CANARIAS, Cuesta de San Jos\'e, s/n, E-38712, Bre\~{n}a Baja, La Palma, Spain
}

\date{Accepted XXX. Received YYY; in original form ZZZ}

\pubyear{2016}

\begin{document}
\label{firstpage}
\pagerange{\pageref{firstpage}--\pageref{lastpage}}
\maketitle

\begin{abstract}

Despite the emerging morphological evidence that luminous quasar-like AGN are triggered in galaxy mergers, the natures of the triggering mergers and the order of events in the triggering sequence remain uncertain. In this work, we present a detailed study of the stellar populations of the host galaxies of 21 type II quasars, with the aim of understanding the sequence of events between the onset of the merger, the triggering of the associated starburst and the initiation of the quasar activity. To this end, we model high quality, wide spectral coverage, intermediate resolution optical spectra of the type II quasars. We find that of the 21 objects, the higher-order Balmer absorption lines, characteristic of young stellar populations (YSP) , are directly detected in $\sim 62\%$ of the sample. We also fit these spectra using a number of combinations of stellar and/or power-law components, representative of viable formation histories, as well as including the possibility of scattered AGN light. We find that $\sim 90 \%$ of the type II quasar host galaxies require the inclusion of a YSP to adequately model their spectra, whilst 71 \% of the sample require the inclusion of a YSP with age < 100 Myr. Since the ages of the YSP in most type II quasar host galaxies are comparable with the expected lifetimes of the AGN activity, these results provide strong evidence that the quasars are triggered close to the peaks of the merger-induced starbursts.

\end{abstract}




\section{Introduction}
\label{intro}

In recent years the role of active galactic nuclei (AGN) in influencing the evolution of massive galaxies has come to the fore. This is because strong correlations have been found between the masses of the super massive black holes (SMBH) which lay at the heart of nearly all massive galaxies \citep{magorrian98}, and the galaxy bulges they inhabit \citep{ferrarese00,gebhardt00}. In spite of their apparent significance in shaping local correlations, we still do not have a clear understanding of the processes which trigger this activity.

One credible scenario is that AGN, and in particular high luminosity (quasar-like) AGN, are triggered by mergers. The attraction of this mechanism is that it helps to solve one of the major problems associated with triggering luminous AGN, namely, how to the gas in a galaxy sheds over 99\% of its angular momentum so that it can move from galaxy-wide scales down to the sub-parsec scale where it can accrete onto the central SMBH (e.g. \citealt{heckman86,hopkins06}). A number of studies investigating the importance of mergers have shown that they can play a significant role \citep{bennert08,ramos11,bessiere12}, indeed \citet{treister12} find a steep increase in merger rate with increasing AGN luminosity. It is also interesting to note that both \citet{ramos12} and \citet{bessiere12} find that the surface brightnesses of the tidal features associated with the mergers are up to two magnitudes brighter than those associated with a matched sample of quiescent galaxies. As discussed in \citet{bessiere12}, this could indicate that the types of mergers being witnessed in luminous AGN and their matched quiescent control samples are fundamentally different (i.e. wet vs. dry/major vs. minor).

Tidal disruption of the host galaxies will not be the only consequence of mergers. Simulations suggest that, at the point of coalescence, not only is there a peak in the accretion rate onto the SMBH, but the gas driven into the nuclear regions also triggers a burst of star formation \citep{springel05,hopkins06}. If this is the case, using stellar synthesis modelling techniques to date the starburst (e.g. \citealt{canalizo00, canalizo01,tadhunter02,holt07}), we can  determine a fiducial time for the peak of the galaxy merger. In this way it is possible to gain insights into the order of events and timing of the quasar activity in relation to the merger -- key information for understanding the co-evolution of quasars and their host galaxies.

Stellar population synthesis modelling can also be used to test evolutionary scenarios for the triggering of quasars in mergers. For example, \citet{sanders88} propose a scenario in which mergers between gas-rich galaxies trigger prodigious star formation, which will manifest itself as an ultraluminous infrared galaxy (ULIRG). As the merger progresses and the two nuclei coalesce, the quasar activity begins to dominate the bolometric luminosity, eventually driving out the envelope of dust and gas, and allowing the quasar to become clearly detectable in the optical \citep{springel05,hopkins06}. The progression of such a sequence means that the burst of star formation will occur while the quasar is shrouded by its natal cocoon. Therefore, by the time the quasar emerges, we would expect to observe that the starburst stellar population has already aged. Both \citet{ramos11} and \citet{bessiere12} find this scenario to be somewhat over simplistic as both studies show that the AGN activity can be triggered before, during, or well after the coalescence of the SMBH. However, such morphological studies do not reveal any fine detail in terms of the nature of the merger and the relative timing of the starburst and AGN activity.

Previous investigations into the stellar populations of AGN host galaxies have produced mixed results. While some find that the triggering of the AGN activity and the starburst are quasi-simultaneous  \citep{heckman97,canalizo00,brotherton02,davies07,wills08,tadhunter11,bessiere14}, and the ages of the young stellar populations are less than the expected maximum lifetime of an AGN of $\sim 100 $ Myr \citep{martini01}, other works find a clear delay in the triggering of the AGN relative to the starburst, with the ages of the stellar populations ranging from a few 100 Myr to a few Gyr \citep{tadhunter05,holt07,wills08,wild10, canalizo13}.

In order to try to understand the root of these differences, we have carried out a campaign of long-slit optical spectroscopic observations of a 95 \% complete sample of type II quasars selected from the catalogue of candidate type II objects presented by \citet{zakamska03} with [OIII]$\lambda5007$ emission line luminosities ($L_{ [OIII]}$) $L_{[OIII]} > 10^{8.5} L\sun$. We exclusively consider type II quasars in this work because they are ideal objects for making detailed studies of the host galaxies of luminous AGN. In type II quasars, the central engine, which would otherwise overwhelm the host galaxy light, is obscured from view by intervening material. Therefore, there is no need to attempt to model and subtract the central point source before embarking on a detailed study of the underlying host galaxy.

A full morphological analysis of this sample has already been presented in \citet{bessiere12}, and a detailed analysis of the stellar populations of one of the type II quasars (J0025-10) has also been presented in \citet{bessiere14}. J0025-10 provides strong evidence for young YSP ages and the simultaneous triggering of the quasar and the merger-induces starburst, but is this true of type II quasar host galaxies in general? Here we present the analysis of the remaining 18 objects from the original sample for which we have spectra, as well as two further type II quasars for which high quality long-slit spectra have subsequently been obtained. We have used the spectra to determine the ages and reddenings of the young to intermediate-age stellar populations (YSPs/ISPs; $0.001 \lid t_{YSP}\mbox{(Gyr)} \lid 5$) that may be present in the host galaxies of these type II quasars. 

Throughout this paper we assume a cosmology with $H_0 = 70 ~\mbox{km s}^{-1}\mbox{Mpc}^{-1}$, $\Omega_M=0.3$ and $\Omega_\Lambda = 0.7$.

\section{Sample, observations and data reduction}
\label{data}

\begin{table*}

	\begin{minipage}{105mm}
	\centering
\caption{Full classification of the sample objects: Columns 1 and 2 give the SDSS identifier and abbreviated identifier that will be used throughout this work. Columns 3 and 4 list the spectroscopic redshift and cosmology corrected scale taken from the NED database. Column 5 corresponds to the value of log$(L_{[OIII]}/L_\odot)$ taken from \citet{zakamska03} upon which the original sample selection was based, while the values in brackets are updated values taken from \citet{reyes08}.}
    \label{sample_table}
 	\begin{tabular}{l c c c c }
	\hline
	Name 	& Abbreviated 	& z 	& Scale 				& log$(L_{[OIII]}/L_\odot)$		\\ 
		 	&	name 		&		& (kpc arcsec$^{-1}$) 	&	\\
	\hline 
	J235818-000919 & J2358-00 & 0.402 & 5.39 & 9.32(9.29)	\\  
	J002531-104022 & J0025-10 & 0.303 & 4.48 & 8.73(8.65) 	\\
	J011429+000037 & J0114+00 & 0.389 & 5.28 & 8.66(8.46) 	\\
	J012341+004435 & J0123+00 & 0.399 & 5.36 & 9.13(9.14) \\
	J014237+144117 & J0142+14 & 0.389 & 5.28 & 8.76(8.87)\\	
	J021757-011324 & J0217-01 & 0.375 & 5.16 & 8.55(8.37) 	\\
	J021758-001302 & J0217-00 & 0.344 & 4.88 & 8.75(8.81) 	\\
	J021834-004610 & J0218-00 & 0.372 & 5.13 & 8.85(8.62) \\
	J022701+010712 & J0227+01 & 0.363 & 5.05 & 8.90(8.70) \\
	J023411-074538 & J0234-07 & 0.310 & 4.55 & 8.77(8.78) 	\\
	J024946+001003 & J0249+00 & 0.408 & 5.44 & 8.63(8.75) 	\\
	J032029+003153 & J0320+00 & 0.384 & 5.23 & 8.52(8.40) \\	
	J033248-001012 & J0332-00 & 0.310 & 4.55 & 8.50(8.53) \\
	J033435+003724 & J0334+00 & 0.407 & 5.43 & 8.61(8.75) \\ 
	J084856+013647 & J0848-01 & 0.350 & 4.95 & 8.56(8.46) \\
	J090414-002144 & J0904-00 & 0.353 & 4.98 & 8.93(8.89)	\\
	J092318+010144 & J0923+01 & 0.386 & 5.27 & 8.94(8.78)  \\
	J092356+012002 & J0924+01 & 0.380 & 5.22 & 8.59(8.46) 	\\
	J094836+002104 & J0948+00 & 0.324 & 4.71 & 8.52(8.57)	\\	
	J130740-021455 & J1307-02 & 0.425 &	5.63 & 8.92(9.01)	\\
	J133735-012815 & J1337-01 &	0.328 &	4.78 & 8.72(8.73)	\\	  	
	
	\hline 	  
	\end{tabular}
	\end{minipage}
\end{table*}

The original sample of 20 type II quasars was derived from the catalogue of candidate objects established by \citet{zakamska03}, who selected galaxies from the Sloan Digital Sky Survey (SDSS; \citealt{york00}) with high ionisation, narrow emission lines, but no indication of broad permitted lines. To characterise the nature of the ionising source, they used diagnostic emission line ratios to demonstrate the energetic dominance of the AGN. \citet{zakamska03} provide full information on the criteria used to select type II objects based on emission line ratios. 

Because the AGN is itself obscured in type II objects, the luminosity of the [OIII]$\lambda5007$ emission line is used as a proxy for its bolometric luminosity (e.g. \citealt{heckman05,lamassa09,dicken14}). It is assumed that a type II object will have the same intrinsic AGN luminosity for a given $L_{[OIII]}$ as a type I object. \citet{zakamska03} make their cut between objects that are considered to be quasars, and those with Seyfert-like luminosities at $L_{[OIII]} >10^{8.5} L_\odot$, which is roughly equivalent to an AGN absolute magnitude $M_B < -23$ mag.

The sample presented in this work comprises a 95\% complete sample of 19 of the 20 objects from \citet{zakamska03} with RAs in the range $23 < RA < 10$ hr, $\delta$  $<+20$ degrees, redshifts in the range $0.3<z<0.41$ and [OIII] emission line luminosities $L_{[OIII]} > 10^{8.5}L_\odot$ (see Table \ref{sample_table}). This includes results for J0025-10 which have already been presented in \citet{bessiere14}. Unfortunately, for one of the 20 objects in the full sample (J0159+14) we were unable to obtain spectroscopic data of sufficient quality attempt modelling of the stellar populations. Given the high level of completeness of our sample, we are not biased towards a particular luminosity or optical morphology in terms of host galaxy properties.

When determining our selection criteria, the $L_{[OIII]}$ limit was chosen to ensure that the objects do indeed contain AGN with quasar-like bolometric luminosities. However, subsequent to the publication of candidate objects by \citet{zakamska03}, the SDSS spectrophotometric calibration pipeline has been updated and improved \citep{abazajian04,adelman07}. Using the same sample selection criteria and employing the same line fitting technique, \citet{reyes08} have since updated the emission line luminosities presented by \citet{zakamska03} using data calibrated with this improved pipeline. These updated luminosities suggest that six objects from the original sample selection fall marginally below our $L_{ [OIII]}$ cut. However, because the original sample selection was based on \citet{zakamska03}, we choose to retain the full sample of 19 objects, although the updated $L_{ [OIII]}$ values are included in Table \ref{sample_table} (bracketed values) as a matter of interest. It is also worth noting that were a luminosity cut of log$(L_{[OIII]}/L_\odot)> 10^{8.37}$ imposed (to reflect the lowest $L_{ [OIII]}$ from the sample according to \citet{reyes08}), no extra objects would be included from the \citet{reyes08} catalogue over those selected from the \citet{zakamska03} catalogue using our original selection criteria.

The primary optical spectroscopic data used in this paper were obtained in service mode using the Gemini Multi-Object Spectrograph (GMOS-S) mounted on the 8.1m Gemini South telescope at Cerro Pach\'{o}n, Chile. The GMOS-S detector \citep{hook04} consists of three adjacent $2048\times 4096$ pixel CCDs separated by two gaps of approximately 2.8\arcsec, and a binned pixel scale of 0.292 (0.146 unbinned) arcsec pixel$^{-1}$. 

Long-slit spectra were taken for 18 objects during semesters 2010B and 2011B using a relatively wide slit to enable accurate spectrophotometry. Due to the absence of an atmospheric dispersion compensator on GMOS-S, the objects were mainly observed at low airmass and/or with the slit placed at the parallactic angle, in order to minimise the effects of differential refraction (see Table \ref{obs_spec}). With the exception of J0332-00, the slit position angles of those objects not observed in the above configuration were within 20 degrees of the parallactic angles, and were also mostly observed at low to moderate airmasses (AM $<1.25$). Due to our use of a relatively wide slit (1.5 arcsec), differential refraction is unlikely to have affected these observations. The observations were made predominately in dark time (except for J0227+01, J0320+00, J0848+01, J0948+00 and J2358-00, which were observed in grey time), in photometric conditions. The seeing for each object was measured using the stars in the acquisition images, and in all but one case the seeing was FWHM $< 1.15$ arcsec ($0.53 < \mbox{FWHM} < 1.58$ arcsec with a median of 0.81 arcsec FWHM, see Table \ref{obs_spec}). For most objects the observations consist of $4 \times 675$ s exposures using the B600 grating with 600 groves mm$^{-1}$ blazed at 461 nm, and $3\times 400$ s exposures using the R400 grating with 400 groves mm$^{-1}$ blazed at 764 nm. The exposures in the two wavelength ranges were  interleaved in order to mitigate any changes in the atmospheric conditions during each observational run. The full details of the observations are shown in Table \ref{obs_spec}.
 
\begin{table*}
\begin{minipage}{140mm}	
\centering
\caption{Details of the long-slit spectroscopic observations carried out at Gemini South using the GMOS instrument: Column 1 gives the object name, column 2 and 3 the observation date and slit position angle, columns 4 and 5 give the exposure times in seconds using the blue and red grisms respectively. Column 5 gives the average airmass throughout the exposures and column 6 gives the seeing for the observations in arcsec as measured from the acquisition image. The final column denotes whether the object was observed at the parallactic angle.}
  \label{obs_spec}
    \begin{tabular}{l c c c c c c c}
      \hline
      Name & obs. date & PA 		&  Exposure  & Exposure  & Average & Seeing		& Mean  \\
           &		   & (degrees)	&  Blue(s)	 & Red(s)	 &	airmass	   & (arcsec)	& Parallactic ?\\
      \hline
	  J2358-00 & 2011-09-03 & 331 & $4 \times 675$ & $3 \times 400$ & 1.40 	& 1.58 	& No \\
	  J0025-10 & 2010-11-30	& 230 &	$ 4 \times 675$ & $3 \times 400$ & 1.10	& 0.84 	& No \\
      J0123+00 & 2011-10-24 & 209 & $4 \times 675$ & $3 \times 400$ & 1.18 	& 1.12 	& No	\\
      J0142+14 & 2011-12-23 & 155 & $ 1 \times 675$ $1\times 461$ & -- & 1.57 	& 0.73	& Yes	\\ 
	  J0217-01 & 2010-12-03 & 146 & $4 \times 675$ & $3 \times 400$  & 1.26 	& 0.97	& Yes	\\	      
      J0217-00 & 2011-10-23 & 222 & $4 \times 675$ & $3 \times 400$ & 1.32 	& 0.86 	& Yes	\\
	  J0218-00 & 2010-12-04 & 171 & $4 \times 675$ & $3 \times 400$ & 1.17 	& 1.11	& Yes	\\
      J0227+01 & 2011-01-26 & 138 & $4 \times 675$ & $3 \times 400$  & 1.57 	& 0.53	& Yes	\\
      J0234-07 & 2010-12-06 & 126 & $4 \times 675$ & $3 \times 400$ & 1.57 	& 0.78	& Yes	\\
      J0249+00 & 2010-12-05 & 0   & $4 \times 675$ & $3 \times 400$  & 1.17	& 0.73	& Yes	\\ 
      J0320+00 & 2010-12-05 & 216 & $4 \times 675$ & $3 \times 400$  & 1.27 	& 0.75	& Yes	\\     
      J0332-00 & 2010-12-06 & 32  & $4 \times 675$ & $3 \times 400$  & 1.16 	& 0.83	& No	\\
      J0334+00 & 2010-12-01 & 350 & $4 \times 675$ & $3 \times 400$  & 1.24 	& 1.13	& No	\\ 
      J0848+01 & 2011-01-25 & 174 & $4 \times 675$ & $3 \times 400$  & 1.22	& 0.72	& No	\\
      J0904-00 & 2012-01-20 & 140 & $4 \times 675$ & $3 \times 400$ & 1.46	& 0.71	& Yes	\\
      J0923+01 & 2011-11-30 & 222 & $4 \times 675$ & $3 \times 400$ & 1.37 	& 0.67 	& Yes	\\ 
      J0924+01 & 2011-02-01 & 0   & $4 \times 675$ & $3 \times 400$ & 1.18 	& 0.75	& Yes	\\
      J0948+00 & 2011-01-25 & 145 & $4 \times 675$ & $3 \times 400$ & 1.32 	& 0.89 	& Yes	\\      
      \hline             
      \end{tabular}  
 \end{minipage} 
\end{table*}

The data were reduced using a combination of the Gemini GMOS dedicated data reduction package within the {\sc{iraf}} environment, other standard {\sc{iraf}} routines, and the {\sc starlink figaro}\footnote{http://starlink.jach.hawaii.edu/starlink/WelcomePage} package. Initially, the three CCD images, which comprise each exposure, were mosaicked into a single image which was then bias subtracted and flat fielded. The images were then combined using imcombine ({\sc{iraf}}) to produce the final 2D frame. The wavelength calibration for each image was checked by extracting a 1 dimensional spectrum of the sky emission lines close to the target position along the slit, and fitting the position of the most prominent lines using a Gaussian profile. Using this technique, the uncertainty in the wavelength calibration was found to be $\sim 0.3~\mbox{\AA}$ and $\sim 0.5~\mbox{\AA}$ for the B600 and R400 exposures respectively. The average spectral resolutions measured from the night sky lines were 7.2 and 11.4 \AA, for the blue and red wavelength ranges respectively.

Due to the fact that an accurate flux calibration is essential when attempting to model stellar populations, several standard stars were observed over the course of each semester. Flux calibration curves were derived for each star individually, which were then compared to each other (independently for each semester) across the full wavelength range. Clearly outlying curves (i.e. due to misidentification of the standard star, non-photometric conditions etc.) were rejected before combining the curves (5 for 2010B and 3 for 2011B) into an average flux calibration curve, which was then applied to all the objects observed in that semester. From comparison between the individual flux calibration curves it was found that the accuracy of the relative flux calibration was $\pm 5\%$. There is also a large region of overlap between the red and blue spectra ($\sim 2080$ \AA), allowing a further check that the calibrations of the spectra taken with the two gratings match well. In most cases, the agreement was excellent, and the red and blue spectra agreed in both shape and level across the full region of the overlap to within 5\%. In three cases (J0217-00, J0249+00 and J0924+01), the differences in the flux levels between the blue and red spectra were significant (8\%, 13\% and 22\% respectively), although the shapes of the spectra in all cases agreed well. However, because this occurred in a limited number of cases, it was considered that this was due to changes in the atmospheric conditions between the observations made using the blue and red grisms, rather than being an issue with the flux calibration. Therefore, either the blue or red spectrum was scaled to match its counterpart with the higher continuum flux. The final step was to correct the spectra for any spatial distortions, using the {\sc iraf apall} package. The useful observed wavelength range is 3940 -- 9140 \AA, although wavelengths longer than 8200 \AA\ are strongly affected by residual fringes. 

We were unable to obtain a spectrum of J0114+00 through our Gemini program. However, its SDSS spectrum is of sufficient quality to enable stellar population modelling. This spectrum was taken using the standard 3\arcsec\ fibre covering an observed wavelength range of 3800--9200 \AA\ ($\sim$ 2700 -- 6620 \AA\ rest-frame) with a spectral resolution varying from 1500 at 3800 \AA\ to 2500 at 9000 \AA.

\subsection{Additional type II quasar observations} 

\begin{table*}
\begin{minipage}{120mm}	
\centering
\caption{Details of the long-slit observations carried out using the OSIRIS instrument on the GTC: Column 1 is the abbreviated object name and column 2 gives the dates of observations. The exposure times in each of the three wavelength ranges are given in columns 3 to 5, whilst the average airmass and seeing for the combined observations of each target are given in columns 6 and 7 respectively. }
  \label{obs_spec_osiris}
    \begin{tabular}{l c c c c c c r}
      \hline
      Name & obs. date &   Exposure  & Exposure	&Exposure  & Airmass & Seeing  \\
           &		   & 	  Blue(s)	 & Yellow(s)	&Red(s)	 &		   & (arcsec)\\
      \hline
     
      J0217-00 & 16/17/18-08-2012 &  $3 \times 1800$ & $3 \times 1200$ & $3 \times 1800$  & 1.37 	& 0.95 \\	  
      J1307-02	& 01/13-04-2012	 &  $3 \times 1800$ & $3 \times 1200$ & $3 \times 1800$  & 1.21 & 0.95 \\
	  J1337-01	& 13/26-04-2012	&  $3 \times 1800$ & $3 \times 1200$ & $3 \times 1800$  & 1.25	&	1.2 \\	
      \hline             
      \end{tabular}  
 \end{minipage} 
\end{table*}

Spectroscopic observations of three type II quasars were also taken at the Gran Telescopio Canarias (GTC), La Palma, on 2012 April 1, 13, 26 and 2012 August 16-18. The observations were obtained under good sky conditions (dark nights with clear/photometric transparency), and seeing values ranging from 0.9-1.0 arcsec (FWHM) for J0217-00 and J1307-02, and 1.2 arcsec (FWHM) for J1337-01 respectively. The OSIRIS instrument was used in long slit spectroscopy mode, with each target being observed with the R2000B, R2500R and R2500I grisms for integration times of $3 \times 1800$ s, $ 3 \times 1200$ s and $3 \times 1800$ s, respectively. A slit of width 1.23\arcsec\ was used oriented at the parallactic angle to minimize flux losses, giving a spectral resolution (FWHM) of 4, 5 and 6.5 \AA\ in the three respective grisms. The reduction and calibration of these long slit observations were performed in the standard way, using IRAF routines. Full details of the observations made for these three objects are given in Table \ref{obs_spec_osiris}. Of these three objects, two (J1307-02 and J1337-01) are in addition to the original sample. Both objects fulfil the original [OIII] luminosity selection criteria (Table \ref{sample_table}), although J1307-02 is at slightly higher redshift ($z=0.425$) than the original sample selection. J0217-00 was also observed as part of the original Gemini sample, however, the data presented here are taken at a different position angle, sampling different extended regions of the galaxy, although the flux is likely still dominated by the nuclear regions also sampled by our Gemini spectra.

\section{Analysis}

For the majority of Gemini spectra, only one aperture was extracted from each object, which was centred on the peak continuum flux. In all such cases, the aperture had a metric diameter of $\sim$ 8 kpc and was centred on the nucleus of each target. However, note that the apertures were not exactly 8 kpc because of the finite pixel scale. The sizes of the apertures were 10--12 pixels, depending on the redshift of the object.

 In two cases (J0218-00 and J0332-00), it was possible to extract two apertures, one with a diameter of 8 kpc centred on the nuclear region of the quasar host galaxy (N), and a second aperture extracted from an extended region using a 5 kpc diameter aperture (E). The physical location of the apertures are shown in Figures \ref{apPos}(J0218-00) and Figure \ref{q0332apPos} (J0332-00).  

All spectra were corrected for Galactic extinction using the E(B-V) values of \citet{schlegel98} and the \citet*{cardelli89} extinction law, and were then shifted to the rest-frame before stellar synthesis modelling. The spectra for all the sample objects including the SDSS spectrum of J0114+00 are shown in Figure \ref{all_spec}. 

\begin{figure*}
\centering
\subfloat{
\includegraphics[scale =  0.4]{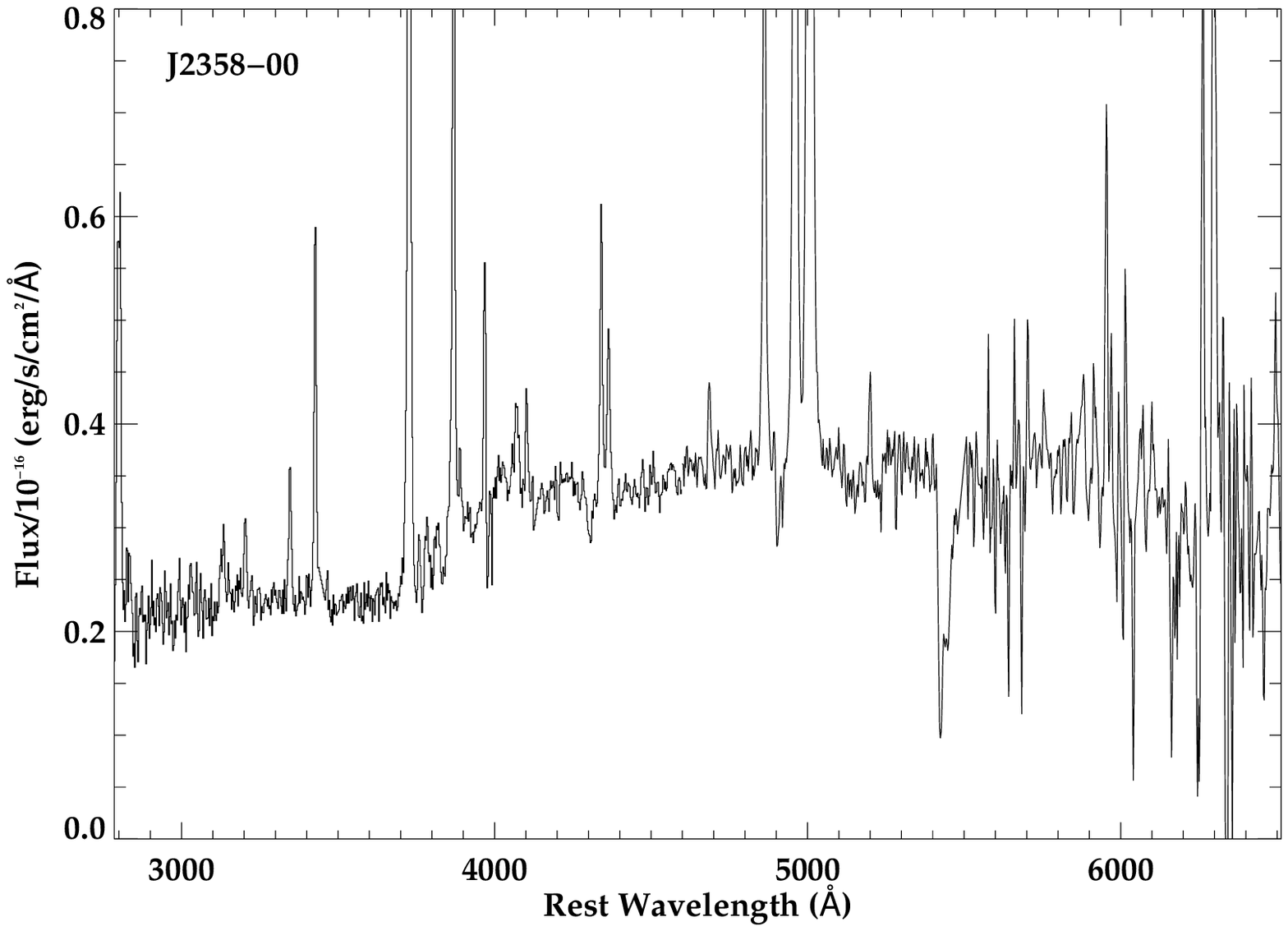}
\label{q2358-spec}}
\subfloat{
\includegraphics[scale =  0.4]{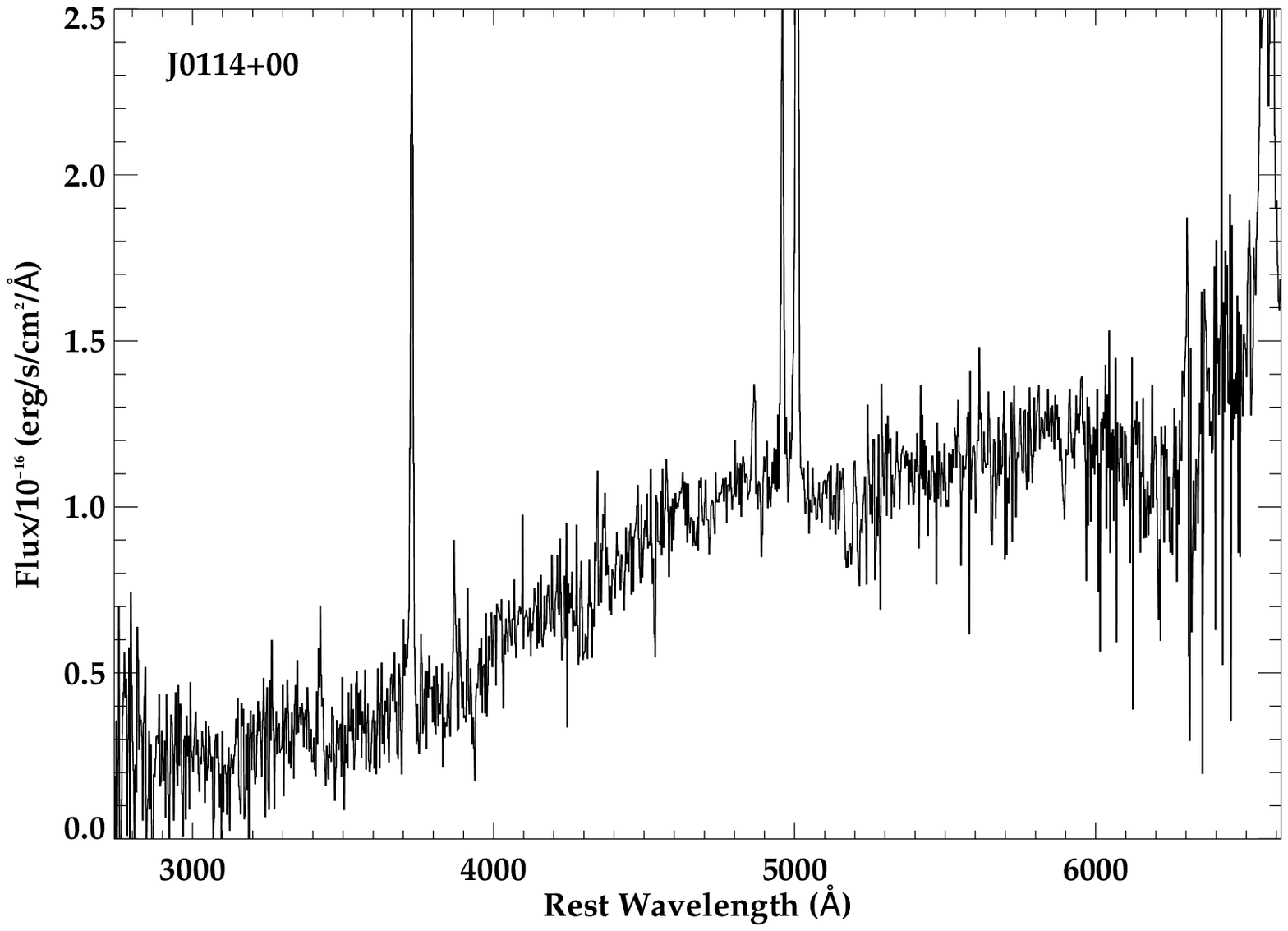}
\label{q0114-spec}}

\subfloat{
\includegraphics[scale =  0.4]{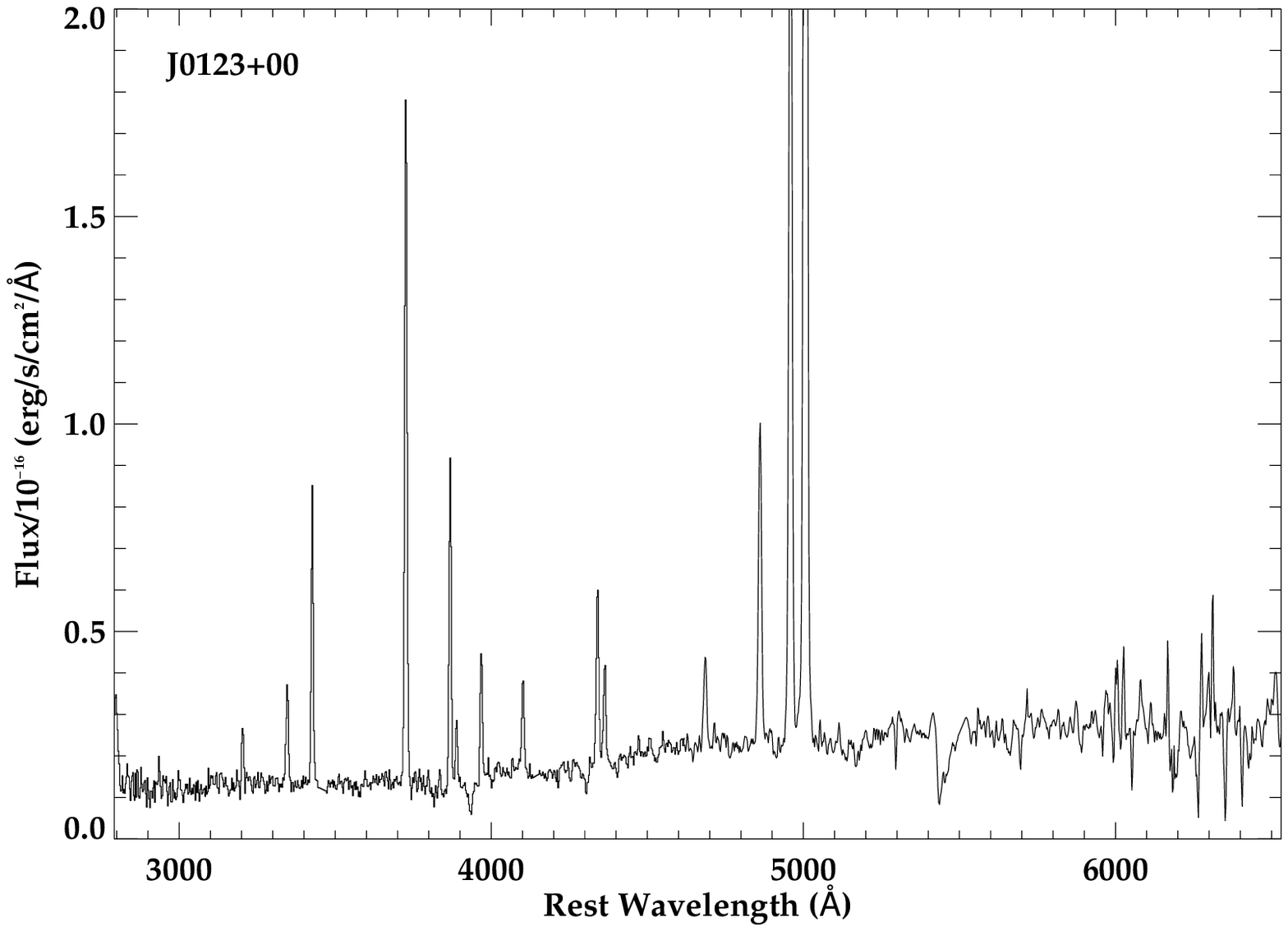}
\label{q0123-spec}}
\subfloat{
\includegraphics[scale =  0.4]{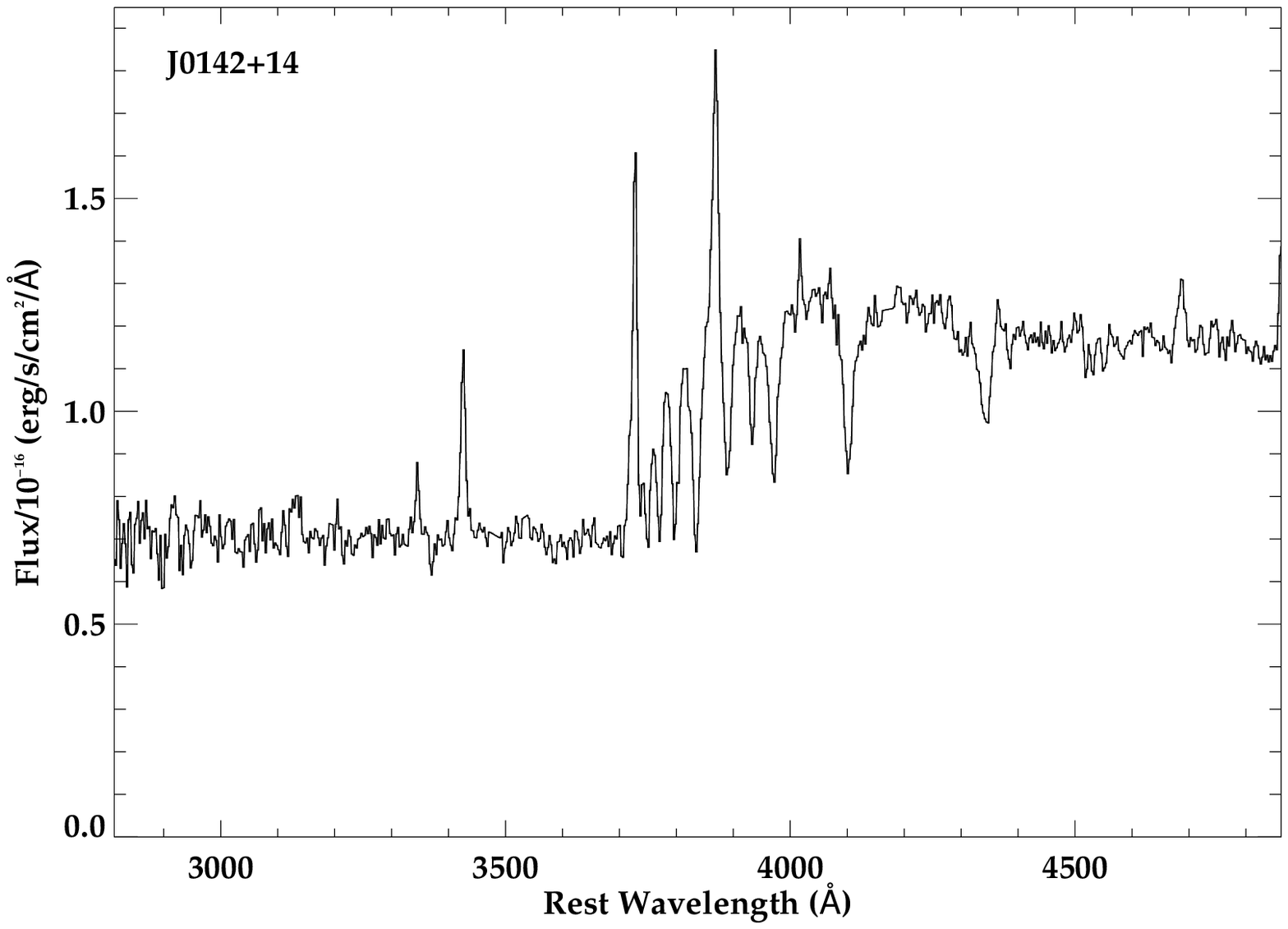}
\label{q0142-spec}}

\subfloat{
\includegraphics[scale =  0.4]{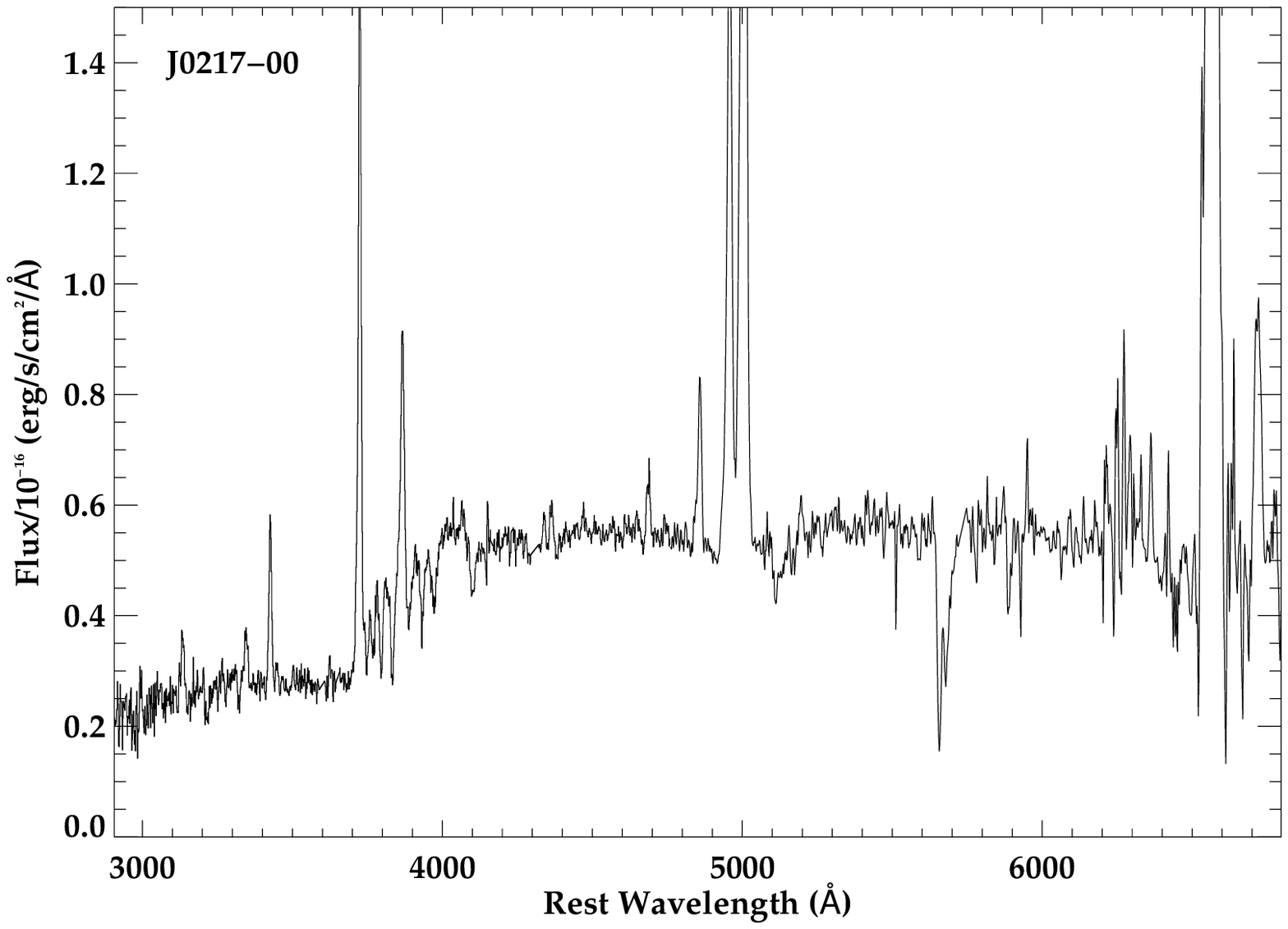}
\label{q021700-spec}}
\subfloat{
\includegraphics[scale =  0.4]{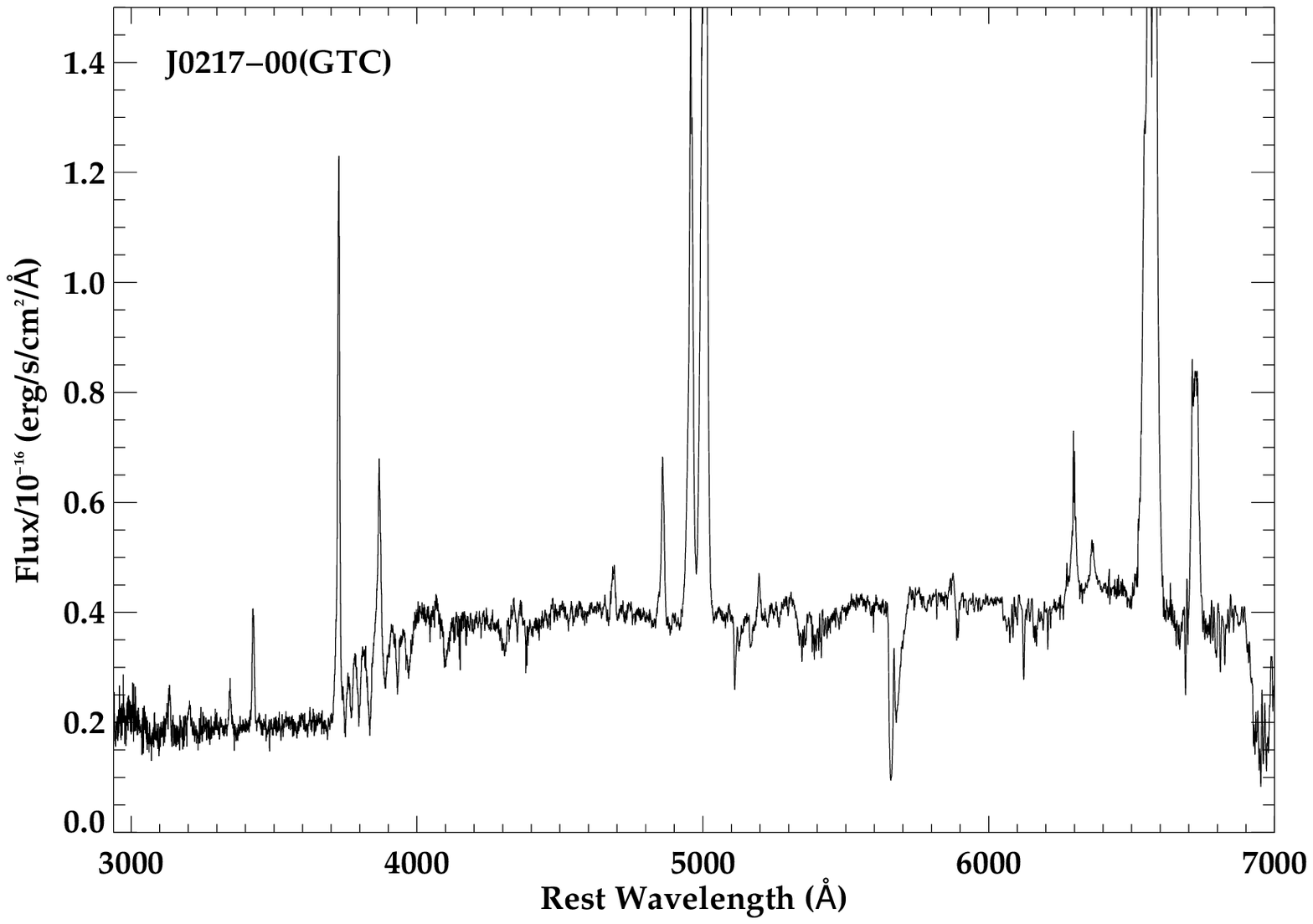}
\label{q0217GTC-spec}}

\subfloat{
\includegraphics[scale =  0.4]{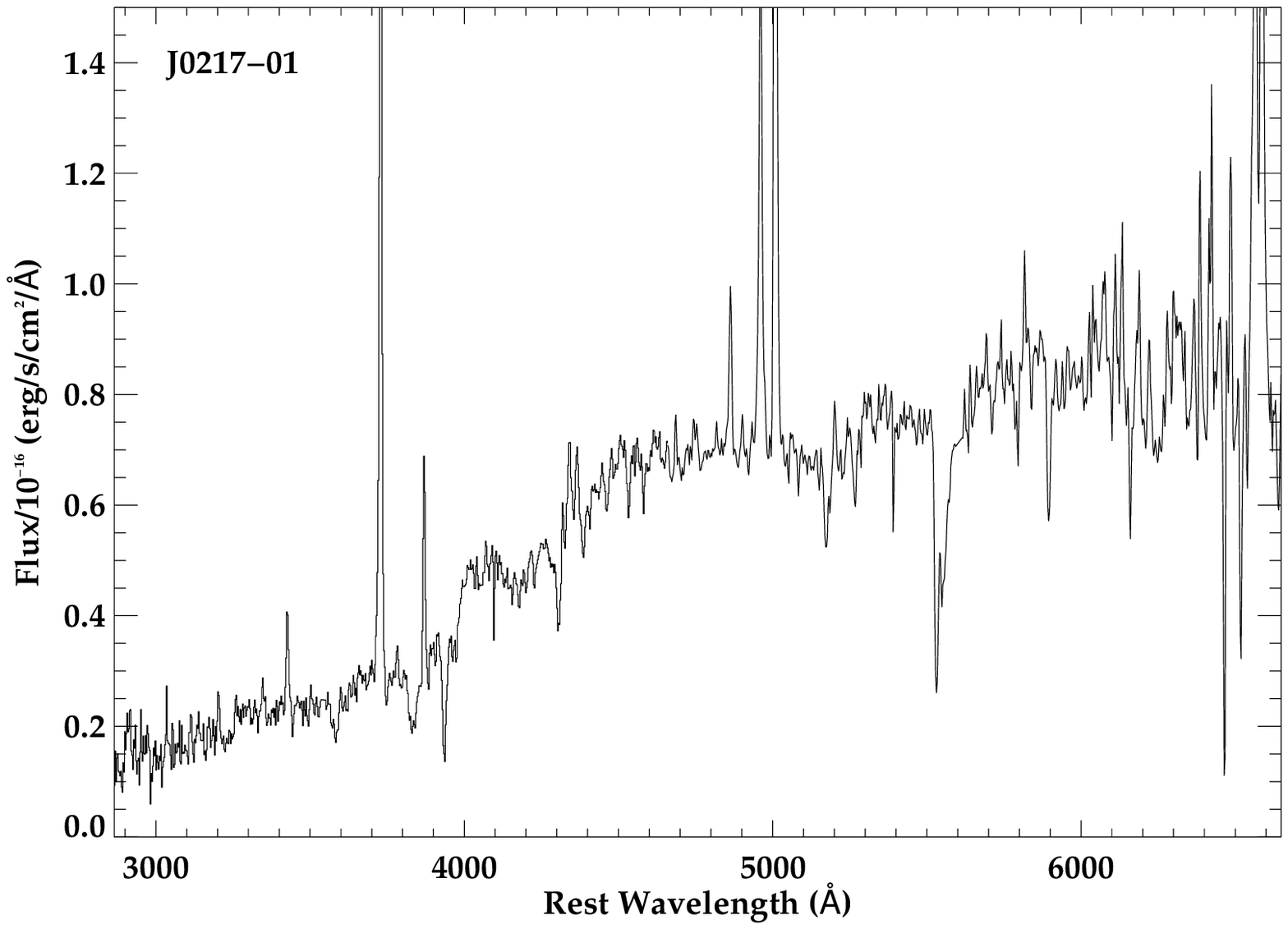}
\label{q021701-spec}}
\subfloat{
\includegraphics[scale =  0.4]{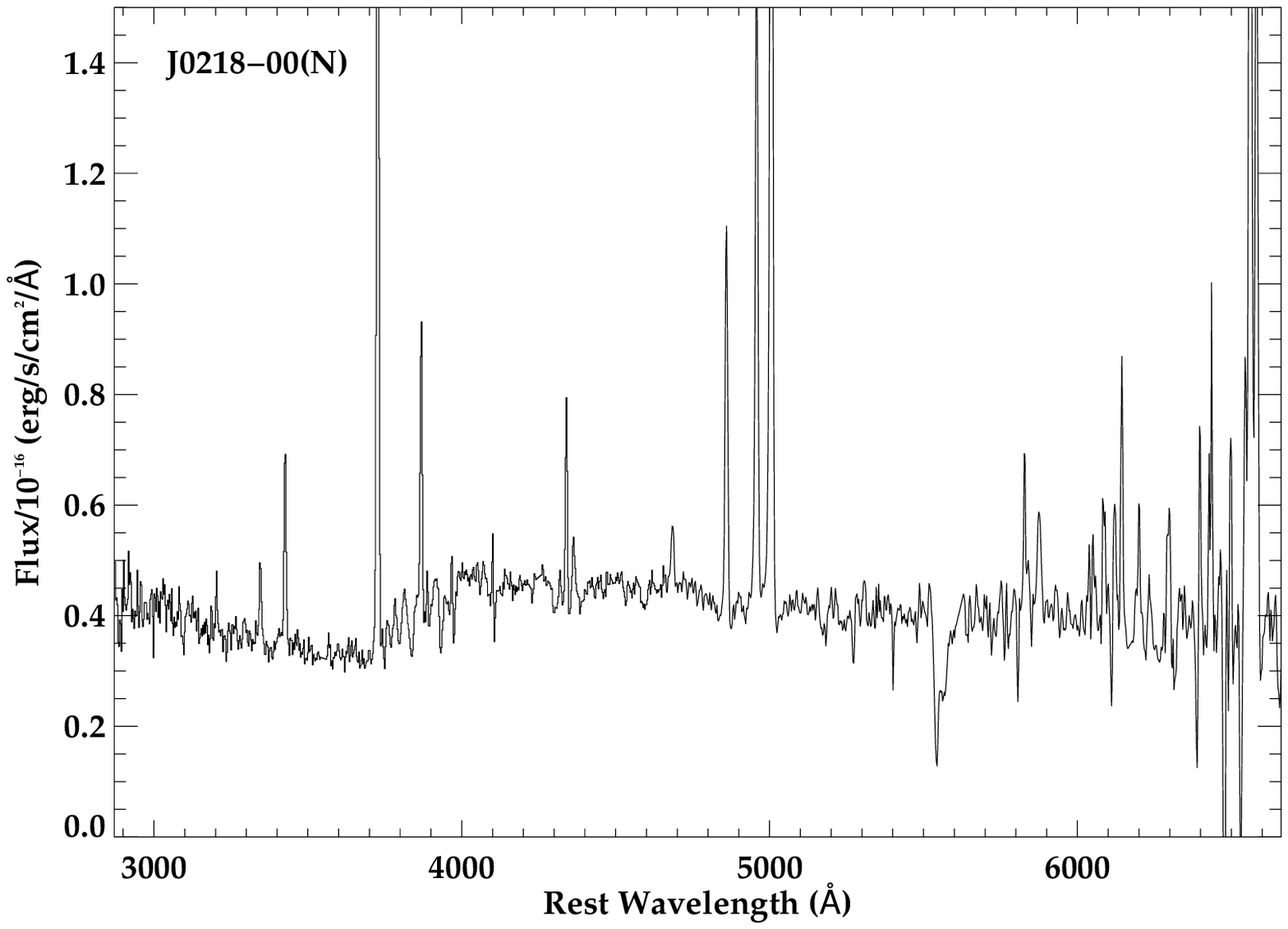}
\label{q02188kpc-spec}}
\caption{The Gemini \gmos\ spectra for 17 of the objects, the SDSS spectrum of J0114+00 and the 3 GTC spectra of J0217-00, J1307-02 and J1337-01 all shifted to the rest frame. The name of each object is given in each panel.}
\label{all_spec}
\end{figure*}
\begin{figure*}
\centering
\subfloat{
\includegraphics[scale =  0.4]{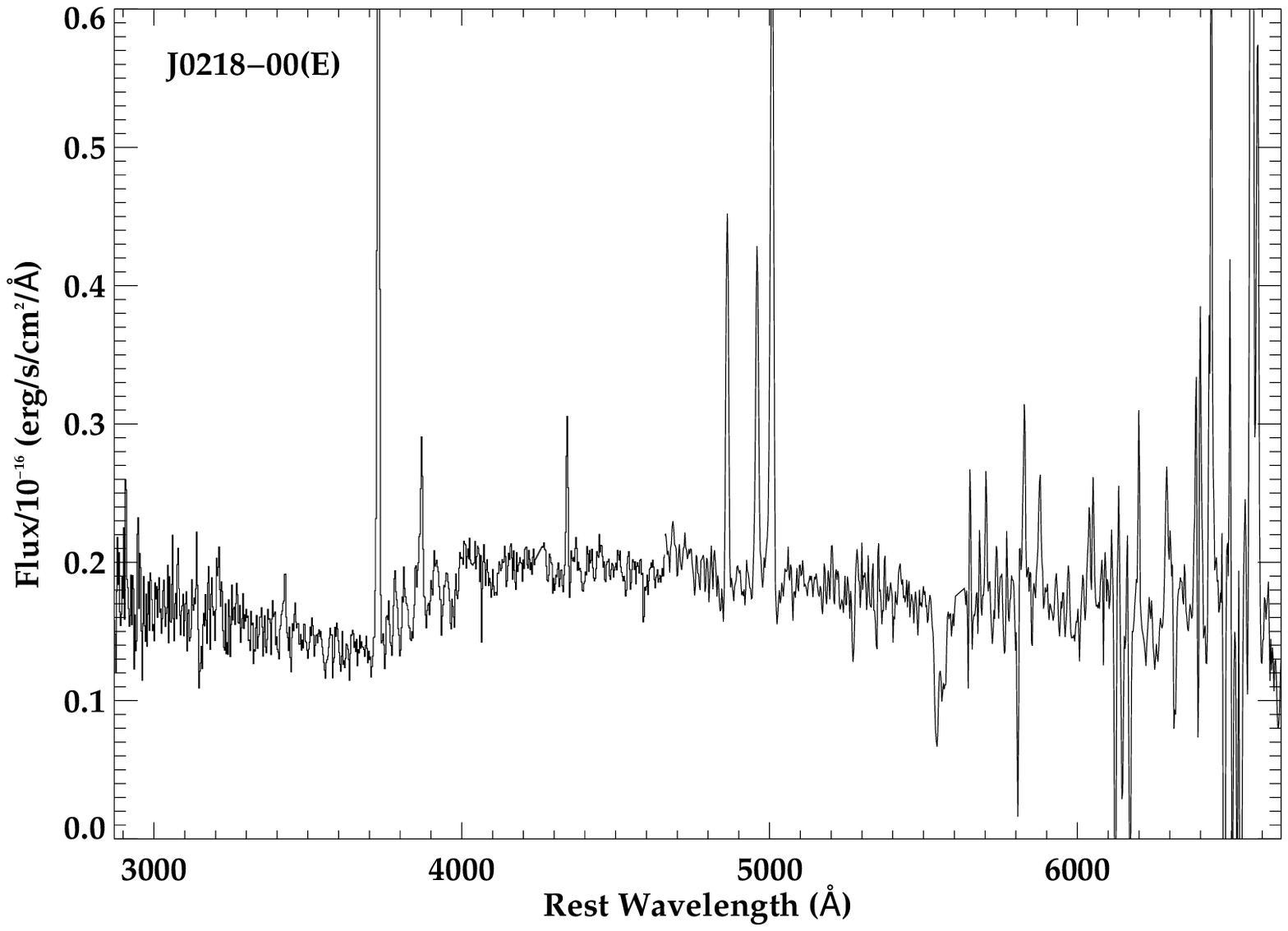}
\label{q02185kpc-spec}}
\subfloat{
\includegraphics[scale =  0.4]{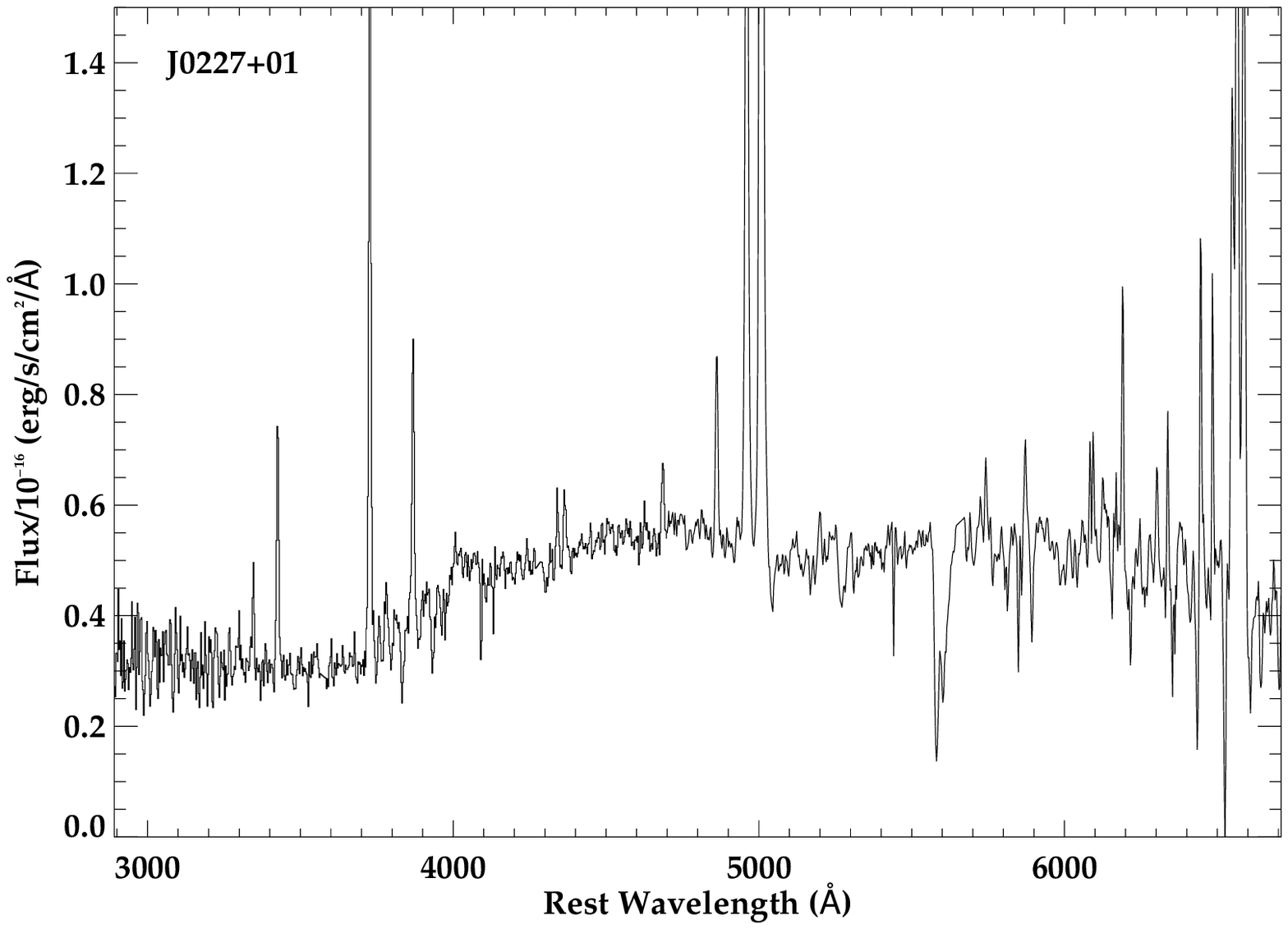}
\label{q0227-spec}}

\subfloat{
\includegraphics[scale =  0.4]{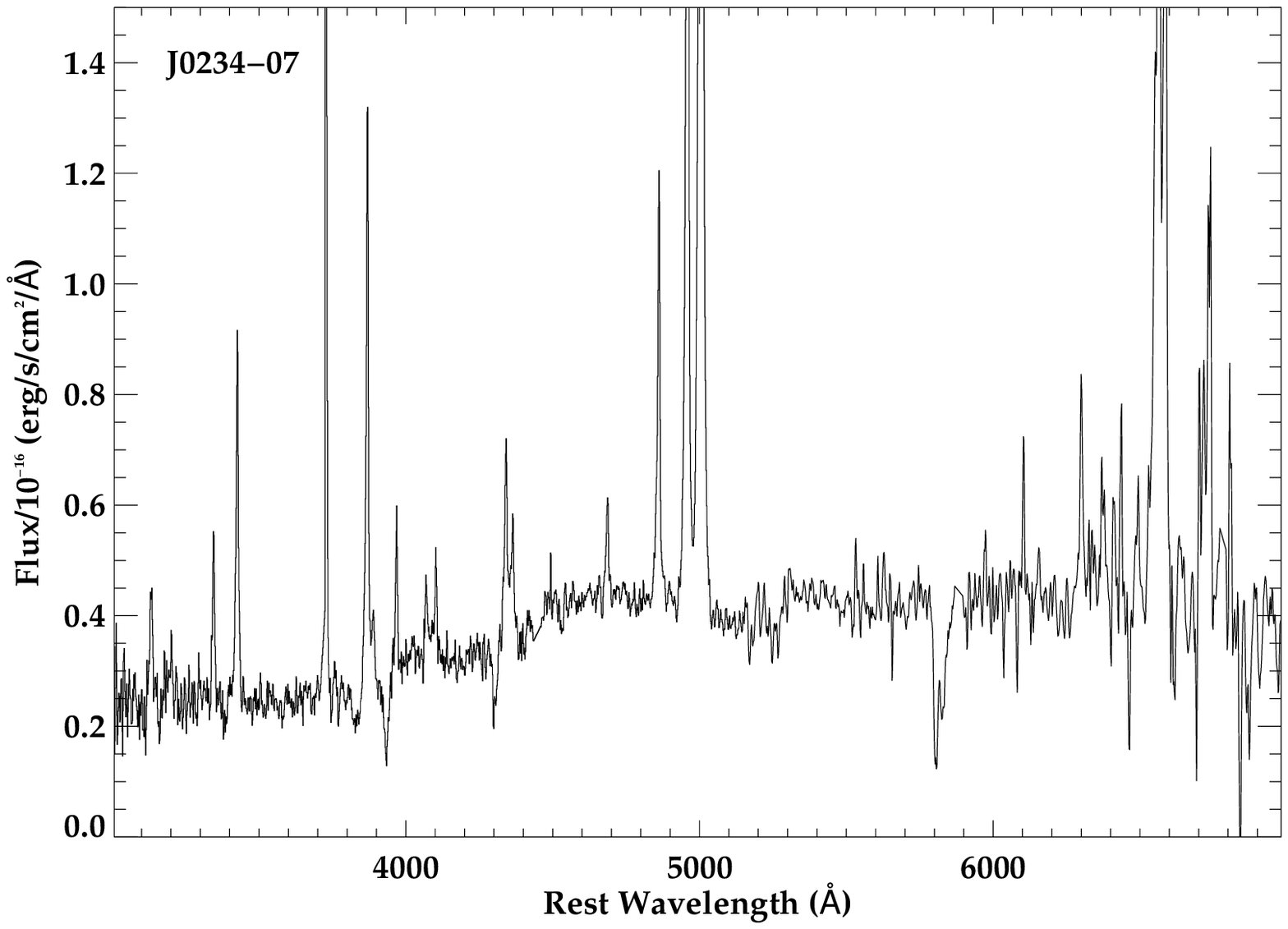}
\label{q0234-spec}}
\subfloat{
\includegraphics[scale =  0.4]{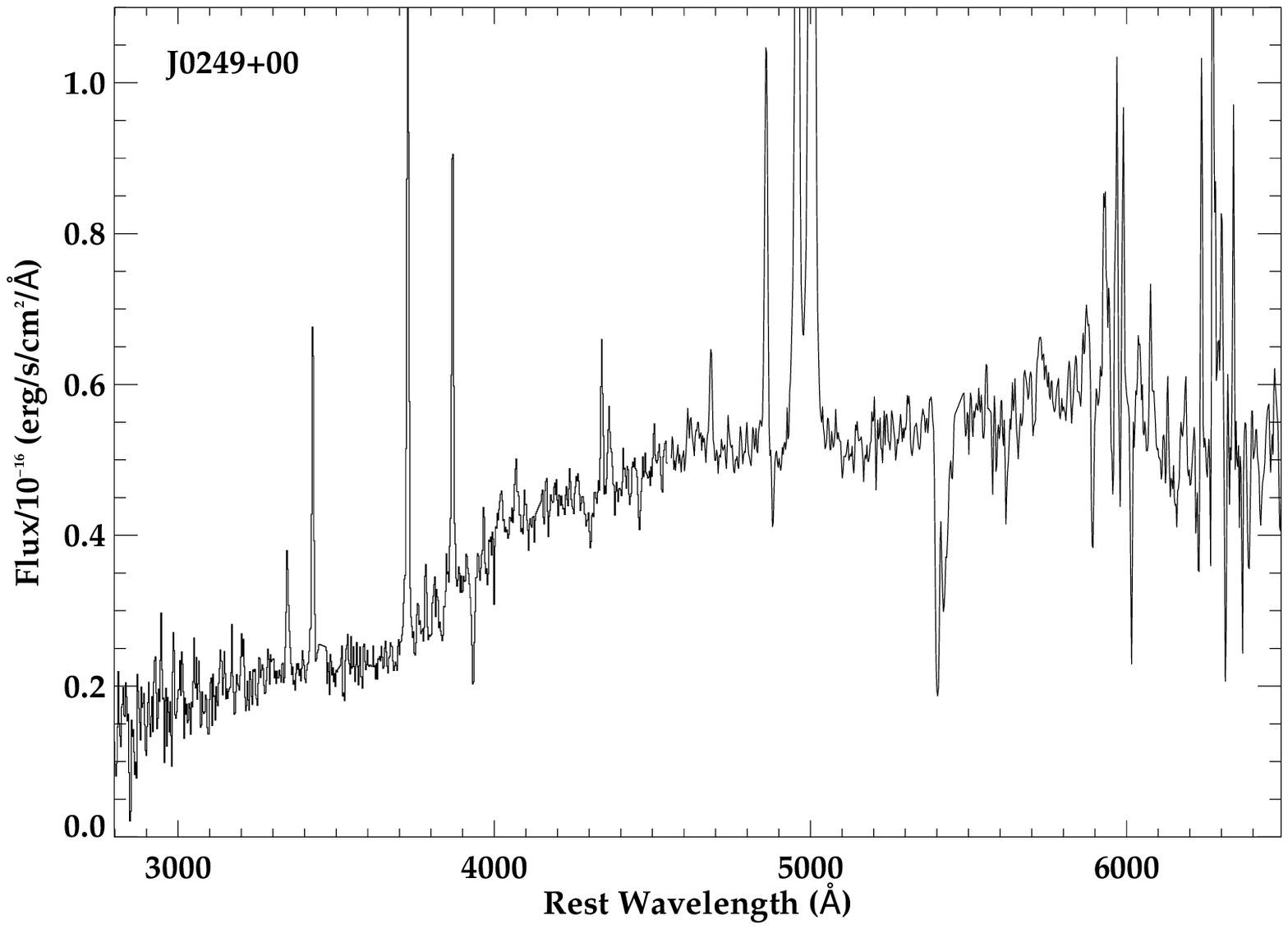}
\label{q0249-spec}}

\subfloat{
\includegraphics[scale =  0.4]{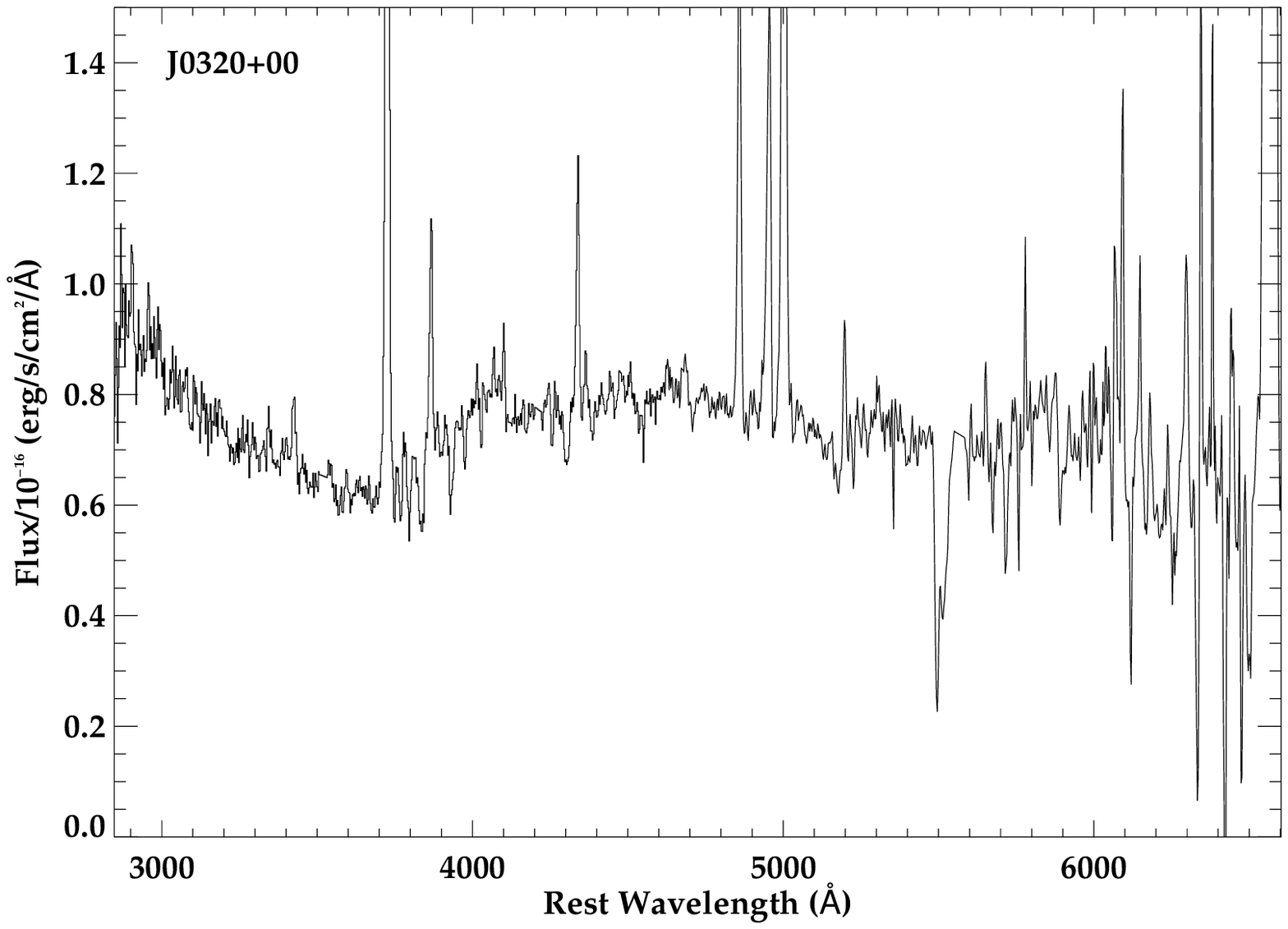}
\label{q0320-spec}}
\subfloat{
\includegraphics[scale =  0.4]{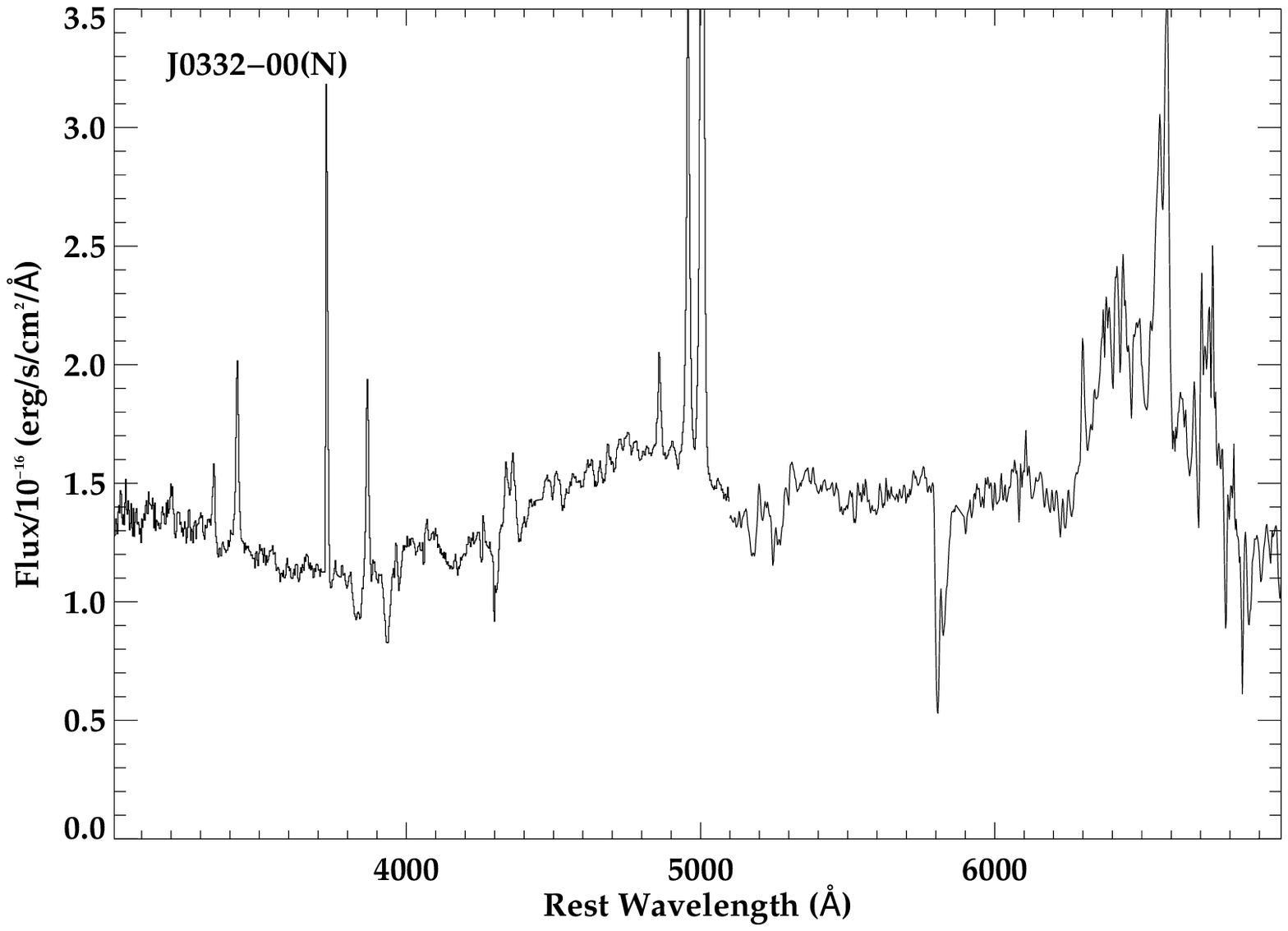}
\label{q03328kpc-spec}}

\subfloat{
\includegraphics[scale =  0.4]{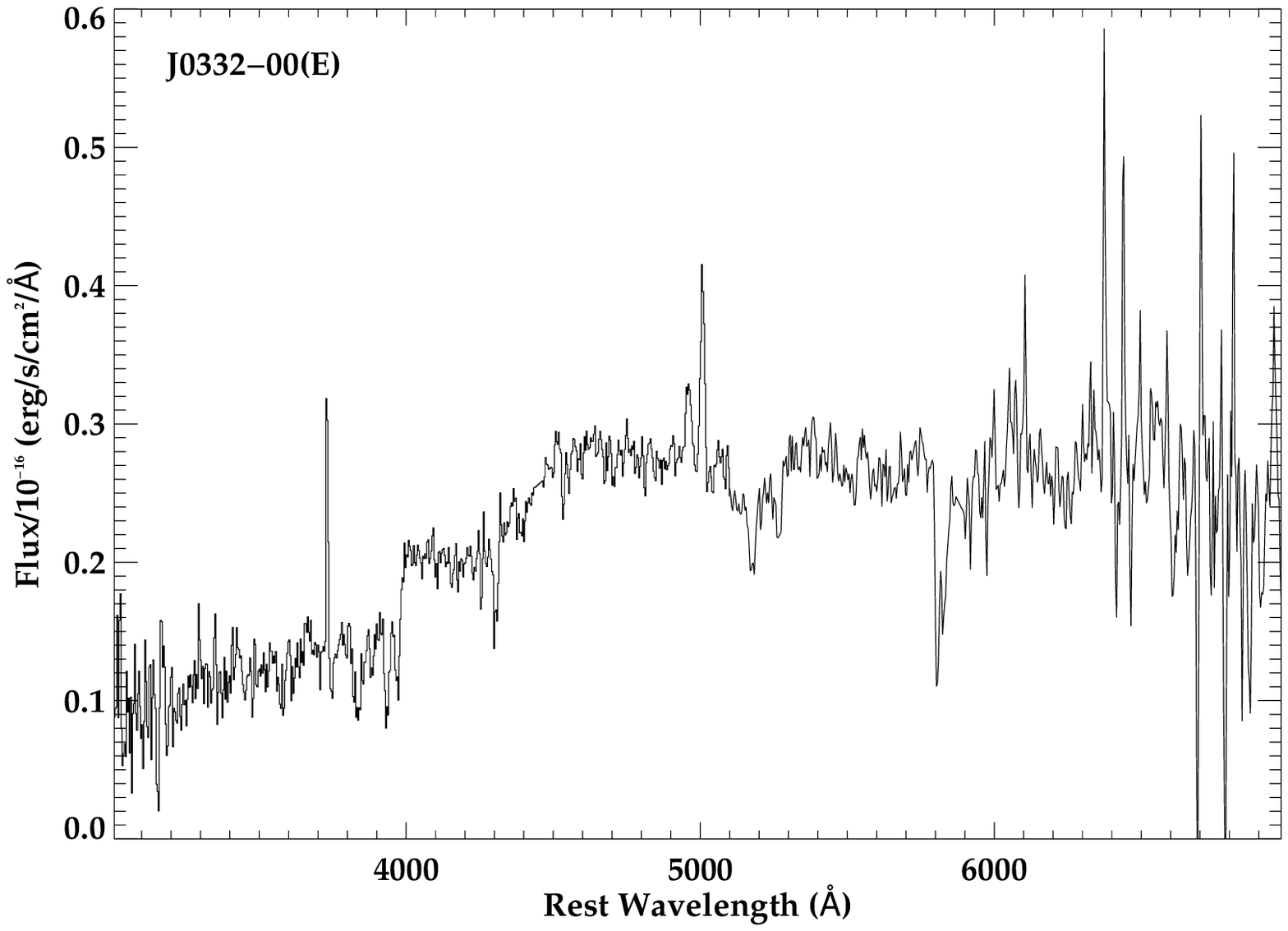}
\label{q03325kpc-spec}}
\subfloat{
\includegraphics[scale =  0.4]{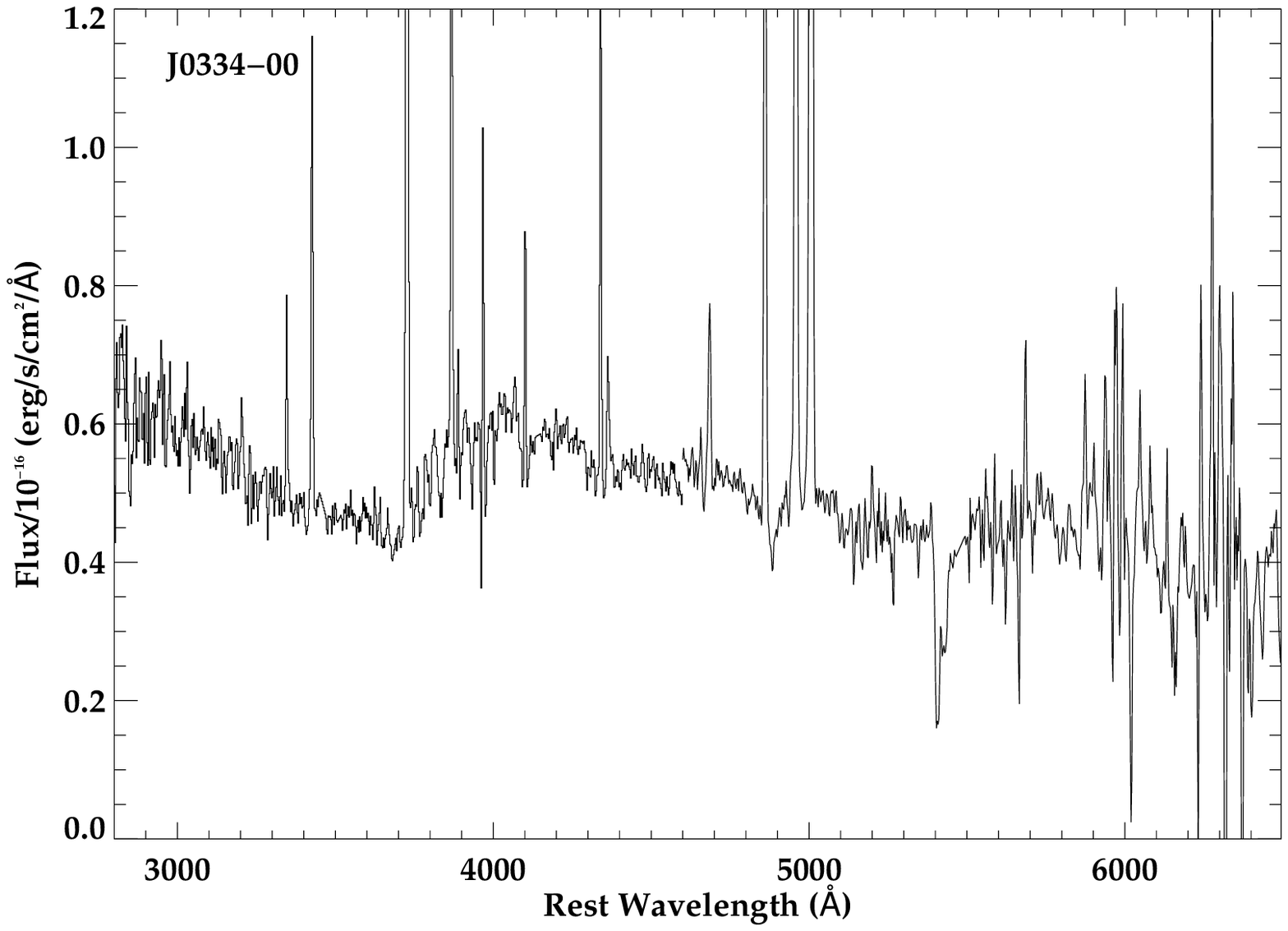}
\label{q0334-spec}}

\contcaption{}
\end{figure*}
\begin{figure*}

\centering
\subfloat{
\includegraphics[scale =  0.4]{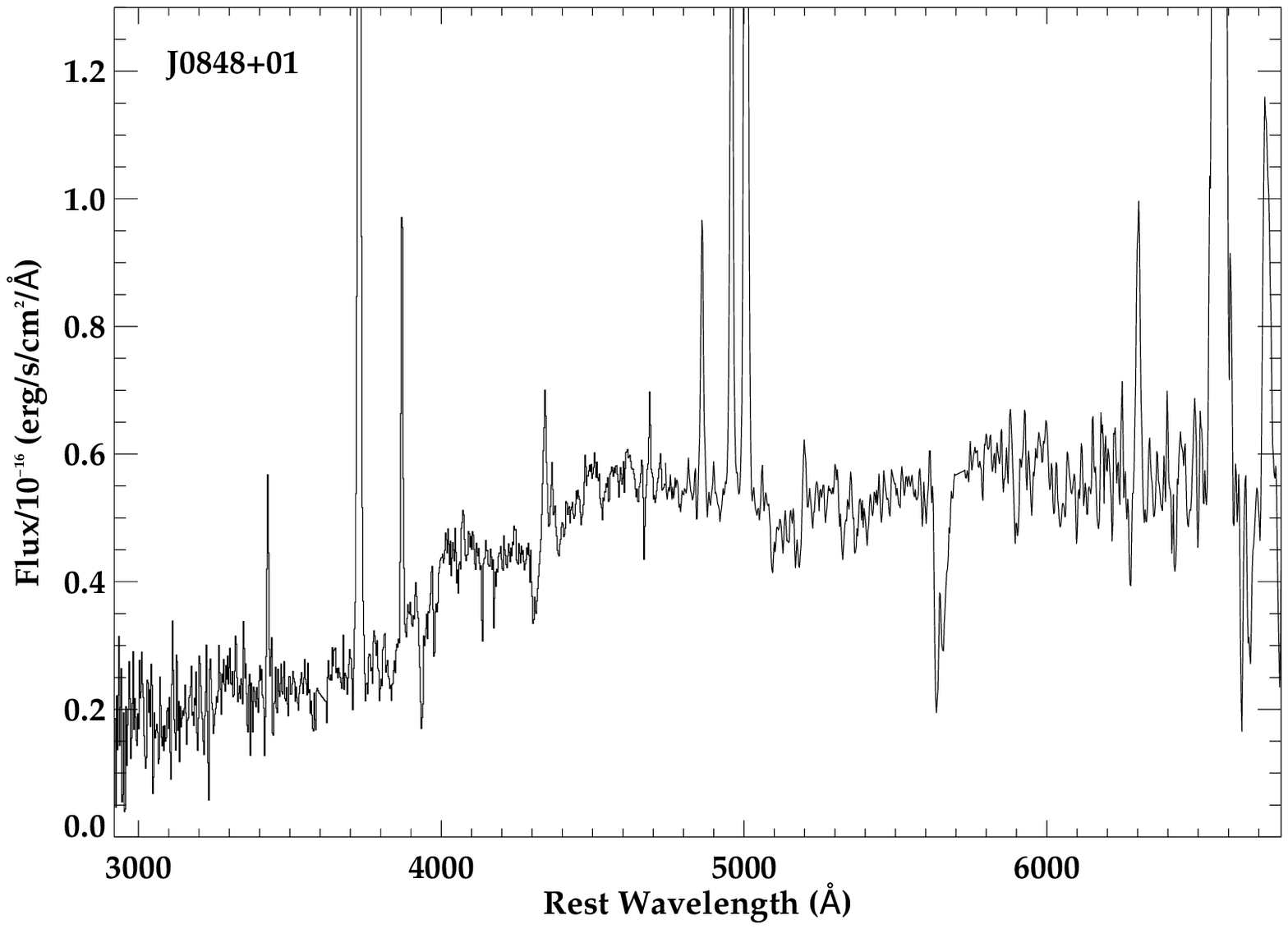}
\label{q0848-spec}}
\subfloat{
\includegraphics[scale =  0.4]{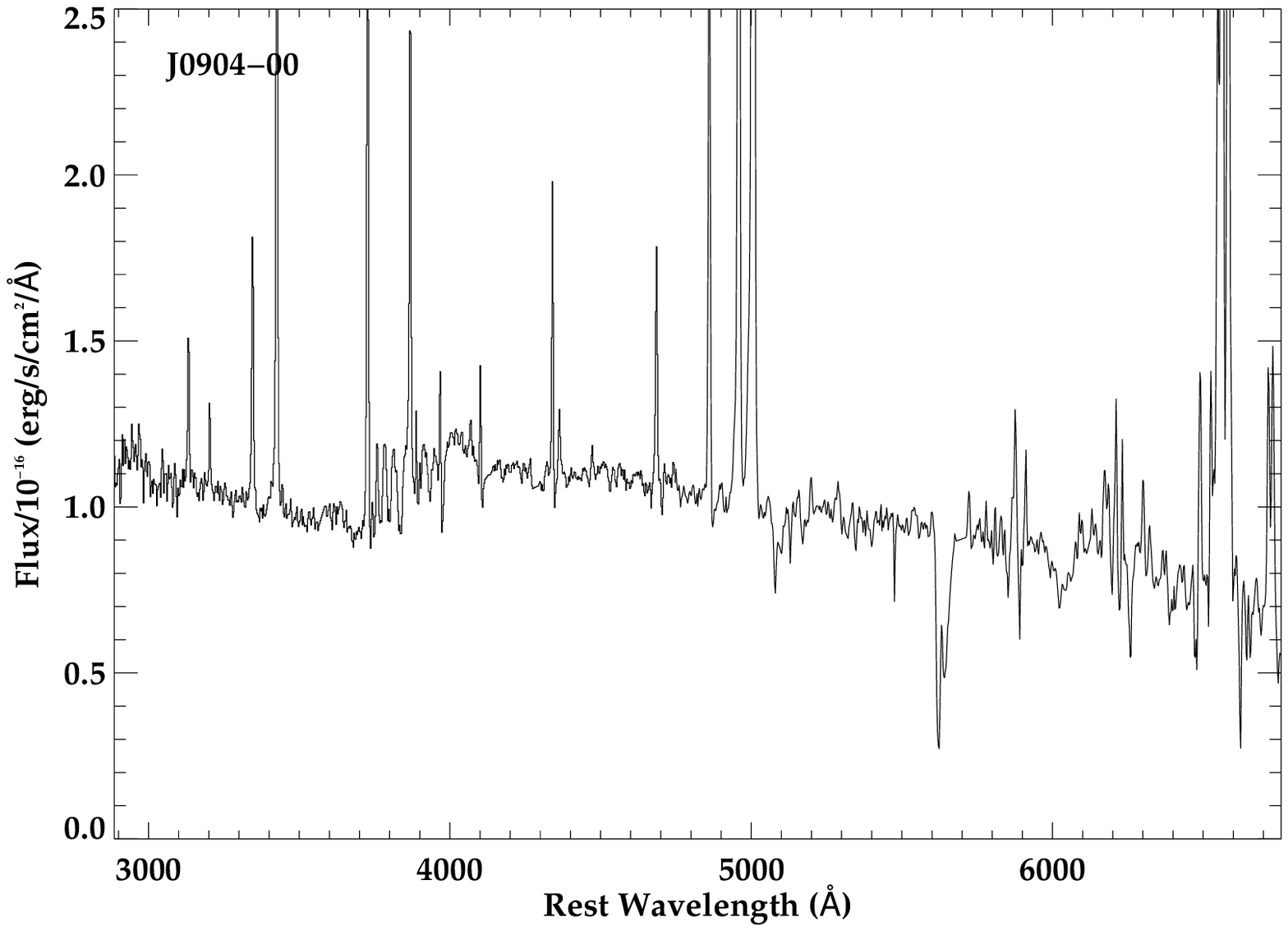}
\label{q0904-spec}}

\subfloat{
\includegraphics[scale =  0.4]{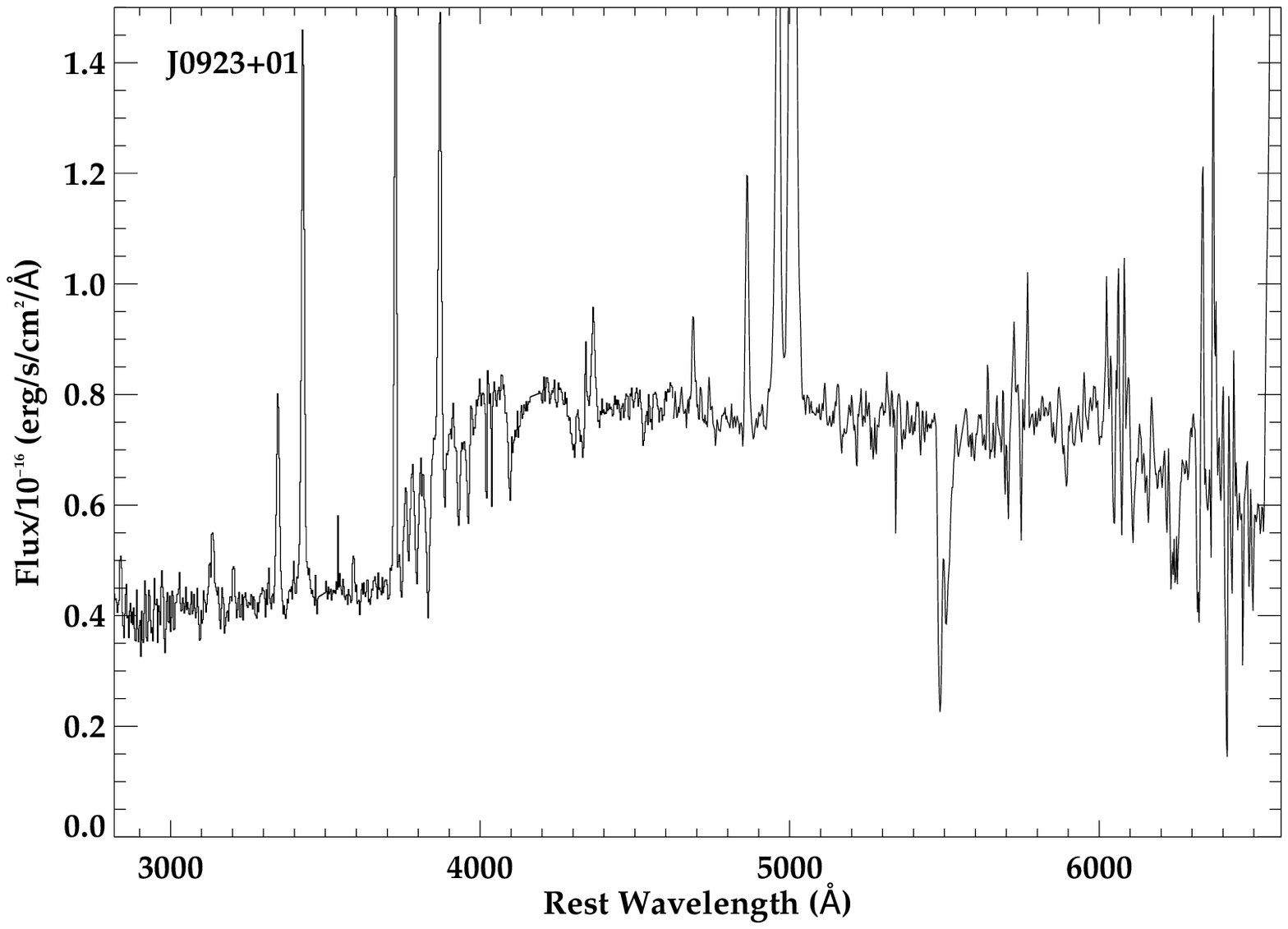}
\label{q0923-spec}}
\subfloat{
\includegraphics[scale =  0.4]{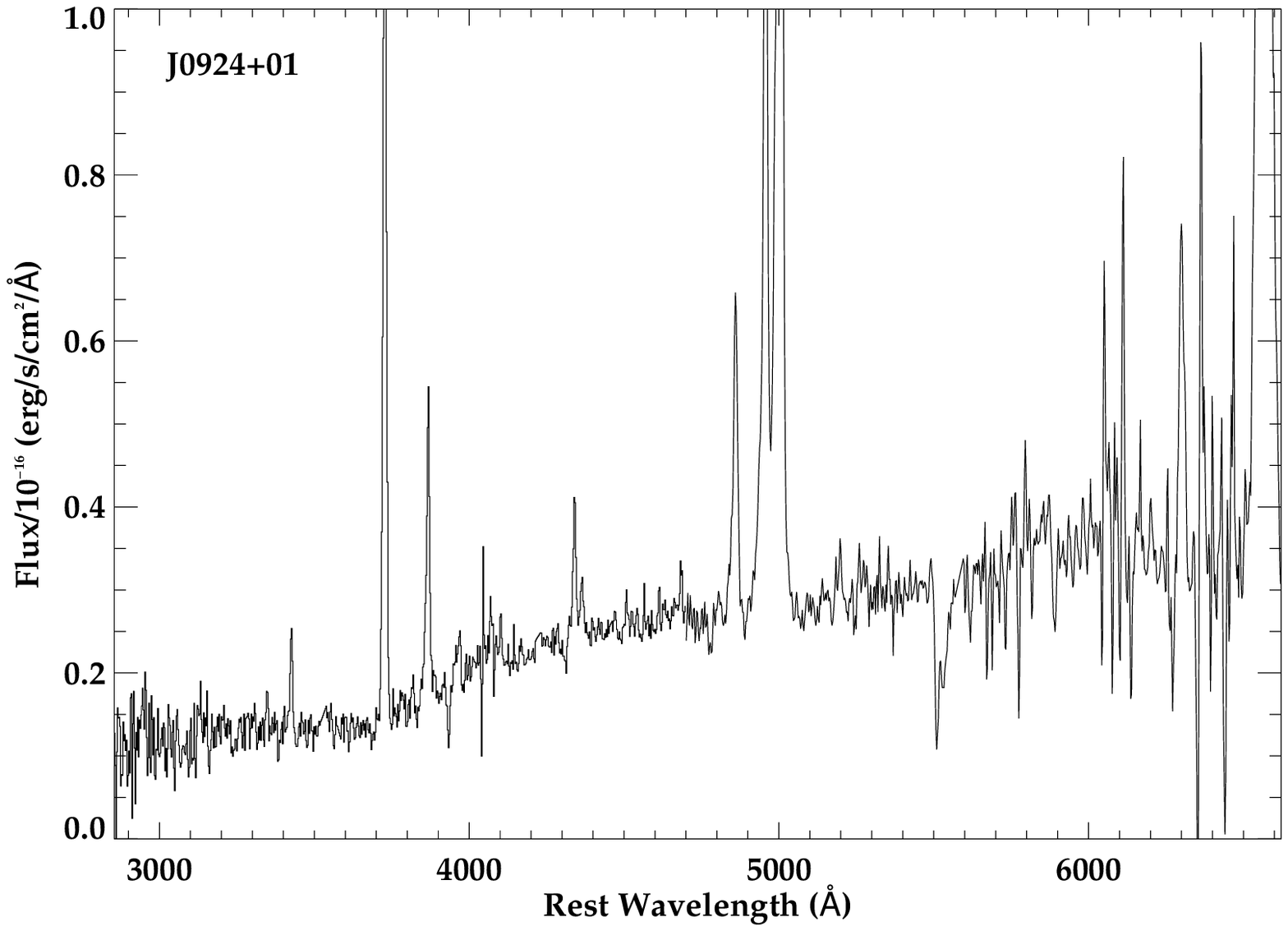}
\label{q0924-spec}}

\subfloat{
\includegraphics[scale =  0.4]{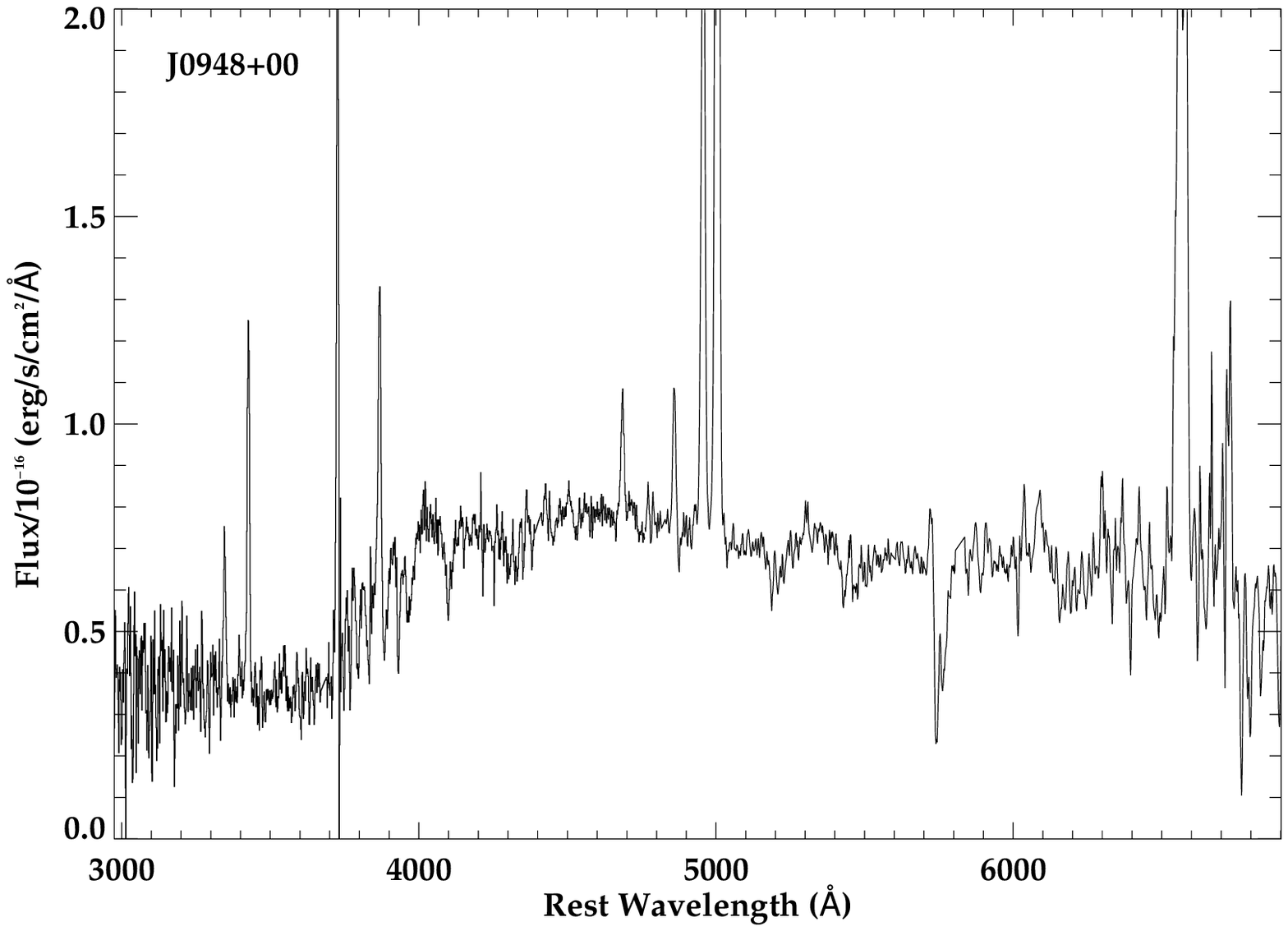}
\label{q0948-spec}}
\subfloat{
\includegraphics[scale =  0.4]{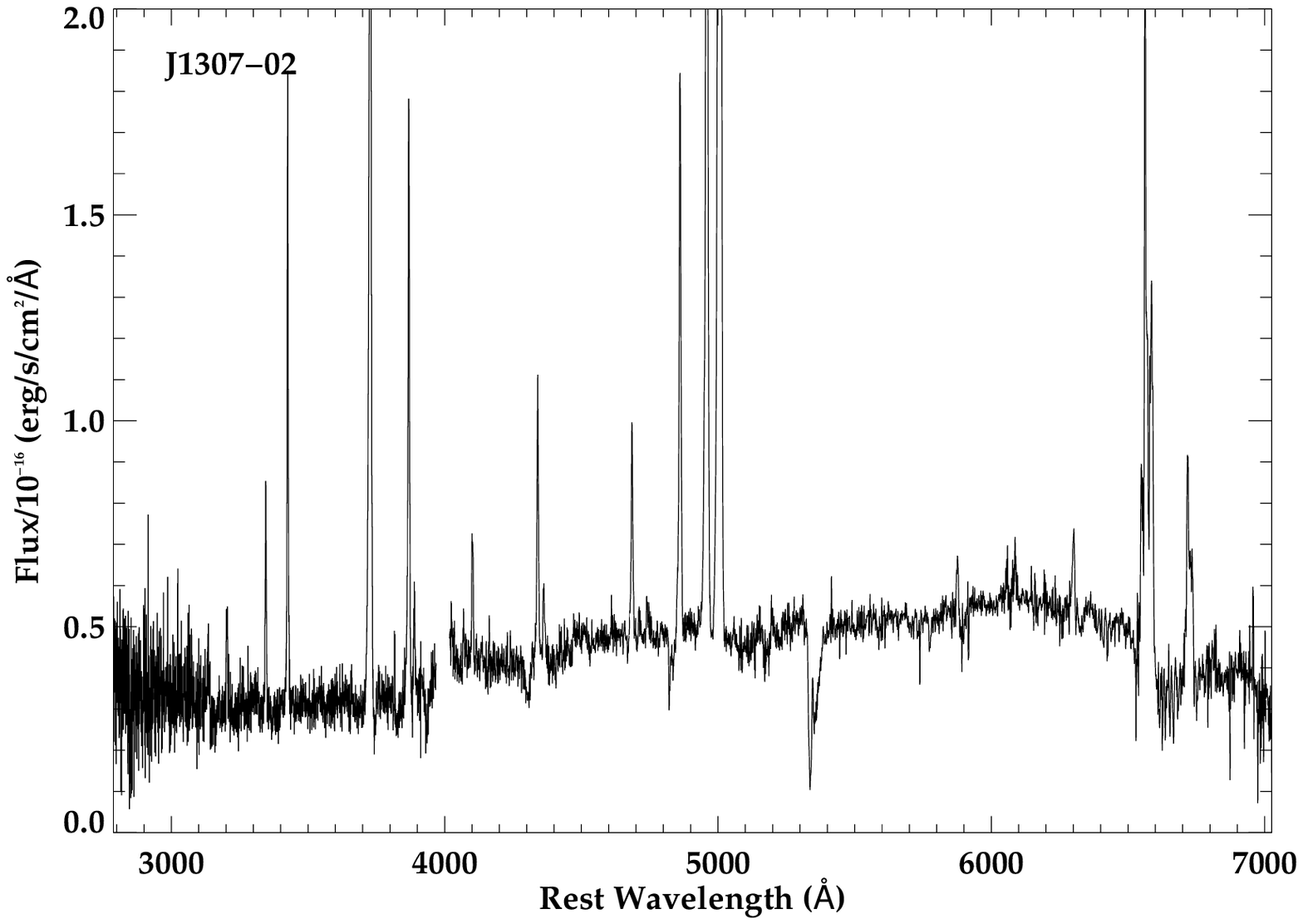}
\label{q1307-spec}}

\subfloat{
\includegraphics[scale =  0.4]{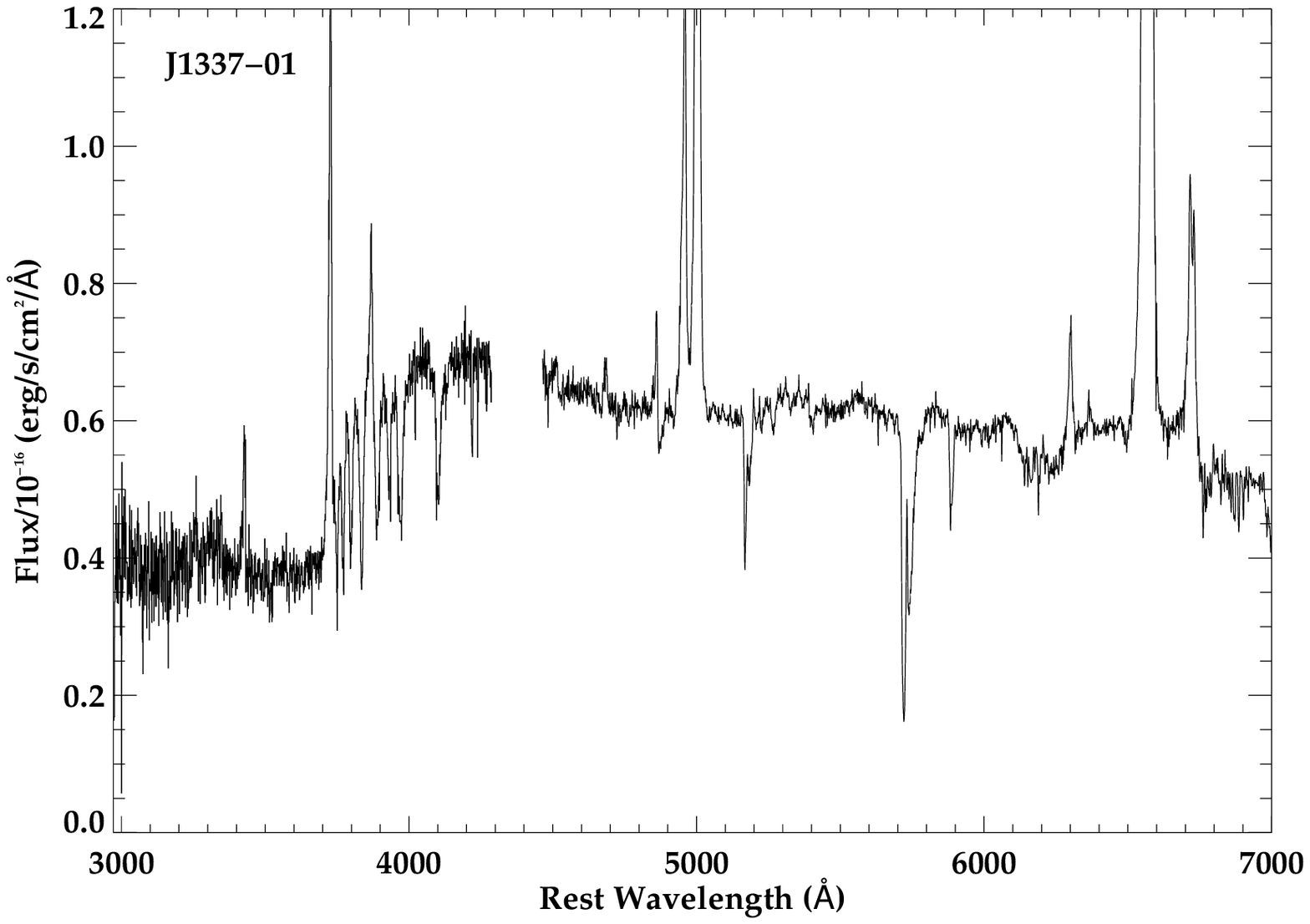}
\label{q1337-spec}}

\contcaption{}
\end{figure*}

\subsection{Continuum Modelling}
\label{contMod}

To model the stellar continuum, we used a purpose written {\sc idl} routine ({\sc confit}; \citealt{robinson00,robinson01}), which is capable of fitting up to three components to the observed spectrum, comprising up to two stellar components (one fixed in age and one varying), and additionally, a power-law component of the form $F_\lambda \propto \lambda^\alpha$ ($-15 < \alpha < 15$). The routine fits the specified number of components to the data using varying combinations of flux in each component to compute the reduced minimum $\chi^2$ (\chisq) value for each combination. The combinations of components are allowed to vary up to 125\% of the measured flux in the normalising bin in order to account for the uncertainties in both the data and the models. Each spectrum was partitioned into a number of wavelength bins ($\sim 70$), avoiding strong emission lines, chip gaps, dead columns and other cosmic defects. In order to perform the modelling, we selected a normalising bin in the region of 4400 -- 4800 \AA\ for each spectrum, and assumed a relative flux calibration error of $\pm5\%$ (as outlined above). \confit\ has been successfully used to carry out spectral synthesis modelling for galaxies with a wide range of properties. Examples can be found in \citet{holt07}, \citet{wills08} and \citet{rodriguez09}.  

In order to better fit the stellar populations, the higher order Balmer emission lines and a nebular continuum were subtracted using the procedure outlined in \citet{dickson95} and \citet*{holt03}. The process of generating the nebular continuum involved several steps. Initially, the best fitting stellar model (generated in an initial \confit\ run) was subtracted from the spectrum to account for any underlying stellar absorption of \hb. Using the {\sc starlink figaro} package, the measured \hb\ flux was then used to generate a spiketrum\footnote{A spectrum where all the elements except for those at the tabulated values are zero. In this case, the tabulated values are the expected fluxes of the higher order Balmer lines, considering the measured \hb\ flux.} of the Balmer higher order lines ($\mbox{n} \ge 8$), assuming case B recombination. This spiketrum was then convolved with Gaussians of the measured FWHM of \hb, producing a model of the Balmer emission lines. 

The next step was to generate a nebular continuum which accounts for the free-free, free-bound and two photon decay continua associated with the nebula. This model was produced using the {\sc starlink dipso} package, assuming an electron temperature of $T=10000~\mbox{K}$ and electron density of $N_e = 10^2~\mbox{cm}^{-3}$. However, it should be noted that the resulting nebular model is not particularly sensitive to the assumed temperature and density. The model of the Balmer emission lines and the continuum nebular emission were then combined to form the final nebular model.

 We then attempted to characterise the internal reddening of the galaxy using the Balmer decrement from the stellar subtracted spectrum. Although it would obviously be preferable to measure any reddening of the nebular continuum by using the \ha/\hb\ ratio, the strong fringing at the longest wavelengths in the Gemini spectra does not allow reliable measurements of the \ha\ flux. In the case of the higher redshift objects in the sample, \ha\ is shifted out of the spectral range altogether. Therefore, in order to make some estimate of the reddening of the nebular continuum, the fluxes of the \hg\ and \hd\ lines were also measured. However, due to the lower equivalent widths of the lines, in some cases the errors were larger than the value of the ratio. In such cases the values were disregarded (Table \ref{blines}).

For the purpose of maintaining a consistent approach throughout, a maximum nebular subtraction (i.e. applying no reddening to the model nebular continuum) was attempted first. The resulting nebular subtracted spectrum was then inspected in order to determine if this resulted in a non-physical step around the region of the Balmer edge (3646 \AA). If this was the case, reddening that was consistent with the Balmer decrement measured from the stellar-subtracted spectrum was applied using the \citet{seaton79} extinction law. However, in the cases where no Balmer decrement could be measured, no reddening was applied to the model nebular continuum. In all cases where this was necessary, this procedure produced an acceptable result, with no clear over-subtraction at the Balmer edge. This is not unexpected, because the fact that the higher order Balmer emission lines are too weak to measure accurately, implies a weaker nebular continuum. Inspection of Table \ref{blines} demonstrates that, when the nebular continuum contributes $<20 \%$ of the flux below the Balmer edge, it was not possible to measure the Balmer decrements. This weaker nebular continuum means that the result of the subtraction will not be too sensitive to any reddening correction applied.

The results presented in column 5 of Table \ref{blines} make it clear that it is important to consider the effects of the nebular continuum before embarking on the modelling process. In the vast majority of cases, the nebular continuum accounts for over 15\% (ranging from 3--28\%) of the total flux just below the Balmer edge, measured in the wavelength range 3540--3640\AA\ (where the contribution from the nebular continuum will be most pronounced). Failure to account for this component could lead to underestimations of the true ages of the YSPs, because the stellar spectrum will appear bluer than it actually is. The proportions of the UV flux attributed to the nebular continuum in these type II quasars host galaxies are in good agreement with those found for the 2 Jy sample of powerful radio galaxies \citep{tadhunter02}, where values also ranging from 3--28\%, were found, with the majority of the sample showing nebular contributions above 15\% over the same wavelength range.

In order to assess the effect of the uncertainties in the nebular continuum subtraction on the modelling results, we have carried out the stellar synthesis modelling of a number of objects with differing YSP ages (J0234-07, J0320+00, J0848+01 and J0948+00) varying the degree of reddening applied to the nebular continuum before subtraction by \ebv $\pm 0.1$. The results produced by these tests conform to those which we would expect. When the reddening applied to the nebular continuum is increased (i.e. less flux is subtracted in the blue part of the spectrum), we see a slight shift to younger ages with slightly higher reddening for the YSP. These shifts tend to extend the YSP age range by $\sim$ 1 to 10 Myr (depending on the increment available in YSP ages due to the input models), and we find a corresponding increase of \ebv = 0.1 at the youngest ages. Decreasing the reddening (i.e. subtracting more flux in the blue part of the spectrum) results in the same behaviour but in the reverse direction, with the ages increasing and the reddening decreasing at the youngest ages. This confirms than an error in the reddening applied to the nebular continuum of \ebv $\sim 0.1$ will not have any significant effect on the range of YSP ages and reddenings determined for each object, and will not impact the conclusions drawn in this work.

\begin{table*}
\begin{minipage}{117mm}
\centering
\caption{The Balmer decrements measured from each of the apertures extracted from the full sample of type II quasars. Column 1 gives the name of the object and if appropriate, which aperture is being referred to. Columns 2, 3 and 4 give the \hg/\hb, \hd/\hb\ and \ha/\hb\ ratios respectively along with the associated fitting error (Case B values: \hg/\hb\ = 0.468, \hd/\hb\ = 0.259 and \ha/\hb \ = 2.86). Column 5 gives the percentage contribution of the nebular continuum to the total flux in the wavelength range 3540--3640\AA\, and the final column gives the E(B-V) value of reddening applied to the model nebular continuum before subtraction from the data. In the cases in which a reddening correction was applied before subtraction, the value shown in column 5 refers to the percentage of the total flux contributed by the model after the reddening correction has been applied. No value was calculated for J0142+14 due to the fact that \hb\ fell outside the spectral range.}
\label{blines}
\begin{tabular}{ l c c c c c }
\hline
Name 	& \hg/\hb\	& \hd/\hb\			&	\ha/\hb\		&	Nebular Continuum	&	E(B-V)\\
			&					&					&					&	(\%)				&		\\
\hline
J2358-00	&	0.39 \pms 0.08	&	0.23 \pms 0.07  &	--				&	28					&	--\\
J0114+00	&	--				&	--				&	--				&	14					&	--			\\	
J0123+00	&	0.43 \pms 0.03	&	0.22 \pms 0.02	&	--				&	20					&	0.2		\\	
J0142+14	&	--				& 	--				& --				&	--					&	-- \\	
J0217-00	&	--				&	--				&	--				&	14					&	--\\		
J0217-01	&	--				&	--				&	--				&	20					&	--	\\
J0218-00(N)	&	0.37 \pms 0.02	&	--				&	--				&	17					&	--	\\
J0218-00(E)	&	0.32 \pms 0.02	&					&	--				&	17					& -- \\
J0227+01	&	--				&	--				&	--				&	11					&	--\\	
J0234-07	&	0.42 \pms 0.12	& --				&	2.41 \pms 0.12	&	25					&	0.25\\		
J0249+00	&	0.35 \pms 0.14	& --				&	--				&	25					&	--\\	
J0320+00	&	0.43 \pms 0.17	&  -- 				&	-- 				&	8					&	0.2\\		
J0332-00(N)	&	--				&--					&	--				&	4.1					& 	-- \\
J0332-00(E)	&	--				&	--				&	--				&	3.3					&	--\\
J0334+00	&	0.38 \pms 0.02	& 0.21 \pms 0.01	&	--				&	23					&	--\\		
J0848+01	&	--				&	--				&	--				&	19					&	-- \\		
J0904-00	& 	0.46 \pms 0.17	&	-- 				& 	3.82 \pms 1.3	&	13					&	--\\		
J0923+01	&	--				&	--				&	--				&	14					&	--	\\	
J0924+01	&	0.33 \pms 0.09	&	--				&	4.26 \pms 1.43	&	14					&	0.3\\	
J0948+00	&	--				&	--				&	--				&	15					&	--\\
J1307-02	&	0.40 \pms 0.07	&					&	--				&	27					&	-- \\
J1337-01	&	--				&	0.20 \pms 0.04	&	4.39 \pms 0.7	&	7.8					&		\\	
	
\hline
\end{tabular}
\end{minipage}
\end{table*}

\subsection{Modelling Strategy}
\label{modelling}

The stellar population models used here were generated using the {\sc{Starburst99}} code of \citet{leitherer99}, assuming an instantaneous burst star formation history, solar metallicity and a Kroupa IMF \citep{kroupa01}.Reddening in the range $ 0 \le E(B-V) \le 2.0$ in steps of $\Delta \ebv = 0.1$ was then applied to these template spectra using the extinction curve of \citet{calzetti00} which is appropriate in the starburst case. To carry out the modelling, we used several different components to account for varying scenarios. In all cases, we assumed that there is an underlying older stellar population of either 8 Gyr (old stellar population; OSP) or 2 Gyr (intermediate stellar population; ISP), and in the majority of cases, we also assumed that there is a YSP component. We chose an 8 Gyr underlying population because this allows for the formation of a population of stars at high redshift. However, because there is very little evolution in the spectra of a simple stellar population at large ages, the modelling is not particularly sensitive to the exact age chosen (e.g. 8 Gyr vs. 12 Gyr).

The 2 Gyr population was used to account for the possibility that the dominant underlying population of stars was formed in more recent times (e.g. in a previous major merger event) and was chosen because experimentation with different possibilities (i.e. 1 Gyr and 0.5 Gyr) showed that the 2 Gyr population produced significantly better results in general. In contrast to the YSP, which are assumed to be the product of a merger induced, dusty, nuclear starburst, these older populations are assumed to be unreddened because it is probable that these stars are distributed throughout the host galaxy, and that those which contribute to the observed spectra are foreground stars.  

It was also possible to include a power-law component in the model, the purpose of which is to account for the possibility of scattered-light from the hidden quasar. Although these are type II quasars, it is possible that scattered light may make a significant contribution to observed spectrum \citep{tadhunter02,obied16}, and thus must be properly accounted for. For the purposes of the modelling, five combinations of components, outlined below, were adopted which reflect plausible combinations of the various components that contribute to the observed spectra\footnote{We do not include an 8 Gyr population combined with a 2 Gyr population and a YSP, which would account for more than one previous episode of star formation, because it would be very difficult to disentangle the contribution of the 8 Gyr population from that of the 2 Gyr population.}.

\begin{enumerate}[1.]
\item An unreddened, 8 Gyr OSP combined with a power-law component, accounting for the case of an OSP plus a scattered or direct AGN component but no YSP.
\item An unreddened, 8 Gyr underlying population, combined with instantaneous burst star formation models, with ages $0.001 < t_{YSP} ~ (\mbox{Gyr}) < 5.0$ and reddening $0 \le E(B-V) \le 2.0$. 
\item An unreddened, 2 Gyr underlying population, combined with instantaneous burst star formation models, with ages $0.001 < t_{YSP} ~ (\mbox{Gyr}) < 2.0$ and reddening $0 \le E(B-V) \le 2.0$. 
\item The same as 2 with the addition of a power-law component.
\item The same as 3 with the addition of a power-law component.
\end{enumerate}

\begin{figure}
\centering
\includegraphics[scale = 0.5]{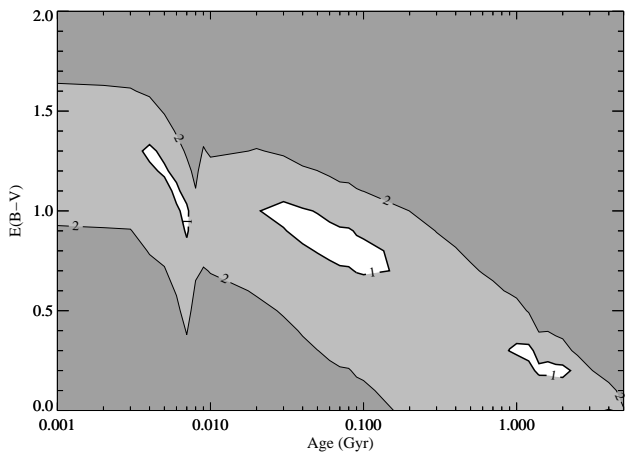}
\caption{The \chisq\ plot produced from modelling J0217-01 assuming combination 2. The plot shows the age of the YSP against the reddening of the YSP. The areas shown in white are the regions in which the modelling produced \chisqlt. This example shows that, in this particular case, three distinct regions produced such values. }
\label{chiplot}
\end{figure}

\begin{figure}
\centering
\subfloat{
\includegraphics[scale = 0.45, trim = 7mm 0mm 0mm 0mm]{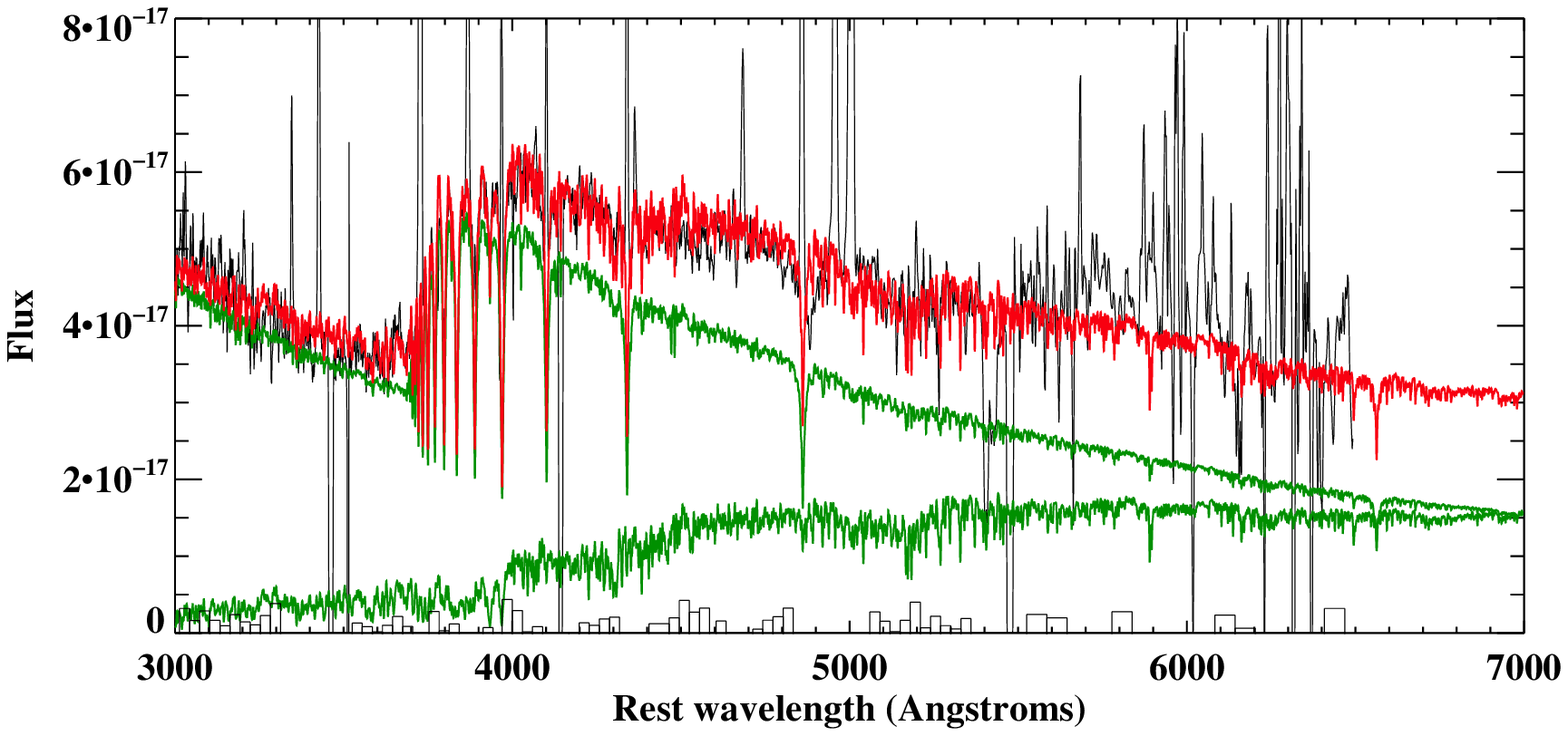}
\label{q0334-8np-b-bf}}

\subfloat{
\includegraphics[scale = 0.45, trim = 7mm 0mm 0mm 0mm]{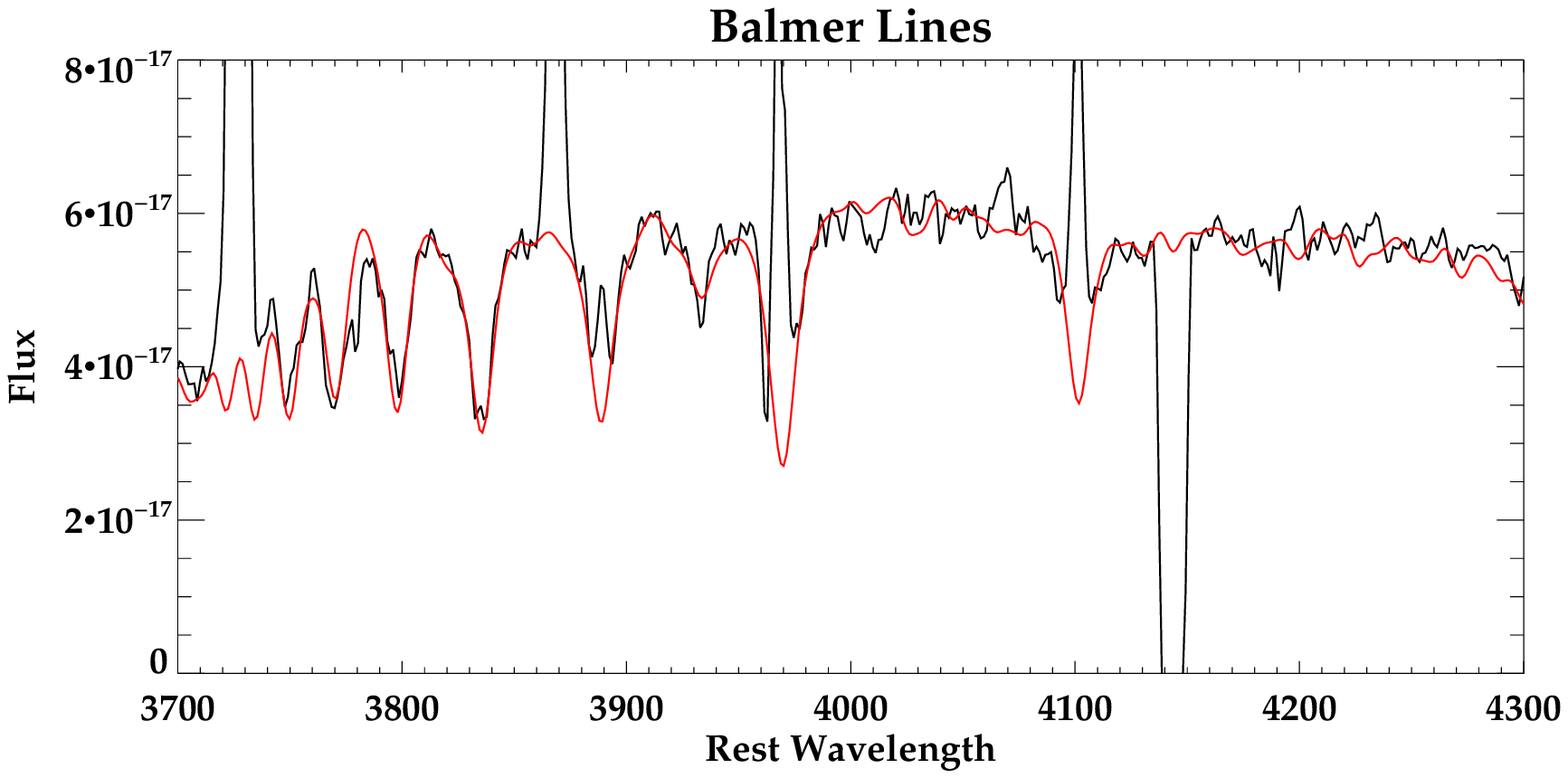}
\label{q0334-8np-b-bl}}

\caption{An example of an acceptable fit produced by \confit for J0334+00. The fit shown here was produced using combination 2 and includes a YSP with \tysp\ = 0.08 Gyr and \ebv\ = 0. The top panel shows the overall fit to the spectrum (red) while the individual components are shown in green. The residual for the fit are shown at the bottom. The bottom panel shows a zoom in of the region including the higher order Balmer lines. The solid black line shows the data while the red line shows our fit. The large drop in the continuum wavelengths $\sim 4140 - 4150$ \AA\ is not intrinsic to the galaxy, but due to a gap between two of the CCD detectors.}
\label{q0334_confit}
\end{figure}

In all cases, the overall shape, higher order Balmer lines, \ca\ and G-band fits for solutions that produce \chisqlt, were visually examined to determine if they were acceptable. When any of these age-sensitive regions produce a very poor fit, the combination of components was rejected as an acceptable solution. In cases where a large number of combinations produce \chisqlt, a representative sample of combinations were examined, taking  account of the allowed values of YSP/ISP ages and associated reddenings. For any combination used, it is possible that more than one minima in \chisq\ will be produced (see Figure \ref{chiplot} for an example). An example of what we consider to be an acceptable fit is shown in Figure \ref{q0334_confit}. The top panel shows the overall fit to the spectrum (red), with the individual contributions for the OSP (8 Gyr) and YSP (0.08 Gyr) shown in green. The residuals of the fit are shown along the bottom for each bin. The bottom panel shows a zoom-in of the Balmer absorption lines, which, in this example, are prominent in the spectrum. This fit assumes combination 2 and therefore, no power-law component is included. 

The exception to the method of inspection outlined above was J0332-00, because it has subsequently become apparent that it is actually a type I object in which the broad Balmer emission lines are directly detected (see section \ref{spec:q0332}). In this case, because of the strong power-law component, for combination 4 all combinations produced \chisqlt\, while for combination 5, large sections of parameter space produced \chisqlt. Therefore, we are not able to offer any meaningful constraint to the age of the YSP (if any) in this object.

\section{Results}

\subsection{The effect of including a power-law component}
\label{spec:plComp}

The significance of the power-law component, particularly in objects that are well fit by combinations that do not \emph{require} the inclusion of one, remains one of the largest sources of uncertainty in this work. To unambiguously detect a scattered quasar component would require spectropolarimetry data, but it should be noted that in all but one case (J0332-00), it is possible to fit the \hb\ line using the same kinematic components as those used for $\mbox{\oiii}\lambda\lambda 4959,5007$, which are, or course, produced in the narrow-line region. As we show next, this strongly suggests that any contamination from scattered quasar light is relatively small because no broad \hb\ component (FWHM $> 2000~ \kms$) is clearly visible in the stellar-subtracted spectrum for any of the objects presented here (except the type I object J0332-00). However, it is important to consider the effect that introducing a power-law has on the results of the stellar population modelling. Therefore, an attempt to place an upper limit on the power-law contribution to the total flux for each object, using the information yielded by the stellar synthesis modelling, has been made. 

\begin{figure*}
\centering
\subfloat{
\includegraphics[scale = 0.45, trim = 7mm 0mm 0mm 0mm]{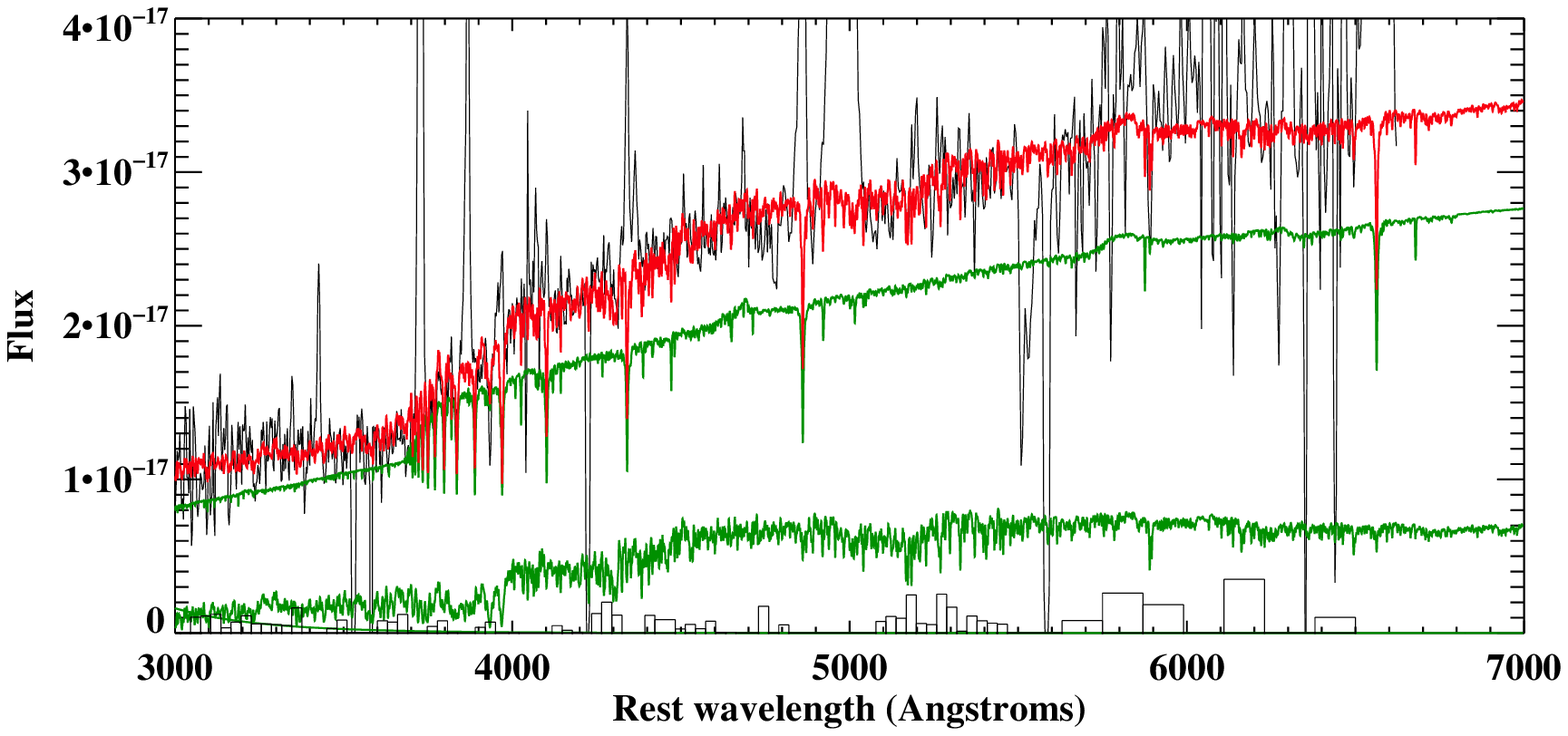}
\label{red_min_pl_example}}
\subfloat{
\includegraphics[scale = 0.45, trim = 7mm 0mm 6mm 0mm]{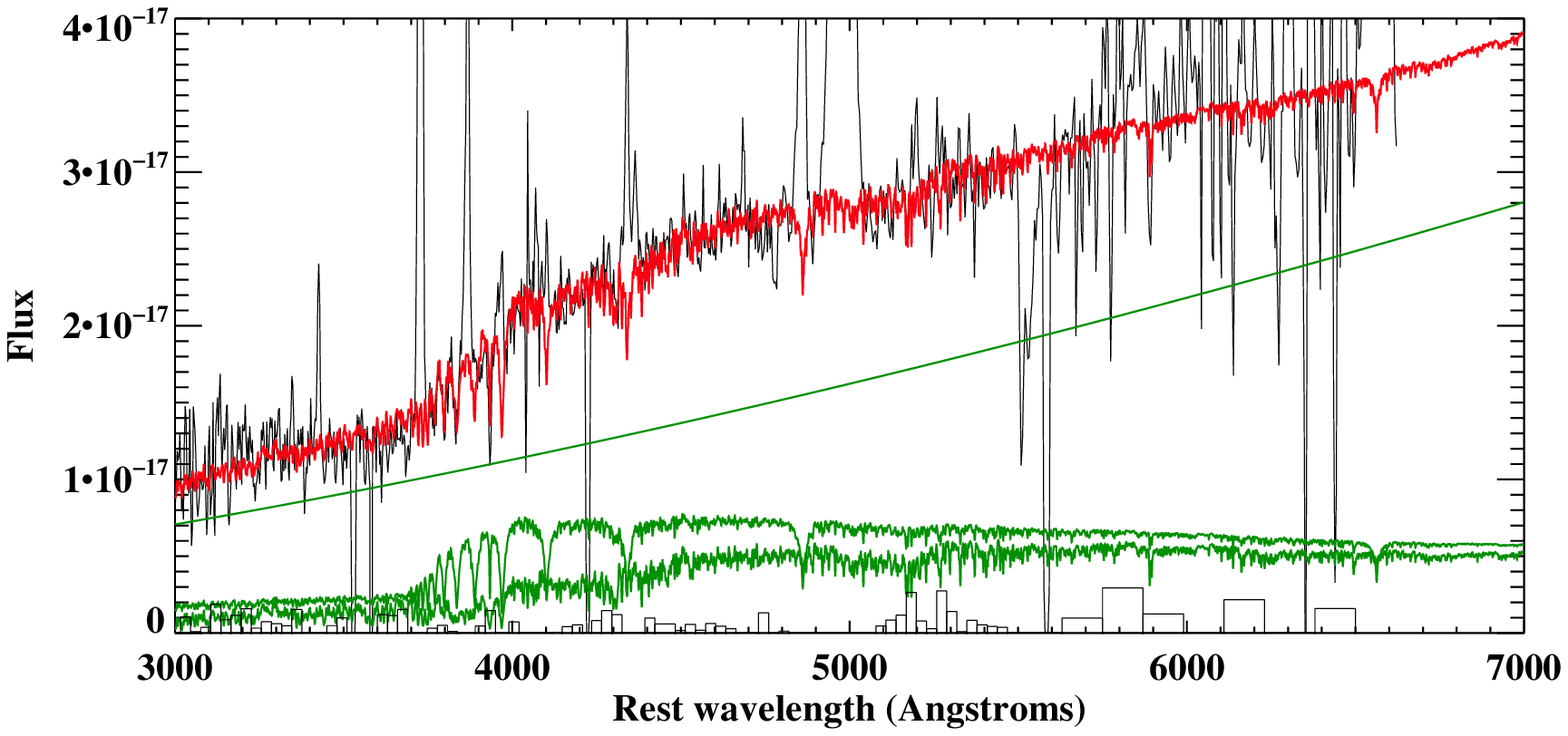}
\label{red_max_pl_example}}

\subfloat{
\includegraphics[scale = 0.45, trim = 7mm 0mm 0mm 0mm]{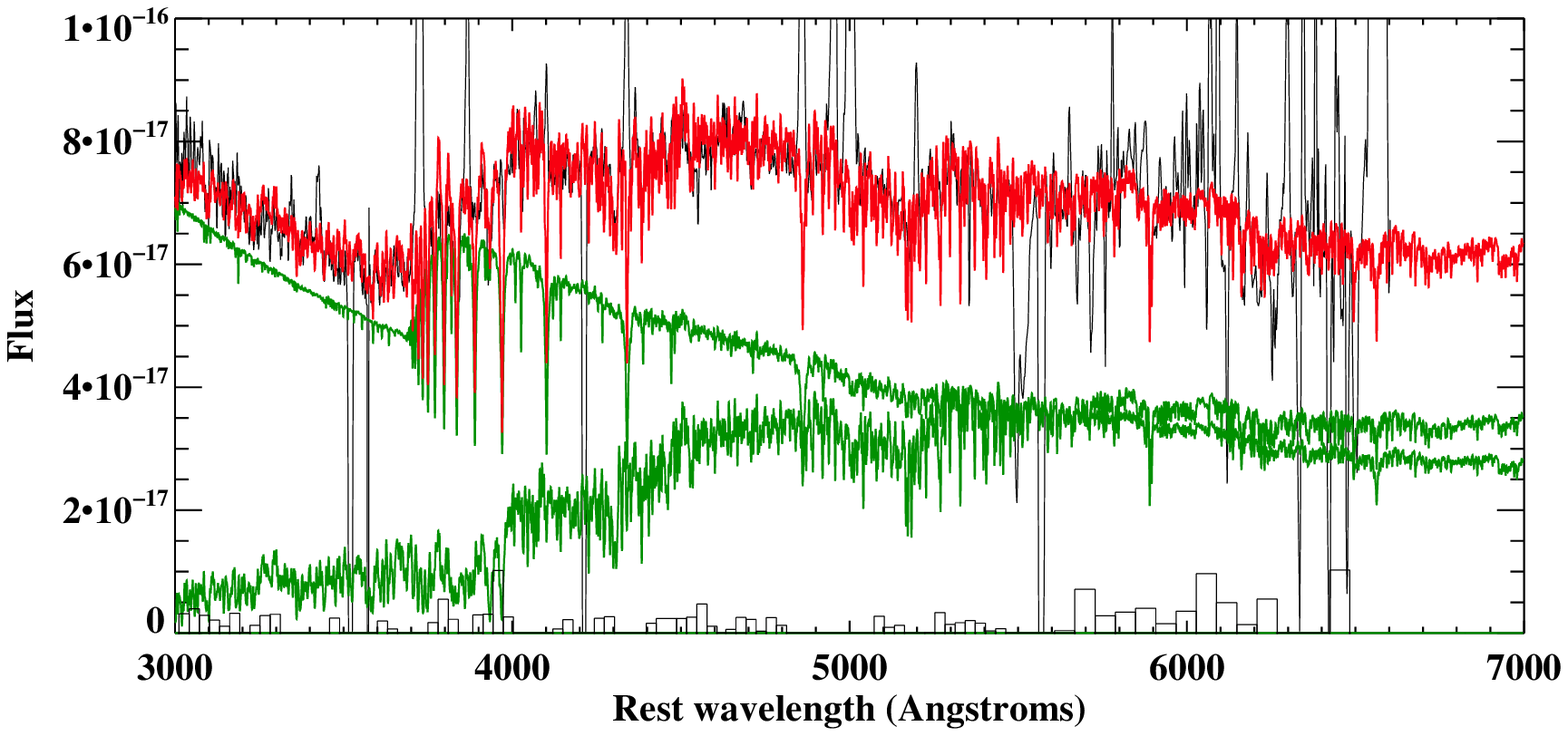}
\label{blue_min_pl_example}}
\subfloat{
\includegraphics[scale = 0.45, trim = 7mm 0mm 6mm 0mm]{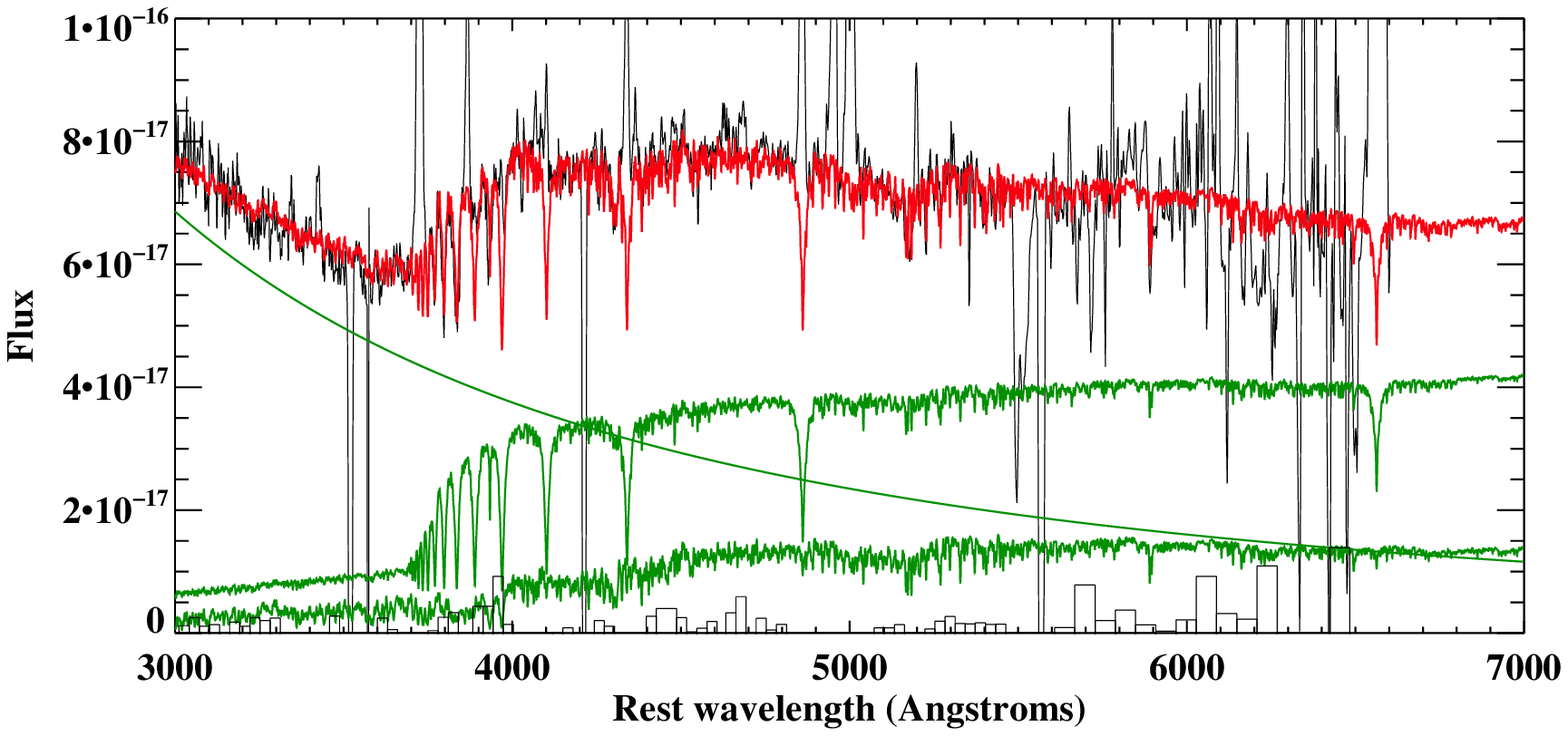}
\label{blue_max_pl_example}}
\caption{Examples of how the power-law contribution is related to the age and reddening of the YSP.  The top left panel shows a fit for J0924+01 which incorporates only an OSP and YSP. The fit shows a YSP of \tysp = 0.005 Gyr and \ebv = 1.0.  The top right panel shows the same object but with the maximum allowable power-law of $\sim 50\%$ included in the fit. In this case, \tysp = 0.5 Gyr with \ebv = 0.3  The bottom panels show the same but for J0320+00, which has a maximum allowable power-law component of $\sim 40 \%$. The left-hand panel shows a YSP with \tysp = 0.02 Gyr and \ebv = 0.2 whilst the right-hand panel shows a YSP with \tysp = 0.3 and \ebv = 0.6.}
\label{age_pl_red_confit_example}
\end{figure*}

One aspect of the stellar synthesis modelling that is common between all the objects is that, when a power-law component is included in the fits, the maximum age that produces solution with \chisqlt\ increases significantly, and the range of reddenings allowed is also extended. In both the cases of objects with increasing and decreasing flux towards the UV, the age of the YSP generally increases with increasing power-law contribution. The reasons for these trends are illustrated in Figure \ref{age_pl_red_confit_example} which shows examples of combination 4 fits for the minimum and maximum cases of power-law flux in the normalising bin. The top two panels show examples of fits for J0924+01, which has one of the most strongly reddened YSPs of the objects in the sample. The left panel shows the minimum power-law contribution allowed by combination 4 (0 \%), whilst the right shows the maximum contribution (54 \%). In both cases, the OSP flux is only a small percentage of the total, and thus does not have a large effect on the overall shape of the stellar model in either case. However, it is clear that in the maximum case, the power-law mimics the shape of the reddened YSP from the minimum case. 

The bottom two panels show the same thing but for J0320+00, which is one of the objects that increases strongly in flux towards the UV. Again the left panel shows the minimum power-law case produced by combination 4 (0\%), and the right panel shows the maximum power-law case (38\%). Once again, it is clear that in the maximum case, the power-law component assumes the general shape of the unreddened YSP from the minimum case, thus allowing older YSPs to provide acceptable fits, as they are not required to contribute as strongly to the overall shape of the continuum.

The fact that the significance of the power-law component has a dramatic impact on the maximum ages allowed by the modelling means that it is important to be able to constrain the amount of the total flux attributable to it. Unfortunately, it is not possible to do this directly from the results of our stellar synthesis modelling. However, using the technique outlined in \citet{bessiere14}, we have made an attempt to place an upper limit on the possible contribution of the power-law to the flux in the normalising bin.

\begin{figure*}
\centering
\subfloat{
\includegraphics[trim = 14mm 0mm 14mm 0mm, width=4.9cm]{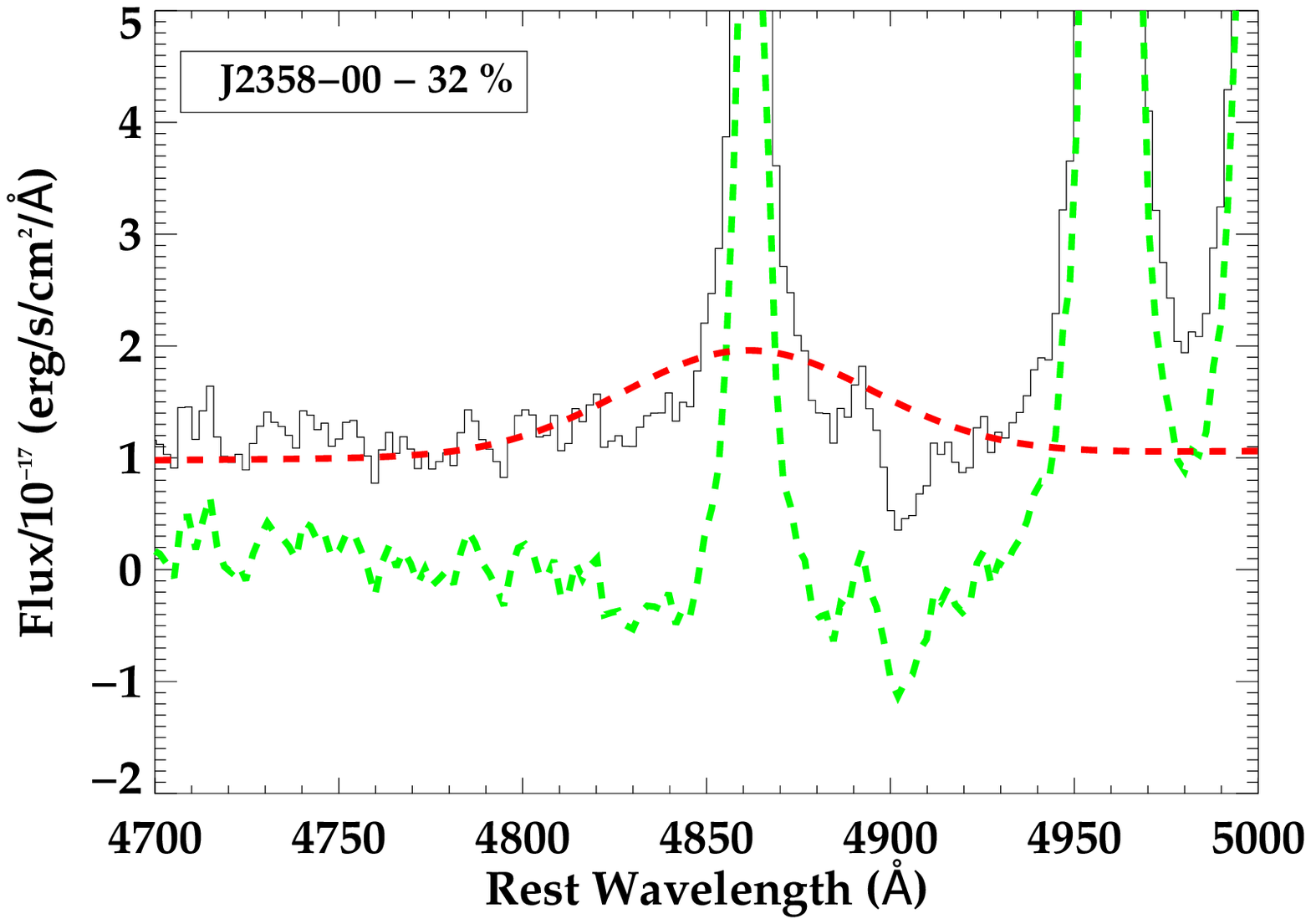}
\label{spec:q2358-ew}}
\subfloat{
\includegraphics[trim = 14mm 0mm 14mm 0mm, width=4.9cm]{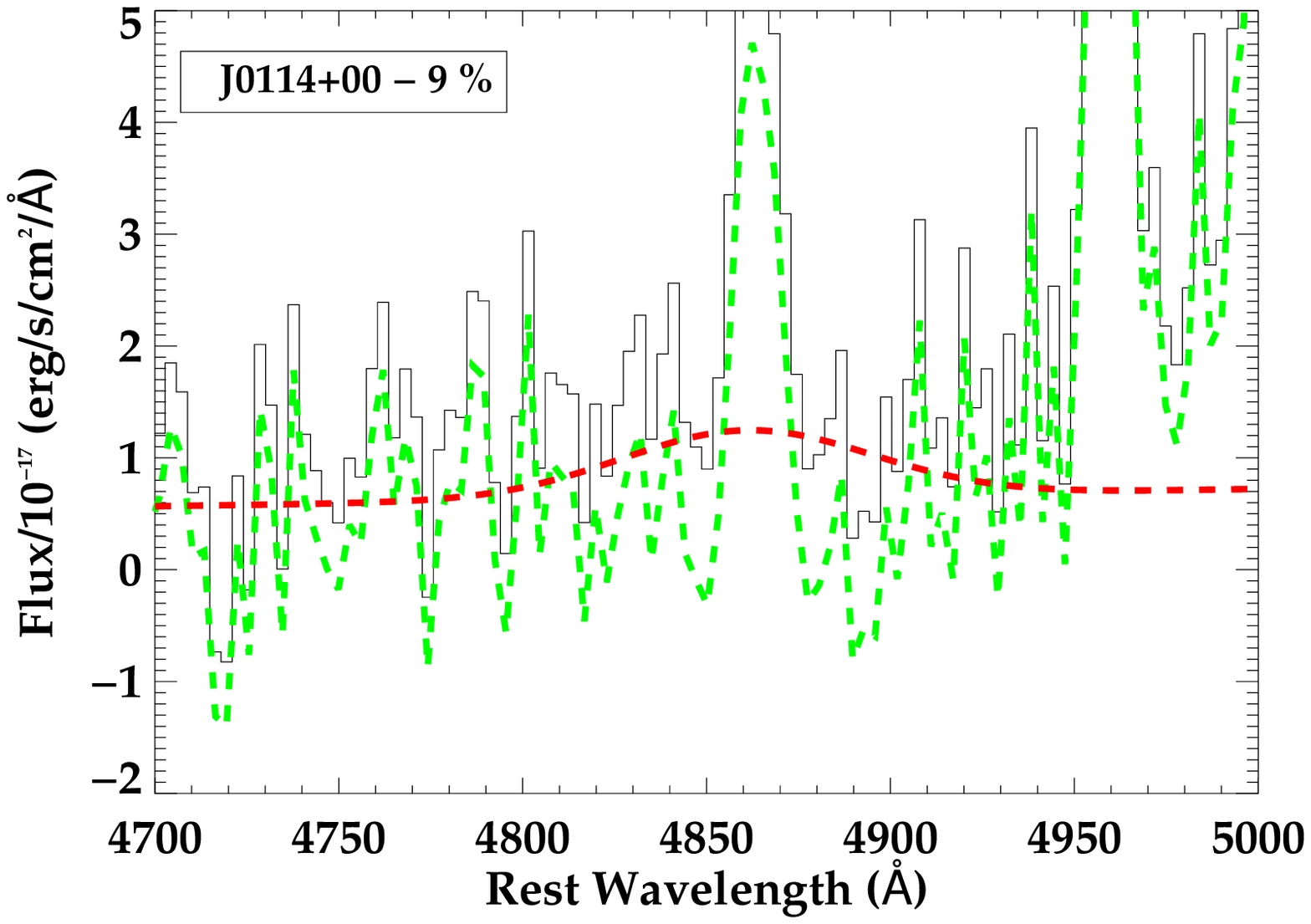}
\label{spec:q0114-ew}}
\subfloat{
\includegraphics[trim = 14mm 0mm 14mm 0mm, width=4.9cm]{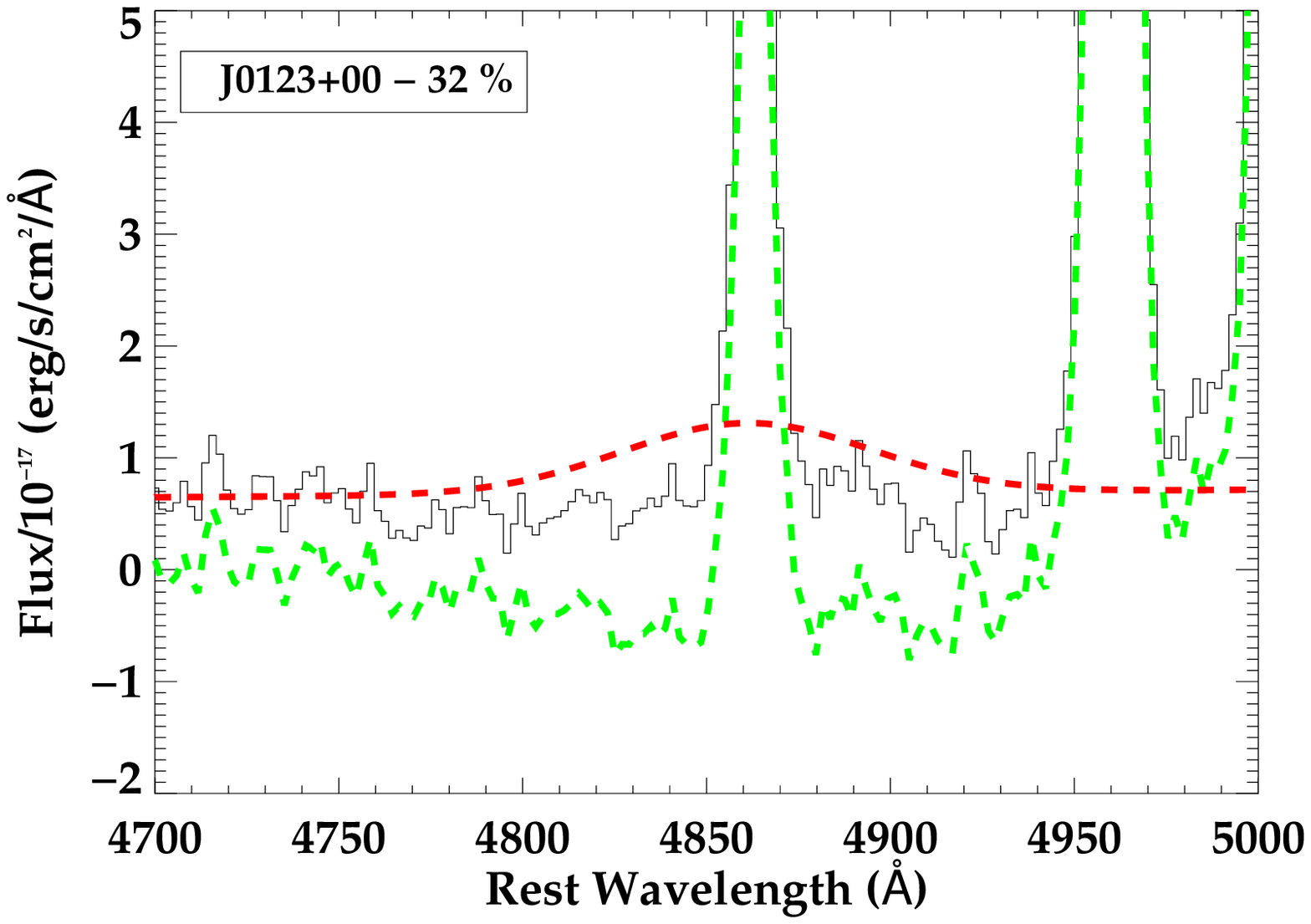}
\label{spec:q0123-ew}}

\subfloat{
\includegraphics[trim = 14mm 0mm 14mm 0mm, width=4.9cm]{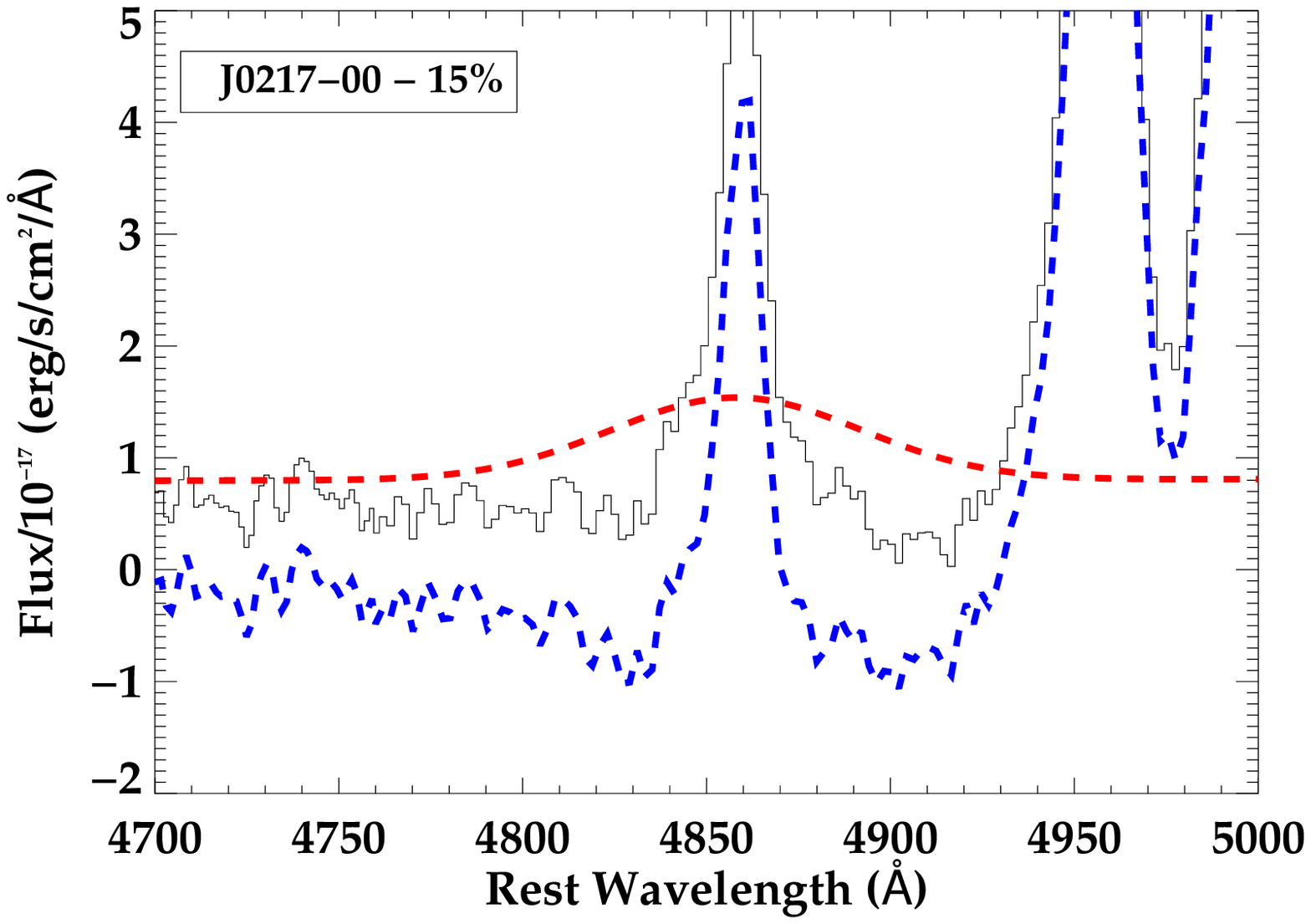}
\label{spec:q021700-ew}}
\subfloat{
\includegraphics[trim = 14mm 0mm 14mm 0mm, width=4.9cm]{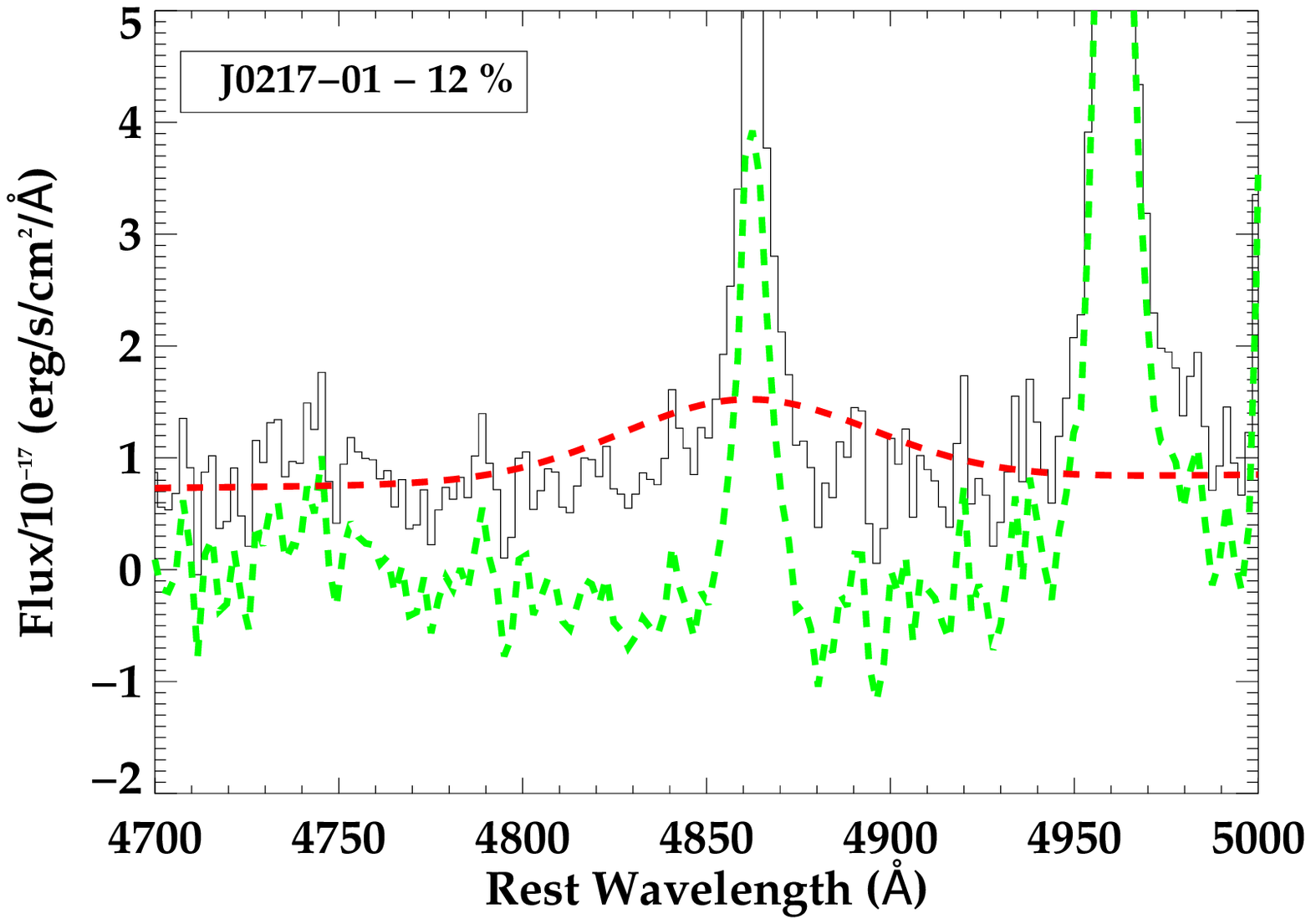}
\label{spec:q021701-ew}}
\subfloat{
\includegraphics[trim = 14mm 0mm 14mm 0mm, width=4.9cm]{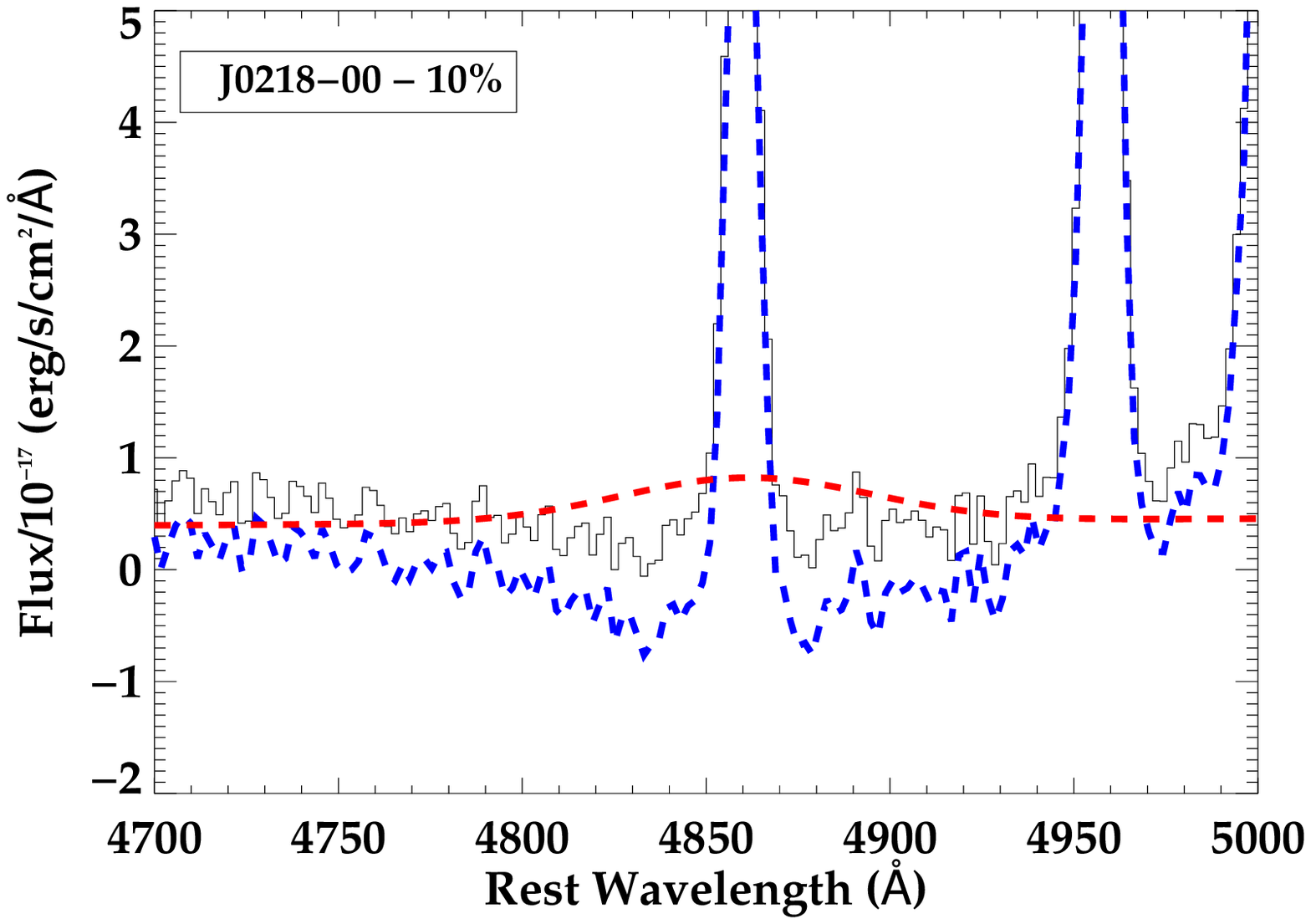}
\label{spec:q0218-ew}}

\subfloat{
\includegraphics[trim = 14mm 0mm 14mm 0mm, width=4.9cm]{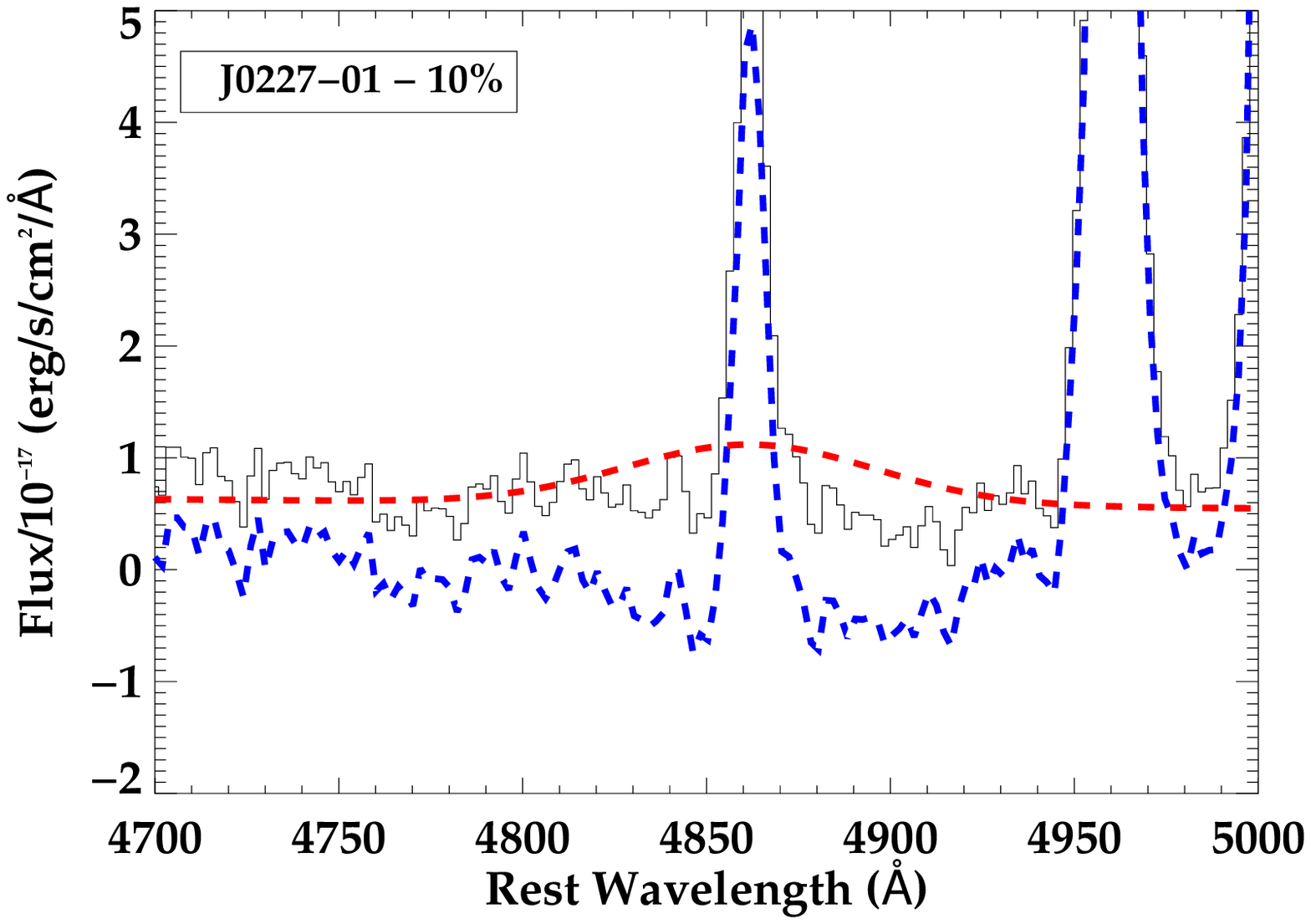}
\label{spec:q0227-ew}}
\subfloat{
\includegraphics[trim = 14mm 0mm 14mm 0mm, width=4.9cm]{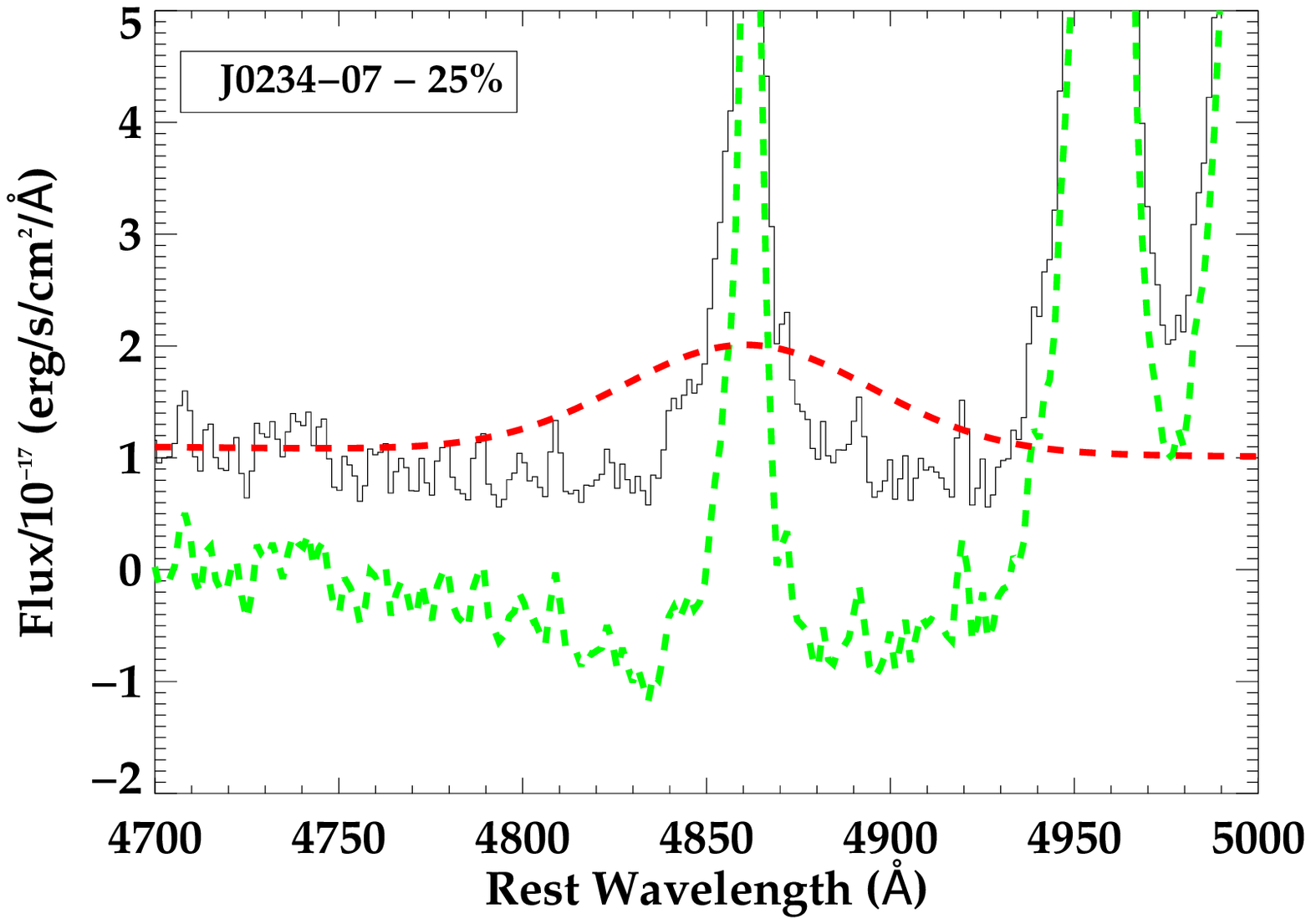}
\label{spec:q0234-ew}}
\subfloat{
\includegraphics[trim = 14mm 0mm 14mm 0mm, width=4.9cm]{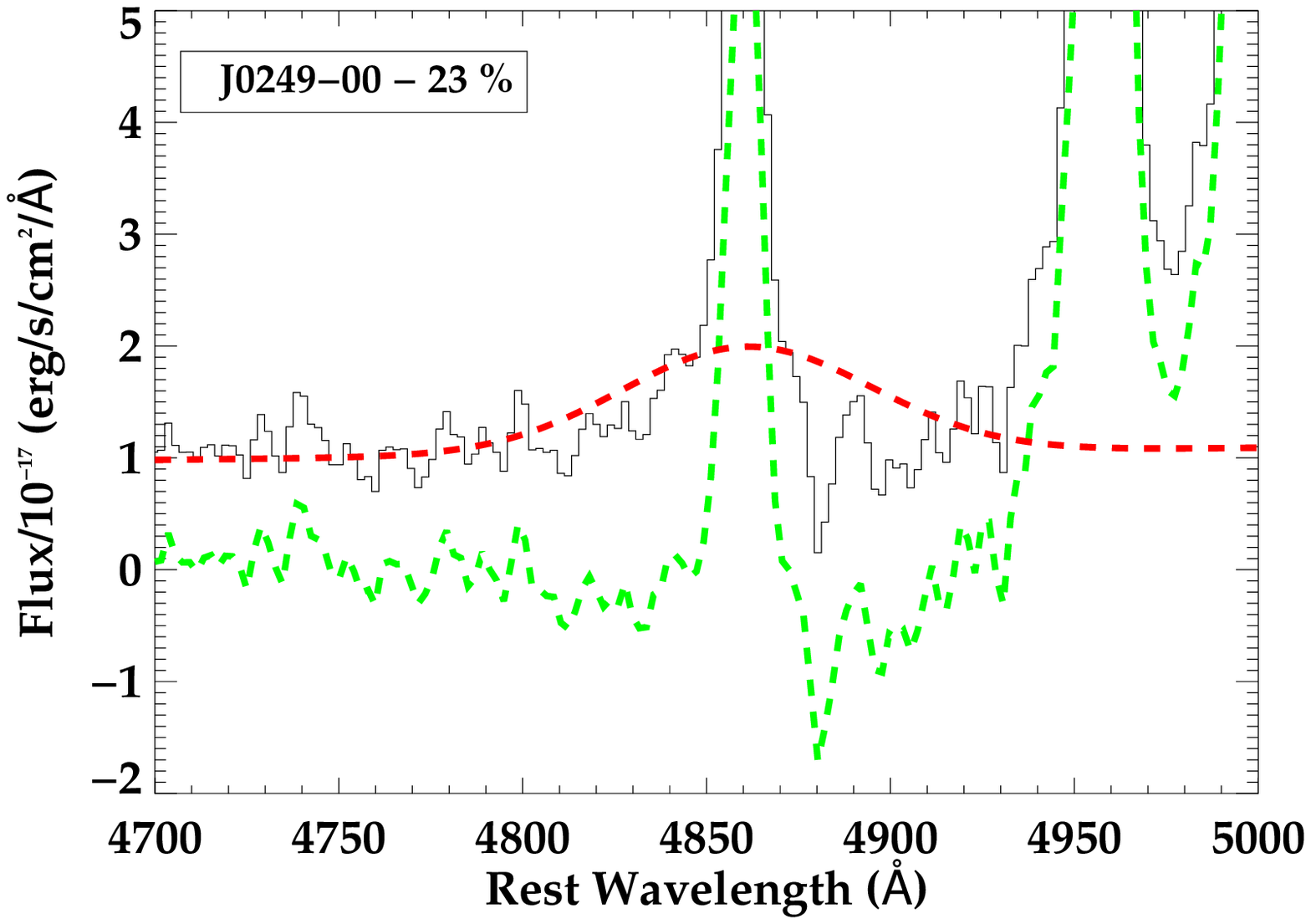}
\label{spec:q0249-ew}}

\subfloat{
\includegraphics[trim = 14mm 0mm 14mm 0mm, width=4.9cm]{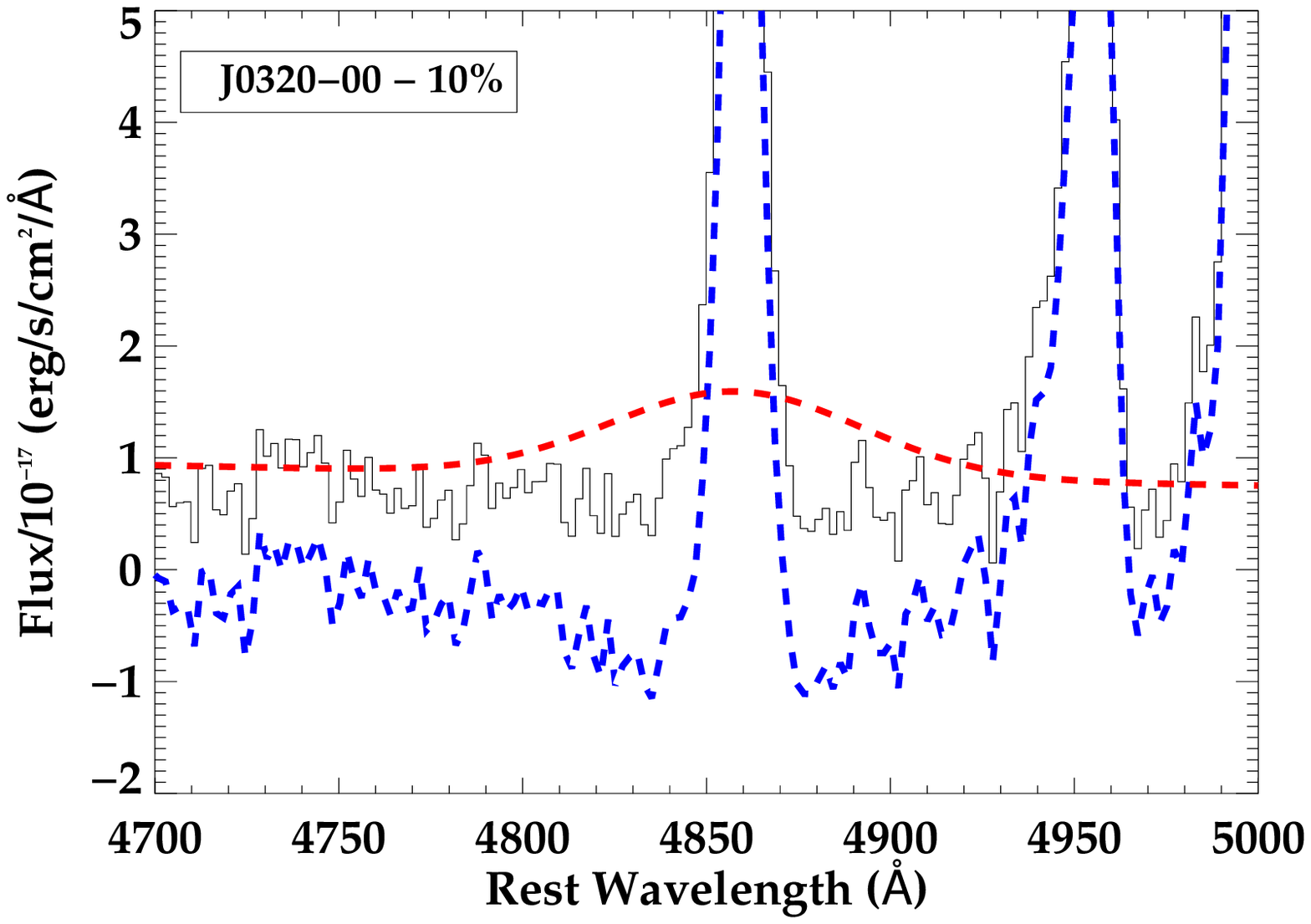}
\label{spec:q0320-ew}}
\subfloat{
\includegraphics[trim = 14mm 0mm 14mm 0mm, width=4.9cm]{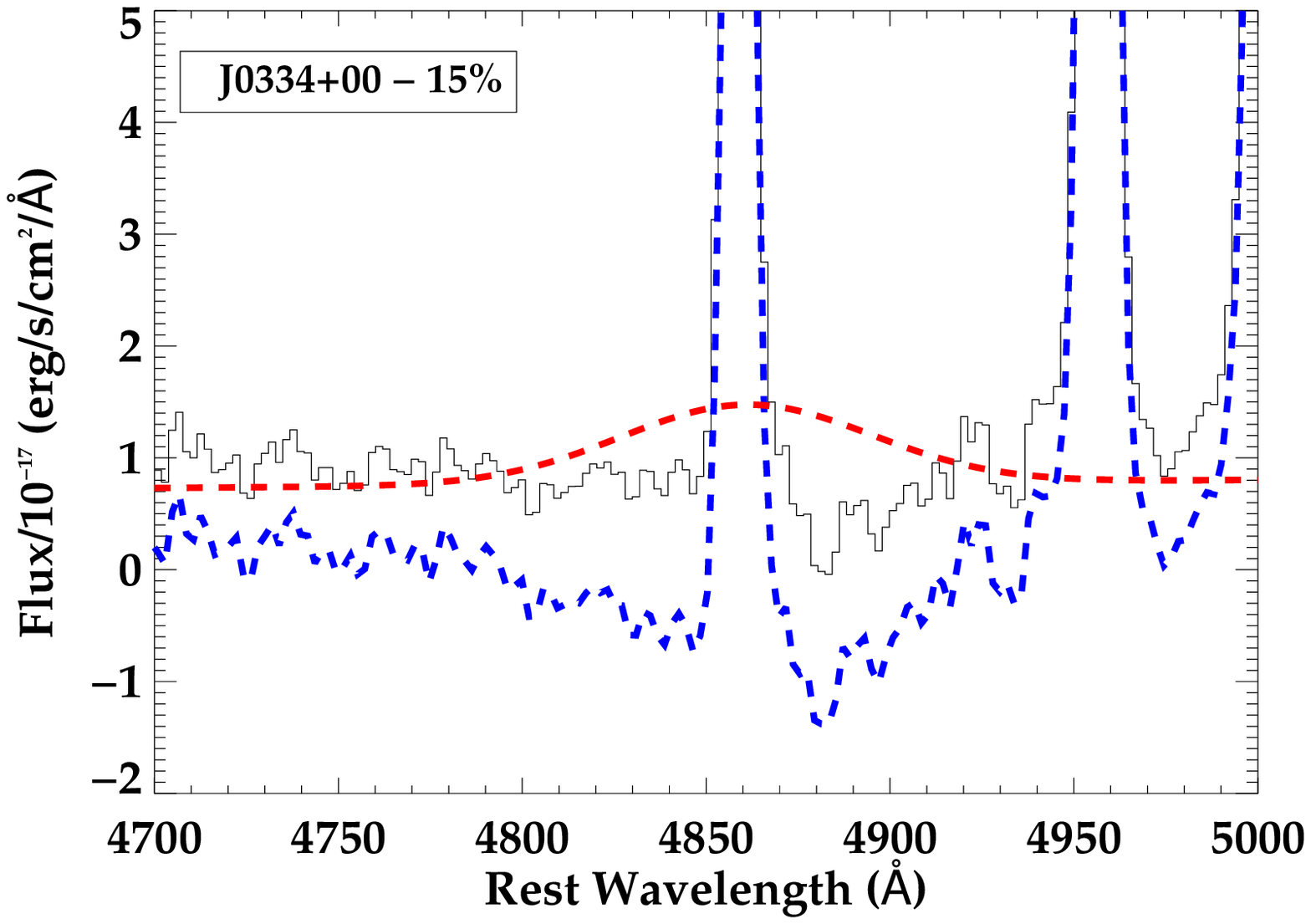}
\label{spec:q0334-ew}}
\subfloat{
\includegraphics[trim = 14mm 0mm 14mm 0mm, width=4.9cm]{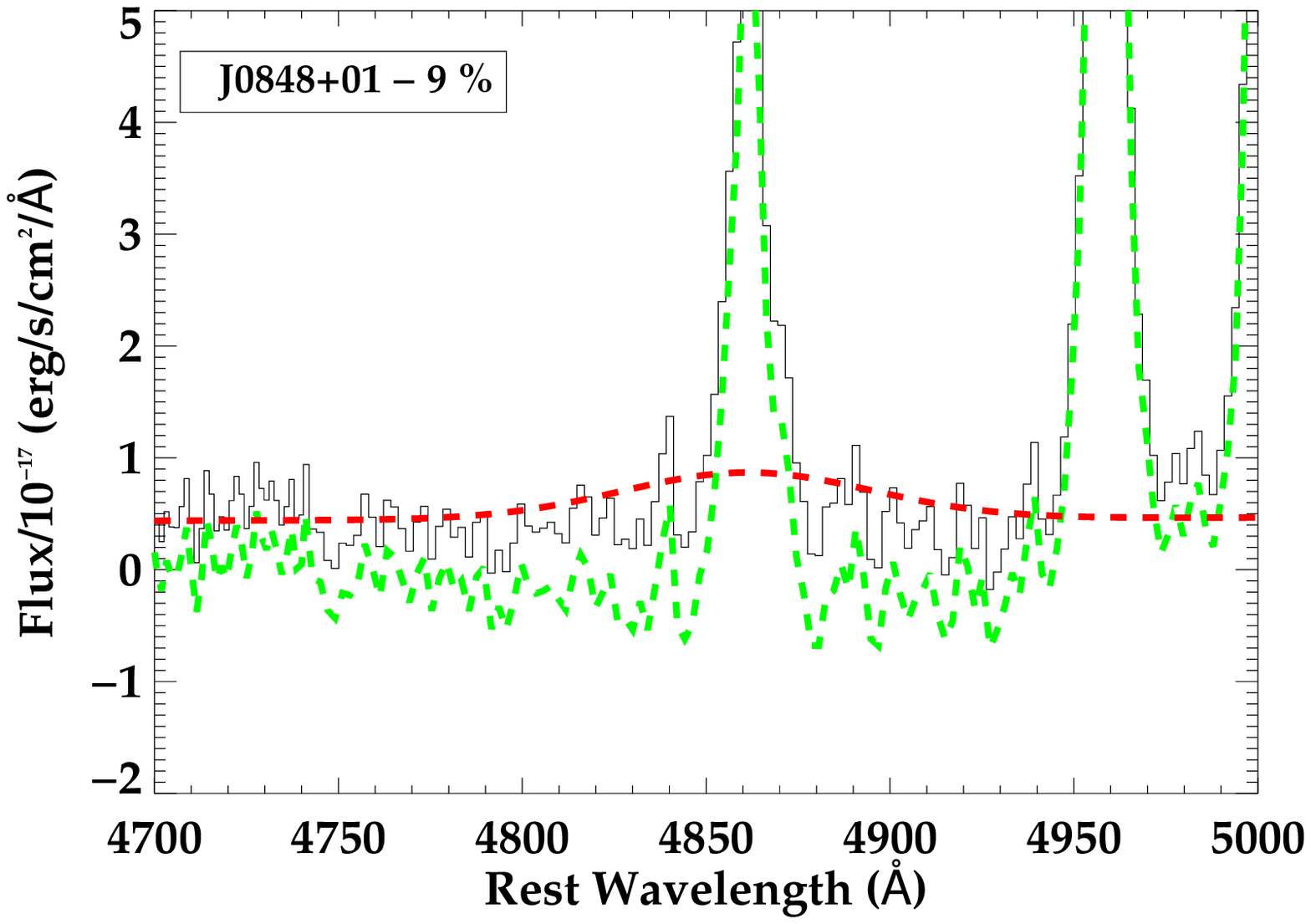}
\label{spec:q0848-ew}}

\subfloat{
\includegraphics[trim = 14mm 0mm 14mm 0mm, width=4.9cm]{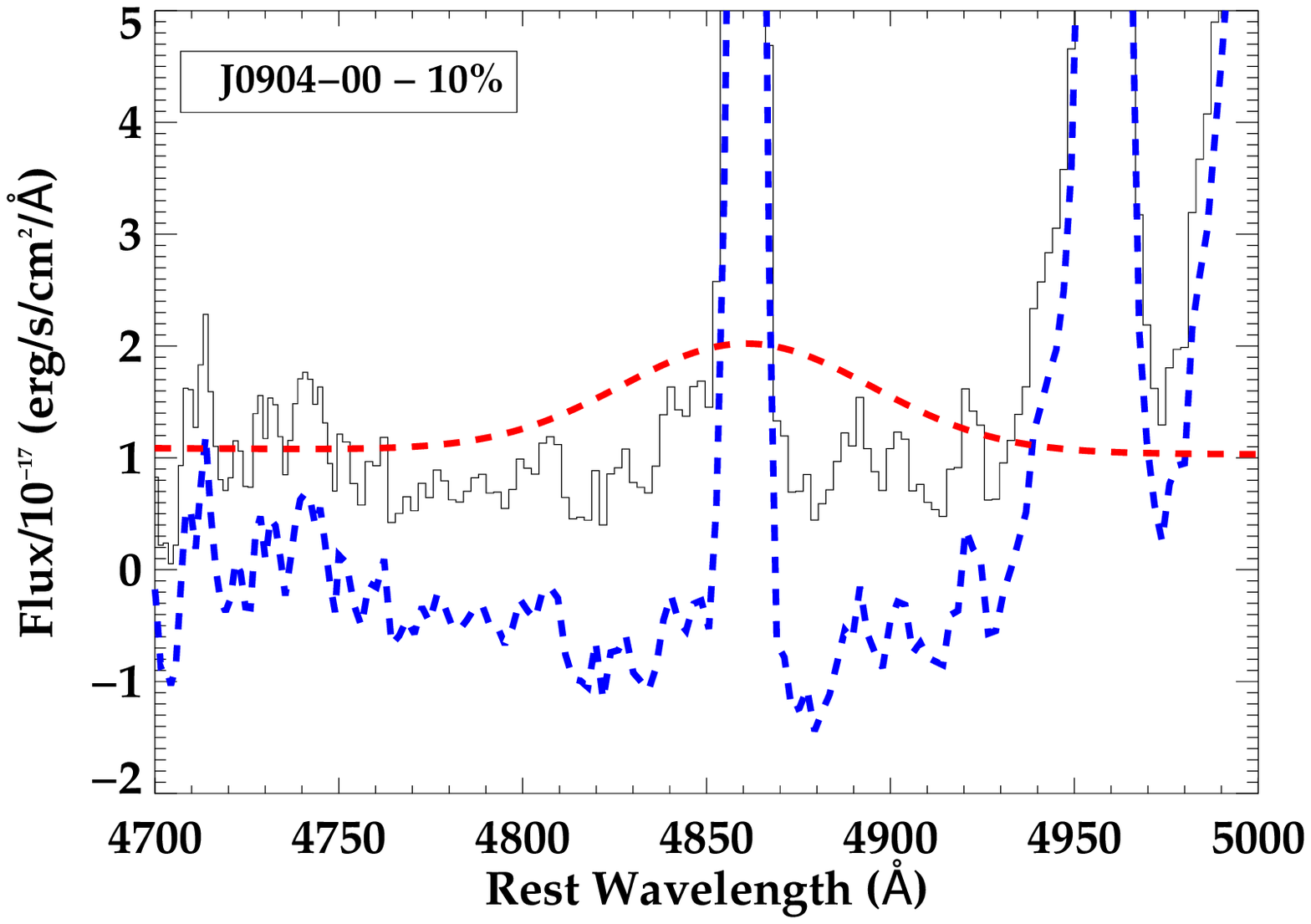}
\label{spec:q0904-ew}}
\subfloat{
\includegraphics[trim = 14mm 0mm 14mm 0mm, width=4.9cm]{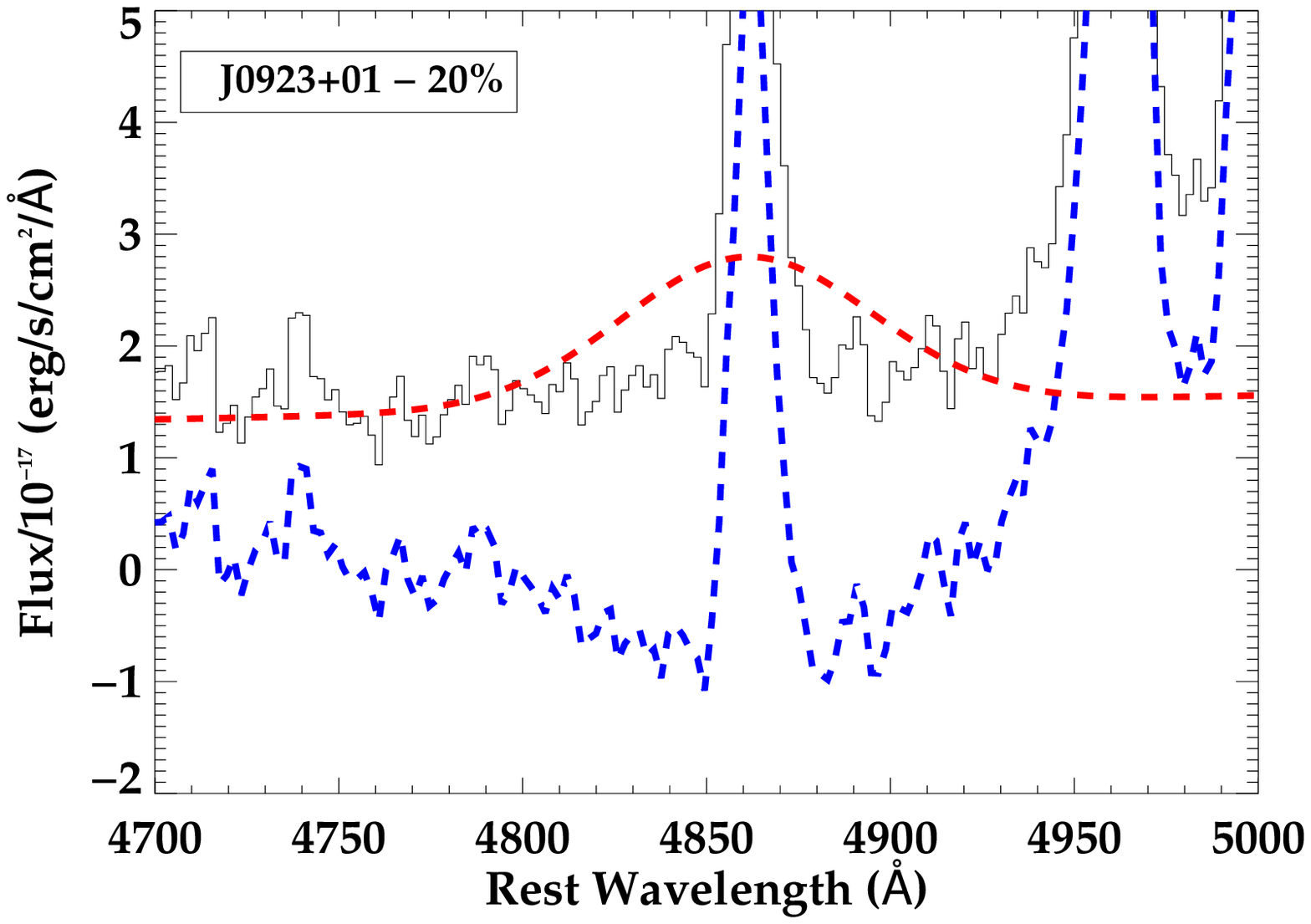}
\label{spec:q0923-ew}}
\subfloat{
\includegraphics[trim = 14mm 0mm 14mm 0mm, width=4.9cm]{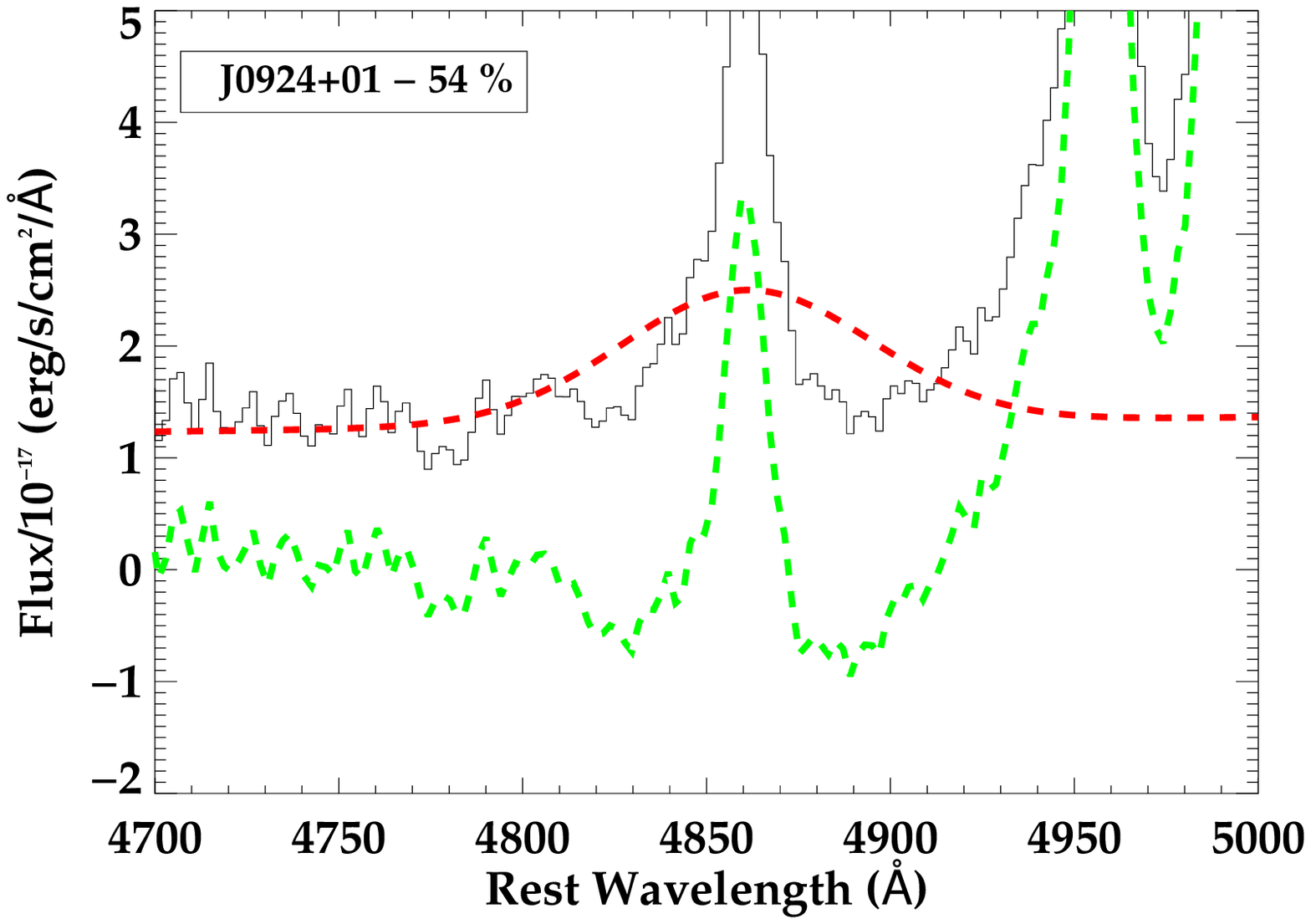}

\label{spec:q0924-ew}}
\caption{A zoom of the stellar subtracted spectrum in the region around the \hb\ line is shown for each object. The solid black line shows the stellar subtracted data, whilst the dashed red line shows the broad \hb\ plus power-law model which has been subtracted from these data. The blue or green dashed line shows the result of the subtraction (explanation in text). The percentage contribution of the power-law to the total flux in the normalising bin from which the broad \hb\ model has been constructed is shown in each panel.}
\label{broadCompPlot}
\end{figure*}
\begin{figure*}
\centering
\ContinuedFloat
\subfloat{
\includegraphics[trim = 14mm 0mm 14mm 0mm, width=4.9cm]{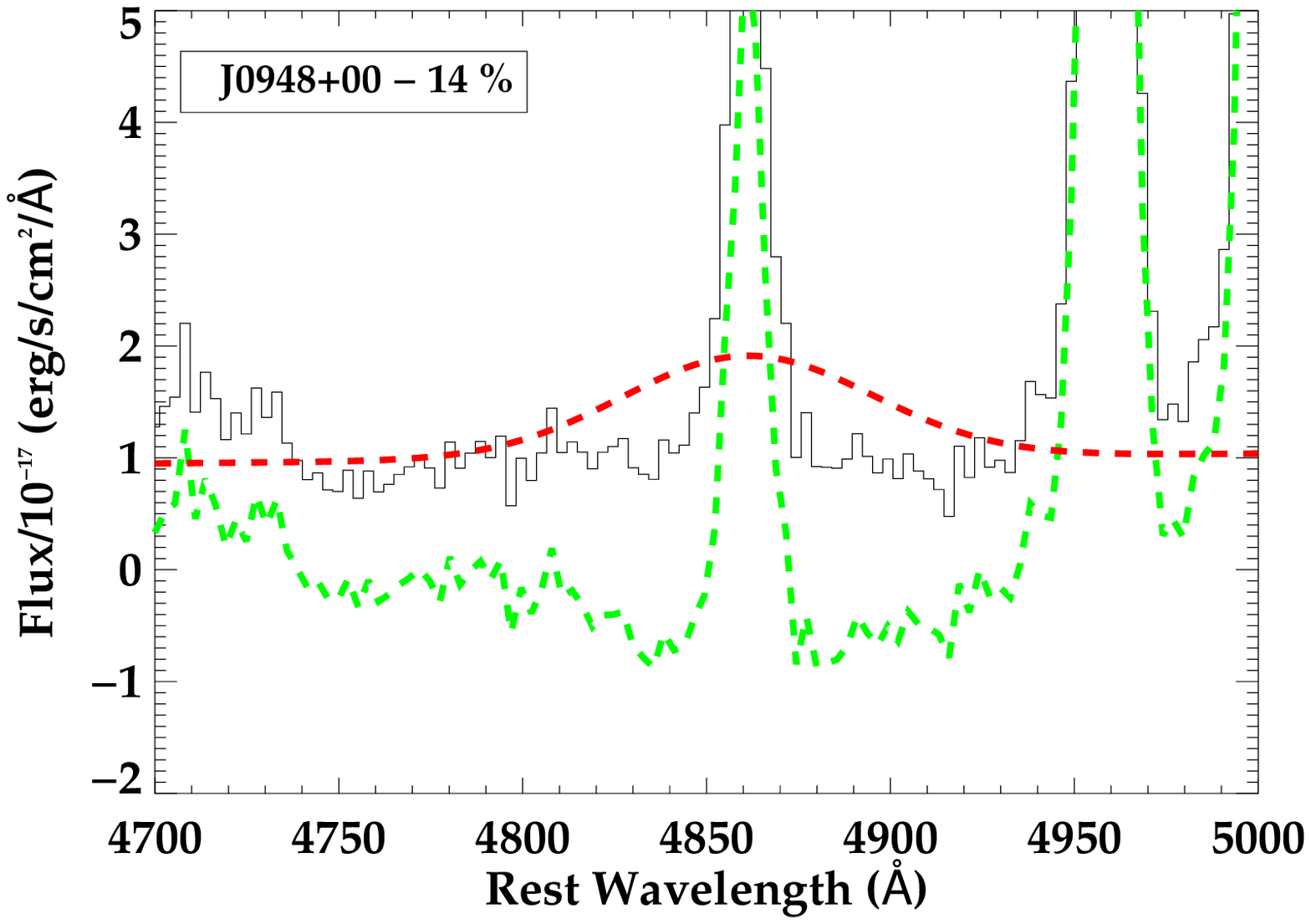}
\label{spec:q0948-ew}}
\subfloat{
\includegraphics[trim = 14mm 0mm 14mm 0mm, width=4.9cm]{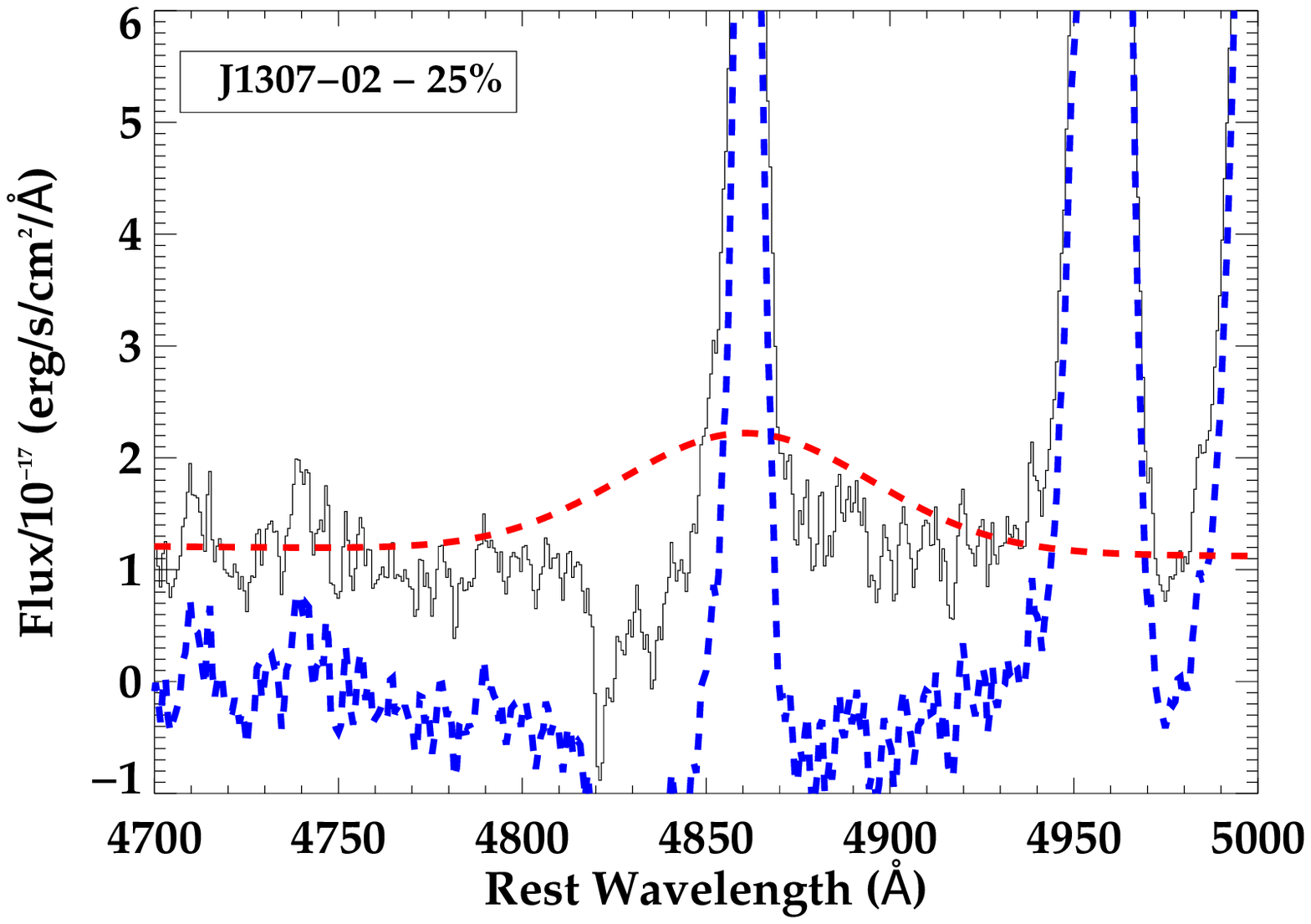}
\label{spec:q1307-ew}}
\subfloat{
\includegraphics[trim = 14mm 0mm 14mm 0mm, width=4.9cm]{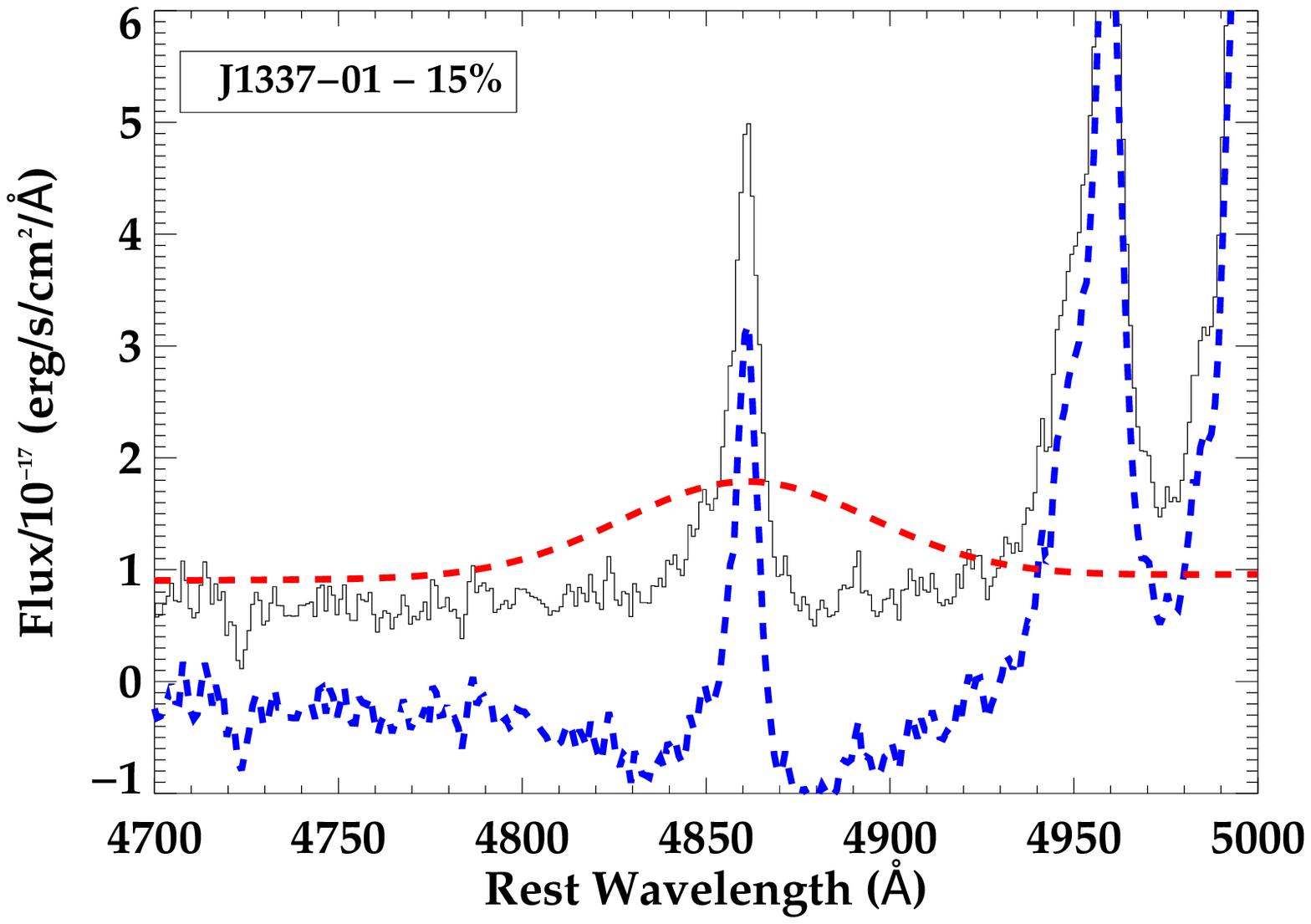}
\label{spec:q1337-ew}}

\contcaption{}
\end{figure*}

To recap, it was assumed that the percentage contributions and the power-law indices ($\alpha$) for every acceptable combination of components (using both combinations 4 and 5) are representative of scattered or directly transmitted quasar light. These contributions were then extrapolated from the normalising bin to the wavelength of \hb\ (using the corresponding value of $\alpha$), where Gaussians were constructed using average values of equivalent width and FWHM (78 \AA\ and 5000 \kms\ respectively\footnote{These values were derived from 'A Catalog of Quasar Properties from SDSS DR7'\citep{schneider10,shen11}, by selecting all objects that have $L_{OIII} > 10^{42}$ erg s$^{-1}$ and taking the median values of broad H$\beta$ equivalent width and FWHM.}) for the broad \hb\ line in quasars of similar luminosity. This produced models for the \hb\ component generated in the the broad line region, that would be present as a scattered component at a given level of power-law flux. As the power-law flux increases, this broad \hb\ component should, at some point, become detectable. In order to test this, broad \hb\ models were constructed assuming different levels of power-law contribution to the total flux, for each object independently (increasing in steps of 5 \% up the the maximum allowed by the modelling). Each level of power-law contribution is associated with a particular stellar model, consisting of a OSP and a YSP. For each level of power-law contribution, the stellar model corresponding to the power-law model was subtracted from the nebular subtracted data. Each Gaussian model was then combined with its power-law component, and then subtracted in turn from their corresponding stellar-subtracted spectrum in an attempt to quantify the point at which the over-subtraction of the continuum becomes clearly detectable on visual inspection.

Figure \ref{broadCompPlot} shows a zoom of the stellar-subtracted spectrum for each object. In each case, the solid black line shows the data with both the stellar model and the nebular continuum subtracted, the dashed red line denotes the `just detected' broad \hb\ model (comprising the Gaussian and power-law component) that was subtracted from the spectrum. In the cases in which the subtraction of the broad \hb\ model, associated with the maximum possible contribution to the total flux by the power-law component, does \emph{not} result in a visible over-subtraction of the continuum, Figure \ref{broadCompPlot} shows this maximum broad \hb\ case (green dashed line). In the cases where the over-subtraction of the continuum becomes clearly detectable \emph{below} the maximum value allowed by the modelling, a constraint is placed on the maximum contribution of the power-law to the total flux in the normalising bin. Where this is the case, a plot of the stellar-subtracted spectrum in which it has been determined that the subtraction of the broad \hb\ model has just become clearly visible is shown (blue dashed line).

Taking these additional constraints on the power-law contribution into account, the results of our stellar synthesis modelling are shown in Tables \ref{results8} and \ref{results2}. Table \ref{results8} shows the results for combinations including an 8 Gyr old underlying population, and Table \ref{results2} shows the results if we assume an underlying population of 2 Gyr.

It should be noted that the fitted power-law does not necessarily represent scattered AGN light, and may instead be associated with a very young stellar population (VYSP). However, it is not possible to clearly distinguish between the two possibilities on the basis of the data presented here because a VYSP ($< 10$ Myr) has only weak age associated features such as the Balmer break and stellar absorption features, so that such features will not contribute strongly to an OSP + VYSP composite. Therefore, to properly constrain the power-law component, not only in terms of its contribution to the observed flux and its spectral index, but also to distinguish between a scattered AGN component or VYSP, spectropolometric observations are essential. This information will allow more stringent constraints to be placed on the ages, reddenings and contributions to the total flux for any YSP/ISP component present in the host galaxies of these type II quasar. 

\begin{table*}
\begin{minipage}{145mm}
\centering
\caption{The results of our stellar synthesis modelling assuming an 8 Gyr underlying stellar population. Columns 1 and 2 give the object name and aperture ((N) nuclear (E) Extended (G) companion galaxy). Column 3 shows whether the higher order Balmer absorption lines were detected in the unsubtracted spectrum. Column 4 gives the nebular continuum contribution in the wavelength range 3540--3640 \AA, whilst column 5 indicates whether a power-law component was included in the model. Columns 6 and 7 show the range of percentage contributions to the total flux of both the OSP and YSP, whilst columns 7 and 8 show the range of ages and reddenings that are acceptable for the YSPs. Finally column 8 gives the range of percentage contribution by a power-law component (if included). }
\begin{tabular}{l c c c c  c c c c r}
\hline
Name		&	Aperture	&	Balmer	&	Nebular	&	PL?	&	OSP	&	YSP	&	YSP age	&	E(B-V)	&	Power-law	\\
			&				&	Lines?	&	(\%)	&		&	(\%)	&	(\%)	&	(Gyr)	&		&	(\%)	\\
\hline	
J2358-00	&	N	&	Y	&	28	&	N	&	0--19	&	88--95	&	0.04--0.05	&	0.5--0.6	&	--	\\
			&		&		&		&	Y	&	0--28	&	44--100	&	0.04--0.2	&	0.2--0.6	&	0--41	\\
																		
J0025-10	&	N	&	Y	&	3	&	N	&	19--37	&	63--79	&	0.02--0.04	&	0--0.2	&	--	\\
			&		&		&		&	Y	&	0--39	&	37--79	&	0.02--0.3	&	0--0.5	&	4--39	\\
																			
J0025-10	&	G	&	Y	&	10	&	N	&	--	&	--	&	--	&	--	&	--	\\
																			
J0025-10	&	E	&	Y	&	9	&	N	&	5-20	&	77-92	&	0.03-0.04	&	0-0.1	&	--	\\
																			
J0114+00	&	N	&	N	&	14	&	N	&	53--82	&	17--44	&	0.004--0.007	&	0.8--1.3	&	--	\\
			&		&		&		&	N	&	51--81	&	19--48	&	0.02--0.4	&	0.6--1.0	&	--	\\
			&		&		&		&	Y	&	50--82	&	17--47	&	0.004--0.007	&	0.9--1.3	&	0--9	\\
			&		&		&		&	Y	&	47--84	&	8--51	&	0.03--0.2	&	0.3--1.1	&	0--9	\\
																			
J0123+00	&	N	&	N	&	20	&	N	&	--	&	--	&	--	&	--	&	--	\\
			&		&		&		&	Y	&	52--81	&	3--26	&	0.001--0.01	&	0.0--0.2	&	12--40	\\
			&		&		&		&	Y	&	0--88	&	1--96	&	0.001--5	&	0.2--2.0	&	2--35	\\

J0142+14	&	N	&	Y	&	--	&	N	&	2--32	&	72--98	&	0.05--0.1	&	0.2--0.4	&	--	\\
			&		&		&		&	Y	&	0--20	&	52--100	&	0.04--0.4	&	0.1--0.5	&	0--48	\\
																			
J0217-00	&	N	&	Y	&	14	&	N	&	0--26	&	72--100	&	0.06--0.1	&	0.5--0.6	&	--	\\
			&		&		&		&	Y	&	0--37	&	55--100	&	0.06--0.3	&	0.1--0.6	&	0--15	\\
			&	GTC	&	Y	&		&	N	&	0--35	&	69--100	&	0.04--0.1	&	0.4--0.6	&	--	\\
			&		&		&		&	Y	&	0--35	&	56--99	&	0.04--0.2	&	0.2--0.6	&	0--15	\\

J0217-01	&	N	&	N	&	20	&	N	&	64--79	&	20--32	&	0.004--0.007	&	0.9--1.3	&	--	\\
			&		&		&		&	N	&	63--82	&	18--34	&	0.03--0.1	&	0.7--1.0	&	--	\\
			&		&		&		&	Y	&	61--82	&	13--32	&	0.004--0.007	&	0.9--1.3	&	0--11	\\
			&		&		&		&	Y	&	33--85	&	11--57	&	0.03-1.4	&	0.3--1.1	&	0--17	\\
																			
J0218-00	&	N	&	Y	&	17	&	N	&	13--39	&	60--87	&	0.03--0.09	&	0.0--0.3	&	--	\\
			&		&		&		&	Y	&	0--37	&	43--99	&	0.03--0.2	&	0.0--0.5	&	0--10	\\
																			
J0218-00	&	E	&		&	17	&	N	&	10--33	&	66--81	&	0.02--0.04	&	0.1--0.4	&	--	\\
																			
J0227+01	&	N	&	Y	&	11	&	N	&	15--52	&	46--82	&	0.03--0.1	&	0.1--0.5	&	--	\\
			&		&		&		&	Y	&	6--49	&	32--89	&	0.03--0.3	&	0.1--0.6	&	0--10	\\
																			
J0234-07	&	N	&	N	&	25	&	N	&	41--85	&	15--55	&	0.001--0.02	&	0--0.7	&	--	\\
			&		&		&		&	Y	&	0--82	&	0--90	&	0.001--5	&	0--2	&	0--33	\\
																			
J0249+00	&	N	&	N	&	25	&	N	&	12--35	&	65--88	&	0.05--0.1	&	0.7--0.8	&	--	\\
			&		&		&		&	Y	&	0--41	&	47--89	&	0.05--0.7	&	0.4--0.8	&	0--30	\\
																			
J0320+00	&	N	&	Y	&	8	&	N	&	37--47	&	50--61	&	0.005--0.006	&	0.1--0.3	&	--	\\
			&		&		&		&	N	&	33--46	&	52--64	&	0.02--0.03	&	0.0--0.2	&	--	\\
			&		&		&		&	Y	&	18--25	&	69--77	&	0.001	&	0.7	&	$<1$	\\
			&		&		&		&	Y	&	12--47	&	35--83	&	0.004--0.006	&	0.1--0.6	&	0--15	\\
			&		&		&		&	Y	&	6--48	&	19--87	&	0.02--0.2	&	0--0.5	&	0--15	\\
																			
J0332-00	&	N	&	N	&	4	&	Y	&	0--76	&	0--51	&	0.001--5	&	0--2	&	0--47	\\
																			
J0332-00	&	E	&		&	3	&	N	&	64--70	&	25--29	&	0.001--0.002	&	0.8	&	--	\\
			&		&		&		&	N	&	59--75	&	22--35	&	0.004--0.006	&	0.4--0.8	&	--	\\
			&		&		&		&	N	&	65--75	&	23--31	&	0.03--0.05	&	0.2--0.4	&	--	\\
			&		&		&		&	Y	&	57--76	&	11--36	&	0.004--0.007	&	0.3--0.8	&	0--18	\\
			&		&		&		&	Y	&	0--78	&	8--85	&	0.03--4	&	0--0.6	&	0--25	\\
																			
J0334+00	&	N	&	Y	&	23	&	N	&	6--27	&	74--97	&	0.04--0.09	&	0--0.2	&	--	\\
			&		&		&		&	Y	&	0--27	&	55--104	&	0.04--0.2	&	0--0.3	&	0--10	\\
																			
J0848+01	&	N	&	Y	&	19	&	N	&	53--66	&	32--43	&	0.2--0.4	&	0.1--0.3	&	--	\\
			&		&		&		&	Y	&	31--65	&	31--55	&	0.3--0.5	&	0.1--0.3	&	0--15	\\
																			
J0904-00	&	N	&	Y	&	13	&	N	&	12--32	&	72--91	&	0.006	&	0.3--04	&	--	\\
			&		&		&		&	N	&	11--31	&	73--92	&	0.03	&	0.2--0.3	&	--	\\
			&		&		&		&	Y	&	0--35	&	51--99	&	0.03--0.04	&	0.1--0.4	&	0--10	\\
																			
J0923+01	&	N	&	Y	&	14	&	N	&	0--19	&	80--100	&	0.05--0.1	&	0.4--0.5	&	--	\\
			&		&		&		&	Y	&	0--22	&	55--100	&	0.05--0.2	&	0.1--0.6	&	0--20	\\
\end{tabular}
\label{results8}
\end{minipage}
\end{table*}

\begin{table*}
\begin{minipage}{170mm}
\centering
\contcaption{}
\begin{tabular}{l c c c c c c c c r}
\hline
Name	&	Aperture	&	Balmer	&	Nebular	&	PL?	&	OSP	&	YSP	&	YSP age	&	E(B-V)	&	Power-law	\\
	&		&	Lines?	&	(\%)	&		&	(\%)	&	(\%)	&	(Gyr)	&		&	(\%)	\\
\hline
J0924+01	&	N	&	N	&	14	&	N	&	10--20	&	77--86	&	0.005	&	1	&	--	\\
			&		&		&		&	N	&	19--28	&	71--78	&	0.02	&	0.8	&	--	\\
			&		&		&		&	Y	&	6--44	&	27--88	&	0.004--0.007	&	0.4--1.1	&	0--44	\\
			&		&		&		&	Y	&	0--46	&	16--86	&	0.02--1.2	&	0--1.1	&	0--60	\\

J0948+00	&	N	&	Y	&	15	&	N	&	30--46	&	53--68	&	0.2--0.3	&	0--0.2	&	--	\\
	&		&		&		&	Y	&	0--47	&	51--94	&	0.2--0.4	&	0.1--0.5	&	0--20	\\

J1307-02	&	N	&	N	&	27	&	N	&	--	&	--	&	--	&	--	&	--	\\
	&		&		&		&	Y	&	36--58	&	24--46	&	0.007	&	0.2--0.3	&	8--25	\\
	&		&		&		&	Y	&	20--46	&	39--71	&	0.006--0.007	&	0.8--0.9	&	1--23	\\
	&		&		&		&	Y	&	35--61	&	21--51	&	0.03--0.1	&	0--0.4	&	4--25	\\
	&		&		&		&	Y	&	0--47	&	33--88	&	0.02--1.0	&	0.4--0.9	&	1--25	\\
																			
J1337-01	&	N	&	Y	&	8	&	N	&	0--30	&	74--100	&	0.04--0.1	&	0.3--0.5	&	--	\\
	&		&		&		&	Y	&	0--39	&	55-100	&	0.04--0.1	&	0.1--0.5	&	1--15	\\
\hline
\end{tabular}
\end{minipage}
\end{table*}

\begin{table*}
\begin{minipage}{150mm}
\caption{The results of our stellar synthesis modelling assuming an 2 Gyr underlying stellar population. Columns 1 and 2 give the object name and aperture ((N) nuclear (E) Extended (G) companion galaxy). Column 3 shows whether the higher order Balmer absorption lines were detected in the unsubtracted spectrum. Column 4 gives the nebular continuum contribution in the wavelength range 3540--3640 \AA, whilst column 5 indicates whether a power-law component was included in the model. Columns 6 and 7 show the range of percentage contributions to the total flux of both the ISP and YSP, whilst columns 7 and 8 show the range of ages and reddenings that are acceptable for the YSPs. Finally column 8 gives the range of percentage contribution by a power-law component.  }
\begin{tabular}{l | c c c c c c c c r}

\hline
Name	&	Aperture	&	Balmer	&	Nebular	&	PL?	&	ISP	&	YSP	&	YSP age	&	E(B-V)	&	Power-law	\\
	&		&	Lines?	&	(\%)	&		&	(\%)	&	(\%)	&	(Gyr)	&		&	(\%)	\\
\hline
J2358-00	&	N	&	Y	&	28	&	N	&	0--13	&	89--100	&	0.04	&	0.6	&	--	\\
	&		&		&		&	Y	&	20--91	&	54--104	&	0.04--0.2	&	0.4--0.7	&	0--29	\\

J0025-10	&	N	&	Y	&	3	&	N	&	--	&	--	&	--	&	--	&	--	\\
	&		&		&		&	Y	&	0--42	&	39--94	&	0.02--0.4	&	0--0.4	&	0--42	\\
																			
J0025-10	&	G	&	Y	&	10	&	N	&	--	&	--	&	--	&	--	&	--	\\
																			
J0025-10	&	E	&	Y	&	9	&	N	&	6--24	&	73--91	&	0.03--0.04	&	0--0.1	&	--	\\

J0114+00	&	N	&	N	&	14	&	N	&	0--71	&	30--98	&	0.5--2	&	0.2--1.0	&	--	\\
	&		&		&		&	Y	&	0--71	&	28--100	&	0.5--2	&	0.2--1.0	&	0--9	\\
																			
J0123+00	&	N	&	N	&	20	&	N	&	--	&	--	&	--	&	--	&	--	\\
	&		&		&		&	Y	&	--	&	--	&	--	&	--	&	--	\\
																			
J0142+14	&	N	&	Y	&	--	&	N	&	0--33	&	70--100	&	0.04--0.1	&	0.2--0.5	&	--	\\
	&		&		&		&	Y	&	0--42	&	62--104	&	0.04--0.1	&	0.2--0.5	&	0--45	\\
																			
J0217-00	&	N	&	Y	&	14	&	N	&	0--37	&	65--100	&	0.06--0.1	&	0.5--0.6	&	--	\\
			&		&		&		&	Y	&	0--40	&	52--100	&	0.06--0.3	&	0.2--0.6	&	0--15\\	
			&	GTC	&		&		&	N	&	0--38	&	65--99	&	0.04--0.1	&	0.5--0.6	&	--	\\
			&		&		&		&	Y	&	0--53	&	47--99	&	0.04--0.3	&	0.2--0.7	&	0--15	\\
																			
J0217-01	&	N	&	N	&	20	&	N	&	--	&	--	&	--	&	--	&	--	\\
	&		&		&		&	Y	&	--	&	--	&	--	&	--	&	--	\\
																			
J0218-00	&	N	&	Y	&	17	&	N	&	--	&	--	&	--	&	--	&	--	\\
			&		&		&		&	Y	&	0--41	&	45--99	&	0.04--0.2	&	0.1--0.4	&	0--10	\\
																			
J0218-00	&	E	&		&	17	&	N	&	24--41	&	56--74	&	0.004--0.006	&	0.3--0.5	&	--	\\
	&		&		&		&	N	&	15--44	&	56--81	&	0.02--0.04	&	0.1--0.4	&	--	\\
																			
J0227+01	&	N	&	Y	&	11	&	N	&	26--52	&	46--72	&	0.03--0.04	&	0.4--0.5	&	--	\\
			&		&		&		&	Y	&	12--71	&	25--80	&	0.03--0.2	&	0.0 -- 0.7	&	0--10	\\
																			
J0234-07	&	N	&	N	&	25	&	N	&	--	&	--	&	--	&	--	&	--	\\
	&		&		&		&	Y	&	68--88	&	6--11	&	0.001--0.003	&	0--0.1	&	3--20	\\
	&		&		&		&	Y	&	59--91	&	4--34	&	0.001--0.3	&	0.7--1.6	&	1--14	\\
	&		&		&		&	Y	&	0--83	&	9--83	&	1.2--2	&	0.2--0.5	&	6--20	\\
																			
J0249+00	&	N	&	N	&	25	&	N	&	19--59	&	44--81	&	0.05--0.2	&	0.6--0.9	&	--	\\
	&		&		&		&	Y	&	1--61	&	41--84	&	0.06--0.6	&	0.5--0.8	&	0--26	\\
																			
J0320+00	&	N	&	Y	&	8	&	N	&	--	&	--	&	--	&	--	&	--	\\
	&		&		&		&	Y	&	10--49	&	27--82	&	0.001--0.006	&	0.6--0.8	&	0--15	\\
	&		&		&		&	Y	&	16--55	&	25--76	&	0.009--0.1	&	0.3--0.6	&	0--15	\\

J0332-00	&	N	&	N	&	4	&	Y	&	46--71	&	19--28	&	0.001--0.003	&	0--0.1	&	3--27	\\
	&		&		&		&	Y	&	40--73	&	14--47	&	0.001--0.004	&	1--1.3	&	5--20	\\
	&		&		&		&	Y	&	26--73	&	13--63	&	0.008--0.1	&	0.5--0.9	&	4--22	\\
																			
J0332-00	&	E	&		&	3	&	N	&	--	&	--	&	--	&	--	&	--	\\
	&		&		&		&	Y	&	--	&	--	&	--	&	--	&	--	\\

J0334+00	&	N	&	Y	&	23	&	N	&	9--32	&	71--95	&	0.04--0.08	&	0--0.2	&	--	\\
	&		&		&		&	Y	&	0--39	&	49--100	&	0.04--0.2	&	0--0.4	&	0--10	\\
																			
J0848+01	&	N	&	Y	&	19	&	N	&	56--72	&	23--38	&	0.2--0.3	&	0.5--0.6	&	--	\\
	&		&		&		&	Y	&	56--79	&	21--42	&	0.3	&	0.5	&	0-8	\\
																			
J0904-00	&	N	&	Y	&	13	&	N	&	16--22	&	82--87	&	0.006	&	0.4	&	--	\\
	&		&		&		&	N	&	26--31	&	72--76	&	0.02	&	0.3	&	--	\\
	&		&		&		&	Y	&	0-42	&	49--98	&	0.02--0.04	&	0.2--0.4	&	0--10	\\
																			
J0923+01	&	N	&	Y	&	14	&	N	&	0--26	&	74--100	&	0.04--0.1	&	0.4--0.6	&	--	\\
	&		&		&		&	Y	&	0--33	&	48--100	&	0.05--0.2	&	0.1--0.6	&	0--20	\\
																			
J0924+01	&	N	&	N	&	14	&	N	&	36--51	&	46--59	&	0.009	&	0.9	&	--	\\
	&		&		&		&	Y	&	13--68	&	4--81	&	0.001--0.007	&	0--1.5	&	0--45	\\
	&		&		&		&	Y	&	0--67	&	5--88	&	0.009--1.2	&	0--1.4	&	0--60	\\
																			
J0948+00	&	N	&	Y	&	15	&	N	&	30--40	&	58--69	&	0.2	&	0.3	&	--	\\
	&		&		&		&	Y	&	0--57	&	41--94	&	0.2--0.4	&	0.1--0.4	&	0--19	\\
																			
J1307-02	&	N	&	N	&	27	&	N	&	37--54	&	47--62	&	0.008--0.01	&	0.6--0.7	&	--	\\
	&		&		&		&	Y	&	44--74	&	25--50	&	0.001--0.002	&	1.2--1.3	&	2--8	\\
	&		&		&		&	Y	&	29--76	&	19--66	&	0.004--0.007	&	0.9--1.2	&	2--23	\\
	&		&		&		&	Y	&	53--80	&	9--25	&	0.03--0.05	&	0--0.2	&	8--25	\\
	&		&		&		&	Y	&	0--79	&	14--88	&	0.009--1.2	&	0.4--1.2	&	1--25	\\
																			
J1337-01	&	N	&	Y	&	8	&	N	&	0--40	&	66--100	&	0.04--0.09	&	0.3--0.5	&	--	\\
	&		&		&		&	Y	&	0--47	&	46--100	&	0.04--0.2	&	0--0.5	&	0--15	\\

\hline
\end{tabular}
\label{results2}
\end{minipage}
\end{table*}

\subsection{The incidence of YSPs in type II quasars}
\label{sec:blines}

One of the most obvious features in an optical spectrum, which can be considered an unambiguous indicator of a strong YSP component, is the detection of the higher-order Balmer absorption lines. Taking into consideration the potential uncertainties with the subtraction of the higher-order Balmer emission lines, the proportion of type II quasar host galaxies (including J0025-10) that show Balmer lines in absorption \emph{before} performing the nebular subtraction was first determined. In this case, it was found that the Balmer absorption lines are detected in 13/21 spectra ($\sim 62\%$). Following subtraction of the higher order Balmer lines (in emission) and nebular continuum, this proportion increased to 14/21 objects ($\sim67 \%$). 

In order to constrain the proportion of type II quasars that contain YSPs more accurately, it is also instructive to consider what percentage \emph{require} a YSP component in order to produce an acceptable fit to their SEDs, that is, the proportion that cannot be fitted with a power-law + 8 Gyr model (combination 1) alone (based on our \chisqlt\ criterion). In this sample, only J0234-07 and the type I object J0332-00 can be adequately fit using \emph{only} an 8 Gyr OSP combined with a power-law component. In all of the remaining cases, this combination of components failed to produce a reasonable fit to the shape of the spectrum around the Balmer break. Therefore, 19/21 (90\%) objects require a YSP in order to fit their overall spectral shape, thus allowing the unambiguous statement that YSPs are present in at least 90\% of this sample of type II quasars.

\subsection{The ages of the YSPs in type II quasar host galaxies}
\label{results:yspAges}

The results presented in Tables \ref{results8} and \ref{results2} show that the assumptions made in the different modelling combinations can have a profound effect on the results produced. Combinations 4 and 5, both of which include a YSP/ISP as well as a power-law component, are successful in producing acceptable fits. Combination 4 ($t_{OSP}= 8~\mbox{Gyr}$) was successful in 23/23 (100 \%) of the apertures to which is was applied, whilst combination 5 ($t_{ISP} = 2~ \mbox{Gyr}$) produced successful fits for 20/23 of the apertures (87 \%). However, given the increased number of free parameters involved, these modelling combinations do not provide any strong constraints on either the ages or the reddenings of the YSP/ISP in the majority of cases (as discussed in Section \ref{spec:plComp}).

\begin{figure}

\subfloat{ 
\includegraphics[scale=0.47, trim = 0mm 2mm 0mm 2mm]{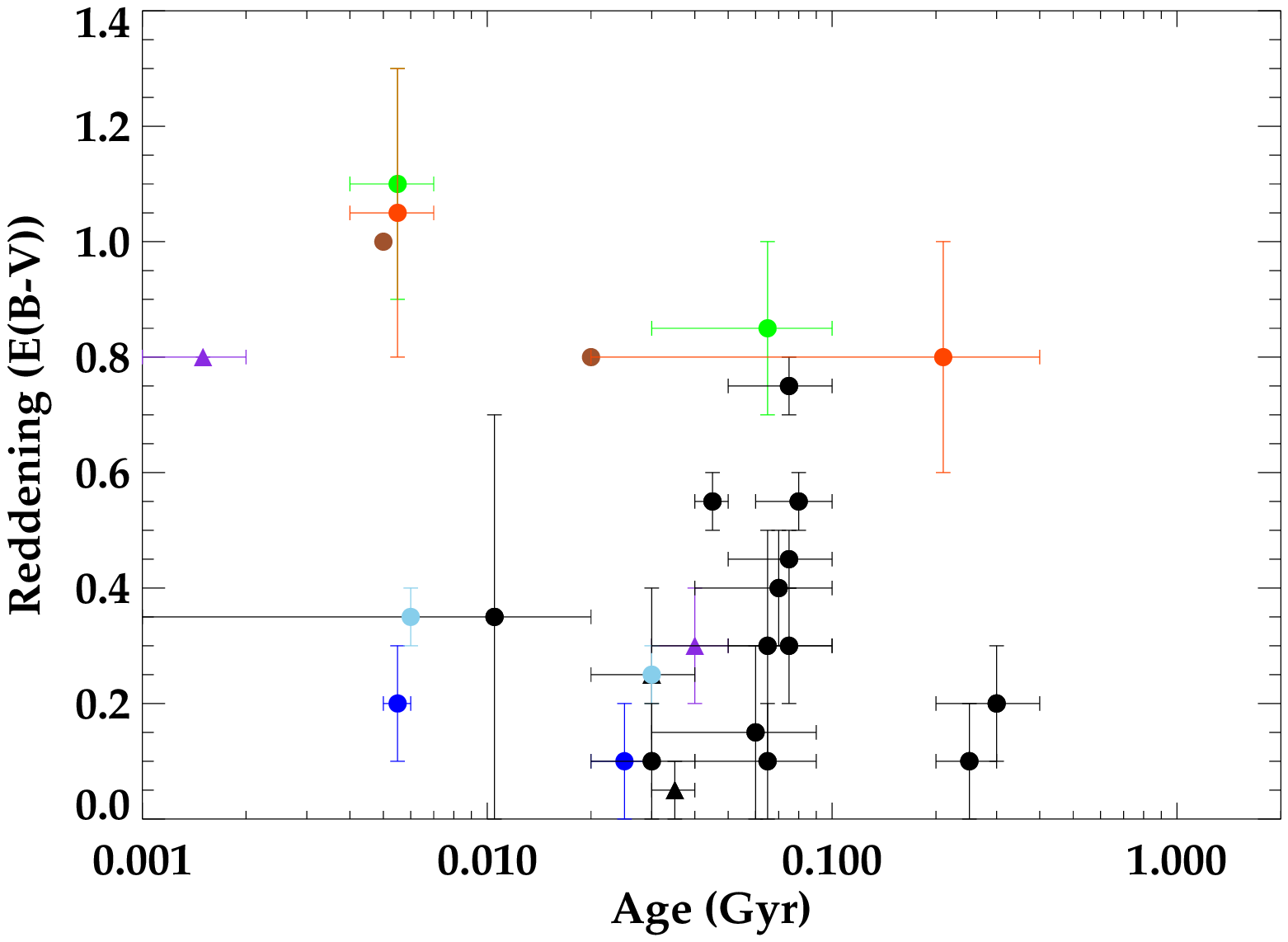} 
\label{8np-b-points}} 

\subfloat{ 
\includegraphics[scale=0.47, trim = 0mm 2mm 0mm 2mm]{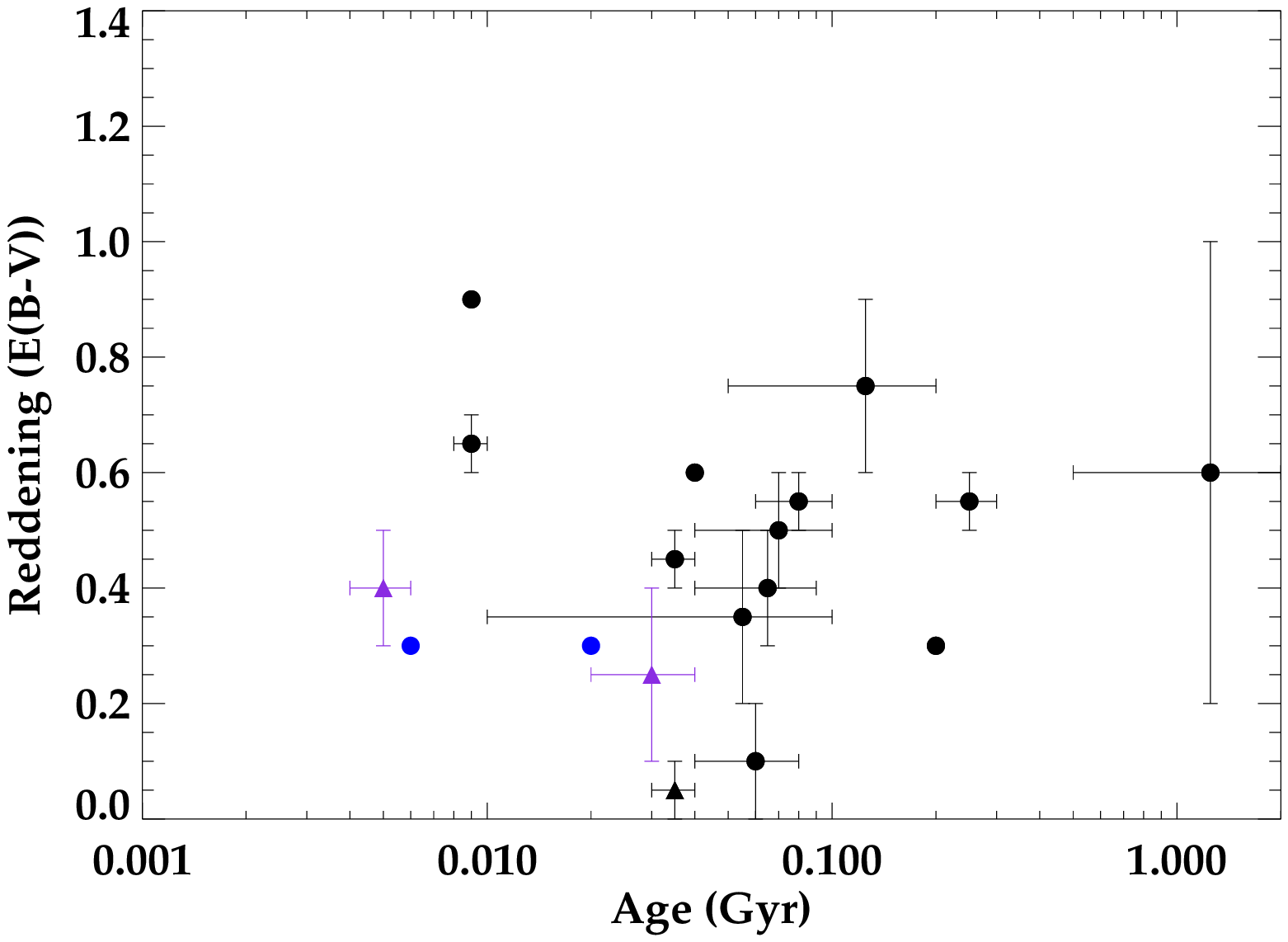} 
\label{2np-b-points}} 

\caption{The mean age and reddening found for each of the spectroscopic apertures. The top panel shows the results for combination 2 and the bottom panel shows the results for combination 3. In both panels, the mean reddening of the YSP is plotted against the mean age found for each aperture. In the cases where, for an individual object, the modelling produced more than one region of acceptable fits with $\chi^{2}_{red} < 1$, both regions are plotted in the same colour (i.e. they refer to the same object), otherwise, the result for each object is plotted in black. The results related to apertures extract from the nuclear regions are plotted as circles, whilst those extract from off nuclear regions a plotted as triangles. The bars show the range of values of both age and reddening that provide acceptable fits.} 
\label{age_red_plot} 

\end{figure}

In light of the above considerations, the remainder of this work will focus on the models which do not include a power-law component. This is justifiable because it is these combinations which give the best constraints on the YSP age, and also allow for direct comparison with other samples of AGN which have taken a similar modelling approach\footnote{However, it must be noted that the conclusions drawn would be significantly altered if a power-law component proved to contribute substantially to the total flux.}. Figure \ref{age_red_plot} shows age in Gyr plotted against E(B-V) for combinations 2 and 3.  For each object, if the modelling for a given combination produced only one region in which $\chi^{2}_{red} < 1$, that value is plotted in black. In the cases where the modelling for a given combination produced two regions of interest, the values for each region are plotted in the same colour. The apertures that were extracted from companion objects or extended regions, rather than the quasar host galaxy nuclei are plotted as triangles. For each point, the bars show the extent of the acceptable ranges of age and reddening, whilst the points are plotted at the median. 

The top panel of Figure \ref{age_red_plot} summarises the results for combination 2, which produced successful results for 22/25 ($\sim 88\%$) of the apertures to which it was applied. However it must be noted that J0217-00 is duplicated in these results, but because it can be clearly seen from Tables \ref{results8} and \ref{results2} that the results from both data sets are consistent with each other, to avoid confusion we will only consider one set of results (the Gemini GMOS-S results) in further discussion. Strikingly, of the 24 unique apertures that are adequately fit by this combination, 18 \emph{require} the inclusion of a YSP component with $t_{max} \le 100~ \mbox{Myr}$, while the majority of acceptable fits for one object (J0114+00) also include a YSP with $t_{YSP} \le 100$ Myr. It is only in two cases (J0848+01 and J0948+00) that a YSP $t_{ysp} > 200$ Myr is \emph{required} in order to achieve acceptable results.

Considering only the unique objects in the sample, the top panel of Figure \ref{age_red_plot} shows that a wide range of reddenings is found, although there does not appear to be any correlation between the YSP age and the reddening associated with the population. If those objects that have stellar populations with average reddenings $E(B-V)> 0.4$ are considered to be strongly reddened, then 8/21 (38\%) of the apertures fall into this category, whilst 13/21 (62\%) are not strongly reddened, of which 38\% have mean $E(B-V) \le 0.1$. 

The bottom panel of Figure \ref{age_red_plot} summarises the results for combinations 3, which produced successful fits for 18/25 ($\sim 72\%$) of the apertures to which it was applied. Again, when considering only unique objects, comparing the two panels of Figure \ref{age_red_plot} shows that, when an underlying population of $t_{ISP}= 2~ \mbox{Gyr}$ is assumed, a higher proportion of the objects for which fits are achieved have $t_{max}>100~ \mbox{Myr}$ (4/17) than in the case where an $t_{OSP}$ = 8 Gyr (3/21) is assumed (24 \% vs. 14 \%), although this difference is marginal. Once again, if YSPs with $E(B-V) > 0.4$ are categorized as significantly reddened, then 9/17 (53\%) of the apertures modelled have a significant average reddening. This proportion is higher than that found for combination 2, which is not unexpected because a significantly reddened YSP can mimic an unreddened OSP in the overall shape of its spectrum. However, it must be noted that for both combination 2 and 3, the majority of the objects \emph{require} the inclusion of a YSP $t_{YSP} < 100~\mbox{Myr}$ in order to adequately fit their spectra.

\section{How do type II quasars compare with other active systems?}
\label{spec:discussion}

\subsection{Comparison with powerful radio galaxies}
\label{spec:qso-pgr}

When considering the properties of YSPs in the host galaxies of type II quasars, it is important to consider them in the wider context of active and merging systems. Are the ages found here typical of those found in other studies?

In \citet{bessiere12}, the 2 Jy sample\footnote{Full details of the 2Jy sample and available data can be found at http://2jy.extragalactic.info/Home.html} \citep{tadhunter93, tadhunter98} was introduced, in order to compare the optical morphologies of powerful radio galaxies (PRG)  with the type II quasars presented here. To recap the selection criteria, the 2Jy objects have radio fluxes $S_{2.7GHz} > 2.0$ Jy, radio spectral indices $\alpha_{2.7}^{4.8} >0.5 ~(F_{\nu} \propto \nu^{-\alpha})$, declinations $\delta < +10^{\circ}$ and redshifts $0.05<z<0.7$. For further details on the criteria used for selection and completeness, refer to \citet{tadhunter93,tadhunter98}. The same stellar synthesis modelling technique has been applied to this sample of PRG facilitating a direct comparison of the incidence of YSPs in two samples.

\cite{tadhunter02} present results for an RA limited sub-sample of that discussed in \citet{bessiere12}, comprising 22 PRG ($0.15 < z < 0.7$). Their modelling showed that only $\sim 15 \%$ of these PRG harbour a clearly detectable, significant YSP. These findings were based on both stellar synthesis modelling and the clear detection of Balmer absorption features in the spectra. The remaining 19 objects were adequately fit by an elliptical galaxy template combined with a power-law component (i.e. combination 1) contributing between 20 \% and 90 \% of the continuum flux just below the 4000 \AA\ break. 

These results have been further updated \citep{holt06,holt07,wills08} with the availability of deeper data with improved S/N in the continuum, higher spectral resolution and, in some cases, wider spectral coverage, combined with the use of higher resolution stellar population models \citep{bruzual03}. This allowed firm detections of YSPs in a further two objects from the original sample, while improving the age estimates for three sources presented in \citet{tadhunter02}. This brings the detection rate of YSPs in this sub-set of the 2 Jy sample to 5/22 (23\%), with 4/5 of these having $\tmax \le 100$ Myr. Therefore, the ages of the YSPs, when detected, are consistent with those found for the type II quasars (Figure \ref{spec:compSamples}), although the detection rate is significantly lower. This detection rate is in stark contrast to the results presented here for the type II quasars, where only two object (J0234-07 and J0332-00) can be adequately fit with an OSP + power-law combination of components alone.

Taking into consideration the results of the comparison of optical morphologies and photometric properties presented in \citet{bessiere12}, this result could be considered somewhat unexpected. This is because the absolute magnitudes, rate of interactions, distribution of host galaxies amongst the morphological groups, and the distribution of the surface brightnesses of the tidal features associated with the mergers were shown to be similar for the two samples, suggesting that both types of AGN reside in similar host galaxies. In contrast, the significant differences in the stellar content of their host galaxies suggests that they are fundamentally different systems.

Although PRG are almost invariably massive ellipticals, the case is not quite so clear cut for radio-quiet quasars. If they are more likely to have a significant disc component than their radio-loud counterparts \citep{floyd04}, this could mean that we would naturally expect star formation to be more prevalent. However, the deep $r^{\prime}$-band imaging presented in \citet{bessiere12} strongly suggests that the majority of these objects are not disc dominated, a finding supported by \citet{wylezalek16}, who also find that the majority of their sample (90\%) of 20 type II quasars (including J0123+00 and J2358-00) are bulge dominated. This suggests that differences in galaxy morphology are unlikely to be the prime origin of the differences in stellar content. \citet{wylezalek16} also derive stellar masses for their sample and find an average of $\mbox{log}(M_{*}/M_{\sun}) = 10.7 \pm 0.3$, consistent with those found for our sample of type II quasars ($\sim 10^{11}$; \citealt{mythesis}), and below that found for PRG with similar stellar populations. This may imply that stellar mass is the more fundamental difference between PRG and radio-quiet type II quasars. The comparison between the stellar masses of the two types of objects will be discussed in more detail in a forthcoming paper.

However, it is important to recognise that the quasar luminosity in a number of these PRG is significantly higher than those of many of the type II quasars presented here. The highest [OIII] luminosity of a type II quasar in our sample is $\mbox{log}(L_{[OIII]}/L_{\sun}) = 9.32$, whilst 6 of the 22 PRG have a higher $L_{[OIII]}$.  This will have the effect of making it more difficult to detect signatures of YSPs in the host galaxies because of the stronger AGN-related continuum components, such as scattered quasar light and nebular continuum. It will also be more difficult to detect the Balmer lines in absorption because there will be a higher degree of contamination by Balmer emission lines. Therefore, the detection rate of 23\% must be taken as a lower limit to the proportion of PRG which host YSPs.

\subsection{Comparison with ULIRGs}
\label{spec:qs-ulirg}
In the context of evolutionary scenarios, such as that proposed by \citet{sanders88}, it is worthwhile to compare the results derived here with those found for a complete sample of ULIRGs presented by \citet{rodriguez09,rodriguez10}. This sample consists of 36 ULIRGs, of which 14 have been classified as Seyferts. The full sample includes a complete RA and Dec limited sample of 26 objects with redshifts $z<0.13$, and in addition, a further 10 objects ($z <0.18$) were included because they are characterised by warm infrared colours. Again, the same stellar synthesis modelling technique as that employed here was used.

\citet{rodriguez09} extracted 5 kpc nuclear apertures for all the objects in their sample, as well as apertures from extended tidal features. Due to the fact that in all but three cases, only one nuclear aperture was extracted from each type II quasar, we will only consider apertures centred on the nucleus, thus allowing a more meaningful comparison.  

Their modelling also utilised a combination of an OSP and YSP, although they use an OSP with \tosp\ = 12.5 Gyr. This use of an older underlying population is unlikely to have a significant effect on the YSP ages because the differences between the stellar templates at large ages is minimal.

The results presented for the ULIRGs in \citet{rodriguez10} are very similar in terms of the YSP ages found for the nuclear apertures, the reddenings associated with the YSPs, and the contribution made by the YSP to the total flux in the normalising bin. They find that 91\% of the nuclear apertures have $\tysp < 100 ~\mbox{Myr}$. In comparison, 89\%\footnote{This proportion excludes J0123+00 and J0332-00, which cannot be modelled without the inclusion of a power-law component.} of type II quasar host galaxies which \emph{require} a YSP component, also have  $\tysp < 100 ~\mbox{Myr}$ for the same modelling combination. Figure \ref{spec:compSamples} shows that the distributions of ages of the type II quasars and ULIRGs are very similar, with similar detection rates. In terms of the reddenings found for the YSPs, \citet{rodriguez10} derived a median value of \ebv = 0.4 for the apertures extracted from the nuclear regions of the ULIRGs. This is the same as the median value found for this sample of type II quasars, with \ebv = 0.4.

\section{Timing the quasar activity in galaxy mergers.}
\label{spec:timing}

Based on the above results, it is clear that in the majority of cases when no power-law component is included, a YSP with $\tmax < 100 ~ \mbox{Myr}$ is a significant flux component of the quasar host galaxies studied here : for combination 2 $\sim 89 \%$  have a significant YSP with $\tmax \le 100~ \mbox{Myr}$, while $\sim 32 \%$  have a significant YSP with $\tmax \le 50~ \mbox{Myr}$

It is interesting to consider these results in the context of an evolutionary sequence such as that of \citet{sanders88}. In such a sequence, the merger will initiate a substantial burst of star formation, leading to the system being classified as a (U)LIRG, which will then evolve into an observable quasar. These kinds of evolutionary models (e.g. \citealt{dimatteo05,springel05,hopkins06}) predict that, directly following the merger, we will observe a system with a YSP that has been heavily reddened by the enshrouding dust. At this stage of the sequence, the quasar activity will be too heavily obscured to be detectable in the optical. As the merger progresses, feedback processes (usually attributed to the AGN activity) will shut down or quench star formation activity, while at the same time removing the material enshrouding the quasar. The observable result would be a single nucleus and an ageing (post) starburst population. If this scenario is correct, then we should observe a delay between the merger driven starburst and the visible quasar activity.

Using various stellar population modelling techniques, some previous studies of AGN host galaxies have found evidence of just such a delay, ranging from a few 100 Myr to a few Gyr \citep{fernandes04,tadhunter05,emonts06,davies07,holt07,wills08,wild10,tadhunter11,canalizo13,ramos13}. This is also true of two of the type II quasar objects presented here, in which delays of up to a few 100 Myr  (assuming combination 2) are found. However, other studies have shown that no such delay exists and that the quasar and starburst activity are triggered quasi-simultaneously \citep{heckman97,canalizo00,canalizo01,brotherton02,holt07,wills08,liu09,tadhunter11, villar12, bessiere14}.

\begin{figure}
\centering
\subfloat{
\includegraphics[trim = 10mm 0mm 10mm 20mm,width=8cm,height = 4.7cm]{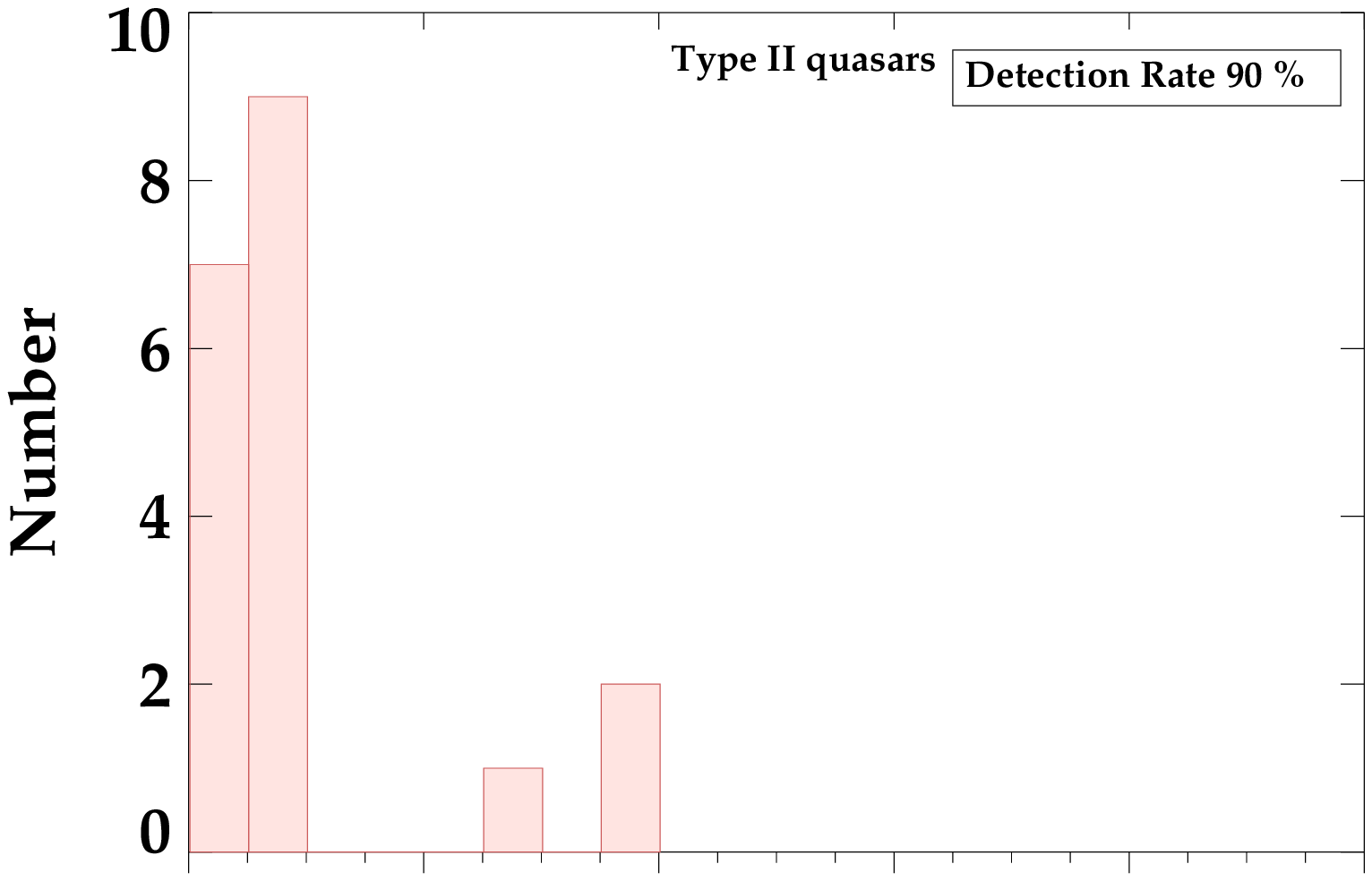}
\label{typeIIhist}}

\subfloat{
\includegraphics[trim =  10mm 0mm 10mm 30mm,width=8cm,height = 4.5cm]{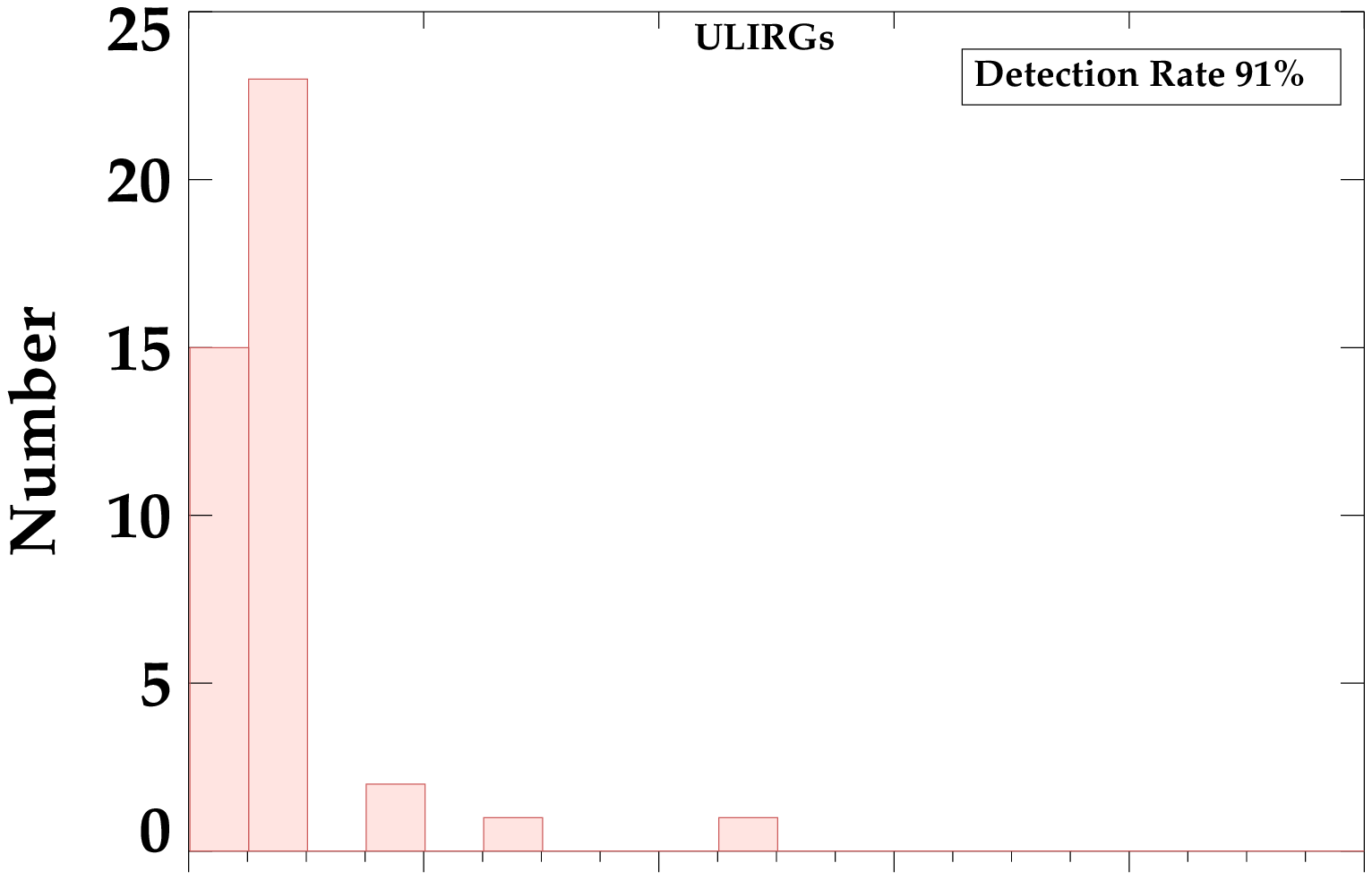}
\label{ulirghist}}

\subfloat{
\includegraphics[trim =  10mm 0mm 10mm 30mm,width=8cm,height = 4.5cm]{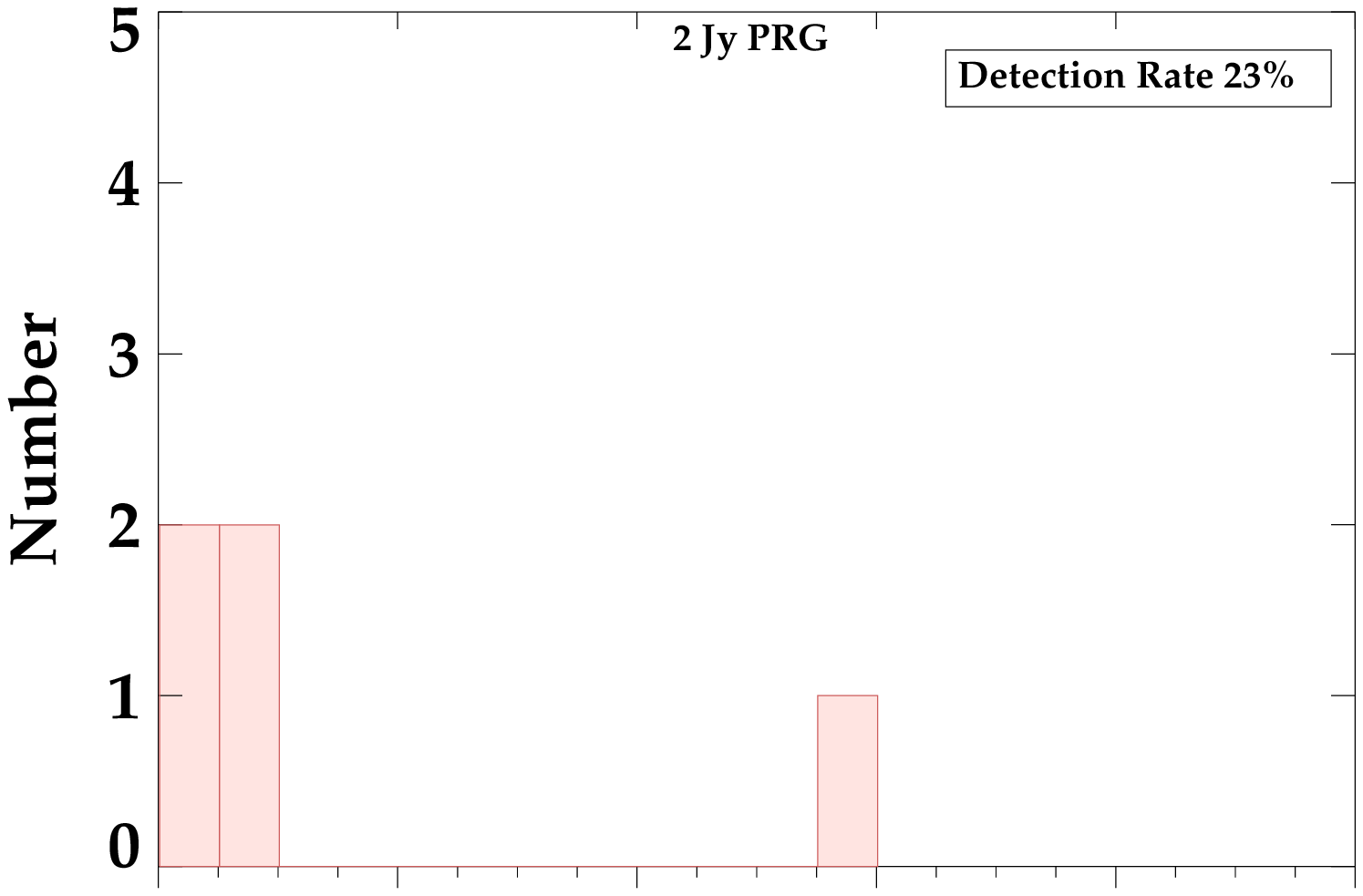}
\label{2jyhist}}

\subfloat{
\includegraphics[trim =  10mm 0mm 10mm 30mm,width=8cm, height = 4.5cm]{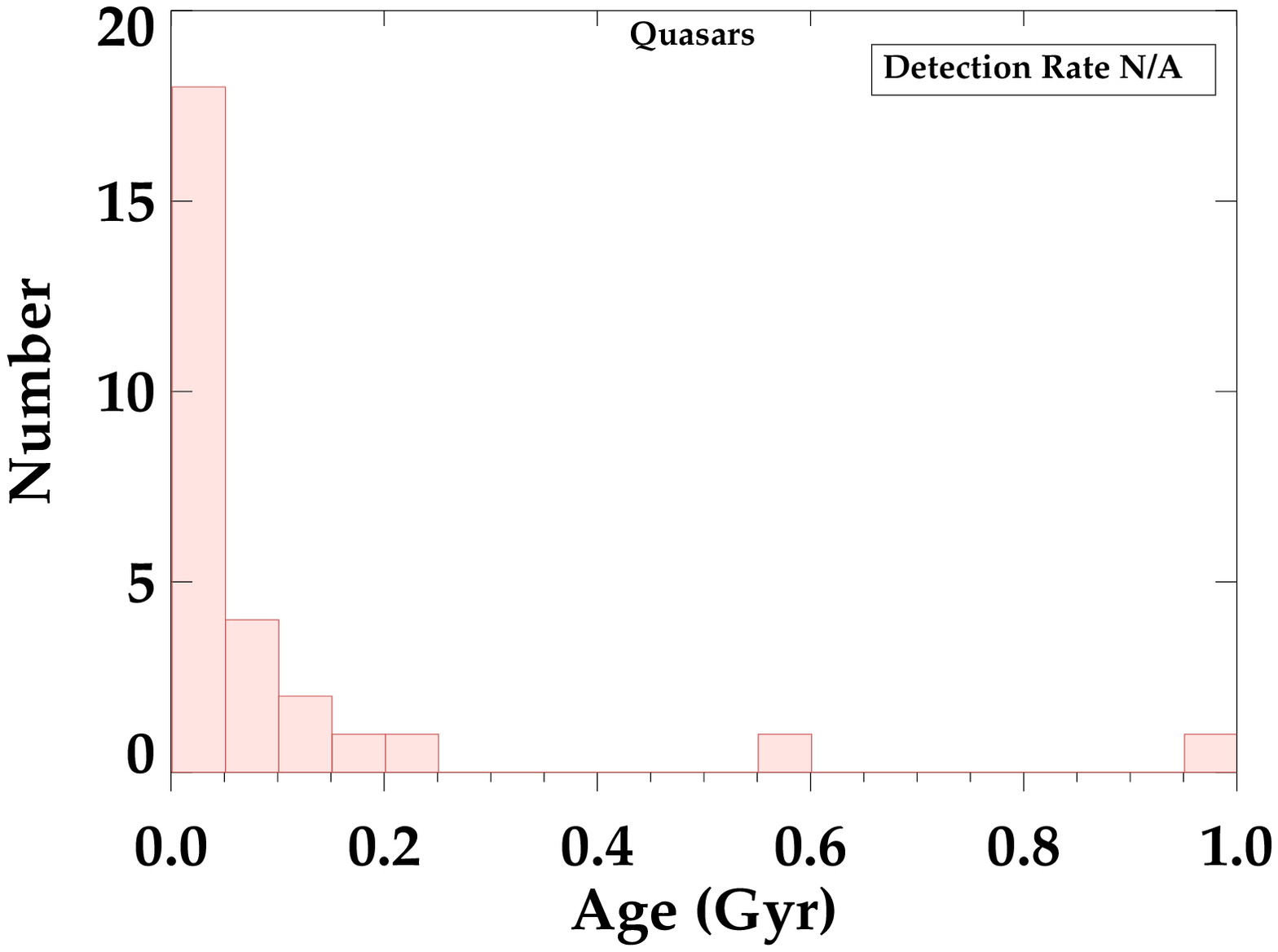}
\label{quasarshist}}

\caption{Comparison of the YSP ages found for the different samples discussed in this work. The top panel shows the distribution of ages for the type II quasars investigated in this paper. The following two histograms show the distribution of ages for the ULIRGs and 2 Jy PRGs respectively. The final panel shows the distribution of ages for a compilation of samples taken from the literature and include results from \citet{heckman97,canalizo00,tadhunter05,emonts06,holt06,holt07,wills08,liu09, tadhunter11, villar12}. The objects shown in this panel include only AGN which have quasar-like luminosities ($L_{[OIII]} > 10^{35}~\mbox{W}$). In each of the panels in which `complete' samples have been considered, the detection rate of YSPs in the host galaxies is also given.}
\label{spec:compSamples}
\end{figure}

\begin{figure}
\centering
\includegraphics[trim = 10mm 0mm 10mm 0mm, scale = 0.5]{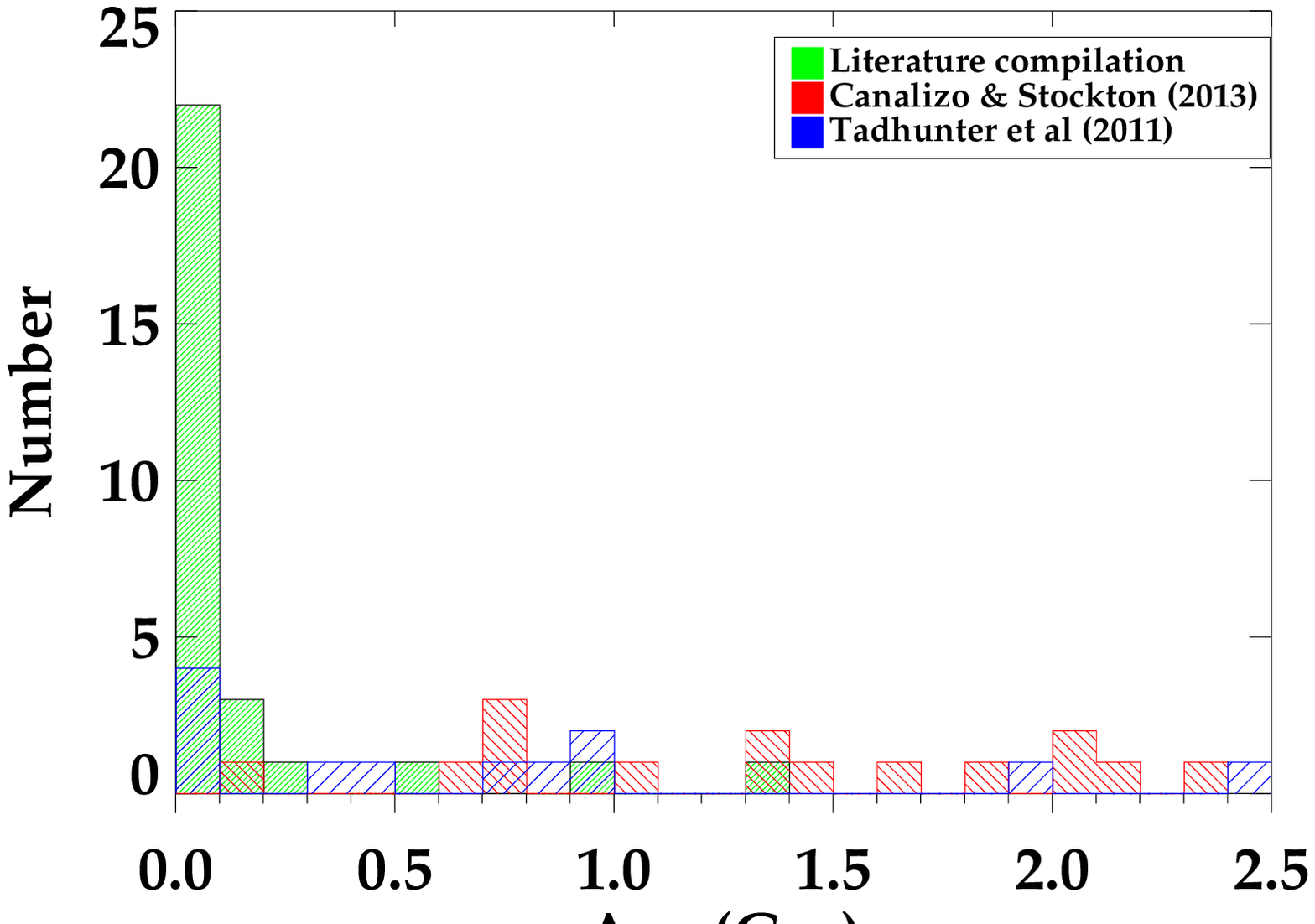}
\caption{The same samples of quasars as shown in the final panel of Figure \ref{spec:compSamples} (green) but also including the lower luminosity AGN discussed by \citet{tadhunter11} ($L_{[OIII]} < 10^{35}~\mbox{W}$; blue) and also the sample of quasars presented in \citet{canalizo13}(red). }
\label{spec:compSamples1}
\end{figure}

We must then consider the causes of these apparently ambiguous results. One explanation may be that the age of the stellar population hosted by an AGN is dependent on the luminosity of the AGN itself. Evidence to support this explanation can be found by examining the results presented for the 2 Jy sample of PRG \citep{tadhunter11}, which show that host galaxies with YSPs with ages $\tysp < 0.1$ Gyr also tend to host AGN of quasar-like luminosity ($\mbox{log}(L_{[OIII]}/L_{\sun})> 8.5$). In contrast, when we consider those objects in the sample which fall below this luminosity threshold, as well as including the results of other studies dominated by lower-luminosity objects \citep{tadhunter05,emonts05,wild10}, YSP ages $0.2 < \tysp < 2$ Gyr are more commonly found. This is effectively demonstrated in Figure \ref{spec:compSamples}, where the distribution of YSP ages is shown for various samples of quasars (see caption for details), and Figure \ref{spec:compSamples1} which shows the same samples but with the inclusion of lower-luminosity PRG (as well as the sample of \cite{canalizo13}). 

Taking this luminosity dependence into consideration, if we concentrate on the quasar-like objects with the best spectroscopic data on the stellar populations \citep{canalizo00,holt07,wills08,tadhunter11,villar12,canalizo13, bessiere14}, in general, ages $t_{ysp} < 0.1$ Gyr are typical in the nuclear regions. Examples of such cases include the powerful radio galaxy 3C459 \citep{wills08}, the type II quasar J0025-10 \citep{bessiere14}, and the nearby quasar Mrk 231 \citep{rodriguez09, canalizo01} in which the populations are found to have ages $t_{ysp} < 0.1$ Gyr. In the latter case, the findings are supported by the detection of the He {\textsc I} absorption features associated with B-type stars \citep{gonzalez99}. The nearby type II quasar Mrk 447 \citep{heckman97} is clearly identified as having a YSP with an age of $t_{YSP} \sim 6$ Myr, supported by the presence of UV absorption features such as Si {\textsc III}$\lambda1417$, which are attributed to the presence of late O and early B super-giants. Figure \ref{spec:compSamples} shows how the results of these various studies compare with those found in this work.

Despite using high-quality spectra, one notable exception to this trend is the study of \citet{canalizo13}, who find that, although 14/15 quasar host galaxies do have evidence for a significant YSP, they are \emph{all} older ($t_{ISP} \sim 1.2$ Gyr) than those found in the majority of other quasar-like systems. Clearly it is important to understand the root of this discrepancy. One possible explanation is that Canalizo \& Stockton do not directly consider reddening of the YSP in their analysis. The importance of accounting for reddening has been underscored by spectroscopic studies of the YSPs of local radio-galaxies and ULIRGs \citep{tadhunter05,rodriguez09,rodriguez10}. If reddening of the YSP is not adequately accommodated in the modelling, a YSP can appear substantially older than it is in reality. However, based on their fits to the stellar absorption line features, they conclude that any YSP/ISP present cannot be significantly younger that their primary modelling suggests. They also perform a double check on their fits which includes a low-order multiplicative polynomial over a spectral window that includes the Balmer absorption features, which can account for reddening (for full details refer to \citealt{canalizo13}), and find that, in the cases where their spectra have sufficient S/N, the results fall within the same range of ages as their primary fits, suggesting that this is unlikely to be the main cause of the difference.

A second source of of discrepancy may be that Canalizo \& Stockton deal with type I quasars, which requires that the apertures used in the modelling are offset from the nucleus by several kpc in order to avoid the overwhelming quasar nucleus. As simulations of galaxy mergers suggest that most of the merger-induced star formation will take place in the central regions of the galaxy on sub-kpc scales (e.g \citealt{dimatteo05, hopkins13a, moreno15}), the inability to model the very central region may mean that the majority of this YSP is missed. However, in this context we note that some of the apertures that give young YSP ages in Mrk231 \citep{canalizo01}, J0025-10 \citep{bessiere14} and J0218-00 (this work) are also significantly offset from the AGN nuclei.

A final basis for this discrepancy may be that type I and type II quasars are at different stages of their evolution. Once again, invoking the evolutionary sequence of \citet{sanders88}, it is possible that the type II objects are being observed earlier in this sequence than the type I quasars, and would therefore have younger merger-induced YSPs. This scenario is supported by the findings of \citet{kocevski15} who have carried out a morphological study of AGN with a range of obscuration from Compton thick to unobscured. They find a significantly higher merger fraction in the obscured objects than the unobscured objects, which cannot be satisfactorily explained exclusively by the orientation based unification scheme \citep{antonucci93}. This might suggest that type I quasars are `older' than type II quasars and thus the difference in the ages of the YSPs can be explained naturally.

Thanks to the quality and spectral range of these data, the type II quasars presented here are included in an expanding group of objects in which the starburst population can be reliably dated, with the clear detection of the strong Balmer absorption features in a large fraction of the sample ($\sim67\%$), and a significant contribution to the total flux by the YSP. These findings strongly emphasise that, in the cases in which the stellar populations can be unambiguously fit, it is found that a significant proportion of the flux is attributable to a YSP $t_{ysp} < 0.1$ Gyr.

One of the goals of the stellar synthesis modelling presented in this work is to determine whether a delay between the peak of star formation and the triggering of AGN activity is observed in luminous AGN. Are the stellar populations of their host galaxies significantly older than the expected lifetime of the AGN? The answer to this question is largely dependent on the assumed lifetime of a typical quasar ($t_{QSO}$), which unfortunately is poorly constrained. Estimates, based on a variety of methods, are in the range $10^6 < t_{QSO} < 10^8 ~\mbox{yr}$ (\citealt{martini04} and references therein). If the typical lifetime of a quasar is in fact closer to the lower limits (1 Myr), then the results presented here provide evidence for significant delay ($\sim 50 -- 100$ Myr) between the peak of star formation and the emergence of the visible quasar activity in many of the sample objects. On the other hand, if the typical quasar lifetime is closer to the upper limit (50--100 Myr), as detailed simulations such as those of \citet{hopkins05} favour for the range for quasars of the luminosities considered here, then it can be concluded that the starburst and AGN activity are concurrent, with the possible exception of two objects (J0848+01 and J0948+00) which require YSP ages $\sim 200$ Myr. Only when tighter constraints are placed on quasar lifetimes, will it be possible to answer this question with a degree of certainty.

\section{Summary}
\label{spec:conclusions}

In this work, the stellar populations of a sample of 21 type II quasar host galaxies (including results presented by \citet{bessiere14} for J0025-10), selected from the SDSS \citep{zakamska03} have been investigated. Our aim was to determine if YSP are present in these host galaxies, and if so, to establish their ages. In this way, we hope to gain valuable insights into the timing of AGN triggering and starbursts, and the sequence of events associated with galaxy mergers.  We have done this using stellar population synthesis modelling, and also by the detection of Balmer absorption lines, which are an unambiguous marker of a YSP. Reddening of the YSP and AGN-related continuum components such as nebular continuum and scattered light have also been taken into account. It is important to consider these components, because failure to account for them may lead to an underestimation of the true ages of the YSPs. The results of the investigation can be summarised as follows.

\renewcommand{\labelitemi}{$\bullet$}

\begin{itemize}
\item The Balmer absorption lines are clearly detected in $\sim 62 \%$ of the objects \emph{before} the nebular continuum and higher order Balmer emission lines are subtracted, and in $\sim 67 \%$ of the objects after the subtraction.
\item $\sim 90\%$ of the type II quasars cannot be adequately fit using a combination of an OSP and power-law only. Therefore, they \emph{require} a YSP component in order to produce an adequate fit.
\item In the majority of cases, a significant proportion of the flux in the normalising bin is attributable to a YSP component. Values in the range 15--100\% have been found for the the objects which can be successfully modelled using an 8 Gyr + YSP combination of components.
\item Of the 24 apertures modelled which can be fit using an 8 Gyr + YSP combination of components, 18 ($ 75 \%$) have YSPs with a maximum acceptable age $\tmax = 100$ Myr, and 7 ($\sim 30 \%$) have a maximum acceptable age of $\tmax = 50$ Myr.
\item Two objects (J0848+01 and J0948) require the inclusion of a YSP component with $\tysp > 0.2$ Gyr.
\item For an 8 Gyr + YSP combination, 38 \% of the apertures are strongly reddened (E(B-V) $> 0.4$), whilst 62 \% of the apertures are not. There does not appear to be any correlation between YSP age and reddening. 
\end{itemize}

We have also compared the stellar properties of type II quasar host galaxies to those of ULIRGs and powerful radio-galaxies, and find that the YSP ages of type II quasars are consistent with those found for both groups. However, the detection rate of YSPs in PRGs is only $\sim 23\%$ (although this is a lower limit), whilst the detection rate for both the type II quasars and ULIRGs is $\sim 90 \%$. This discrepancy in the detection rates of YSPs between PRGs and (in the vast majority of cases) radio-quiet type II quasars could suggest that the host galaxies of these two types of active systems are fundamentally different, whilst ULIRGs and type II quasars may form part of the same evolutionary sequence.

Although quasars selected for this study are type II, so that the direct quasar light is obscured from our direct view, scattered light from the quasar may nonetheless have an impact on the ages derived for the YSPs. Although the results summarised above assume that scattered light does not contribute significantly to the flux,  we have explored the inclusion of a power-law component on the results of our modelling.  We find that the inclusion of a power-law has no bearing on whether a YSP is detected, however, its major impact is to increase the range of ages and reddenings of the YSPs that produce acceptable fits, generally allowing for older ages. Clearly, future spectropolarimety observations -- which explicitly quantify the level and spectrum of the scattered AGN component -- are required to improve the accuracy of the determination of the YSP ages for the type II quasar host galaxies.

Our thorough investigation of the stellar properties of the host galaxies of type II quasars has unambiguously shown that YSPs are a fundamental constituent of a substantial proportion of objects. Moreover, because in most cases the maximum acceptable age of the YSP is comparable with the expected quasar life time, our study provides clear evidence that these type II quasars are being triggered quasi-simultaneously with the starburst activity, suggesting that both phenomena are being triggered at the peak of a galaxy merger.

\section{Acknowledgements}
Firstly, we would like to acknowledge the valuable constructive feedback and suggestions of the referee. PSB acknowledges support from FONDECYT through grant 3160374. CRA acknowledges the Ramn y Cajal Program of the Spanish Ministry of Economy and Competitiveness through project RYC-2014-15779 The main body of this work is based on observations obtained at the Gemini Observatory (acquired through the Gemini Observatory Archive and processed using the Gemini IRAF package), which is operated by the Association of Universities for Research in Astronomy, Inc., under a cooperative agreement with the NSF on behalf of the Gemini partnership: the National Science Foundation (United States), the National Research Council (Canada), CONICYT (Chile), Ministerio de Ciencia, Tecnolog\'{i}a e Innovaci\'{o}n Productiva (Argentina), and Minist\'{e}rio da Ci\^{e}ncia, Tecnologia e Inova\c{c}\~{a}o (Brazil). The data were obtained under programs GS-2010B-Q-83 and GS-2011B-Q-42. We have also made use of observations made with the Gran Telescopio Canarias (GTC), installed in the Spanish Observatorio del Roque de los Muchachos of the Instituto de Astrof\'isica de Canarias, on the island of La Palma.



\bibliographystyle{mnras}
\bibliography{ref_list} 




\appendix

\section{Results For Individual Objects}
\label{spec:obj_results}
In the following appendix a brief description of the stellar synthesis modelling for each of the objects is outlined. We also present examples of acceptable fits for each of the objects, showing the fit to both the overall spectrum, and also in the key Balmer absorption region. J0025-10 is discussed in detail in \citet{bessiere14}.

\subsection{J2358-00}
\label{spec:q2358}

The initial modelling for J2358-00 was performed using combination 2, and resulted in a best fitting model with \tysp = 0.02 Gyr and \ebv = 0.5. The model nebular continuum contributes $\sim 28\%$ of the total flux below the Balmer edge. The reddening derived from the \hg/\hb\ and \hd/\hb\ ratio is consistent with case B (Table \ref{blines}), and therefore, no reddening was applied before subtraction. The resulting spectrum shows no evidence of an unphysical step at the Balmer edge, thus was deemed acceptable.

All five modelling combinations were utilised in order to perform the stellar synthesis modelling, and the results presented in Tables \ref{results8} and \ref{results2} show that all models which include a YSP component produce acceptable solutions.

\begin{figure}
\centering
\subfloat{
\includegraphics[scale = 0.45, trim = 7mm 0mm 0mm 0mm]{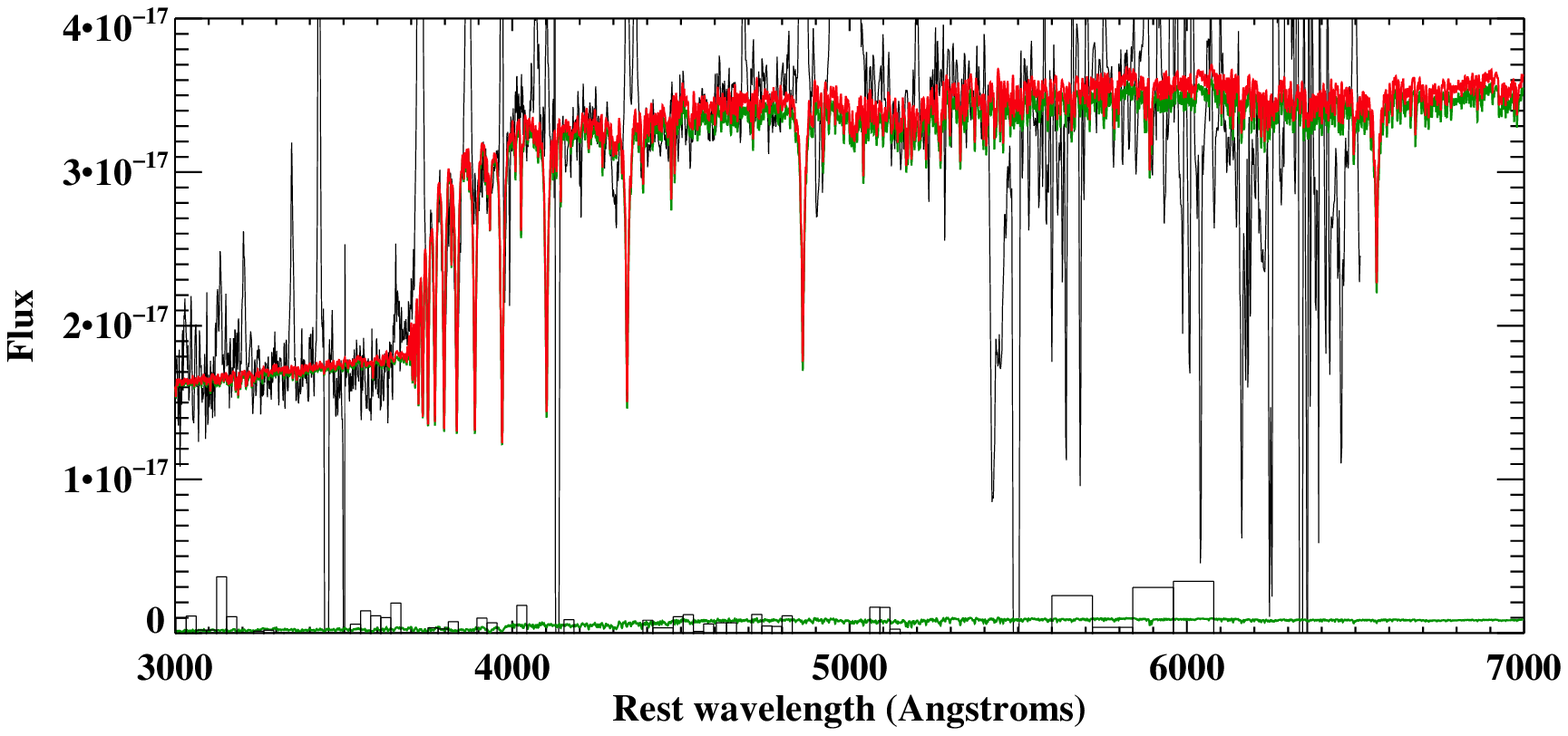}
\label{q2358-8np-b-bf}}

\subfloat{
\includegraphics[scale = 0.45, trim = 7mm 0mm 0mm 0mm]{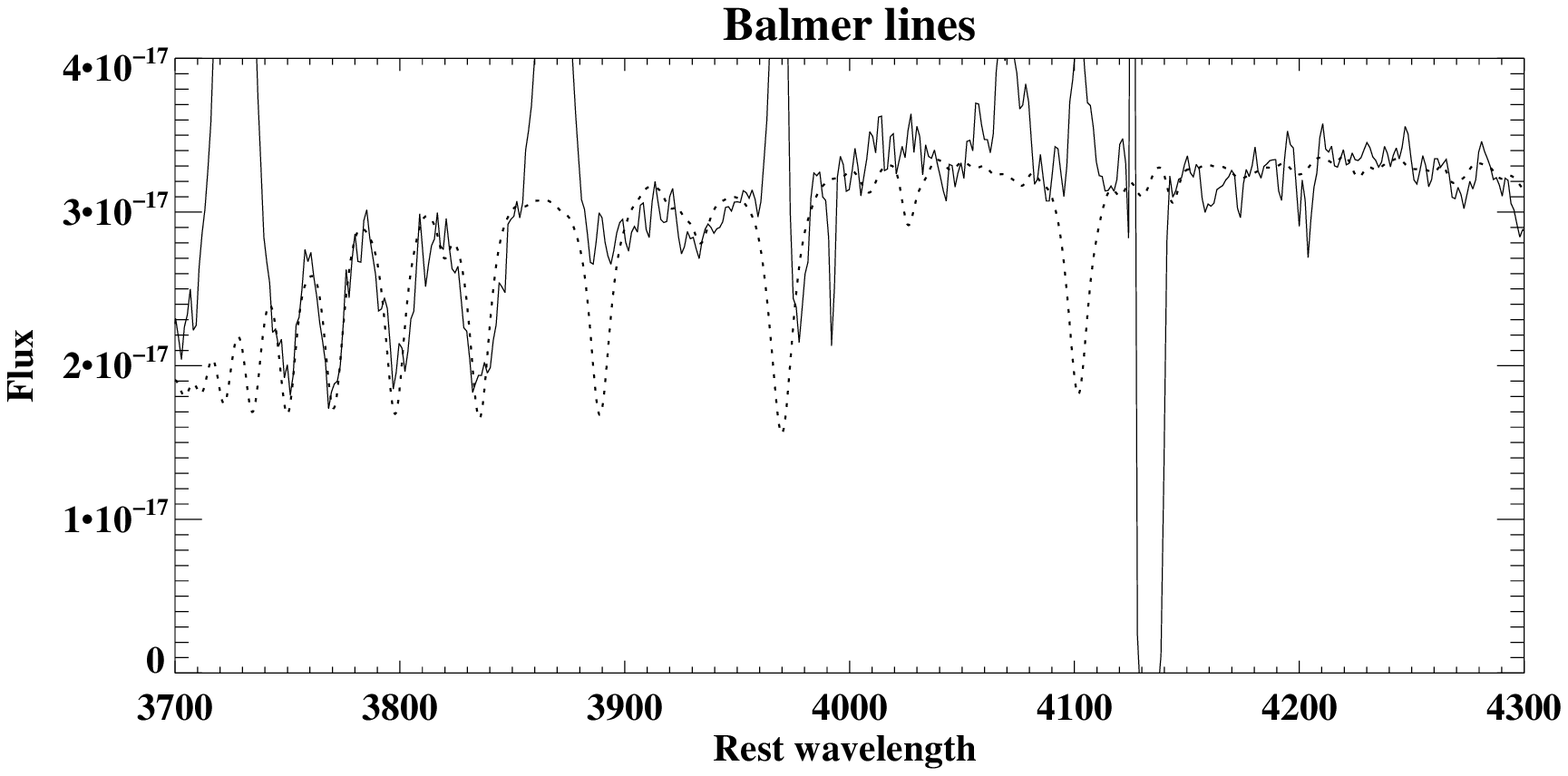}
\label{q2358-8np-b-bl}}

\caption{An example of an acceptable fit produced by \confit for J2358-00. The fit shown here was produced using combination 2 and includes a YSP with \tysp = 0.04 Gyr and \ebv = 0.6.}
\label{spec:q2358_confit}
\end{figure}

It was possible to reject a significant number of potential solutions based on the strong over-prediction of \ca\ absorption. In fact, the bottom panel of Figure \ref{spec:q2358_confit} shows that this feature is unusually weak in the spectrum of J2358-00, although the Balmer lines are clearly detectable. The reason for this is not immediately apparent but is not due to any cosmetic defects in the spectrum at this wavelength. This feature provides a strong constraint on the acceptable ages and reddenings of the YSP when no power-law component is included ($0.04 \le \tysp~\mbox{(Gyr)} \le 0.05 $), with reddenings in the range $0.5 \le \ebv \le 0.6$. However, when a power-law component is included, a much wider range of ages and reddenings becomes permissible, although in a majority of cases, the YSP flux is dominant over the OSP. Figure \ref{spec:q2358_confit} shows an example of an acceptable fit generated using combination 2. 

\subsection{J0114+00}
\label{spec:q0114}

Unfortunately, no Gemini \gmos\ spectrum is available for J0114+00 and therefore, we model the SDSS spectrum instead. Although this spectrum has a lower S/N than the Gemini GMOS-S data, it is still of adequate quality and covers a sufficient wavelength range to perform spectral synthesis modelling.

After performing the initial run (combination 2), the stellar model, which included a YSP with \tysp= 0.04 Gyr and \ebv = 0.8  was subtracted. The model nebular continuum then contributes $\sim 15 \%$ of the total flux blue-ward of the Balmer edge (Table \ref{blines}). In this case, it was not possible to make a reliable estimate of the Balmer decrements, because of the intrinsically low equivalent widths of the \hg\ and \hd\ emission lines, and the low S/N of these data. Therefore, no reddening correction was applied to the model nebular continuum before subtraction.

\begin{figure}
\centering
\subfloat{
\includegraphics[scale = 0.45, trim = 7mm 0mm 0mm 0mm]{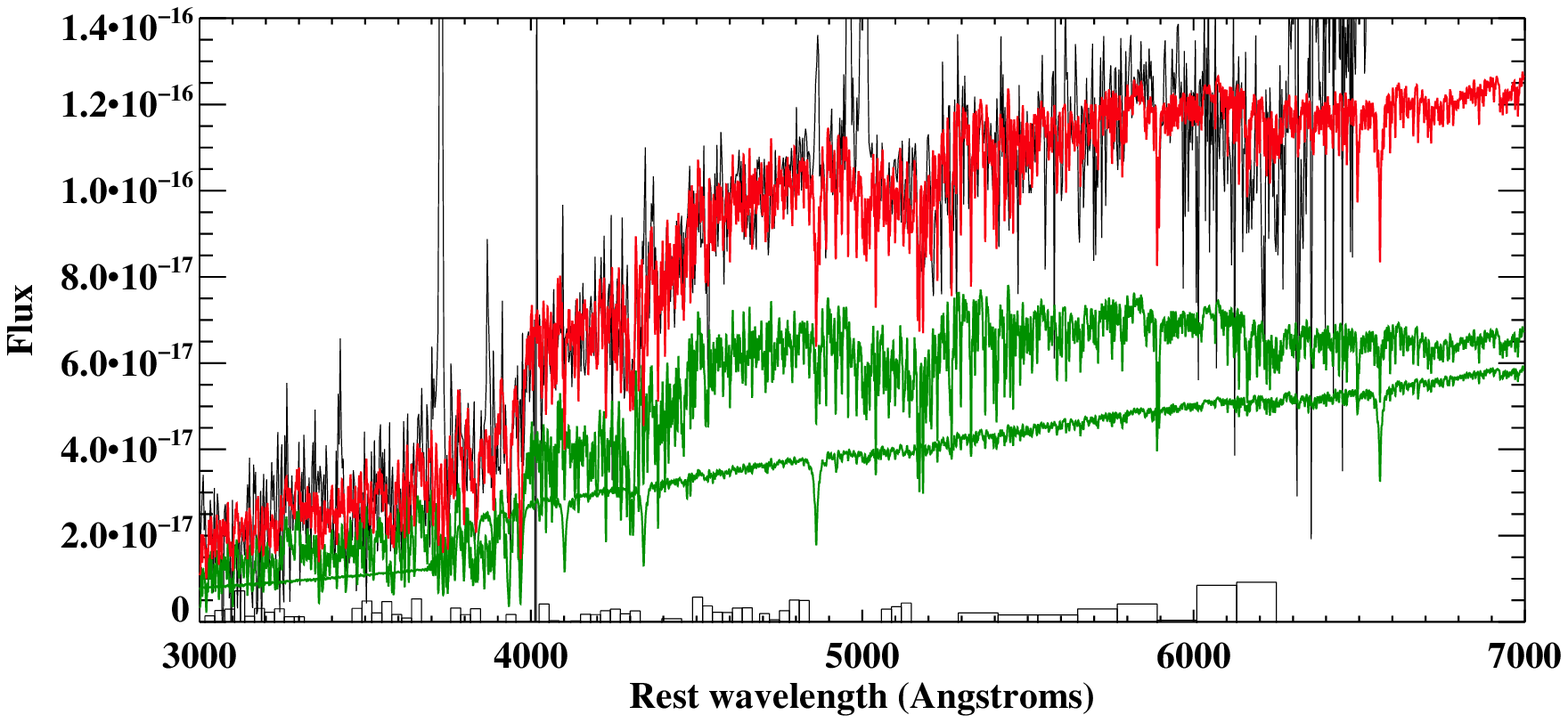}
\label{q0114-8np-b-bf}}

\subfloat{
\includegraphics[scale = 0.45, trim = 7mm 0mm 0mm 0mm]{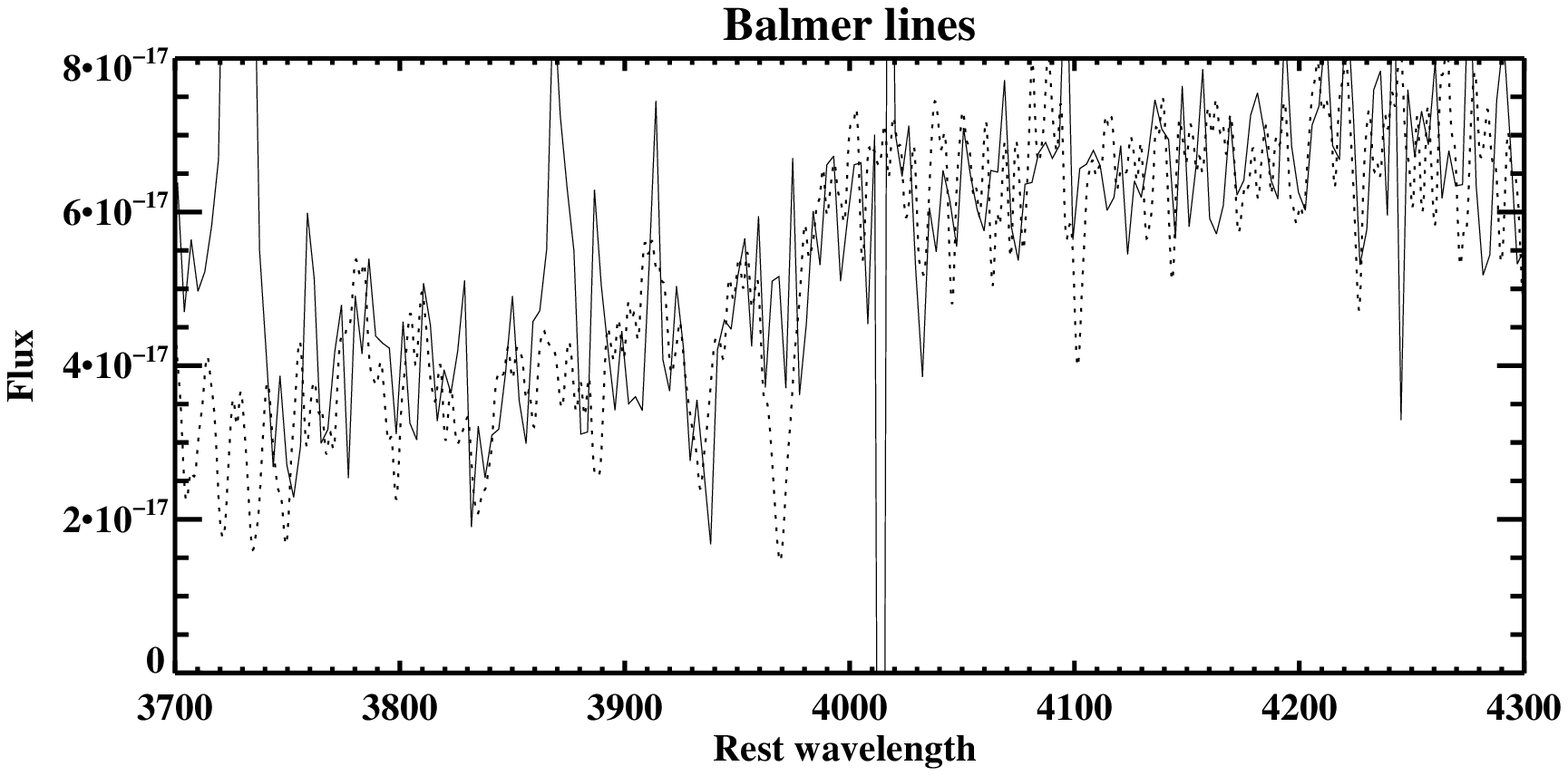}
\label{q0114-8np-b-bl}}

\caption{An example of an acceptable fit produced by \confit for J00114+00. This fit is produced by combination 2 and includes a YSP with \tysp = 0.06 Gyr and \ebv = 0.9. }
\label{spec:q0114_confit}
\end{figure}

Tables \ref{results8} and \ref{results2} demonstrates that for this particular object, the OSP dominates the flux in the normalising bin, which leads to the models being relatively unconstrained. Therefore, the fits produce a wide range of ages and reddenings ($0.004 \le \tysp~\mbox{(Gyr)} \le 0.4$ and $0.6 \le \ebv \le 1.3$) for combination 2. Figure \ref{spec:q0114_confit} shows an example of an acceptable fit generated by combination 2 with \tysp = 0.06 Gyr and \ebv = 0.9.

\subsection{J0123+00}
\label{spec:q0123}

The initial modelling for J0123+00 was performed using combination 4, because it was not possible to obtain acceptable fits without the inclusion of a power-law component. However, this resulted in a wide range of ages and reddenings producing acceptable fits ($ 2 \le \tisp~\mbox{(Gyr)} \le 5$ and $0.4 \le \ebv \le 1.2$), which, based on age-sensitive stellar features, were impossible to differentiate between. Therefore, the stellar model that produced the lowest \chisq\ value was subtracted, which included an ISP with $\tisp=5.0$ Gyr, $\ebv = 0.6$ and a power-law component.

The nebular continuum model produces a large contribution to the total flux blue-wards of the Balmer edge ($\sim 48 \%$), and subtraction of this model without first applying reddening of the value derived from the Balmer line ratios (Table \ref{blines}), results in an unphysical step in the spectrum. Therefore, the model continuum was reddened using a value of \ebv = 0.2, which is the mean value of reddening derived from the \hg/\hb\ and \hd/\hb\ ratios. The reddened nebular continuum contributes $\sim 20\%$ of the flux just below the Balmer edge, and does not result in an over subtraction of the nebular continuum component.

Tables \ref{results8} and  \ref{results2} show that, when an 8 Gyr population is combined with a YSP model and a power-law component, it is possible to achieve an acceptable fit to the spectra, but not with the inclusion of an 2 Gyr population. When attempting to perform the stellar population modelling for J0123+00, it has proven difficult to constrain the ages of the YSP/ISP component. Once again (as for J0114+00), this may be due to the fact that the flux is dominated by the OSP for nearly all possible acceptable solutions, making it difficult to accurately model the lower level YSP contribution.


\begin{figure}
\centering
\subfloat{
\includegraphics[scale = 0.45, trim = 7mm 0mm 0mm 0mm]{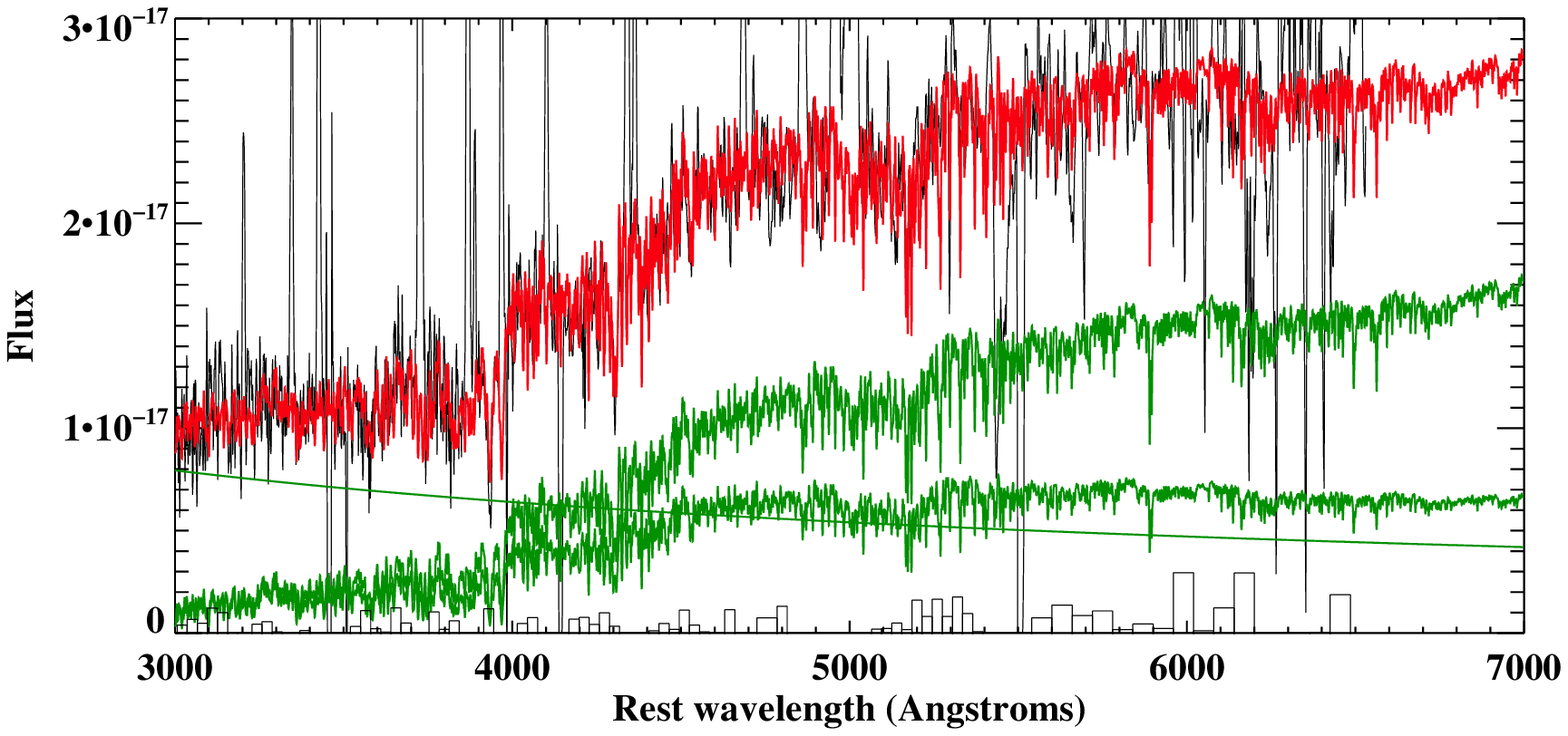}
\label{q0123-8p-b-bf}}

\subfloat{
\includegraphics[scale = 0.45, trim = 7mm 0mm 0mm 0mm]{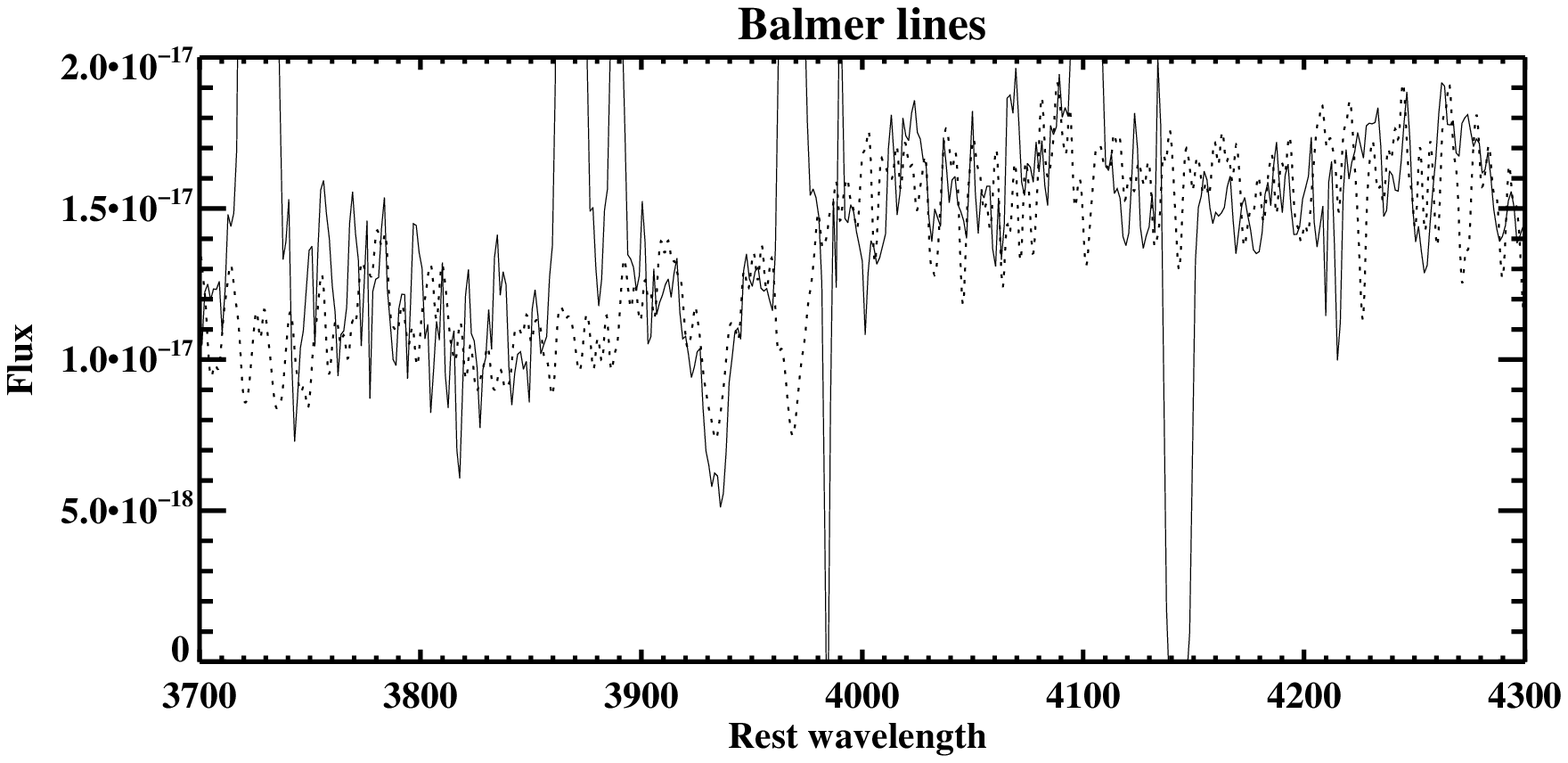}
\label{q0123-8p-b-bl}}

\caption{An example of an acceptable fit produced by \confit for J0123+00. This fit is generated by combination 6 and includes an ISP with \tisp = 4 and \ebv = 0.3.}
\label{spec:q0123_confit}
\end{figure}

Figure \ref{spec:q0123_confit} shows an example of an acceptable fit using combination 4, which in this case include a young component of \tisp = 4 Gyr and \ebv = 0.3.


\subsection{J0142+14}
\label{spec:q0142}

Unlike the other objects for which GMOS-S spectroscopic data is available, for J0142+14 only the blue part of the spectrum was observed (Table \ref{obs_spec}). However, modelling of this object is still well constrained because the spectrum has a high S/N as well as strong Balmer absorption lines. The rest-frame spectral range covered by this spectrum is 2813 -- 4863 \AA, meaning we are unable to recover the \hb\ flux and generate a nebular continuum model.

Tables \ref{results8} and \ref{results2} show that all the modelling combinations including a YSP component were successful. In all cases, where an adequate fit was achieved, the allowable ages are very similar ($0.04 \leq \tysp~\mbox{(Gyr)} \leq 0.1$), except in the case of combination 4, where the maximum allowable age is 0.4 Gyr. In all cases, the reddening of the YSP is $0.1 \leq \ebv \leq 0.5$, and the flux of the YSP dominates the total flux of the system, in some cases accounting for 100 \% of the flux.

\begin{figure}
\centering
\subfloat{
\includegraphics[scale = 0.45, trim = 7mm 0mm 0mm 0mm]{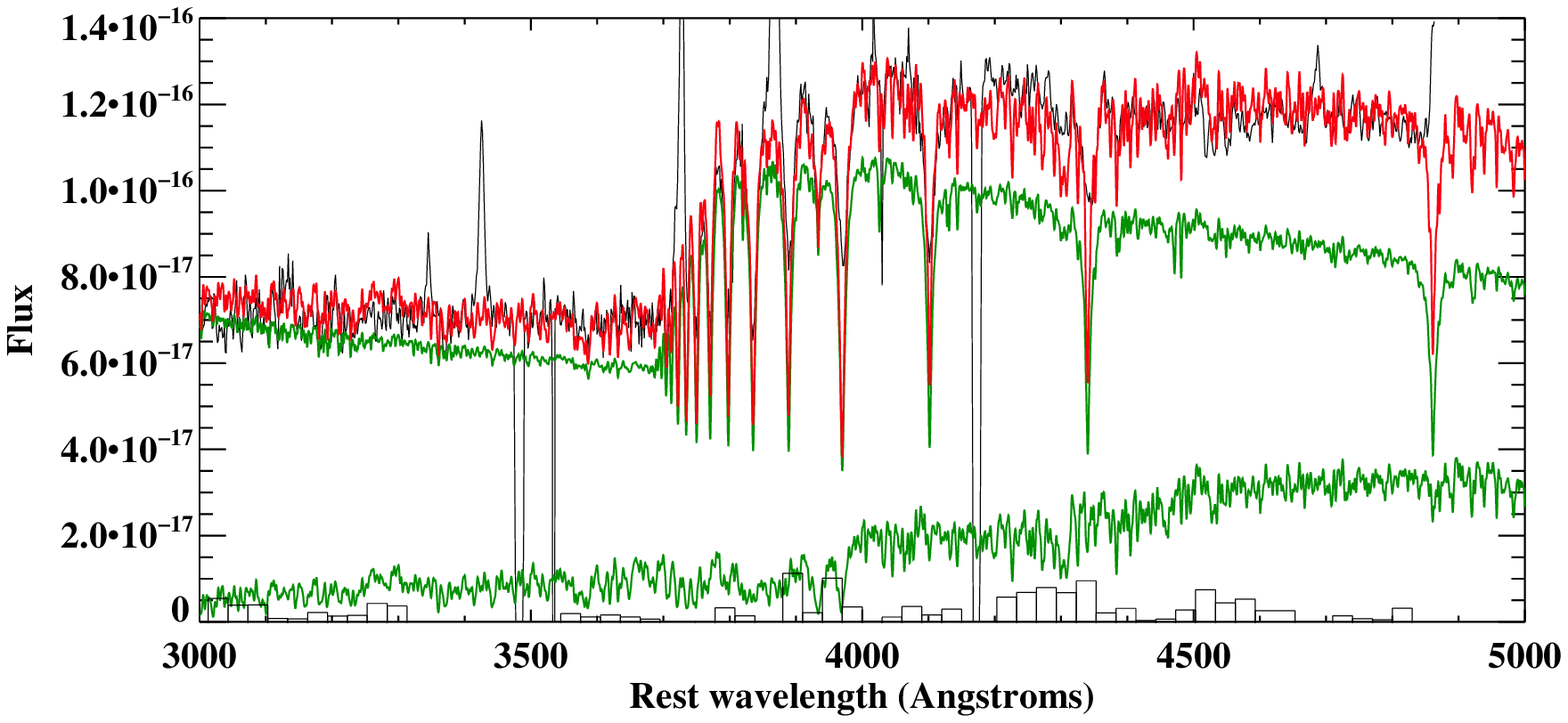}
\label{q0142-8np-b-bf}}

\subfloat{
\includegraphics[scale = 0.45, trim = 7mm 0mm 0mm 0mm]{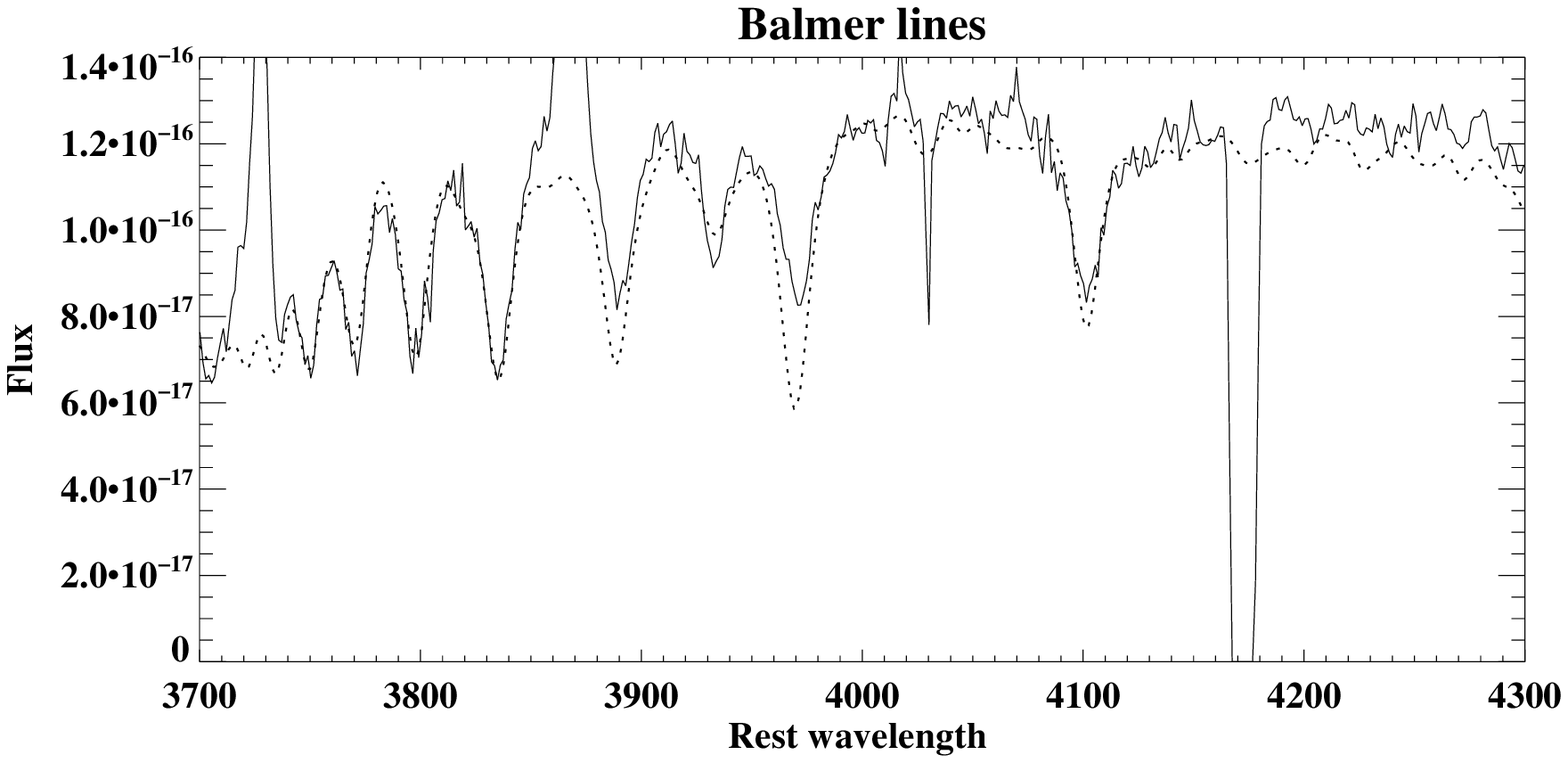}
\label{q0142-8np-b-bl}}

\caption{An example of an acceptable fit produced by \confit for J0142+14. This fit is produced by combination 2 and includes a YSP with $\tysp= 0.08$ Gyr with $\ebv= 0.2$.}
\label{spec:q0142_confit}
\end{figure}

It is interesting to note that J0142+14 is classified as undisturbed in terms of its morphology, and appears to be relatively compact in the deep $r^\prime$-band image presented in \citet{bessiere12}.  However, the modelling results presented here combined with the clearly detectable Balmer absorption lines in the spectrum (illustrated in Figure \ref{spec:q0142_confit}), make it clear that there has been a powerful episode of star formation within the last 0.4 Gyr. This suggests that J0142+14 may be an `E+A' galaxy, which is thought to be a short lived phase in which a galaxy transforms from star-forming to passively evolving \citep{tran03,tran04}. These galaxies are characterised by weak nebular lines (associated with star formation) such as [OII]$\lambda 3727$ and strong absorption lines, such a \hd,  characteristics which are clearly visible in the spectrum of J0142+14 (Figure \ref{all_spec}).

\citet{swinbank12} present integral field unit (IFU) observations for 11 E+A galaxies ($0.07 < z < 0.12$) and find that 90\% of their sample have residual rotational motion, with the A type stars spread throughout the galaxy, a finding which they attribute to previous minor/major merger activity. It is plausible that this is also the case for J0142+14 implying that, even though no evidence of a recent merger is detected, it is likely that one has occurred fairly recently. Figure \ref{spec:q0142_confit} shows an example of a fit produced using combination 2, which includes a YSP of $\tysp= 0.08$ Gyr with $\ebv= 0.2$.


\subsection{J0217-00}
\label{spec:q021700}

After the initial modelling of J0217-00 (using combination 2),  a stellar model with a YSP $\tysp= 0.05$ Gyr and $\ebv = 0.5$ was selected for subtraction, resulting in a nebular continuum that contributes $\sim14\%$ of the total flux just below the Balmer edge (3540 - 3640 \AA). In this case, it was not possible to determine any values for the Balmer decrements because \hg\ falls directly adjacent to a chip gap, and the low equivalent width of \hd\ results in uncertainties which are larger than the measured values. Therefore, no reddening correction was applied to the nebular continuum before subtraction, however this did not result in an unphysical step at the Balmer edge.

\begin{figure}
\centering
\subfloat{
\includegraphics[scale = 0.45, trim = 7mm 0mm 0mm 0mm]{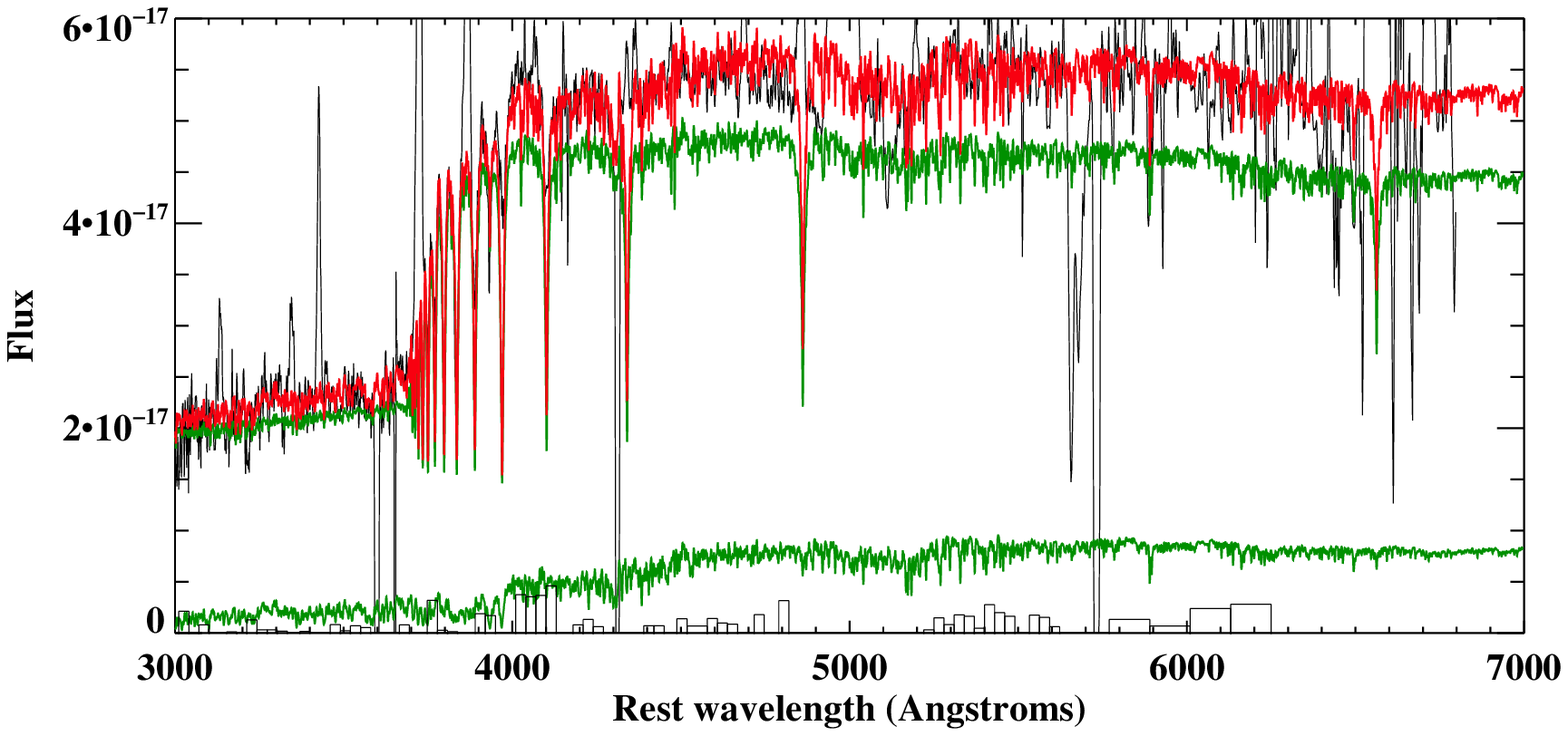}
\label{q021700-8np-b-bf}}

\subfloat{
\includegraphics[scale = 0.45, trim = 7mm 0mm 0mm 0mm]{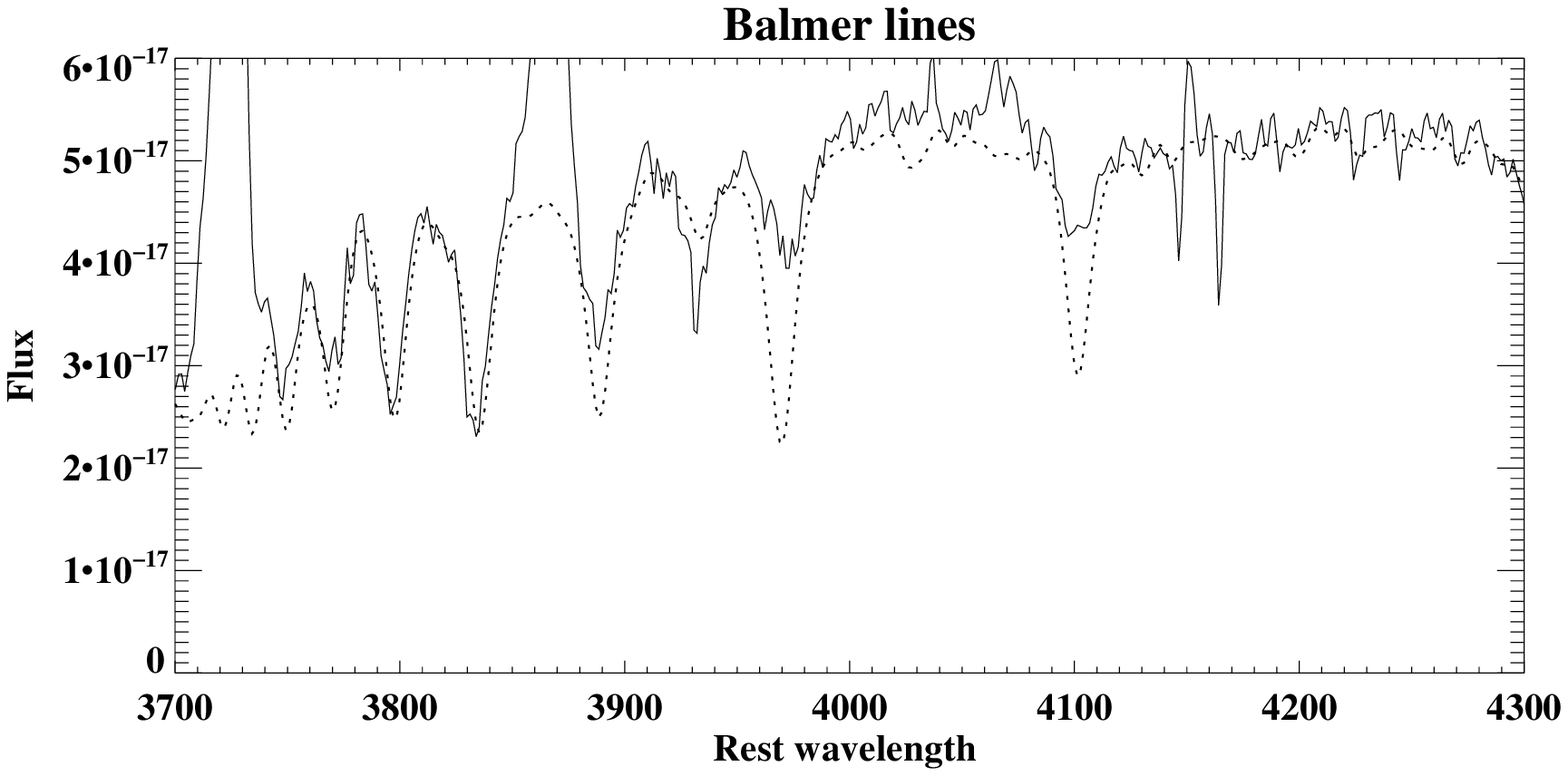}
\label{q021700-8np-b-bl}}

\caption{An example of an acceptable fit produced by \confit for J0217-00. This fit was produced by using combination 2 and includes a YSP with $\tysp = 0.1$ and $\ebv = 0.5$.}
\label{spec:q021700_confit}
\end{figure}

Tables \ref{results8} and \ref{results2} show that all combinations which include a YSP are successful, and that for both cases in which the spectrum was modelled with no power-law component (combinations 2 and 3), the acceptable ages are in the range $ 0.06 \le \tysp~\mbox{(Gyr)} \le 0.1$ and the associated values of reddening are in the range $0.5 \le \ebv \le 0.6$. When a power-law component was also included, the range of acceptable ages in both cases is extended to 0.4 Gyr. Figure \ref{spec:q021700_confit} shows an example of an acceptable fit generated by combination 2.

\subsection{J0217-01}
\label{spec:q021701}

The initial modelling of J0217-01 was performed using combination 2, and produced a best fitting model which included a YSP of age $\tysp = 0.2$ and $\ebv = 0.9$, which resulted in a nebular continuum model that contributes $\sim 5 \%$ of the total flux just below the Balmer edge. Attempting to derive the Balmer decrements proved unsuccessful because of the large errors associated with the \hg\ and \hd\ measurements, which are ultimately due to this weak nebular continuum. 

\begin{figure}
\centering
\subfloat{
\includegraphics[scale = 0.45, trim = 7mm 0mm 0mm 0mm]{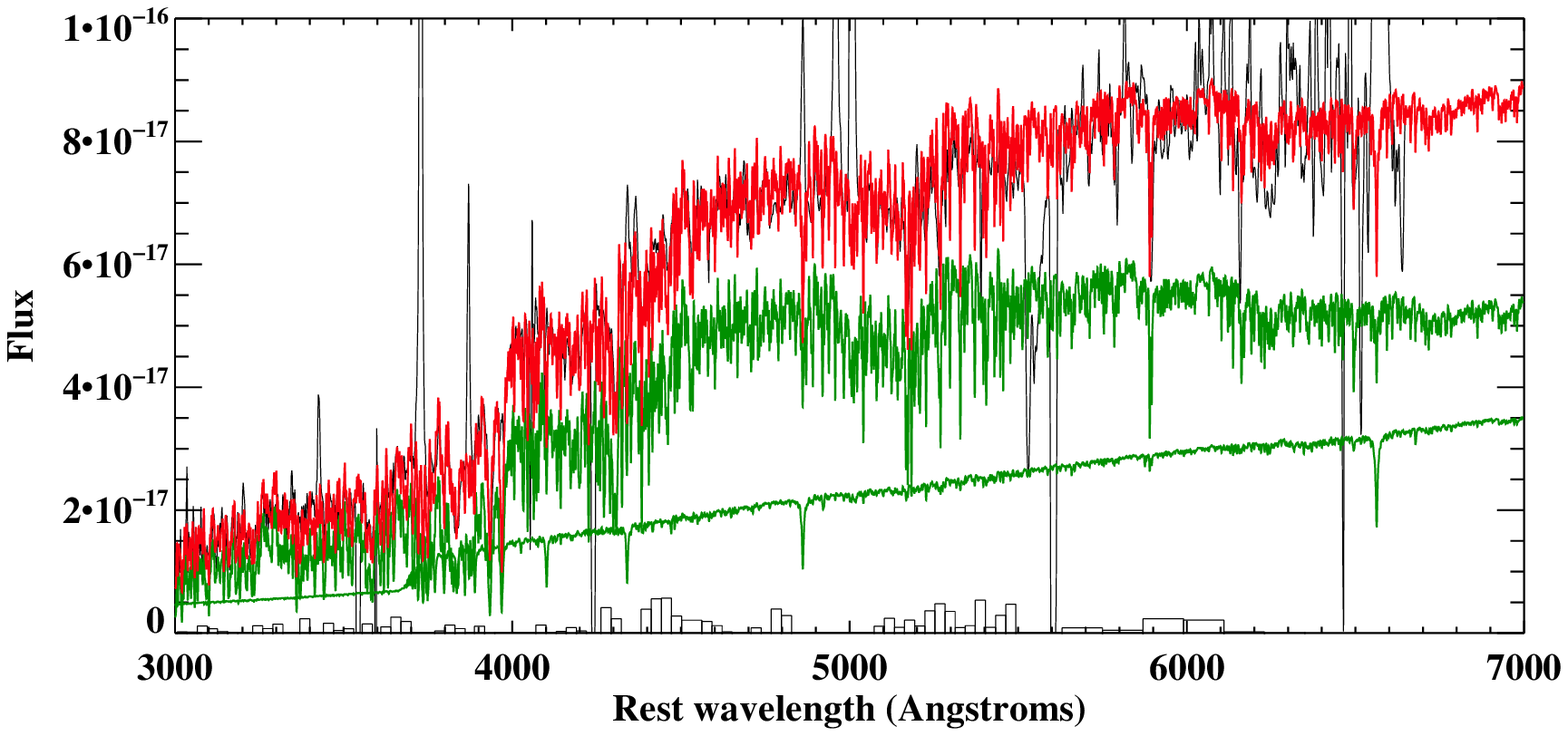}
\label{q021701-8np-b-bf}}

\subfloat{
\includegraphics[scale = 0.45, trim = 7mm 0mm 0mm 0mm]{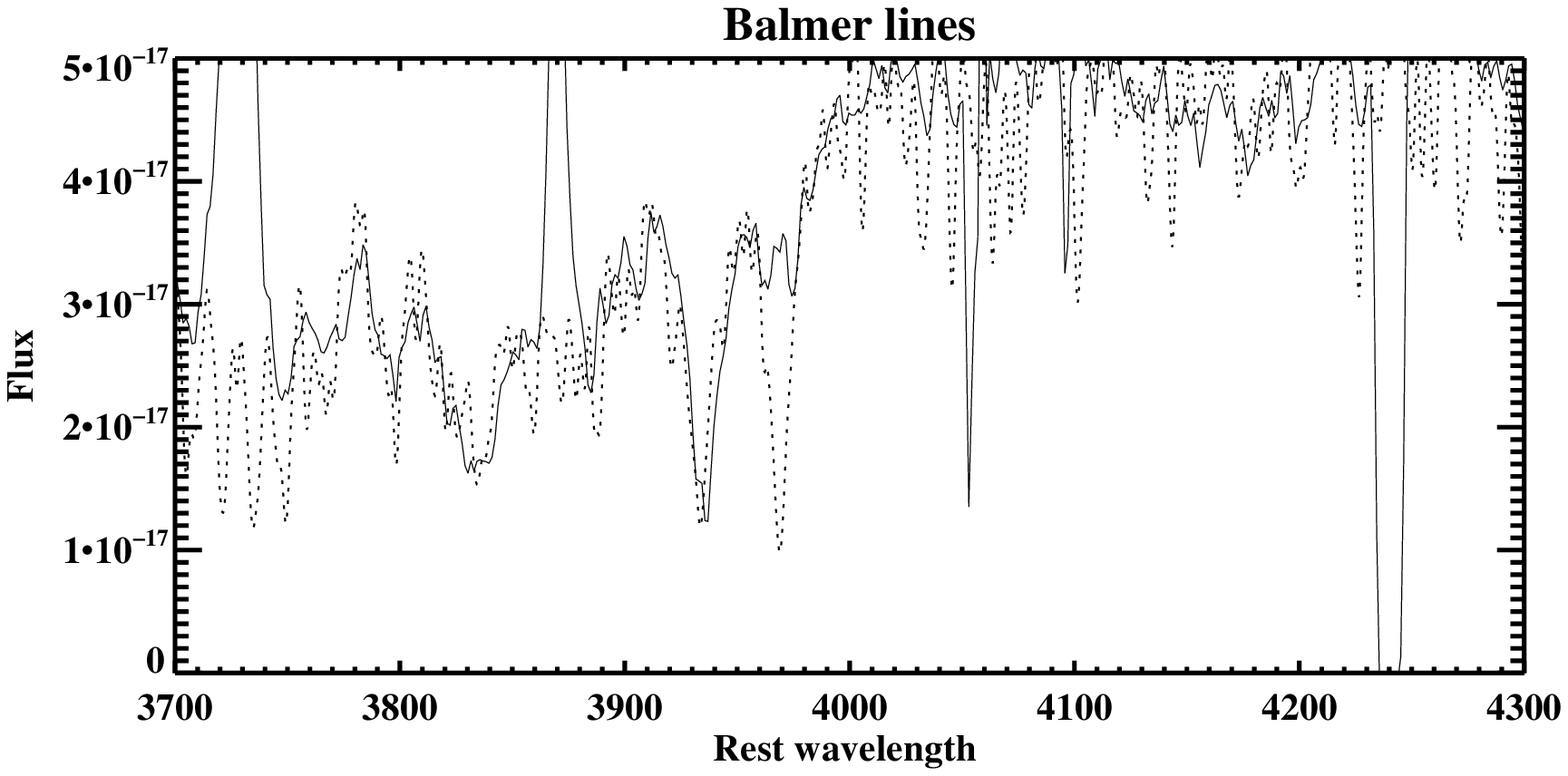}
\label{q021701-8np-b-bl}}

\caption{An example of an acceptable fit produced by \confit for J0217-01. This fit was produced by combination 2 and incorporates a YSP with \tysp = 0.007 Gyr and \ebv = 0.9.}
\label{spec:q021701_confit}
\end{figure}

All five combinations of components were applied, and of these, only combinations that include an 8 Gyr unreddened component and a YSP (combinations 2 \& 4) produced acceptable fits (Table \ref{results8}). In the majority of cases where acceptable solutions are found, the flux is dominated by the OSP component. Figure \ref{spec:q021701_confit} shows an example of an acceptable fit to J0217-01, produced by combination 2, which includes a YSP of $\tysp = 0.007$ Gyr and $\ebv = 0.9$

\subsection{J0218-00}
\label{spec:q0218}

\begin{figure}	
\centering
\subfloat{
\includegraphics[width = 3.9cm, height = 3.9cm]{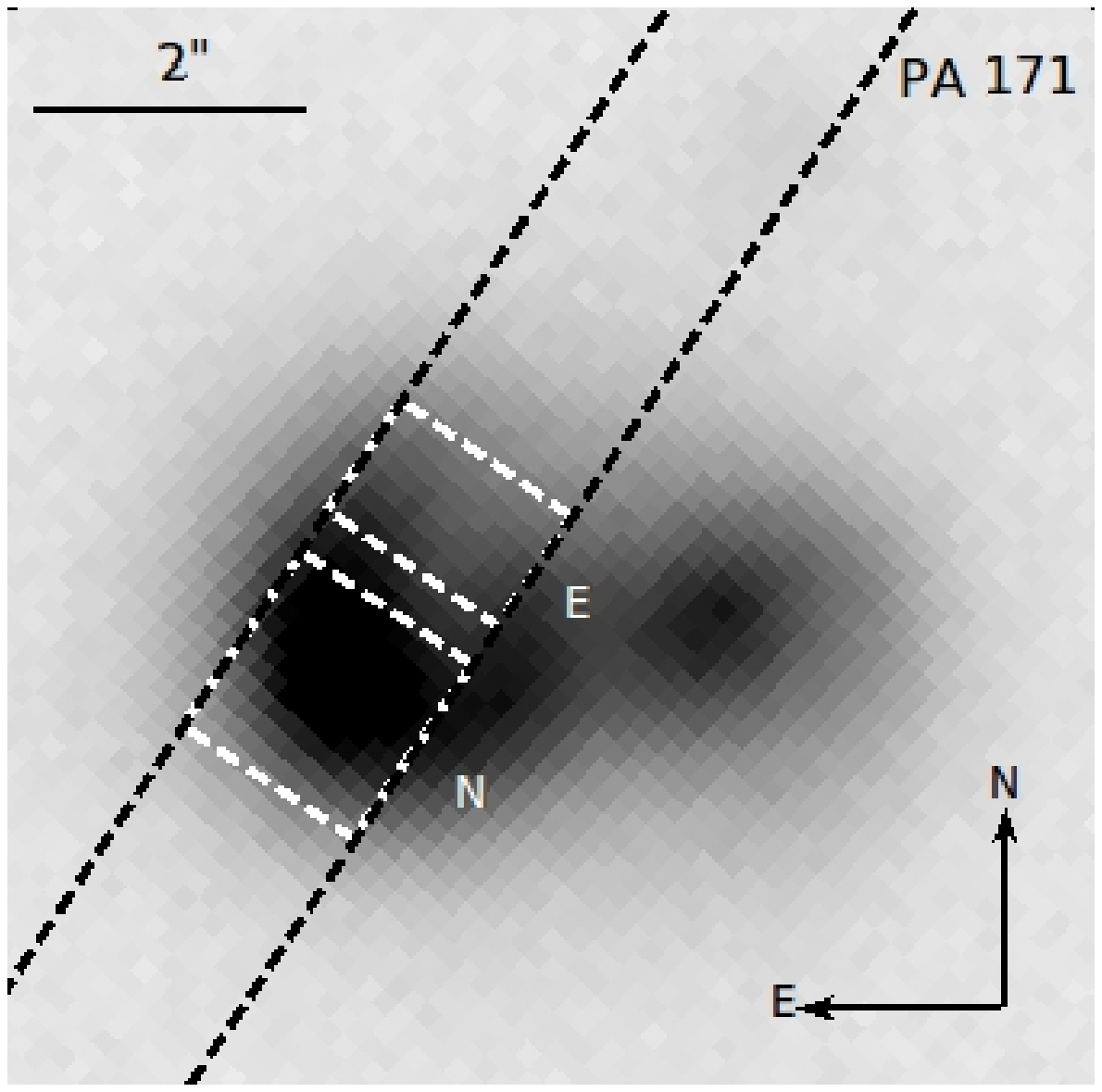}
\label{q0218OnImSlit}}
\subfloat{
\includegraphics[trim = 5mm 7mm 7mm 7mm, width = 3.9cm, height = 3.9cm]{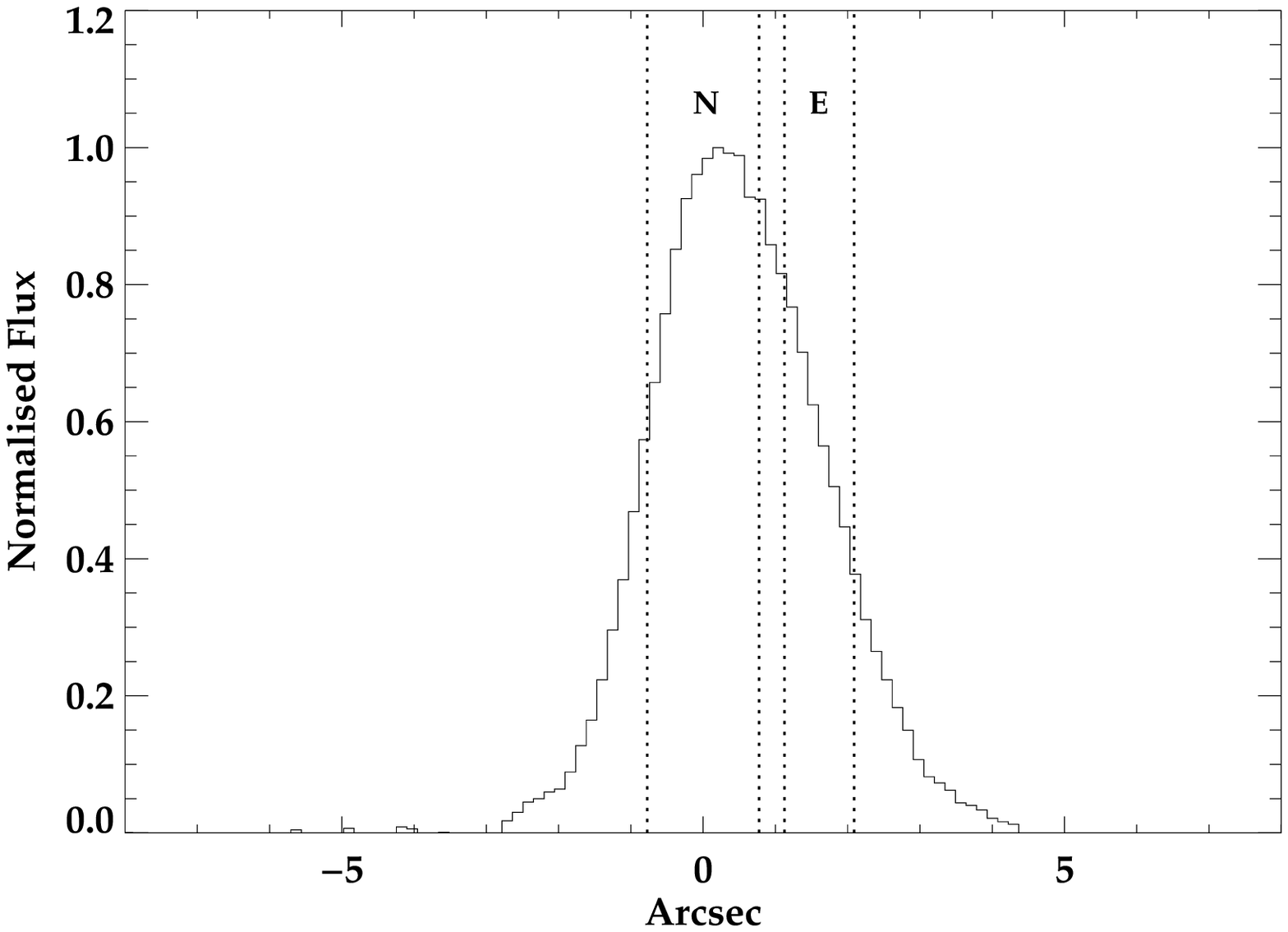}
\label{q0218-ysec}}
\caption{The left panel shows our Gemini GMOS-S image of J0218-00 with the spectroscopic slit overlayed. The positions of the two extracted apertures are also shown. The right panel shows a cross-section of the 2D spectrum of J0218-00 showing the position of each extracted aperture. The cross-section was extracted at an observed wavelength of 6450--6600 \AA\ (4700--4810 \AA\ rest-frame.)}
\label{apPos}
\end{figure}

J0218-00 is a complex system, with two distinct nuclei which have a projected separation of $\sim 12 \mbox{kpc}$ \citep{bessiere12}. This interaction has left the quasar host galaxy in what appears to be a ring-like configuration. The spectroscopic slit does not pass through both nuclei, yet it was still possible to extract two distinct apertures from the quasar host galaxy. The main aperture is centred on the region containing the active nucleus (with a spatial scale of 8 kpc) and a second, narrower aperture is centred on an extended region of the ring like structure (with a spatial scale of 5 kpc). The approximate positions of the apertures is shown in Figure \ref{apPos} where they have been overlayed on the image. A cross-section of the 2D spectrum is also shown, showing the exact position of each aperture. The centres of the two apertures are separated by $\sim 8.2$ kpc. Each aperture was modelled independently, and the results are presented separately below.

\subsubsection{Nuclear aperture}
\label{spec:q0218:8kpc}
The initial fit for the nuclear aperture was performed using combination 2, and a best fitting model, which included a YSP $\tysp = 0.03$ Gyr and $\ebv = 0.1$, was selected. This was used to generate a nebular continuum which contributes $\sim 16 \%$ of the total flux in the 3540--3640 \AA\ range. The \hg/\hb\ ratio was measured to be 0.37 \pms 0.02 (Table \ref{blines}), and the reddening derived from this ratio was initially applied to the nebular continuum before subtraction. However, this resulted in an under-subtraction in the region of the Balmer edge, and therefore, an unreddened nebular continuum was subtracted instead, which produced a reasonable result. The failure of the technique in this case is most likely due to the relatively low equivalent width of the Balmer emission lines and strong underlying Balmer absorption lines. It is probably the case that the true value of reddening is somewhere between these extreme cases. However, it is difficult to judge what the most appropriate value is by visual inspection alone. 

\begin{figure}	
\centering
\subfloat{
\includegraphics[scale = 0.45, trim = 7mm 0mm 0mm 0mm]{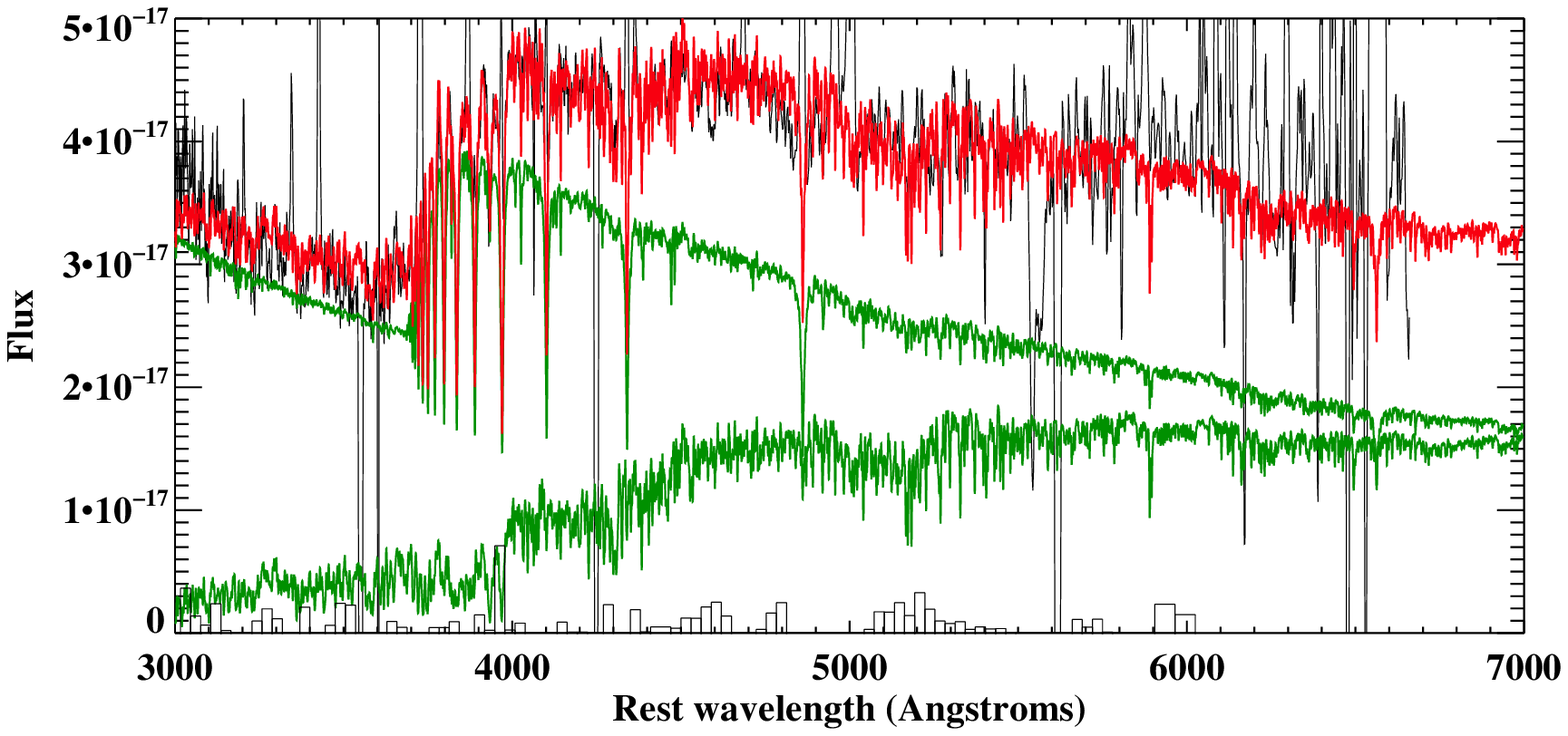}
\label{q0218-8kpc-8np-b-bf}}

\subfloat{
\includegraphics[scale = 0.45, trim = 7mm 0mm 0mm 0mm]{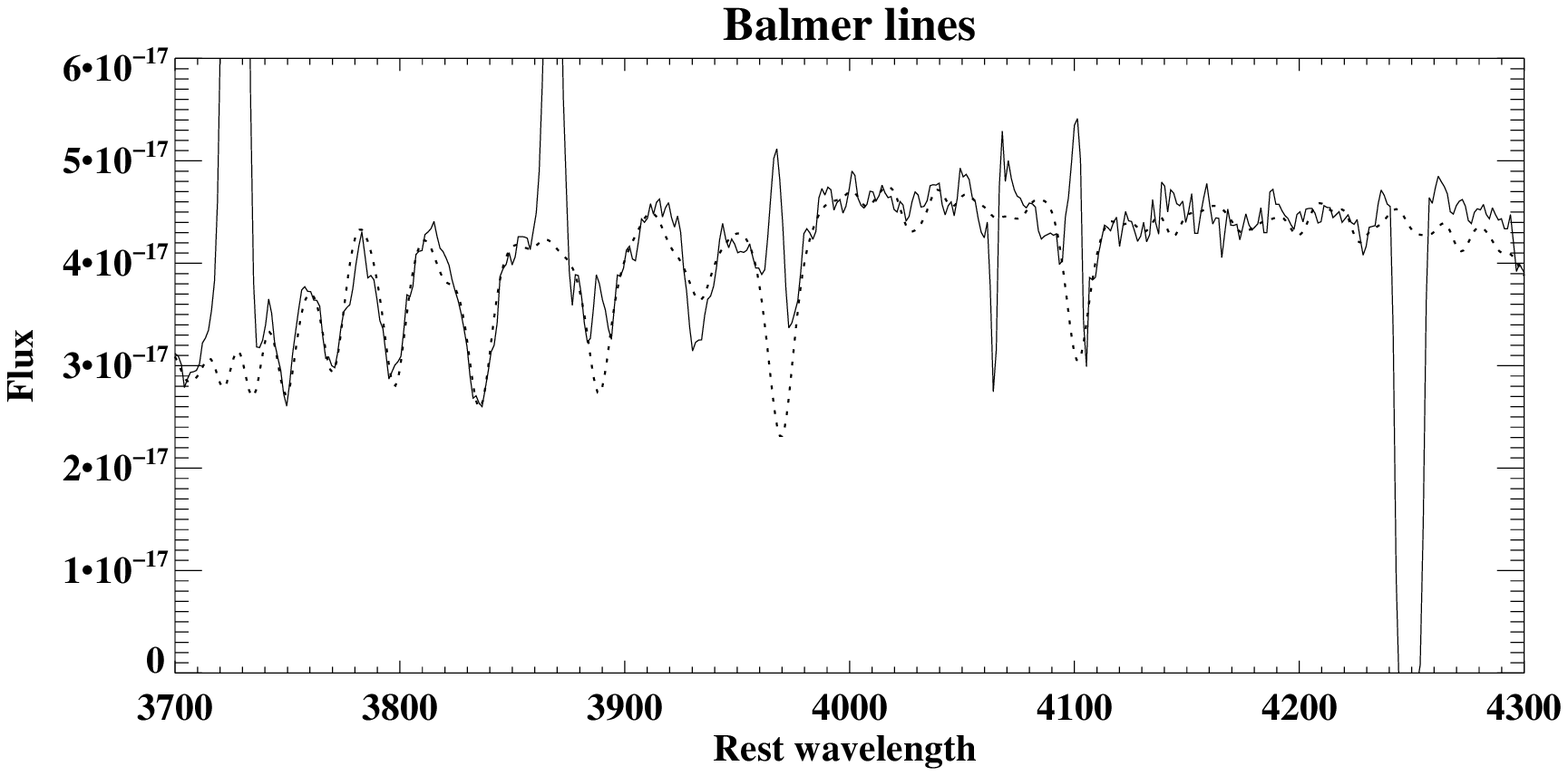}
\label{q0218-8kpb-8np-b-bl}}

\caption{An example of an acceptable fit produced by \confit for J0218-00 -- Nuclear aperture. This fit was generated using combination 2 and includes a YSP with $\tysp = 0.04$ Gyr and $\ebv = 0.2$.}
\label{spec:q0218:8kpc_confit}
\end{figure}

All five combinations of components were run for the nuclear aperture, the results of which are given in Tables \ref{results8} and \ref{results2}. These result show that, when no power-law component is included, it was only possible to achieve acceptable fits for combination 2. Combinations that included a power-law component and a YSP all produced acceptable fits (combinations 4 \& 5). Figure \ref {spec:q0218:8kpc_confit} shows an example of an acceptable fit produced by combination 2 which has a YSP of $\tysp = 0.04$ Gyr and $\ebv = 0.2$. 

\subsubsection{Extended aperture}
\label{spec:q0218:5kpc}

The initial modelling for the extended aperture extracted from J0218-00 was performed using combination 2, which resulted in a best-fitting model with $\tysp = 0.03$ and $\ebv = 0.2$. This is very similar in terms of the components used, and their contribution to the total flux as for the nuclear aperture. This resulted in a model nebular continuum which contributes $\sim 16\%$ of the total flux in the 3540\AA\ -- 3640 \AA\ range. The measured Balmer decrement produced a value of \hg/\hb\ = 0.32 \pms 0.02, suggesting that the nebular continuum is slightly more reddened in the extended region than in the nuclear region (see Table \ref{blines}). However, when the continuum was reddened before subtraction, the same issues was encountered as described for the nuclear aperture. Therefore, an unreddened continuum was subtracted before continuing with the full modelling process.

\begin{figure}
\centering
\subfloat{
\includegraphics[scale = 0.45, trim = 7mm 0mm 0mm 0mm]{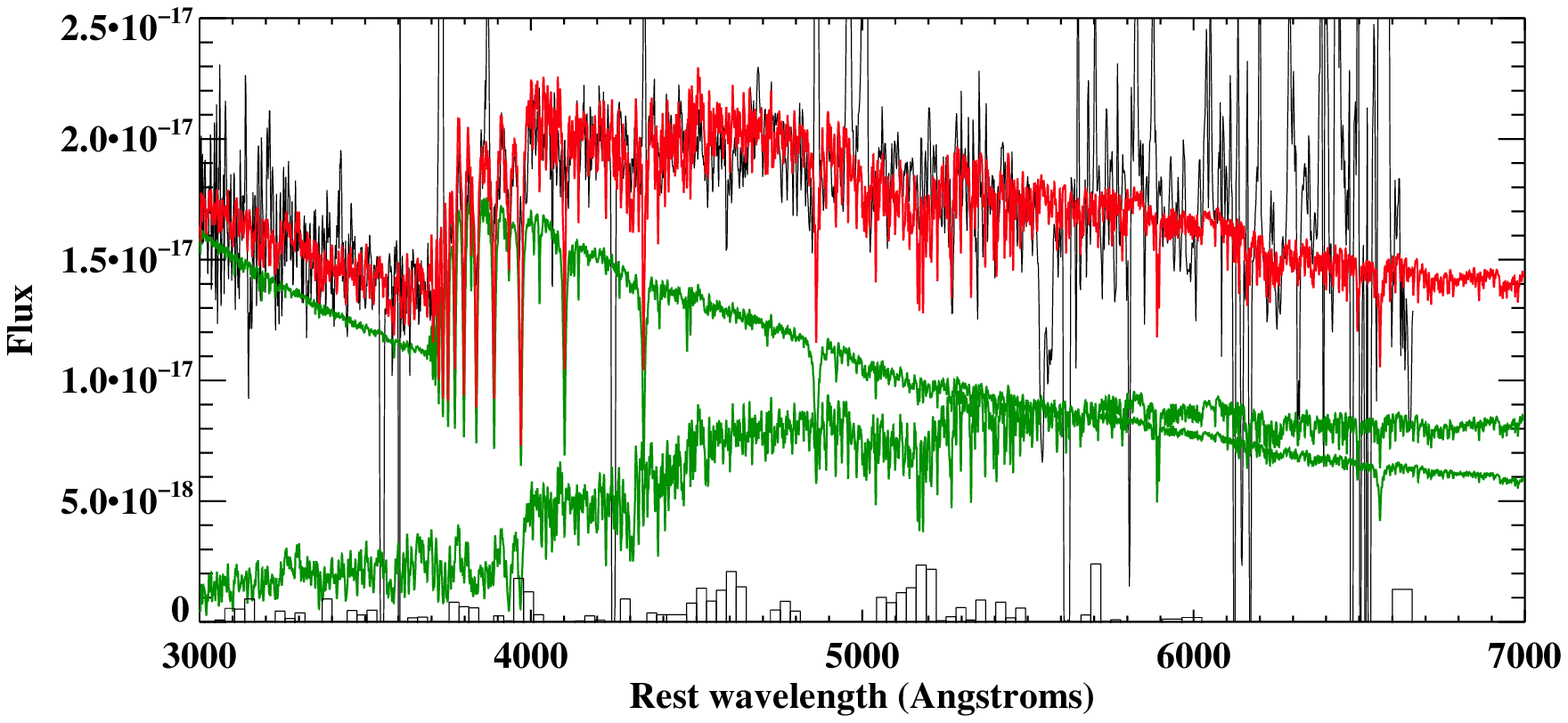}
\label{q0218-8np-b-5kpc-bf}}

\subfloat{
\includegraphics[scale = 0.45, trim = 7mm 0mm 0mm 0mm]{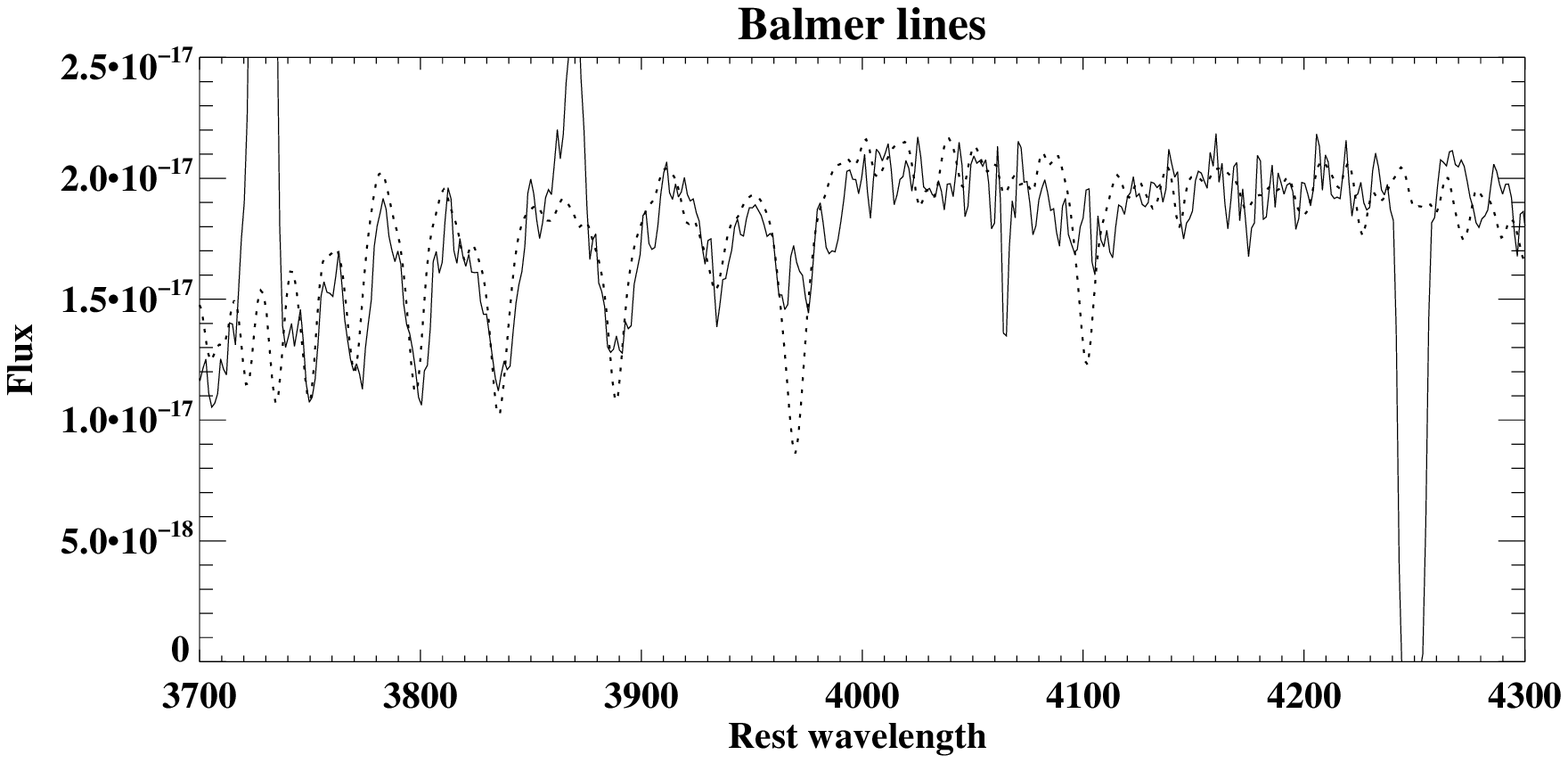}
\label{q0218-8np-b-5kpc-bl}}

\caption{An example of an acceptable fit produced by \confit for J0218-00 -- Extended aperture. This model was generated using combination 2 and includes a YSP with \tysp = 0.04 and \ebv = 0.1.}
\label{spec:q02185kpc_confit}
\end{figure}

The extended aperture is centred on a non-active, extended region of the galaxy and therefore, only combinations 2 and 3 (no power-law component) were used for the modelling process. Figure \ref{spec:q02185kpc_confit} shows an example of an acceptable fit for combination 2, which includes a YSP \tysp\ = 0.04 Gyr and \ebv\ = 0.1. It can be seen by comparison with Figure \ref{spec:q0218:8kpc_confit} that the overall shape and the age sensitive absorption lines in the region of the higher order Balmer lines are very similar, which may suggest that the burst of star-formation induced by this obvious merger was initiated across the entire galaxy.

\subsection{J0227+01}
\label{spec:q0227}

The initial modelling for J0227+01 was performed using combination 2, which resulted in a best fitting model with $\tysp = 0.04~ \mbox{Gyr}$ and $\ebv = 0.6$.  This resulted in a nebular continuum model that contributes $\sim 11 \%$ of the total flux  just below the Balmer edge (Table \ref{blines}). Unsurprisingly, with such a comparatively weak nebular continuum, both the \hg\ and \hd\ emission lines are weak, preventing a reliable measurement of the nebular reddening, via the Balmer decrements, being made. In light of this, it was not possible to apply a reddening correction to the nebular component before subtraction. However, the resulting nebular subtracted spectrum produced an acceptable result.

\begin{figure}
\centering
\subfloat{
\includegraphics[scale = 0.45, trim = 7mm 0mm 0mm 0mm]{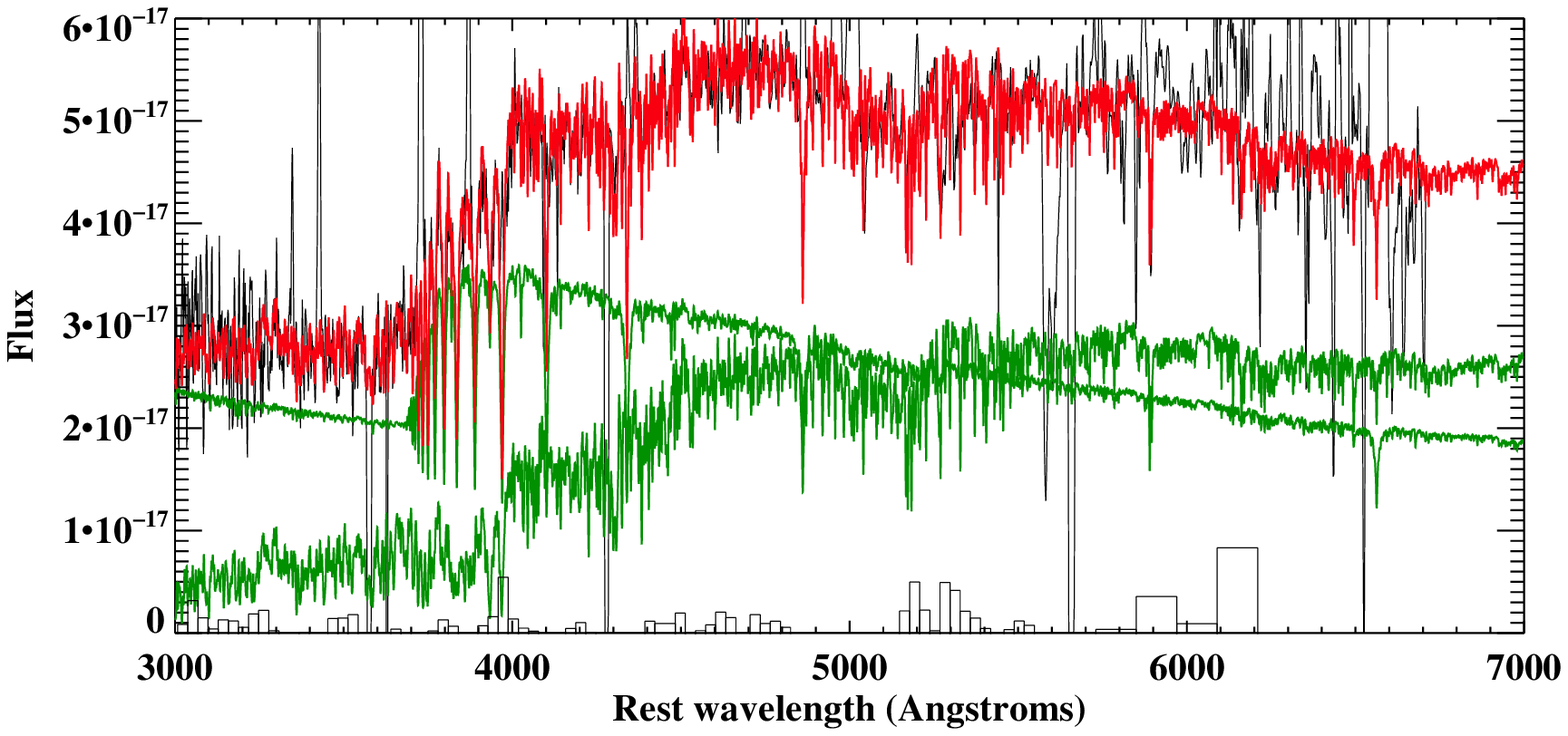}
\label{q0227-8np-b-bf}}

\subfloat{
\includegraphics[scale = 0.45, trim = 7mm 0mm 0mm 0mm]{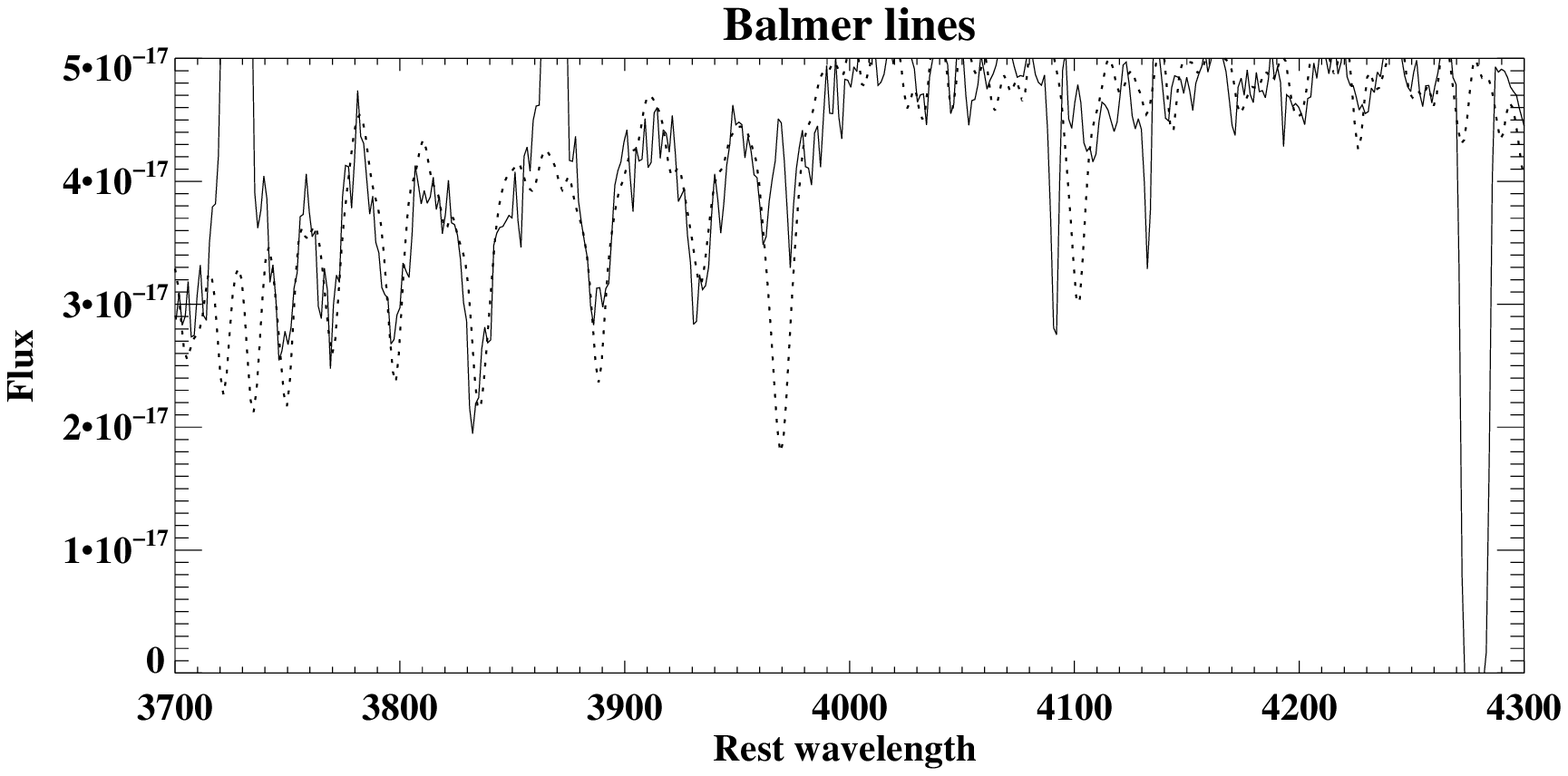}
\label{q0227-8np-b-bl}}

\caption{An example of an acceptable fit produced by \confit forJ0227+01. This fit was generated using combination 2 and consists of and 8 Gyr OSP combined with a YSP of \tysp = 0.05 Gyr and \ebv = 0.3.}
\label{spec:q0227_confit}
\end{figure}

All five combination of components were then run for J0227+01, and the results, presented in Tables \ref{results8} and \ref{results2}, show that the full range of combinations which include a YSP produce an acceptable fit. When combinations which do not include a power-law component are used, the range of ages that produce acceptable fits is $0.03 \le \tysp \le 0.1 $, with reddenings in the range $ 0.1 \le \ebv \le 0.5$ and the YSP component contributing up to 85\% of the flux in the normalising bin. When a power-law component is also included, the ages move to older values ($0.09 \le \tysp \le 0.2$) and a larger range of reddenings is allowed ($0 \le \ebv \le 0.6$). Figure \ref{spec:q0227_confit} shows an example of an acceptable fit, using combination 2, which includes a YSP with $\tysp = \mbox{0.05}$ Gyr and $\ebv = 0.3$. 

\subsection{J0234-07}
\label{spec:j0234}

The initial modelling for J0234-07 was performed using combination 2 and resulted in a wide range of possible solutions that have \chisqlt. An inspection of the overall shape of the continuum and regions around age sensitive stellar absorption lines showed the best fit to have a YSP \tysp\ = 0.001 and \ebv\ = 0.1. However, Table \ref{results8} shows that the OSP dominates the flux for all other possible solutions for combination 2 and therefore, the resulting stellar model is not strongly dependent on the choice of YSP component. A nebular continuum model that contributes $\sim 24\%$ of the flux in the 3540--3640 \AA\ range was constructed. Initially, to remain consistent with the technique of maximum nebular subtraction applied in the majority of cases, no reddening was applied to the nebular continuum before subtraction. However, this resulted in a step in the resulting stellar continuum at 3646 \AA. To resolve this issue, reddening of \ebv\ = 0.25 was applied to the nebular continuum before subtraction (Table \ref{blines}), which produced a much improved result.

\begin{figure}
\centering
\subfloat{
\includegraphics[scale = 0.45, trim = 7mm 0mm 0mm 0mm]{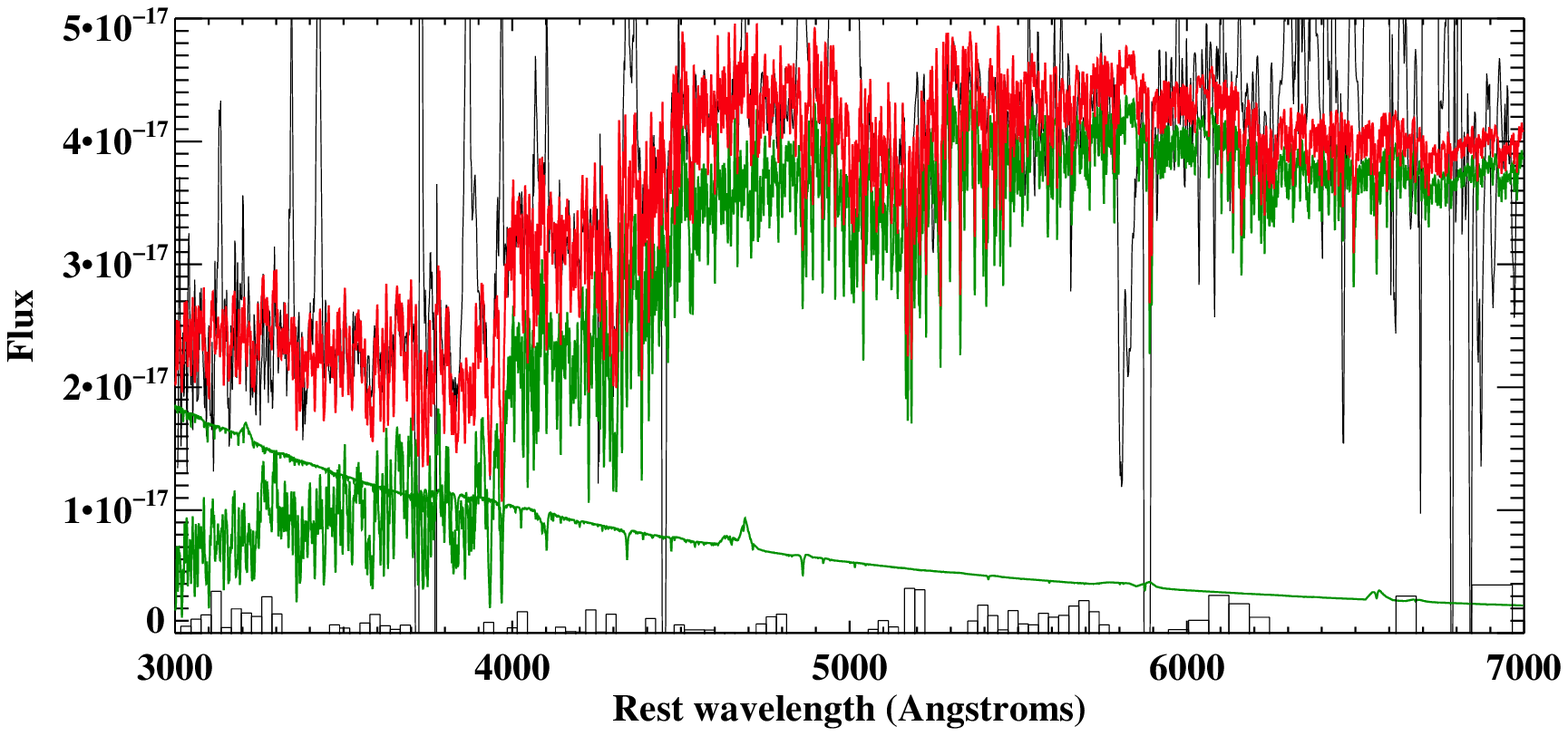}
\label{q0234-8np-b-bf}}

\subfloat{
\includegraphics[scale = 0.45, trim = 7mm 0mm 0mm 0mm]{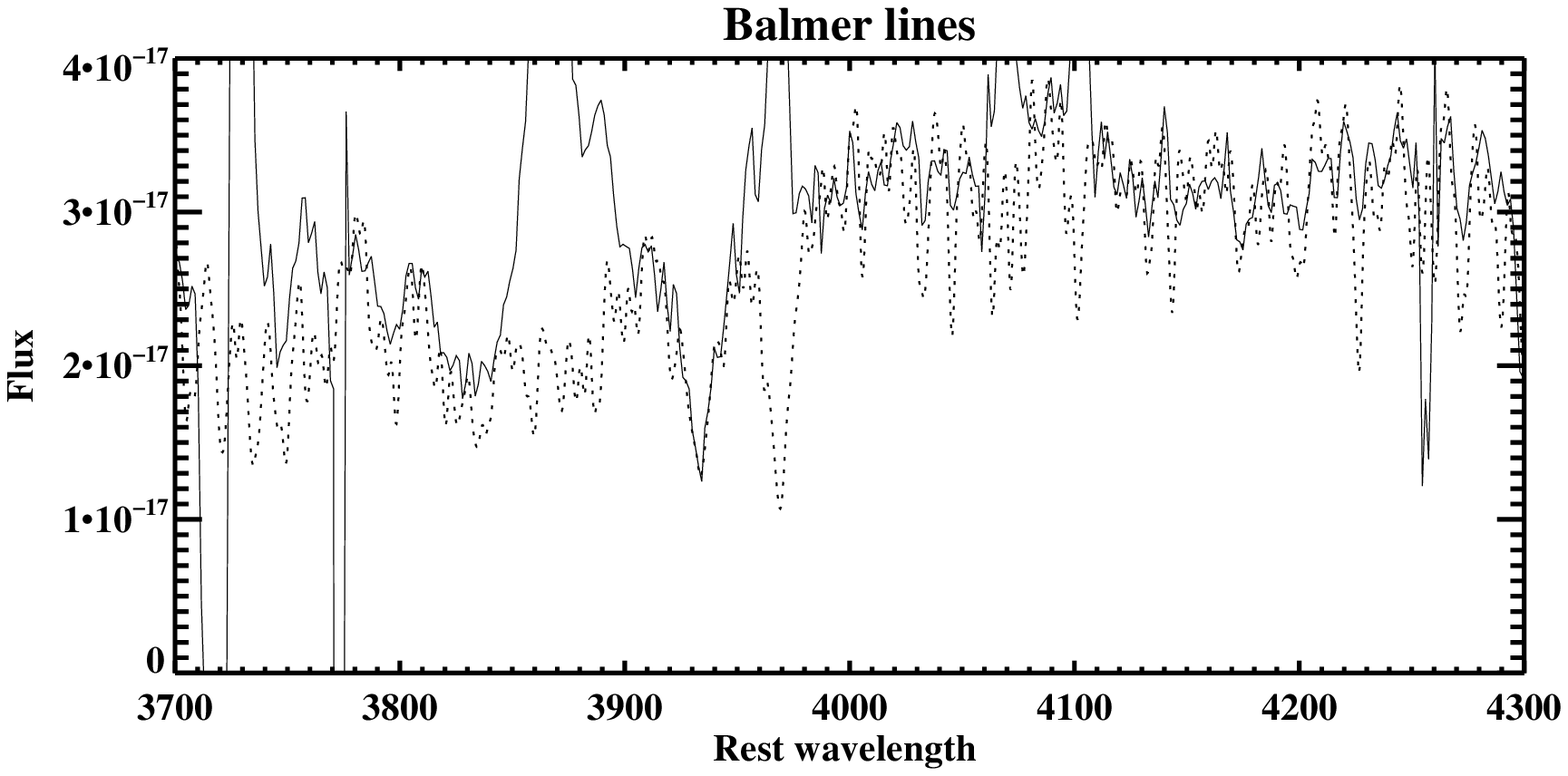}
\label{q0234-8np-b-bl}}

\caption{An example of an acceptable fit produced by \confit for J0234-07. The fit shown here was generated using combination 2 and includes a YSP with \tysp = 0.003 Gyr and \ebv= 0.2.}
\label{spec:q0234_confit}
\end{figure}

All five combinations were run, and all but combination 3 produce acceptable results. It should be noted however that all possible YSP and reddening values for combination 4 produce acceptable results, and therefore, do not allow us to discriminate between solutions. In addition to combinations which include a YSP, J0234-07 can also be fit using only an 8 Gyr, unreddened population combined with a power-law component, making it unclear whether a YSP component is at all necessary. Tables  \ref{results8} and \ref{results2} show that the combinations which do not include a power-law component, tend to include very young or young components at a low level of contribution to the total flux in the normalising bin. This means that in these cases, the VYSP/YSP could, in fact, be taking the form of a power-law, as illustrated in Figure \ref{spec:q0234_confit}, where the VYSP ($\tysp\ = 0.003$, $\ebv\ =0.2$) has an approximately power-law form. 

With such a wide range of possible solutions and such young ages producing acceptable fits to these data, as well as an 8 Gyr plus power-law combination, it is not possible to discriminate between the various possibilities. However, there is little evidence for the Balmer absorption lines, associated with young stars (Figure \ref{q0234-8np-b-bl}), although these are also weak at the youngest model ages.



\subsection{J0249-00}
\label{spec:q0249}

In order to produce a stellar model, the initial run was performed using combination 2. This resulted in a best fitting model with \tysp\ = 0.03 and \ebv\ = 0.7 and produced a model nebular continuum which contributes $\sim 25\%$ of the flux in the 3540--3640\AA\  range. No reddening was applied to this continuum because, as is shown in Table \ref{blines}, the uncertainty associated with measuring the \hg/\hb\ ratio is large in comparison to the measured value, and thus is consistent with reddenings of $ 0 \le \ebv \le 1.7$. Therefore, because the value of reddening is poorly constrained, and also because subtracting an unreddened nebular continuum provides an acceptable result, a maximum nebular subtraction was performed before proceeding with the fitting.

\begin{figure}
\centering
\subfloat{
\includegraphics[scale = 0.45, trim = 7mm 0mm 0mm 0mm]{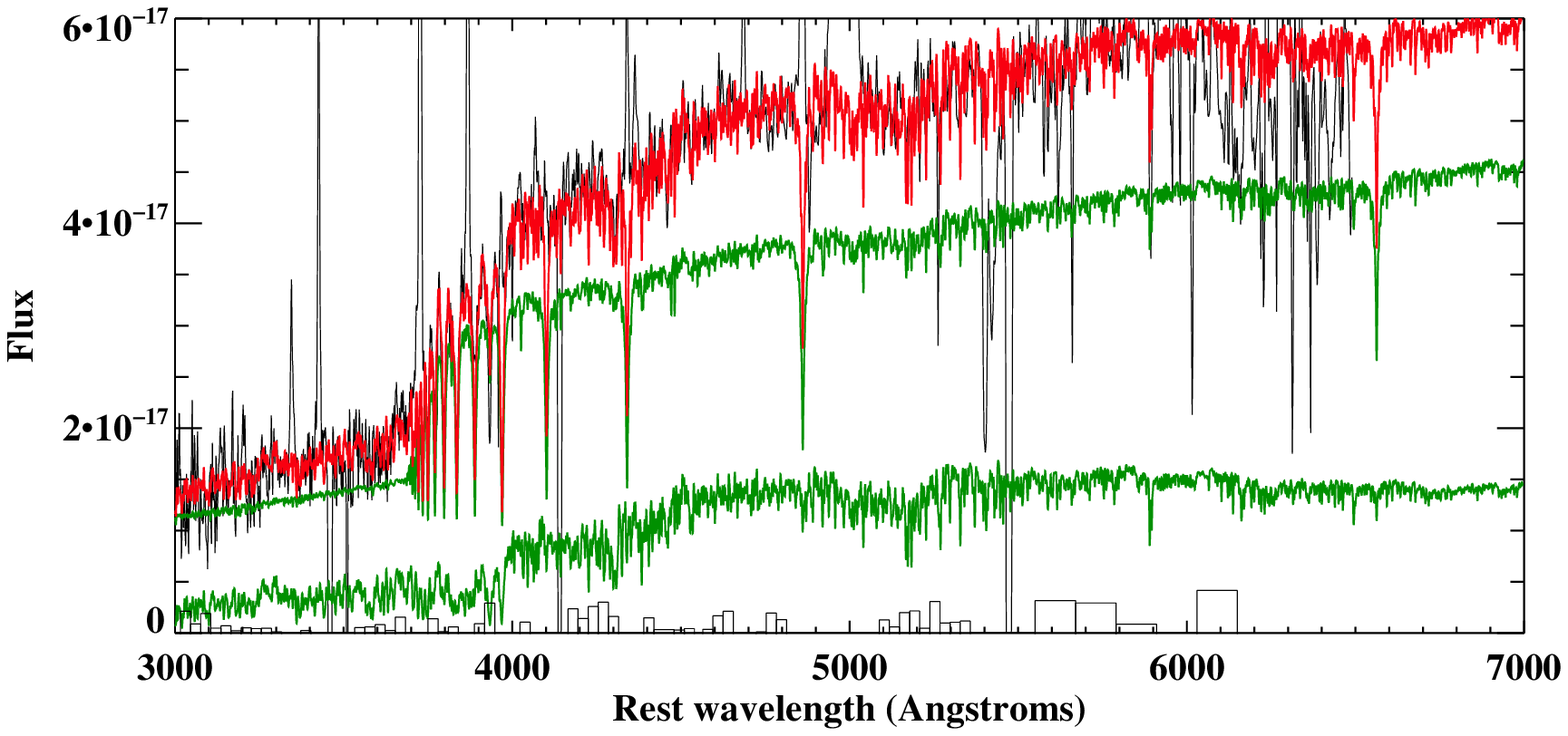}
\label{q0249-8np-b-bf}}

\subfloat{
\includegraphics[scale = 0.45, trim = 7mm 0mm 0mm 0mm]{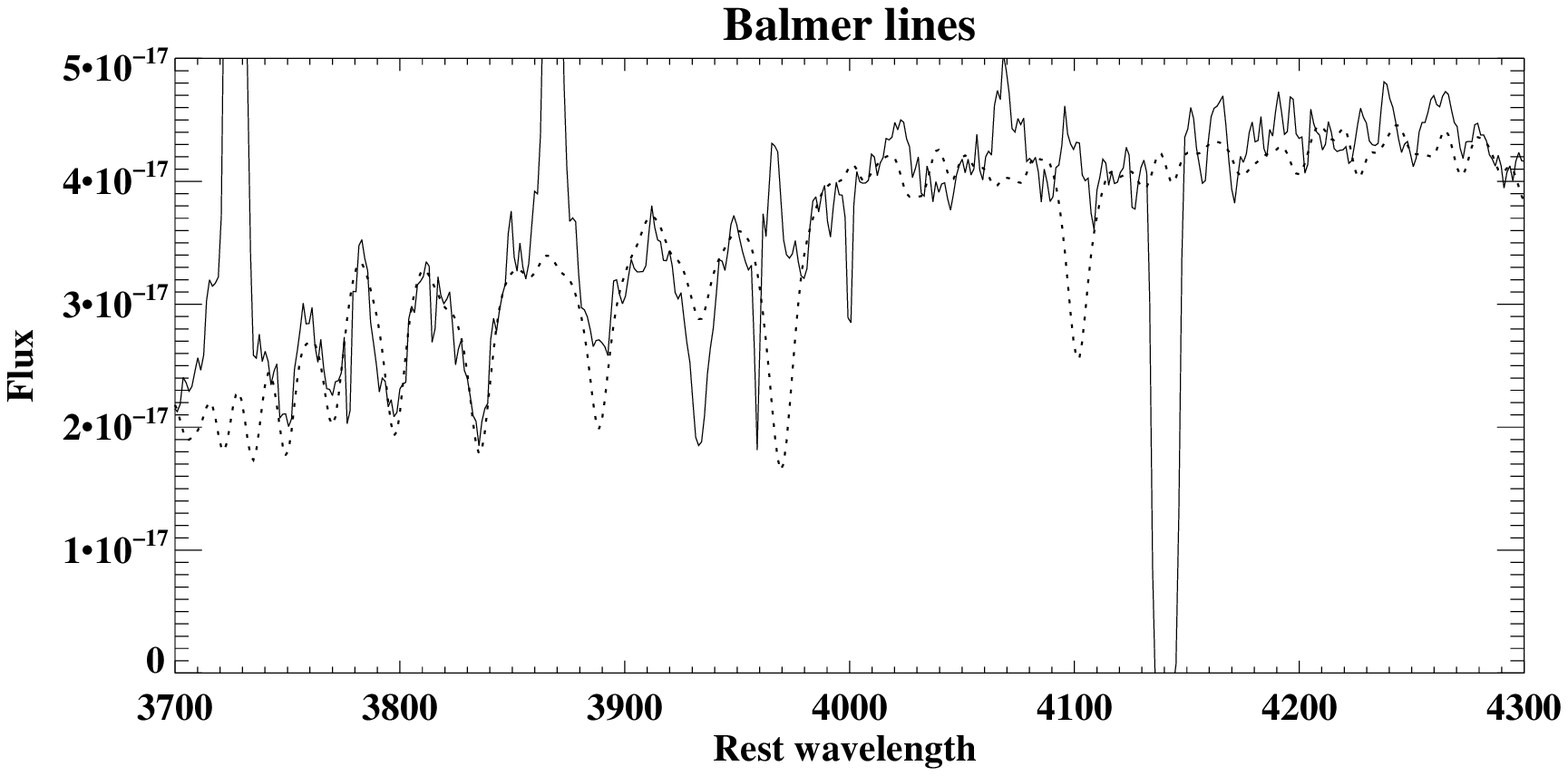}
\label{q0249-8np-b-bl}}

\caption{An example of an acceptable fit produced by \confit for J0249-00. The fit shown here was generated by combination 2 and shows a YSP with $\tysp=0.07$ Gyr and $\ebv=0.7$.}

\label{spec:q0249_confit}
\end{figure}

Of the five combinations of components, all those which include a YSP produce acceptable results. Tables \ref{results8} and \ref{results2} show that, in the cases of combinations 2 and 3, the ages and reddenings allowed are broadly similar, with combination 3 allowing a slightly older age and larger range of reddenings. As is typical, when a power-law component is included, significantly older ages provide acceptable fits at a wider range of reddenings. Figure \ref{spec:q0249_confit} shows an example of an acceptable fit using combination 2, and includes a YSP of $\tysp\ = 0.07$ Gyr and $\ebv\ = 0.7$.

\subsection{J0320+00}

The initial modelling for J0320+00 was performed using combination 2, and the resulting best-fit model includes a YSP of \tysp\ = 0.02 Gyr and \ebv\ = 0.1 and a model nebular continuum which contributes $\sim 14 \%$ of the flux between in the 3540\AA\ -- 3640\AA\ range was generated. The Balmer decrements were measured using \hb\ and \hg\ (Table \ref{blines}) but because of the large uncertainty associated with the measured value, which allows for reddenings in the range $ 0 \le \ebv \le 1.3$, and the fact that the maximum subtraction produces a good result around the Balmer edge, no reddening correction has been applied to the nebular continuum model before subtraction.

\begin{figure}
\centering
\subfloat{
\includegraphics[scale = 0.45, trim = 7mm 0mm 0mm 0mm]{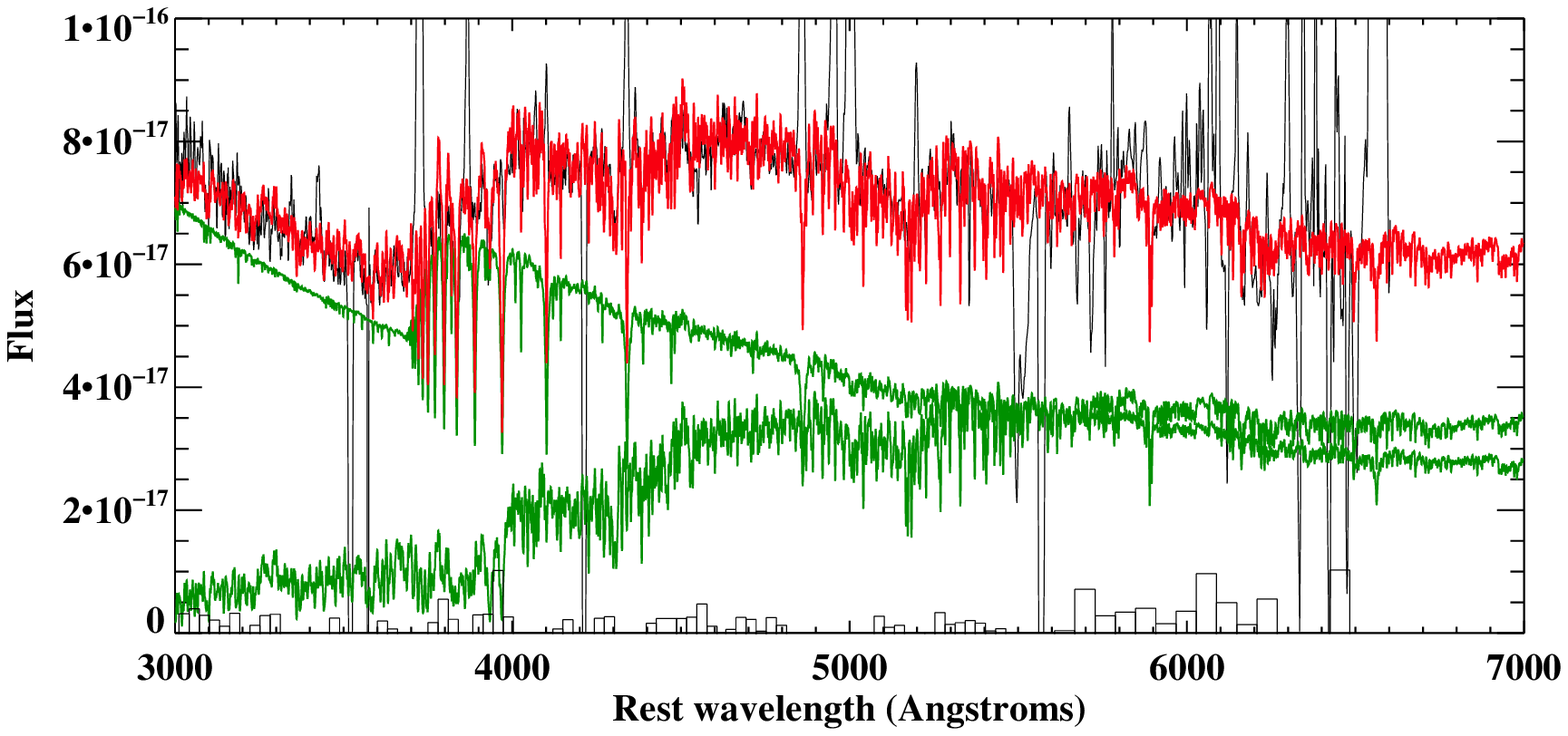}
\label{q0320-8np-b-bf}}

\subfloat{
\includegraphics[scale = 0.45, trim = 7mm 0mm 0mm 0mm]{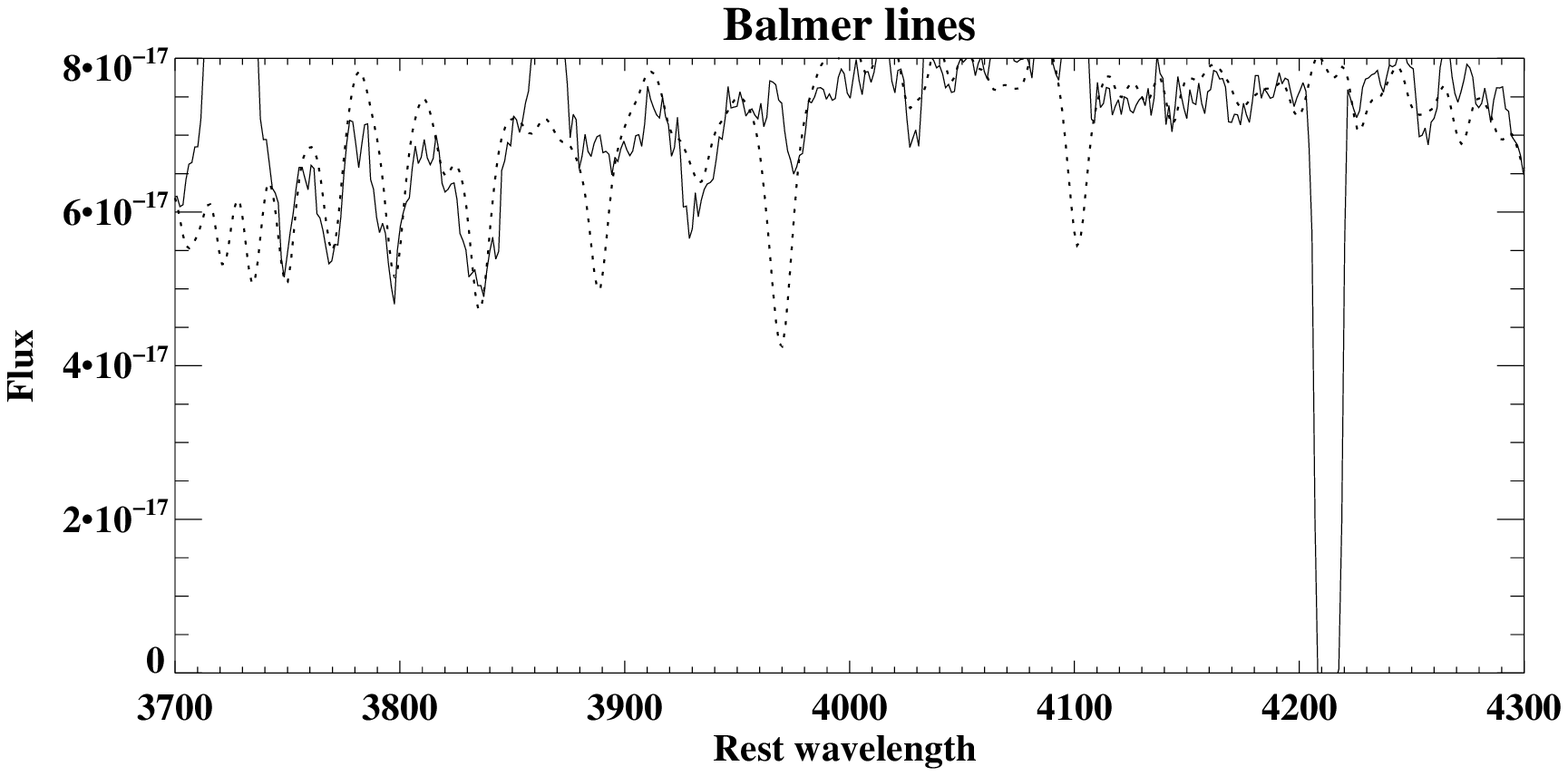}
\label{q0320-8np-b-bl}}

\caption{An example of an acceptable fit produced by \confit for J0320-00. The fit shown here was generated using combination 2, and includes a YSP with \tysp\ = 0.02 and \ebv\ = 0.2.}
\label{spec:q0320_confit}
\end{figure}

The results of running combinations 2--5 are presented in Tables \ref{results8} and \ref{results2}, which show that only combination 3, which includes an underlying 2 Gyr ISP, does not provide acceptable solutions. Columns 5 and 6 of Tables \ref{results8} and \ref{results2} demonstrate that, when no power-law is included in the model, the YSP contributes a slightly higher proportion of the flux than the OSP, and that when a power-law component is included, this difference becomes more pronounced. Figure \ref{spec:q0320_confit} shows an example of an acceptable fit using combination 2, including a YSP of \tysp\ = 0.02 and \ebv\ = 0.2.


\subsection{J0332-00}
\label{spec:q0332}

Although this object was originally classified by \citet{zakamska03} as a type II quasar, the very broad \ha\ component detected in  the Gemini \gmos\ spectrum presented in Figure \ref{all_spec} clearly shows that J0332-00 is, in fact, a type I AGN. This finding is confirmed by \citet{barth14}, who also find clear evidence for a broad \ha\ component in their data. Despite this, an attempt was made to fit the stellar populations in the same manner as the other objects presented in this section. 

\begin{figure}	
\centering
\subfloat{
\includegraphics[width = 3.9cm, height = 3.9cm]{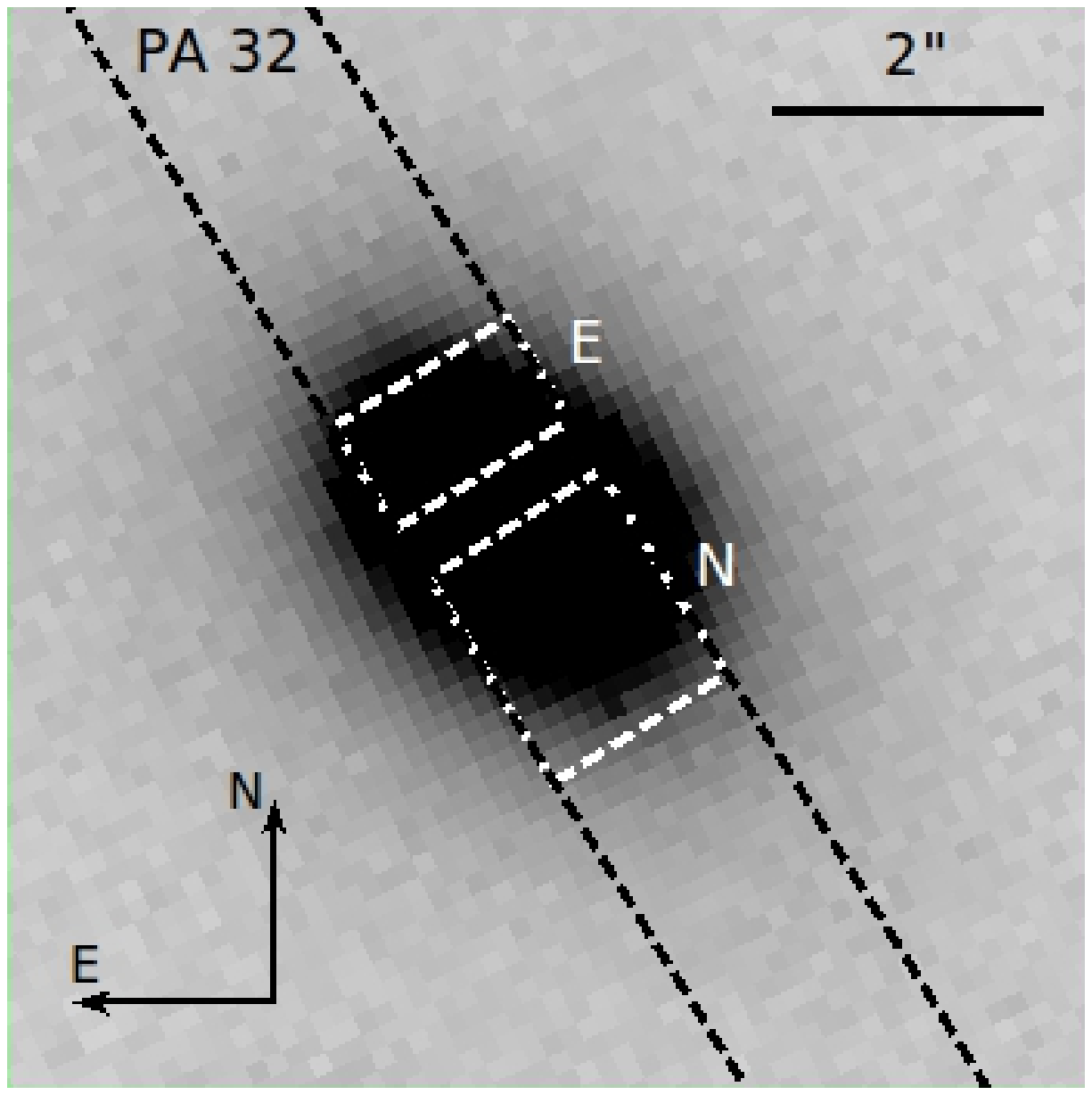}
\label{q0332OnImSlit}}
\subfloat{
\includegraphics[trim = 5mm 7mm 7mm 7mm, width = 3.9cm, height = 3.9cm]{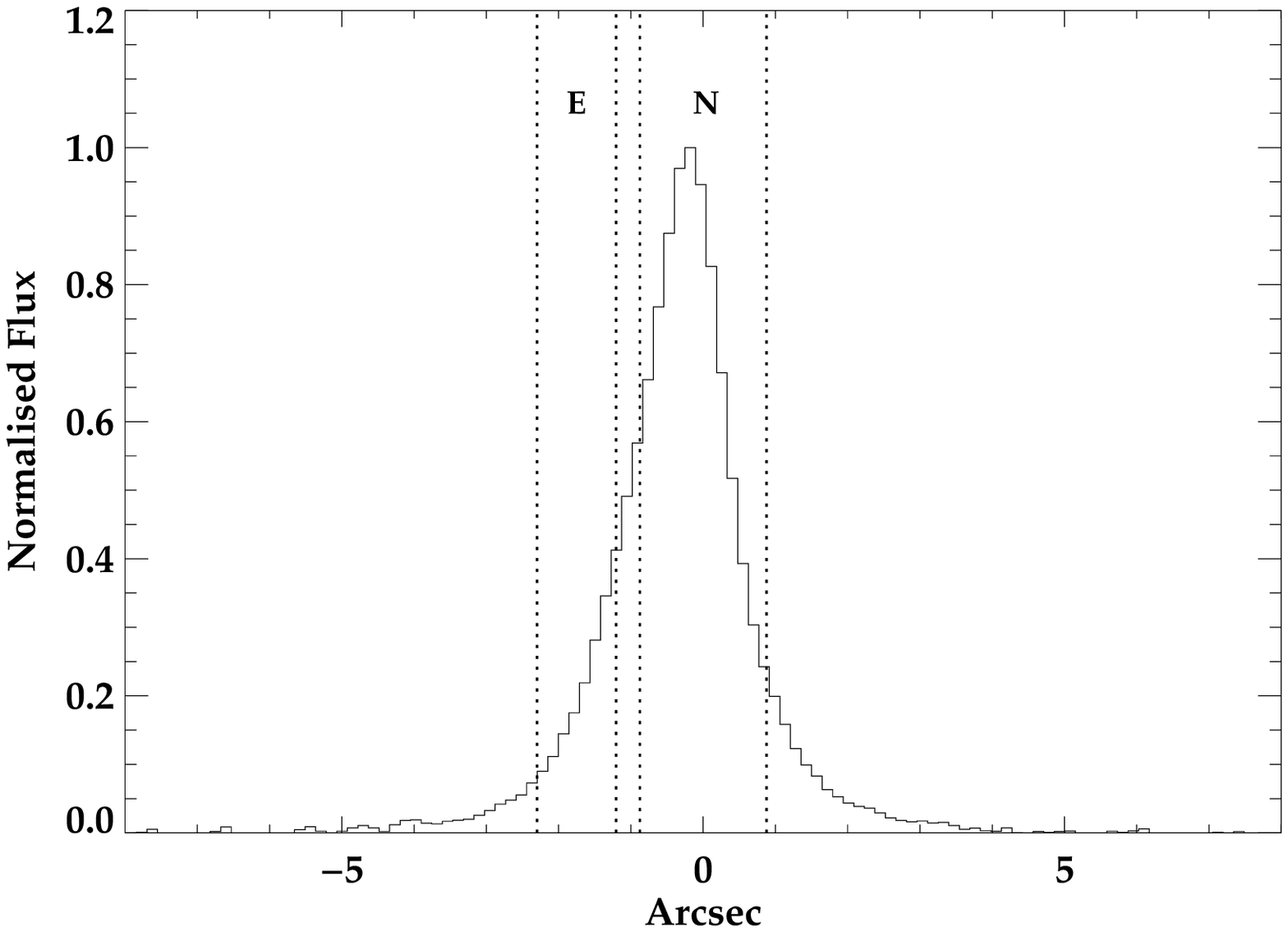}
\label{q0332-ysec}}
\caption{The left panel shows our Gemini GMOS-S image of J0332-00 with the spectroscopic slit overlayed (PA 32). The positions of the two extracted apertures are also shown. The right panel shows a cross-section of the 2D spectrum of J0332-00 showing the position of each extracted apertures. The cross-section was extracted at an observed wavelength of 6185--6215 \AA\ (4720--4745 \AA\ in the rest-frame). }
\label{q0332apPos}
\end{figure}

In \citet{bessiere12}, J0332-00 was classified as a double nucleus, with the two components separated by 4 kpc ($\sim 0.9\arcsec$). In this case, the 1.5 arcsec slit passes directly through both nuclei, allowing the extraction of two distinct spatial apertures centred on each of the components (i.e. the quasar host galaxy and the companion nucleus). Figure \ref{q0332apPos} shows the positioning of the spectroscopic slit overlayed on our Gemini image, with the positions of the two extracted apertures highlighted. The figure also shows a cross-section of the 2D spectrum with the position of both apertures highlighted. The centres of the two apertures are separated by $\sim 8.0$ kpc. Each aperture was modelled independently and discussed individually below.

\subsubsection{Quasar nucleus}
\label{spec:q0332-8kpc}

\begin{figure}
\centering
\subfloat{
\includegraphics[scale = 0.45, trim = 7mm 0mm 0mm 0mm]{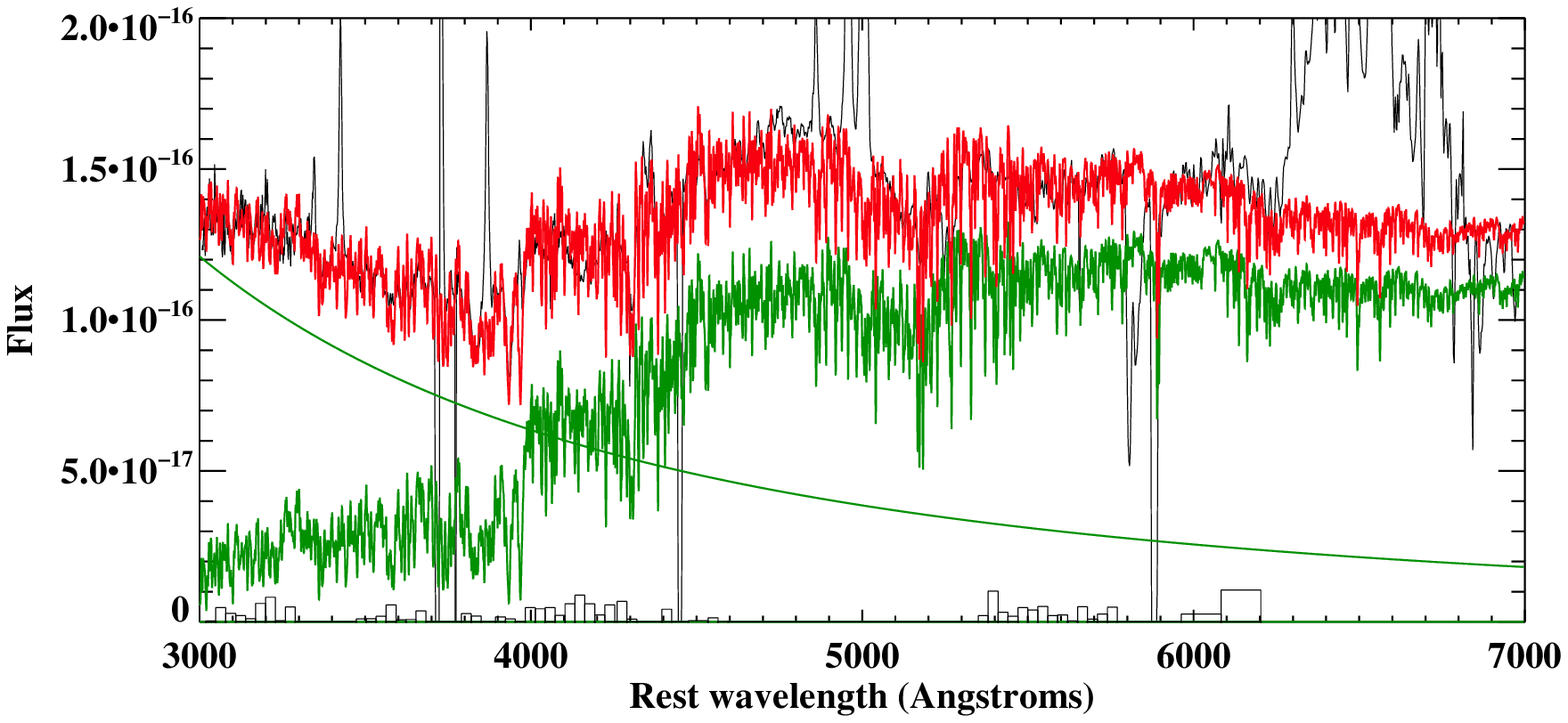}
\label{q0332-8kpc-8p-b-bf}}

\subfloat{
\includegraphics[scale = 0.45, trim = 7mm 0mm 0mm 0mm]{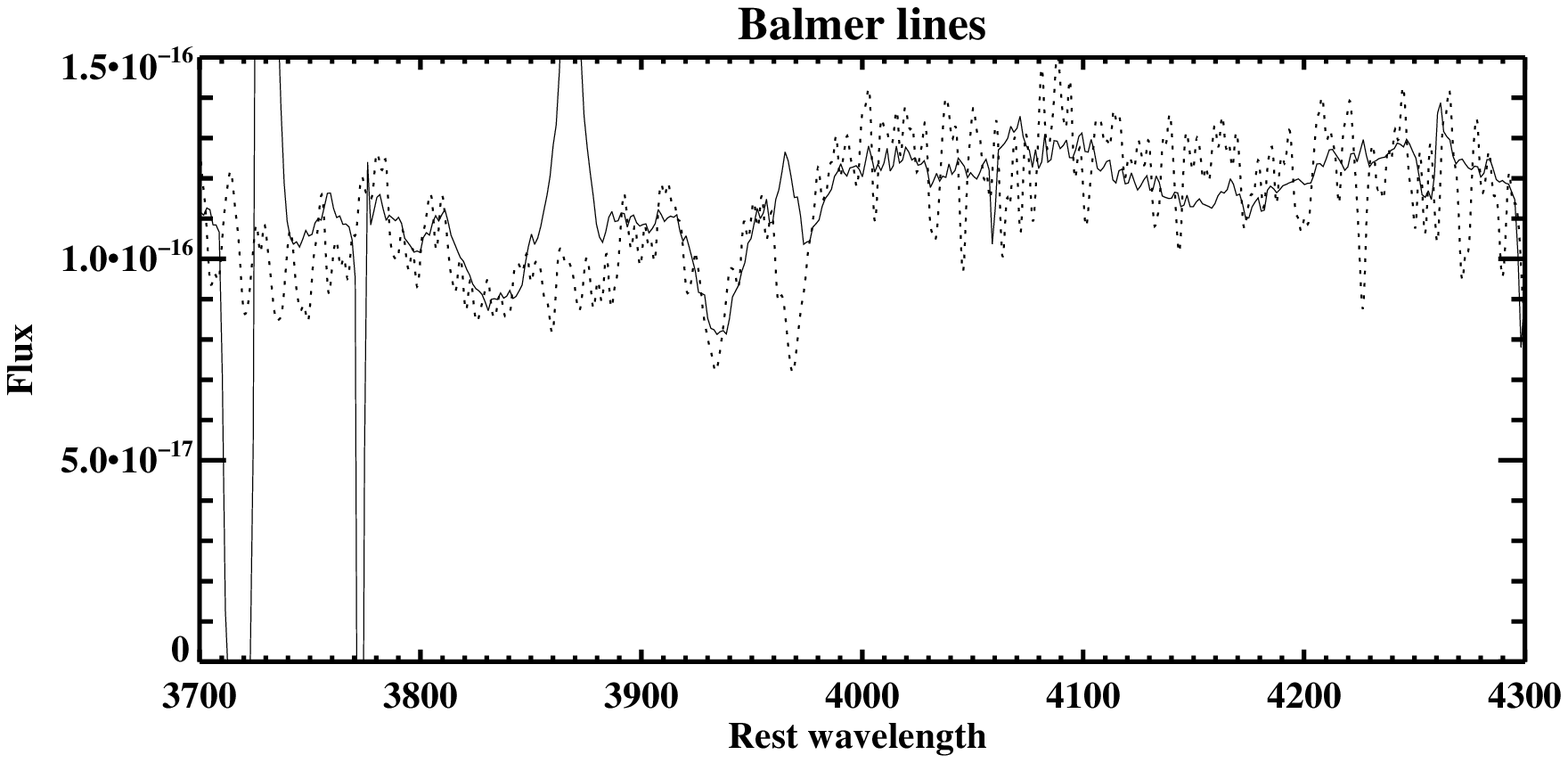}
\label{q0332-8kpc-8p-b-bl}}

\caption{An example of an acceptable fit produced by \confit for J0332-00 -- Quasar Host Galaxy. The fit shown here was produced using combination 1 and comprises an 8 Gyr population and a power-law component with $\alpha = -2.24$.  }
\label{spec:q0332-8kpc_confit}
\end{figure}

It is interesting to note that the quasar appears to be relatively unreddened given the slope of the power-law component ($\alpha = -2.24$)  in the combination 1 fit shown in Figure \ref{spec:q0332-8kpc_confit} (top panel). This would imply that, if no YSP component is present, the quasar nucleus has a relatively low luminosity considering the high [OIII] luminosity. In order to determine if this is the case, an attempt to fit the broad \ha\ and \hb\ components has been made, in order to determine the Balmer decrements in the BLR. This is complicated by the fact that, due to the strong fringing and residuals from the sky subtraction, the spectrum is very noisy at wavelengths greater than 6000 \AA. This object also has a double peaked broad emission line profile which requires a minimum of two broad components, further complicating the fitting process. We measure the blue shifted component to have FWHM = $6683 ~\mbox{km s}^{-1} \pm 992 ~\mbox{km s}^{-1}$ and the redshifted component to have FWHM = $12873 ~\mbox{km s}^{-1} \pm 1427 ~\mbox{km s}^{-1}$. The Balmer decrements were derived for each of these components separately, and in both cases it was found that $\ha/\hb \sim 10$, which is substantially larger than expected for case B recombination. This is likely due to the poor quality of the data in the region of \ha, as well as the uncertainty in the measurement of the \hb\ flux, due to the associated uncertainty in the continuum subtraction. An alternative explanation may be that the BLR reddening is, in fact, large, however the shorter wavelengths may be dominated by the YSP. Figure \ref{spec:q0332-8kpc_confit} shows the fit generated by combination 1, and it can be seen that the combination of an 8 Gyr OSP and a power-law component provides and adequate fit. 

\subsubsection{Companion nucleus}
\label{spec:q0332-5kpc}

The initial modelling for the companion was performed using combination 2,  and resulted in a best fitting model which included an unreddened YSP with \tysp = 0.9 Gyr. Measuring the \hb\ flux from the resulting stellar-subtracted spectrum generated a nebular continuum model which contributes $\sim3\%$ of the flux just below the Balmer edge. This is consistent with the value found for the aperture centred on the quasar nucleus. Once again, it was not possible to obtain a reliable estimate for the Balmer decrements, and therefore, an unreddened nebular continuum was subtracted before proceeding with the stellar synthesis modelling.

\begin{figure}
\centering
\subfloat{
\includegraphics[scale = 0.45, trim = 7mm 0mm 0mm 0mm]{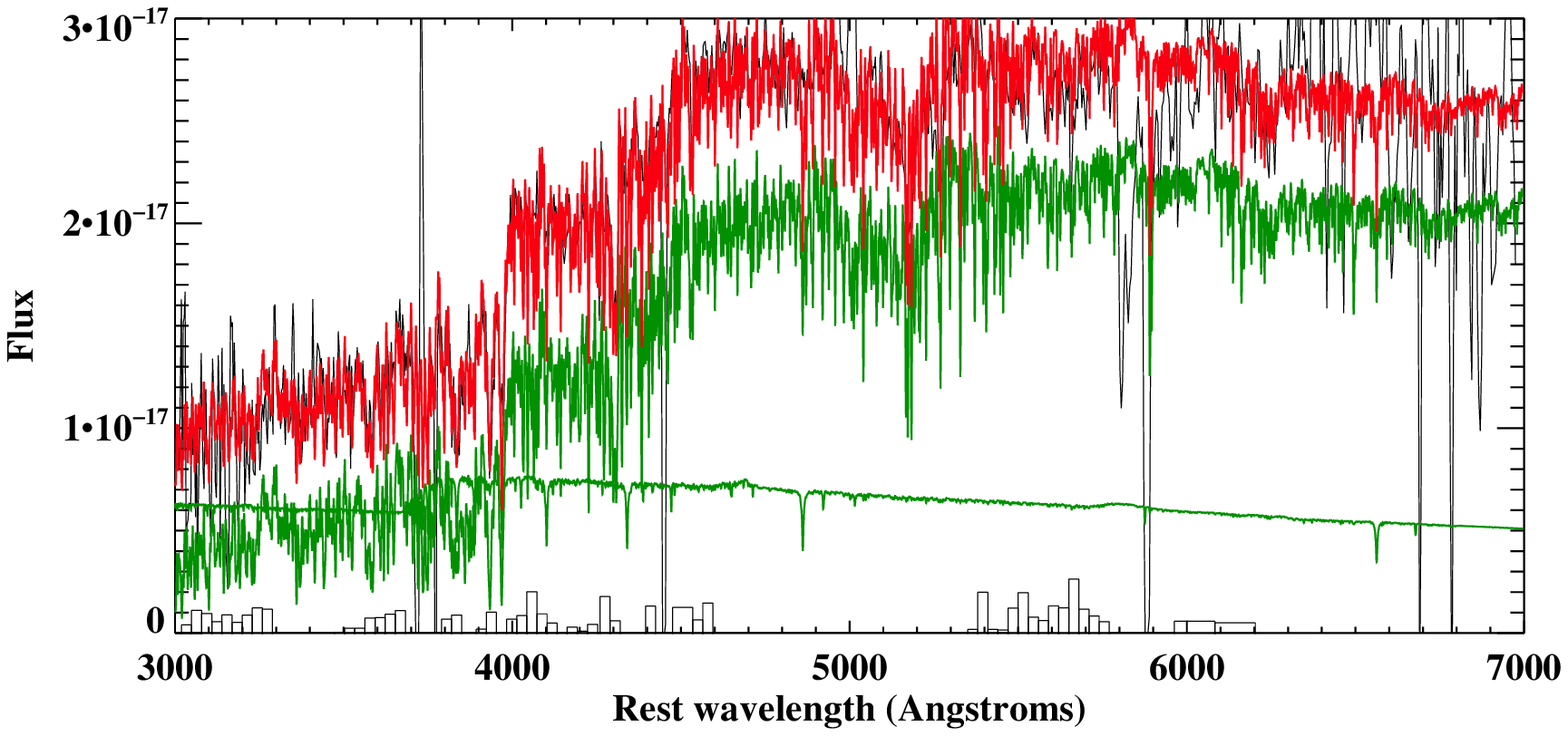}
\label{q0332-5kpc-8np-b-bf}}

\subfloat{
\includegraphics[scale = 0.45, trim = 7mm 0mm 0mm 0mm]{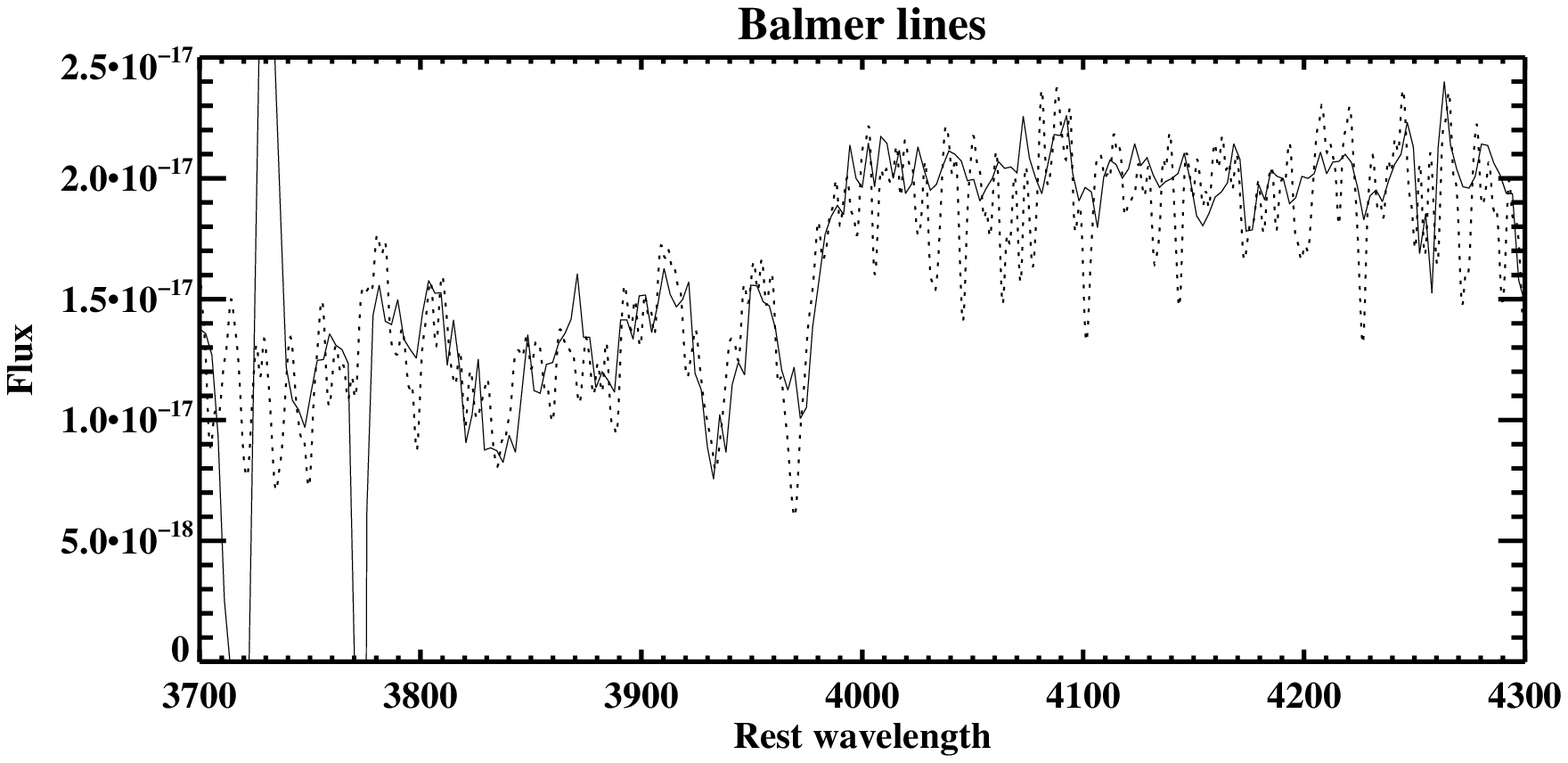}
\label{q0332-5kpc-8np-b-bl}}

\caption{An example of an acceptable fit produced by \confit for J0332 -- Companion nucleus. This particular fit was generated using combination 2 and incorporates a YSP with \tysp = 0.005 Gyr and \ebv = 0.6.}

\label{spec:q0332-5kpc_confit}
\end{figure}

Even though the companion does not host an AGN, its proximity to the type I quasar means that the extent to which this aperture will be contaminated by scattered light is unclear, and therefore, all five combination of components were run. Tables \ref{results8} and \ref{results2} shows that only combinations which include an 8 Gyr component produce acceptable fits. All combinations that include a 2 Gyr ISP do produce solutions with \chisqlt, however, these invariably over-predict the \ca\ absorption feature and were therefore rejected. Figure \ref{spec:q0332-5kpc_confit} shows an example of an acceptable fit produced using combination 2. In this case, the YSP has age \tysp = 0.005 Gyr and \ebv = 0.6.


\subsection{J0334+00}
\label{spec:q0334}
The initial modelling for J0334+00 was performed using combination 2, and  resulted in a best fitting model with \tysp\ = 0.03 Gyr and \ebv\ = 0.1. The nebular continuum model then contributes $\sim 22\%$ of the flux in the 3540 \AA\ -- 3640 \AA\ region. The  Balmer decrements were measured using both the \hg/\hb\ and \hd/\hb\ ratios (Table \ref{blines}). The two measurements result in different ranges of possible reddenings, and therefore a value of \ebv\ = 0.36 was assumed because this is the average value of the region in which the values overlap. However, this resulted in slightly poorer fits than was achieved in the initial run, and therefore an unreddened nebular model was subtracted before proceeding with the modelling.

\begin{figure}
\centering
\subfloat{
\includegraphics[scale = 0.45, trim = 7mm 0mm 0mm 0mm]{arxivSub/q0334-8np-b-bf.eps}
\label{q0334-8np-b-bf}}

\subfloat{
\includegraphics[scale = 0.45, trim = 7mm 0mm 0mm 0mm]{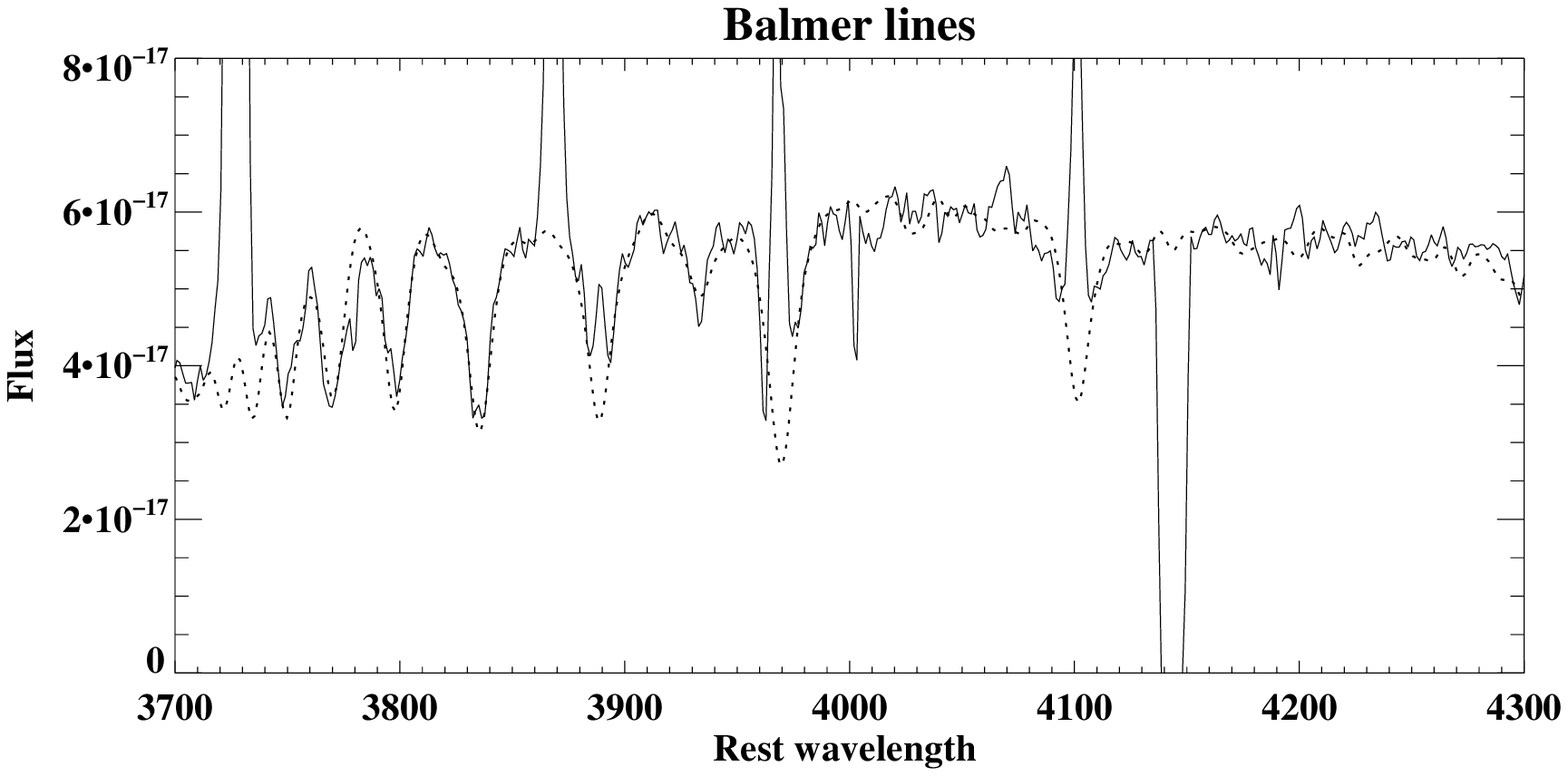}
\label{q0334-8np-b-bl}}

\caption{An example of an acceptable fit produced by \confit for J0334+00. The fit shown here was produced using combination 2 and includes a YSP with \tysp\ = 0.08 Gyr and \ebv\ = 0.}
\label{spec:q0334_confit}
\end{figure}

Tables \ref{results8} and \ref{results2} show that all four combinations that include a YSP component are successful.  When no power-law component is included, the upper age is always less that 100 Myr, with low values of reddening for the YSP population ($0 \le \ebv \le 0.2$).  The inclusion of a power-law allows a wider spread of ages and reddenings with a maximum age of 0.4 Gyr and reddening in the range $0 \le \ebv \le 0.4$. In all cases for which successful fits are achieved, the minimum allowable age is 0.04 Gyr. Figure \ref{spec:q0334_confit} shows an example of an acceptable confit fit produced by combination 2. This particular fit has \tysp\ = 0.08 Gyr and \ebv\ = 0.


\subsection{J0848+01}
\label{spec:q0848}

The initial modelling for J0848+01 was carried out using combination 2, and the resulting best fitting model comprised a YSP \tysp\ = 0.05 with \ebv\ = 0.4. The nebular continuum generated for this object contributes $\sim19 \%$ of the flux just below the Balmer edge. J0848+01 has intrinsically low equivalent width \hg\ and \hd\ emission lines, and therefore it was not possible to make a reliable measurement of the Balmer decrements in order to estimate the reddening, and an unreddened nebular continuum was subtracted. 

\begin{figure}
\centering
\subfloat{
\includegraphics[scale = 0.45, trim = 7mm 0mm 0mm 0mm]{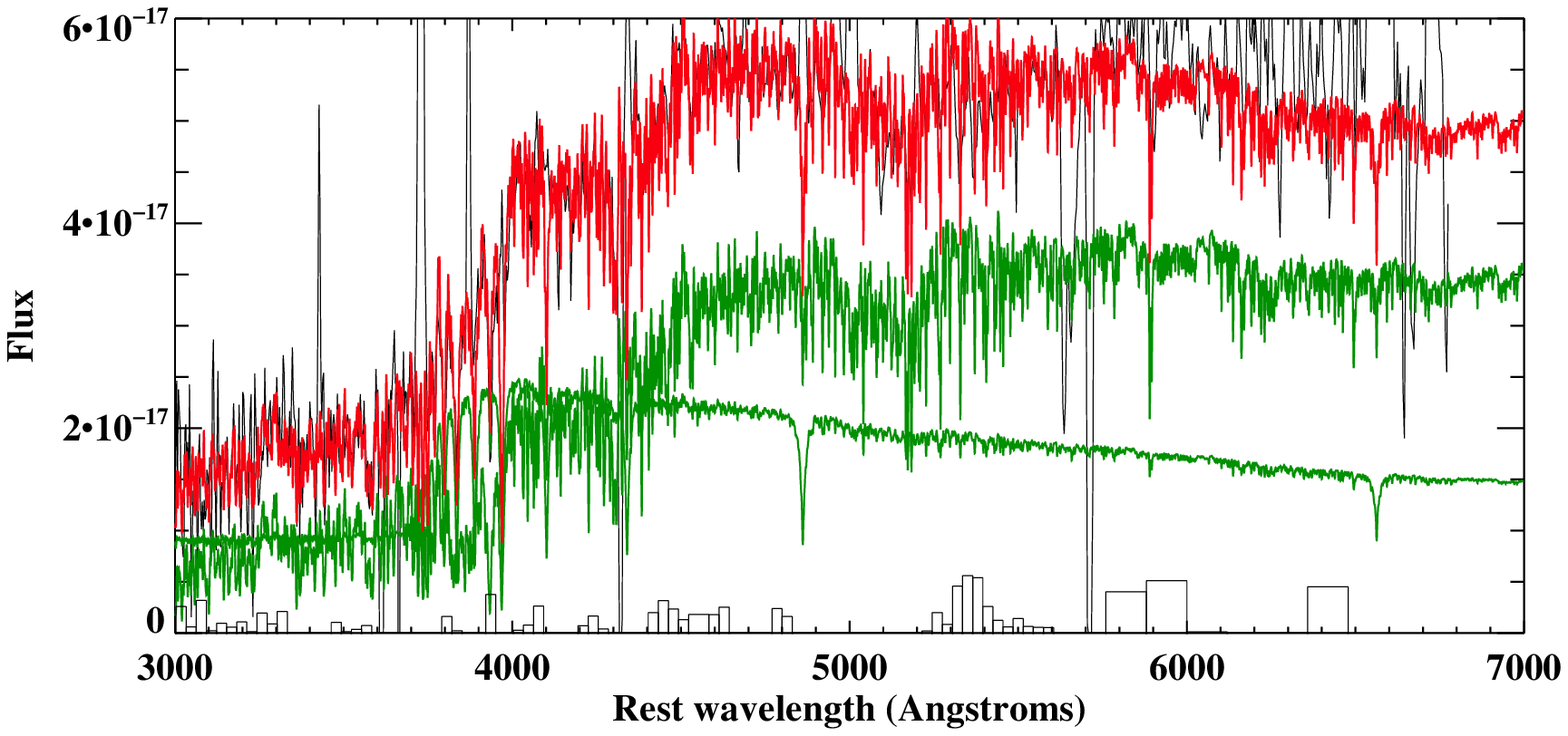}
\label{q0848-8np-b-bf}}

\subfloat{
\includegraphics[scale = 0.45, trim = 7mm 0mm 0mm 0mm]{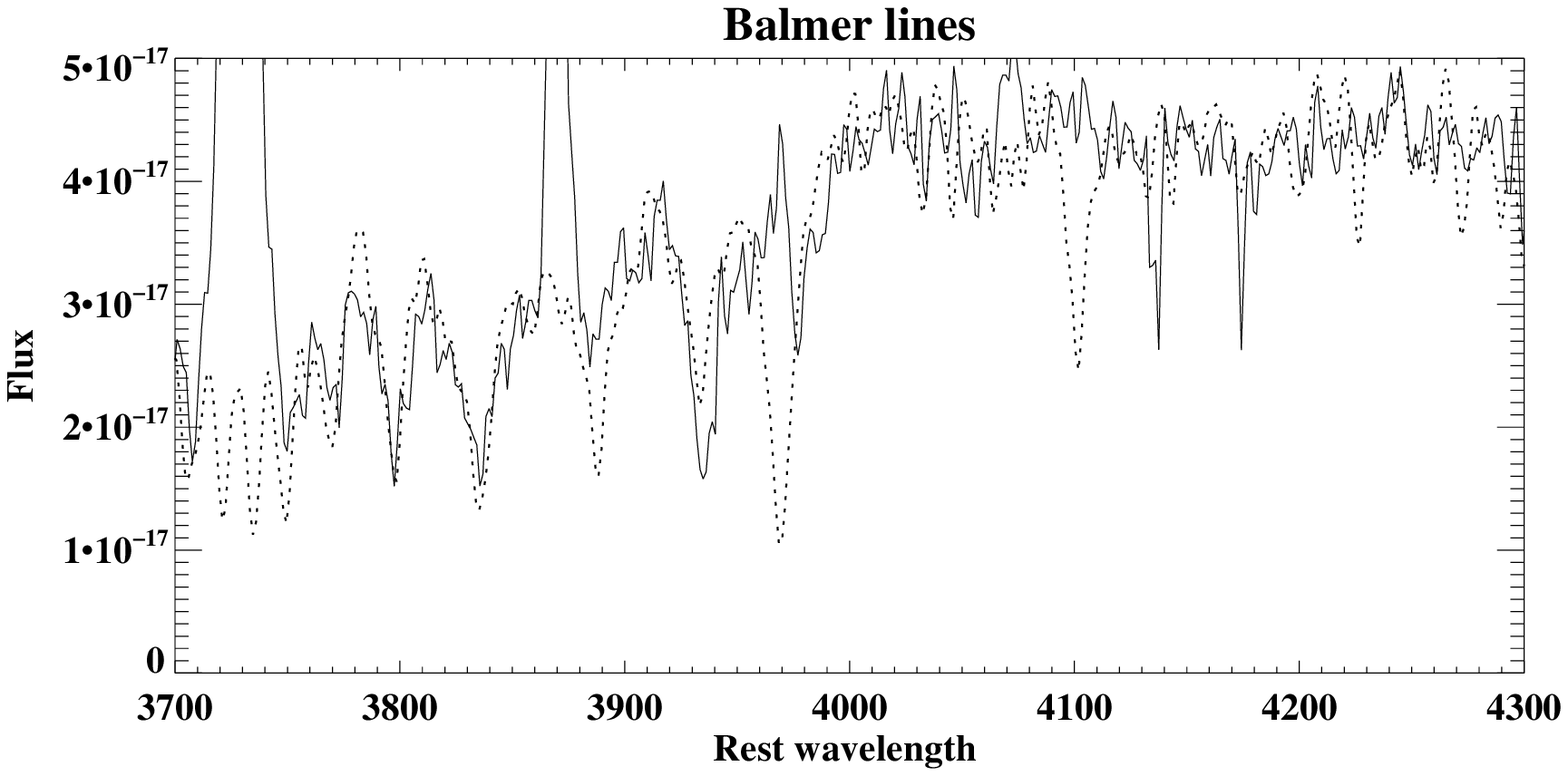}
\label{q0848-8np-b-bl}}

\caption{An example of an acceptable fit produced by \confit for J0848+01. The fit shown here was generated using combination 2 and include a YSP with \tysp\ = 0.2 Gyr and \ebv\ = 0.3.}
\label{spec:q0848_confit}
\end{figure}

Tables \ref{results8} \& \ref{results2} show that all the models which included a YSP component produced successful fits with fairly consistent ages of $0.2 \le \tysp \le 0.5$ Gyr. In nearly all the successful fits, the YSP contributes $ < 50 \% $ of the total flux of the system. Figure \ref{spec:q0848_confit} shows an example of an acceptable fit generated by combination 2, which utilises a YSP \tysp\ = 0.2 Gyr and \ebv\ = 0.3.

\subsection{J0904-00}
\label{spec:q0904}

The initial modelling for J0904-00 was carried out using combination 2, which resulted in a best fitting model with \tysp = 0.02 Gyr and \ebv =0.3. After subtracting this model, the flux of \hb\ was measured from the resulting spectrum in order to construct the nebular model, which contributes $\sim 13\%$ to the total flux just below the Balmer edge. No reddening was applied to the nebular continuum before subtraction, because the \hg/\hb\ ratio is consistent with case B recombination (Table \ref{blines}), and thus it was deemed unnecessary to do so.

\begin{figure}
\centering
\subfloat{
\includegraphics[scale = 0.45, trim = 7mm 0mm 0mm 0mm]{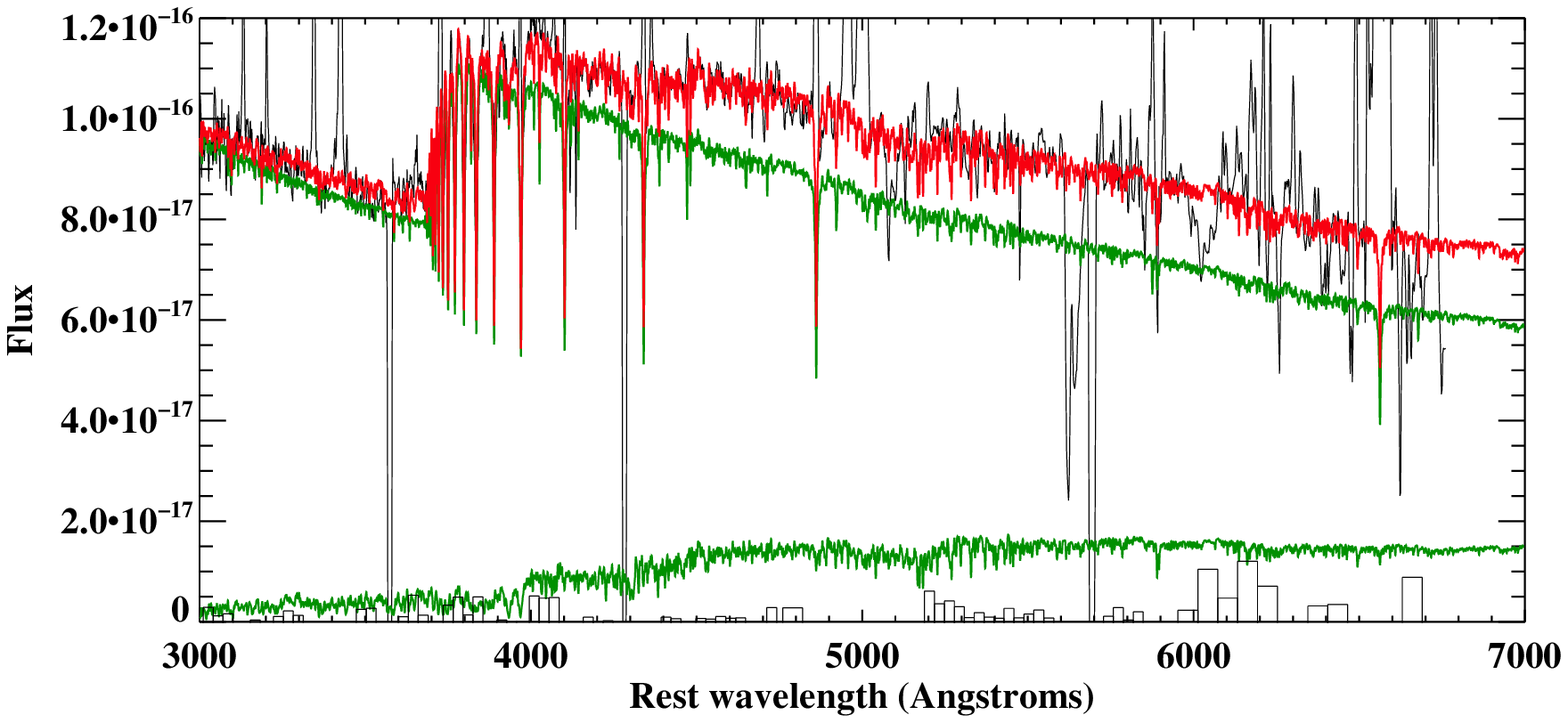}
\label{q0904-8np-b-bf}}

\subfloat{
\includegraphics[scale = 0.45, trim = 7mm 0mm 0mm 0mm]{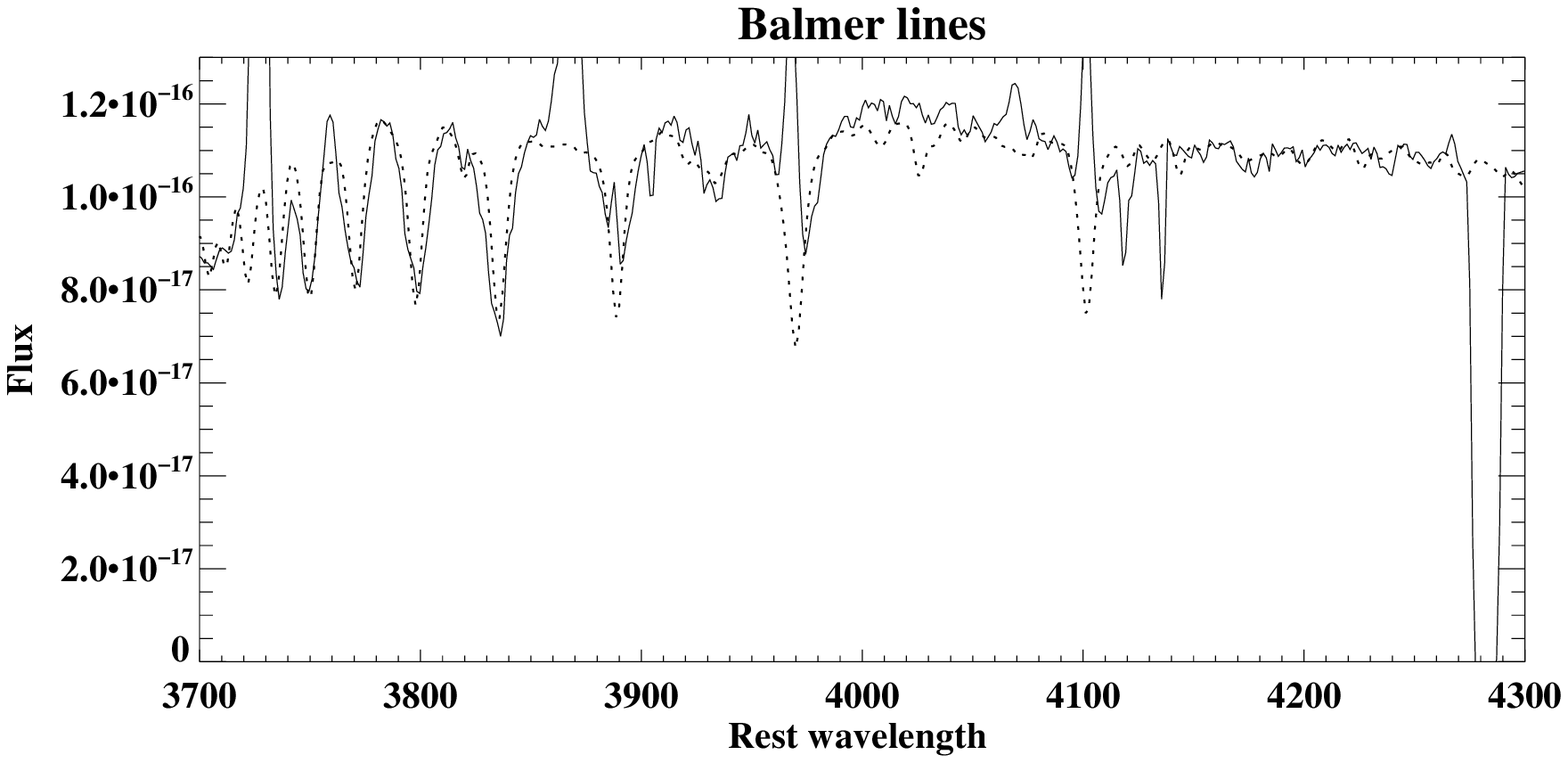}
\label{q0904-8np-b-bl}}

\caption{An example of an acceptable fit produced by \confit for J0904-00. This fit was produced using combination 2 and includes a YSP with \tysp = 0.006 Gyr and \ebv = 0.4.}
\label{spec:q0904_confit}
\end{figure}

Tables \ref{results8} \& \ref{results2} show the results of the stellar synthesis modelling performed for J0904-00,  and shows that all combinations which require two stellar components produce an acceptable fit. However, combination 1 does not produce an adequate result in this case. This is reinforced in Figure \ref{spec:q0904_confit} (bottom panel) which demonstrates that the Balmer lines are clearly detectable in the spectrum, meaning that a YSP component is \emph{required}. Figure \ref{spec:q0904_confit} shows an example of an acceptable combination 2 fit for J0904-00, which includes a YSP with \tysp = 0.006 Gyr and \ebv = 0.4.


\subsection{J0923+01}
\label{spec:q0923}

The initial modelling for J0923+01 was performed using combination 2, which resulted in a best fitting model that included a YSP with \tysp = 0.04 Gyr and \ebv = 0.5. This model was subtracted from the observed data, and the \hb\ flux was then measured from the stellar-subtracted spectrum, which resulted in a nebular continuum model which contributes $\sim 14\%$ of the total flux in the 3540--3640 \AA\ range. It was not possible to obtain a reliable estimate of the Balmer decrements in this case because the errors in the ratios of \hd/\hb\ and \hg/\hb\ were larger than the measured value. Therefore, an unreddened nebular continuum was subtracted from the data, which did not result in an unphysical step in the spectrum at the Balmer edge, and was thus considered acceptable.

\begin{figure}
\centering
\subfloat{
\includegraphics[scale = 0.45, trim = 7mm 0mm 0mm 0mm]{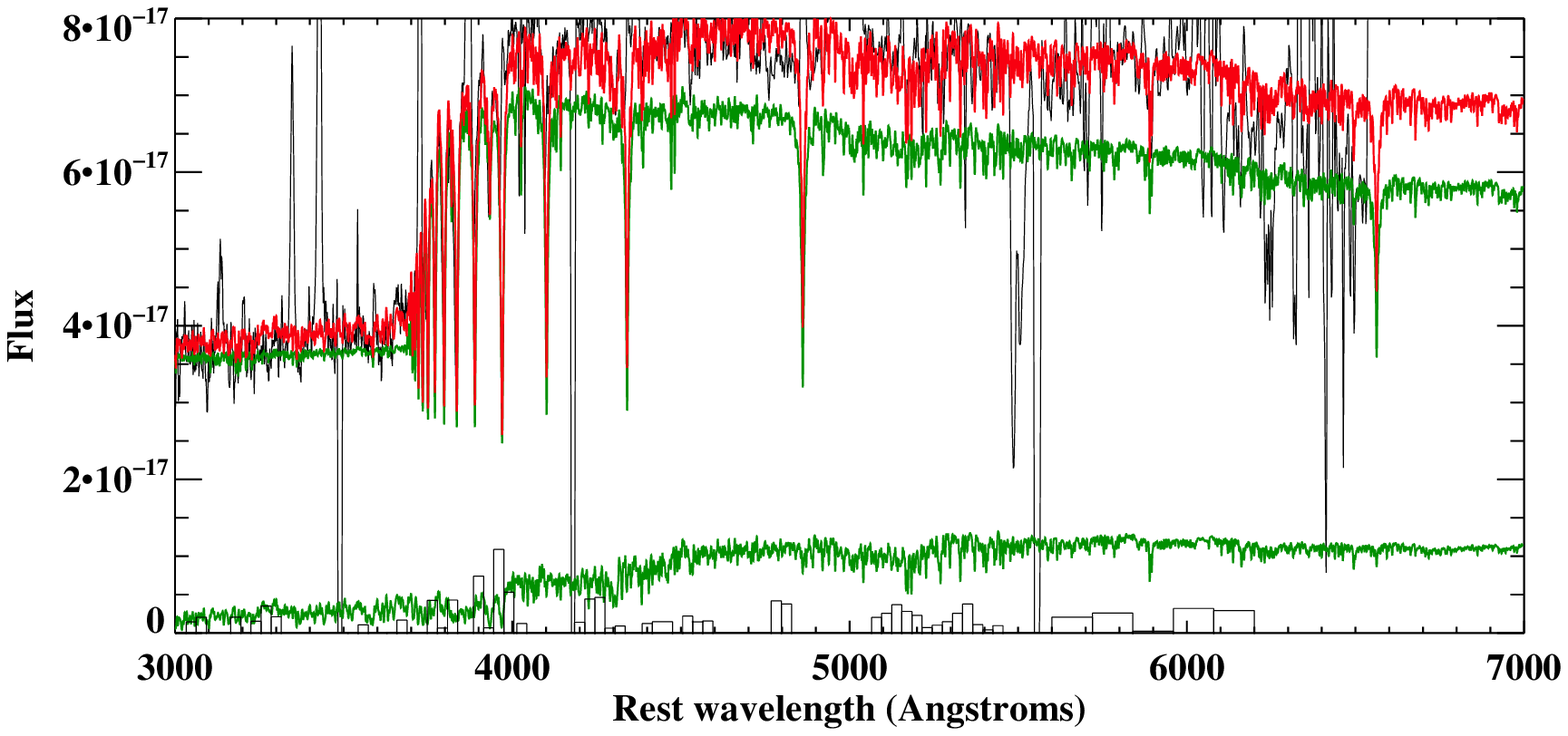}
\label{q0923-8np-b-bf}}

\subfloat{
\includegraphics[scale = 0.45, trim = 7mm 0mm 0mm 0mm]{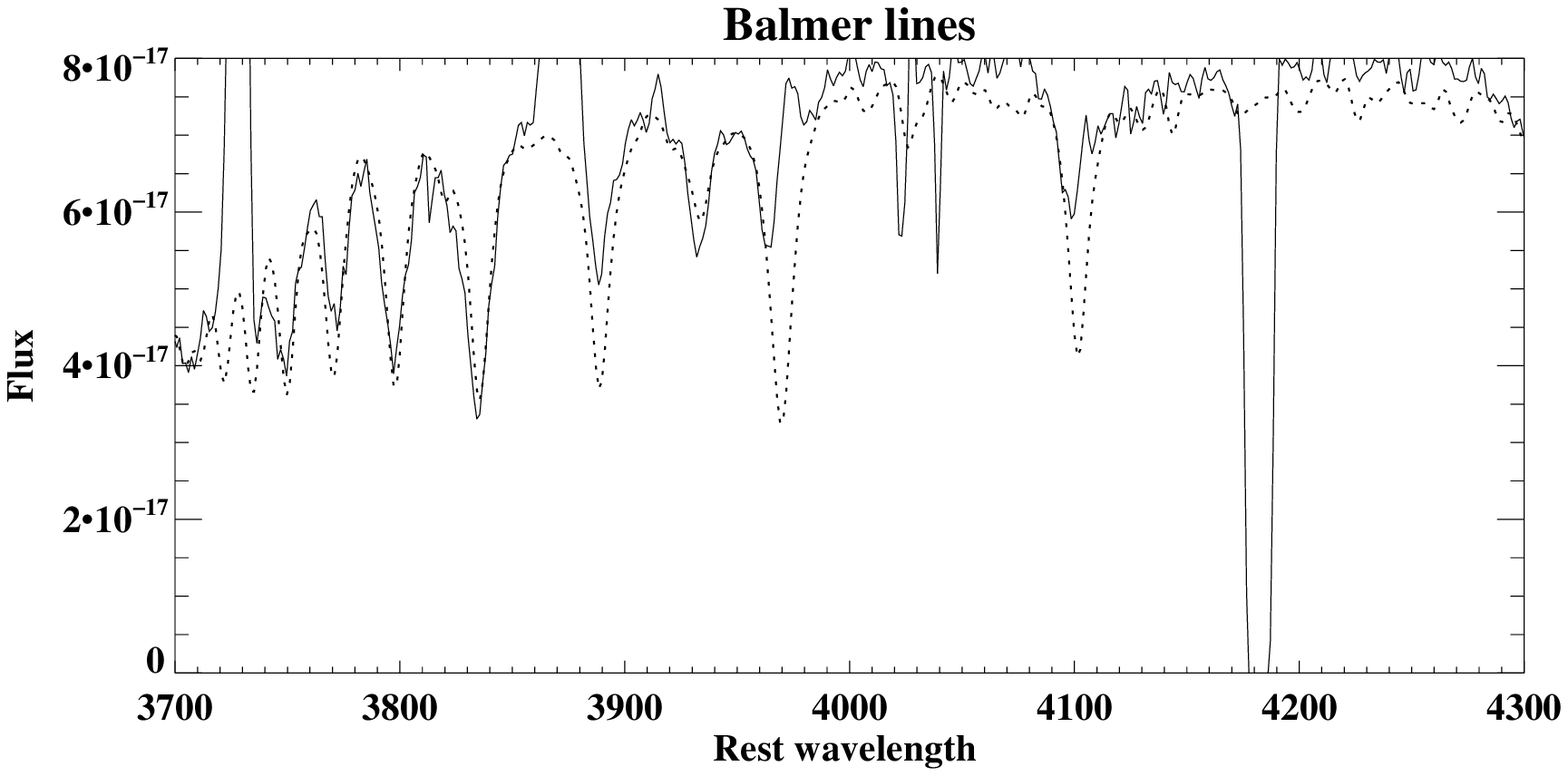}
\label{q0923-8np-b-bl}}

\caption{An example of an acceptable fit produced by \confit for J0923+01. The fit shown here was generated using combination 2 and incorporates a YSP with \tysp = 0.05 Gyr and \ebv = 0.5.}
\label{spec:q0923_confit}
\end{figure}

Tables \ref{results8} \& \ref{results2} show that, all the combinations which include a YSP component (2--5) produce acceptable fits, all of which allow for a contribution of up to 100 \% of the flux in the normalising bin from the YSP.  Figure \ref{spec:q0923_confit} shows an example of an acceptable fit generated using combination 2. In this case, the fit includes a YSP with age \tysp = 0.05 Gyr and \ebv = 0.5.


\subsection{J0924+01}
\label{spec:q0924}


The initial modelling of J0924+01 was performed using combination 3, which produced a best fitting model with \tysp = 0.003 and \ebv = 1.1. The resulting nebular continuum model then contributes $\sim 37\%$ of the total flux below the Balmer edge. The Balmer decrements were measured using both the \hg/\hb\ and \ha/\hb\ ratios, however, both values produce large errors. Therefore, the nebular continuum was initially reddened using a value of \ebv = 0.7, which resulted in an under subtraction of the nebular component. In order to test various values of reddening for the model nebular continuum, reddening beginning with the lowest value determined from the Balmer decrements (\ebv=0.2) was applied, and then increased in increments of $\Delta \ebv = 0.05$, until the point at which the continuum subtraction no longer resulted in a step in the spectrum at the Balmer edge (\ebv = 0.3, Table \ref{blines}). The nebular continuum contributes $\sim 14\%$ of the total flux just below the Balmer edge, after the reddening correction has been applied.

J0924+01 was modelled using all five combinations of components outlined in Section \ref{modelling}, and Tables \ref{results8} and \ref{results2} show that all combinations which include a YSP/ISP component produce acceptable fits. When considering combinations 2 and 3, the range of acceptable fits is very limited, with combination 2 producing two solutions with \chisqlt, in which the YSP dominates the flux, and combination 3 producing only one solution.  When using combination 4 and 5, which both include a power-law component, a wide range of solutions produce \chisqlt, and in both cases all solutions are acceptable, meaning that these combinations do not provide any real constraint on the age and/or reddening of any YSP present.

\begin{figure}
\centering
\subfloat{
\includegraphics[scale = 0.45, trim = 7mm 0mm 0mm 0mm]{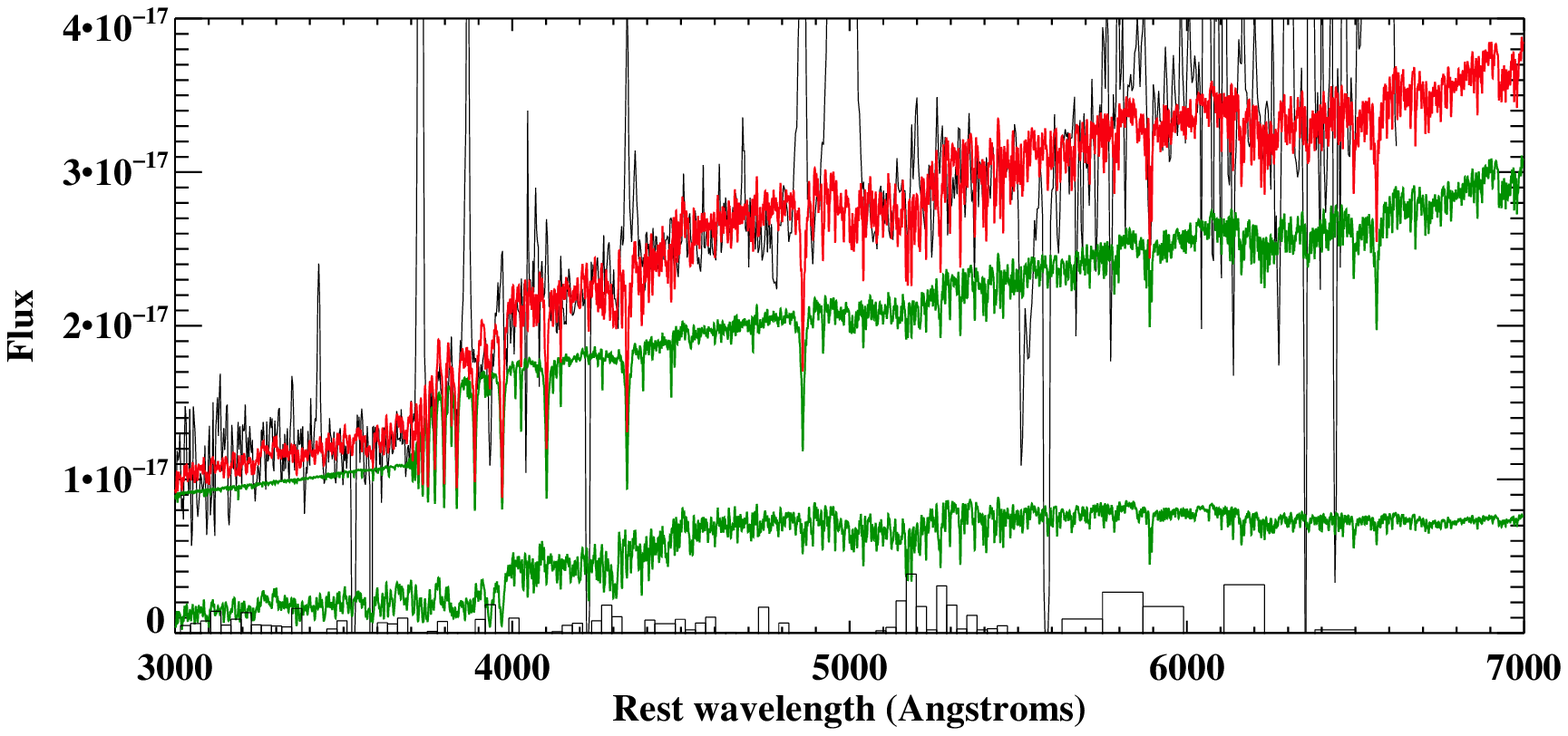}
\label{q0924-8np-b-bf}}

\subfloat{
\includegraphics[scale = 0.45, trim = 7mm 0mm 0mm 0mm]{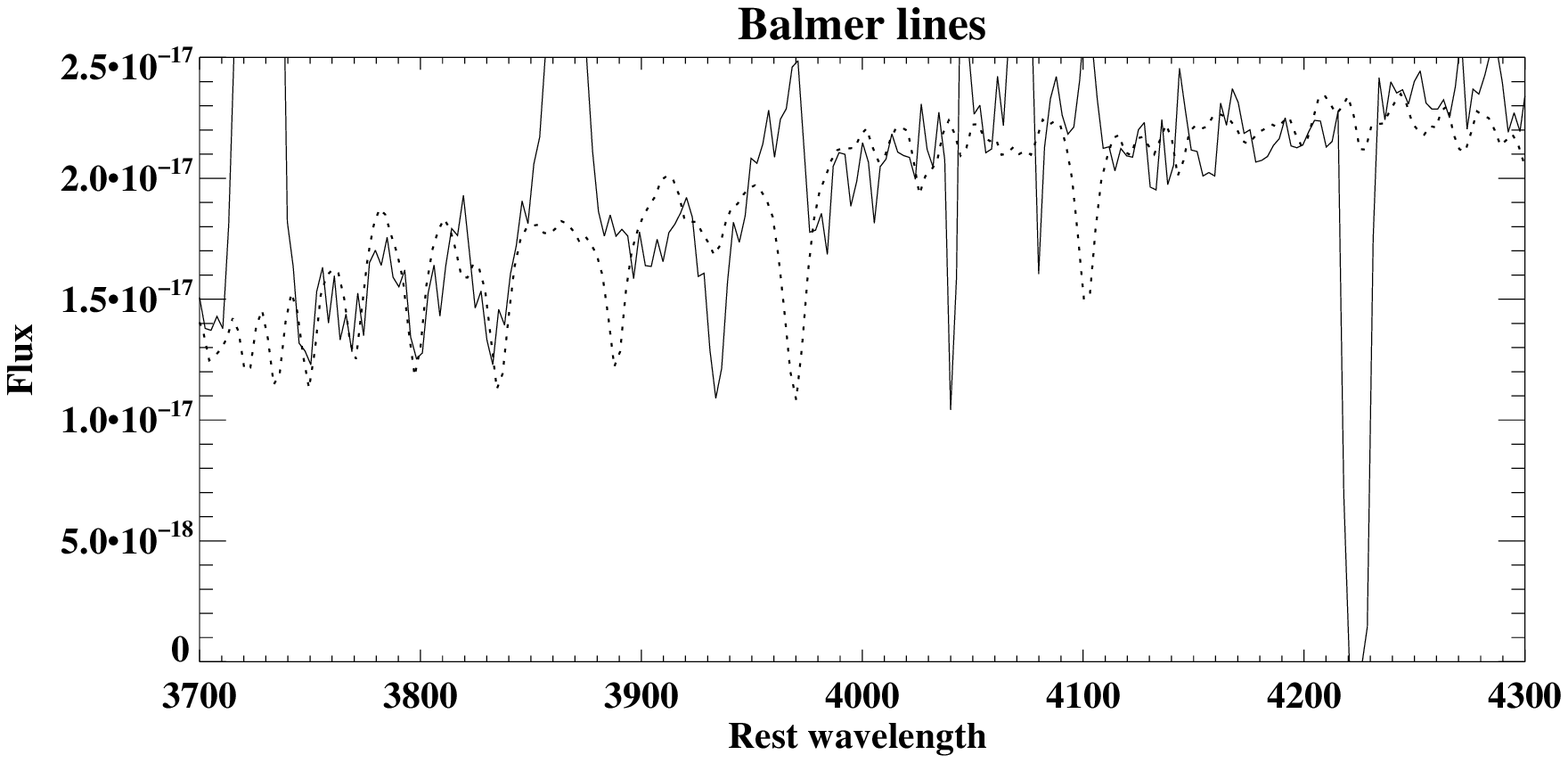}
\label{q0924-8np-b-bl}}

\caption{An example of an acceptable fit produced by \confit for J0924+01. The combination shown here was produced using combination 2 and includes a YSP with \tysp = 0.02 Gyr and \ebv = 0.8.}
\label{spec:q0924_confit}
\end{figure}

Figure \ref{spec:q0924_confit} shows an example of an acceptable fit produced by \confit, which was generated using combination 2. This fit includes a YSP with \tysp = 0.02 Gyr and \ebv = 0.8. Figure \ref{q0924-8np-b-bl} shows that the strength of the \ca\ line is strongly under predicted by the model, which is consistent with the high value of reddening determined here. This is because, the gas and dust which is responsible for the reddening will enhance this feature. In this case, it is relatively easy to search for \ca\ absorption due to the reddening dust column because the YSP dominates the flux, however, when the YSP contributes a lower proportion of the flux, this effect would be more difficult to detect. 

\subsection{J0948+00}
\label{spec:q0948}

The initial modelling for J0948+00 was performed using combination 3 and resulted in a best fitting model which included a YSP with \tysp = 0.1 Gyr and \ebv=0.3. Having subtracted this model from the spectrum, the nebular continuum model generated from the measured \hb\ flux contributes $\sim 15\%$ of the total flux below the Balmer edge, and is thus relatively weak. It was not possible to make a reliable measurement of the Balmer decrements because of the low equivalent widths of the \hg\ and \hd\ lines. Therefore, an unreddened nebular continuum model was subtracted, which did not result in an unphysical step at the Balmer edge.

All five modelling combinations, which are outlined in Section \ref{modelling}, were applied to the spectrum and the results, given in Tables \ref{results8} and \ref{results2}, show that all combinations which include a YSP component provide acceptable fits.



\begin{figure}
\centering
\subfloat{
\includegraphics[scale = 0.45, trim = 7mm 0mm 0mm 0mm]{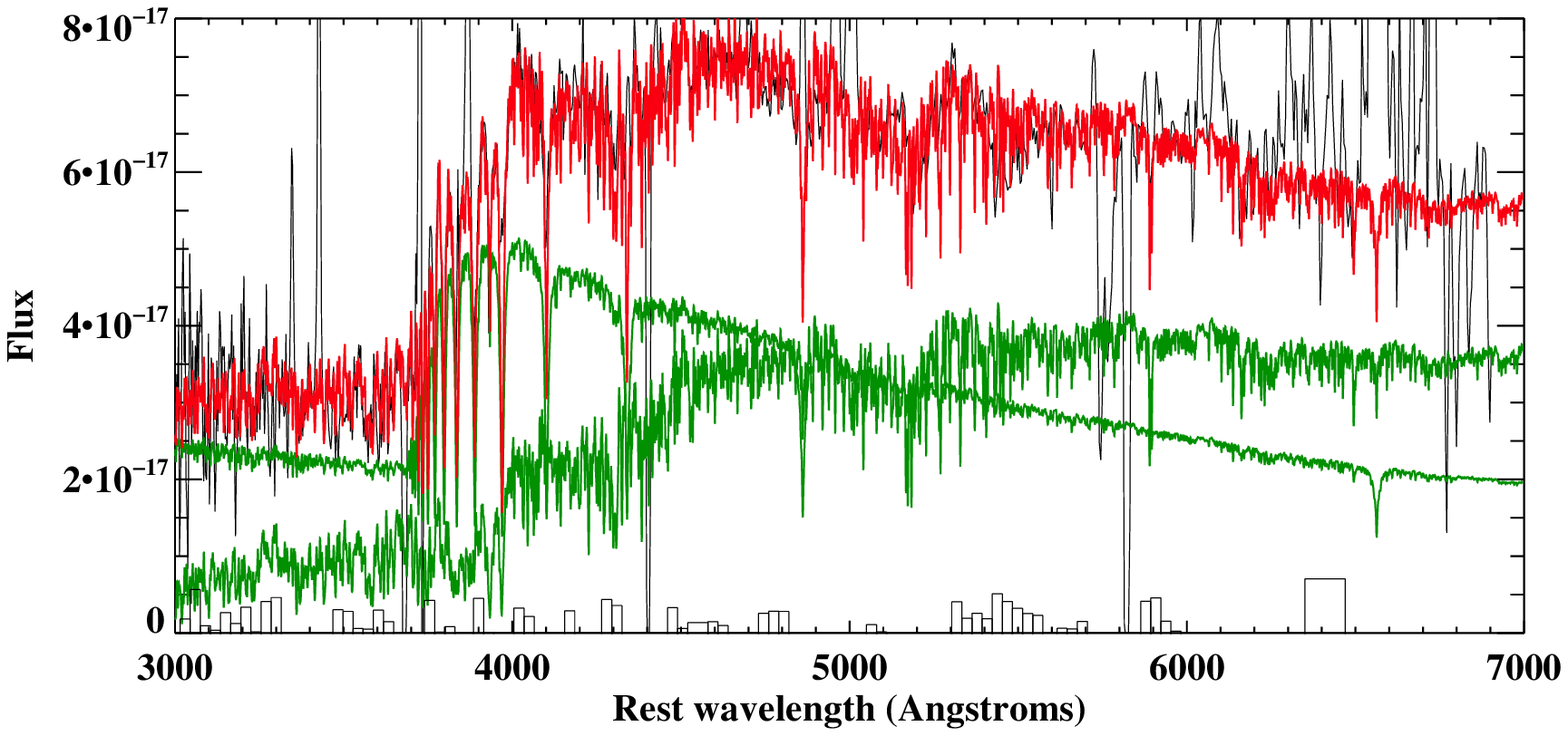}
\label{q0948-8np-b-bf}}

\subfloat{
\includegraphics[scale = 0.45, trim = 7mm 0mm 0mm 0mm]{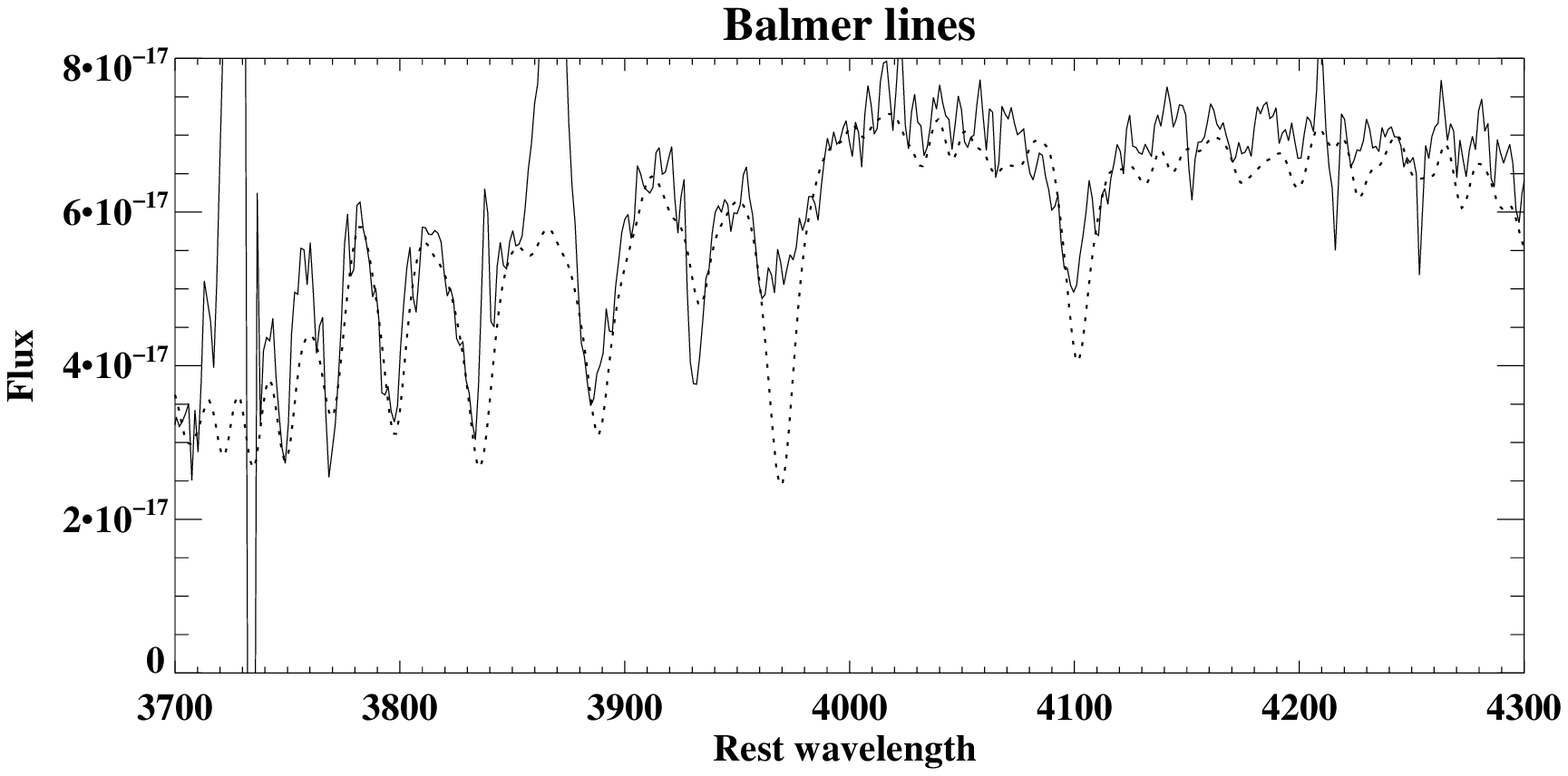}
\label{q0948-8np-b-bl}}

\caption{An example of an acceptable fit produced by \confit for J0948+00. The fit shown here was generated using combination 2 and includes a YSP with \tysp = 0.2 Gyr and \ebv = 0.1.}
\label{spec:q0948_confit}
\end{figure}

On the basis of the deep $r^\prime$-band Gemini \gmos\ image \citep{bessiere12}, J0948+00 was classified as an undisturbed galaxy, however, the spectrum shows clear evidence of an episode on star formation within the last 0.4 Gyr. This means that like J0142+14 (Section \ref{spec:q0142}), J0948+00 could also be a candidate E+A galaxy. Figure \ref{spec:q0948_confit} shows an example of an acceptable fit generated using combination 2, and shows that the OSP and YSP components contribute roughly equal proportions of the total flux in the normalising bin (4520--4580 \AA). This fit includes a YSP with \tysp = 0.2 Gyr and \ebv = 0.1.

\subsection{J1307-02}
\label{spec:q1307}

\begin{figure}
\centering
\subfloat{
\includegraphics[scale = 0.45, trim = 7mm 0mm 0mm 0mm]{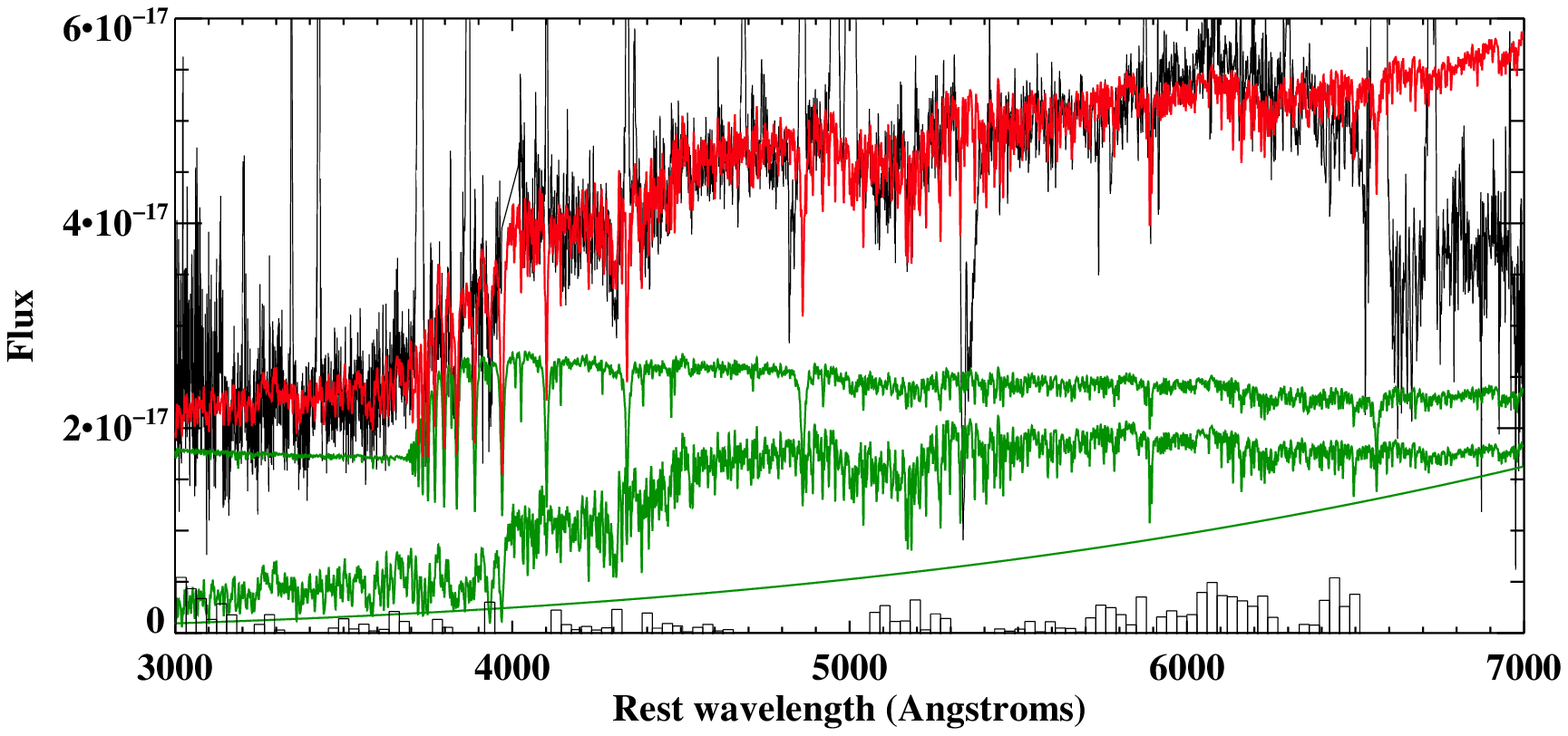}
\label{q1307-8p-b-bf}}

\subfloat{
\includegraphics[scale = 0.45, trim = 7mm 0mm 0mm 0mm]{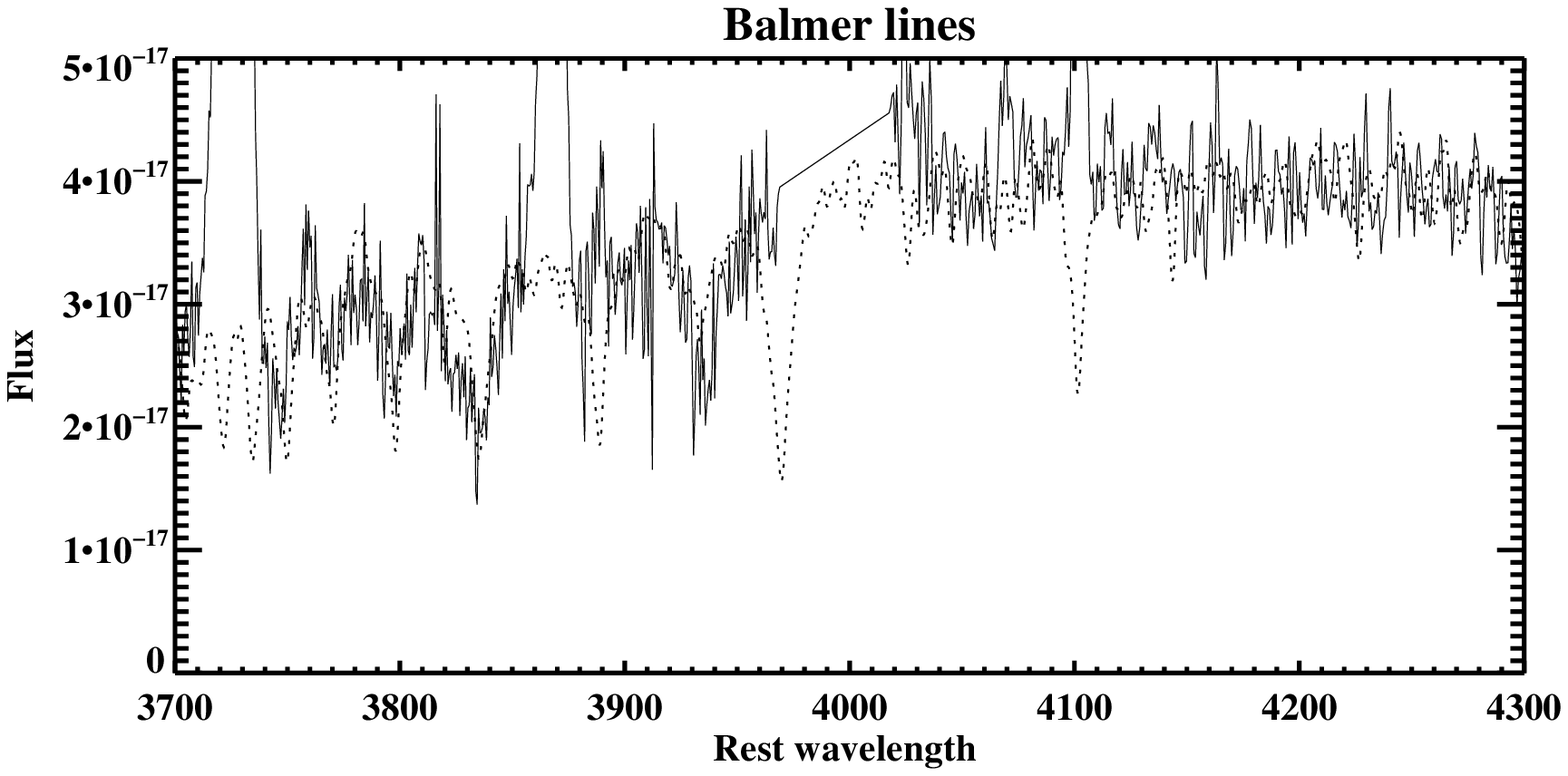}
\label{q1307-8p-b-bl}}

\caption{An example of an acceptable fit produced by \confit for J1307-02. The fit shown here was generated using combination 4 and includes a YSP with \tysp = 0.08 Gyr and \ebv = 0.1.}
\label{spec:q1307_confit}
\end{figure}

The initial modelling for J1307-01 was carried out using combination 4. Although acceptable fits could be achieved with combination 2, those which included a power-law component were found to provide better fits around the Balmer absorption features and G-band. A stellar model including a YSP \tysp= 1.8 Gyr and \ebv = 0.3 was subtracted from the data and the nebular continuum generated from the stellar-subtracted spectrum contributes 27\% of the total flux just below the Balmer edge. The Balmer decrement was determined from \hg/\hb\ to be 0.40 $\pm$ 0.07, however, when this reddening correction was applied to the nebular continuum, this resulted in an under subtraction around the Balmer edge in the data. Therefore, the unreddened nebular continuum was subtracted from the data before proceeding with the modelling. All five modelling combinations were attempted, the results of which are shown in Tables \ref{results8} and \ref{results2}. We can see that when an 8 Gyr OSP is assumed, it is only with the inclusion of a power-law component that acceptable fits are achieved. However, in the case where a 2 Gyr underlying population is assumed, acceptable fits can be achieved both with and without a power-law component. Figure \ref{spec:q1307_confit} shows an example of an acceptable combination 4 fit for J1307-02, which includes a YSP with \tysp = 0.08 Gyr and \ebv = 0.1. 

\subsection{J1337-01}
\label{spec:q1307}

\begin{figure}
\centering
\subfloat{
\includegraphics[scale = 0.45, trim = 7mm 0mm 0mm 0mm]{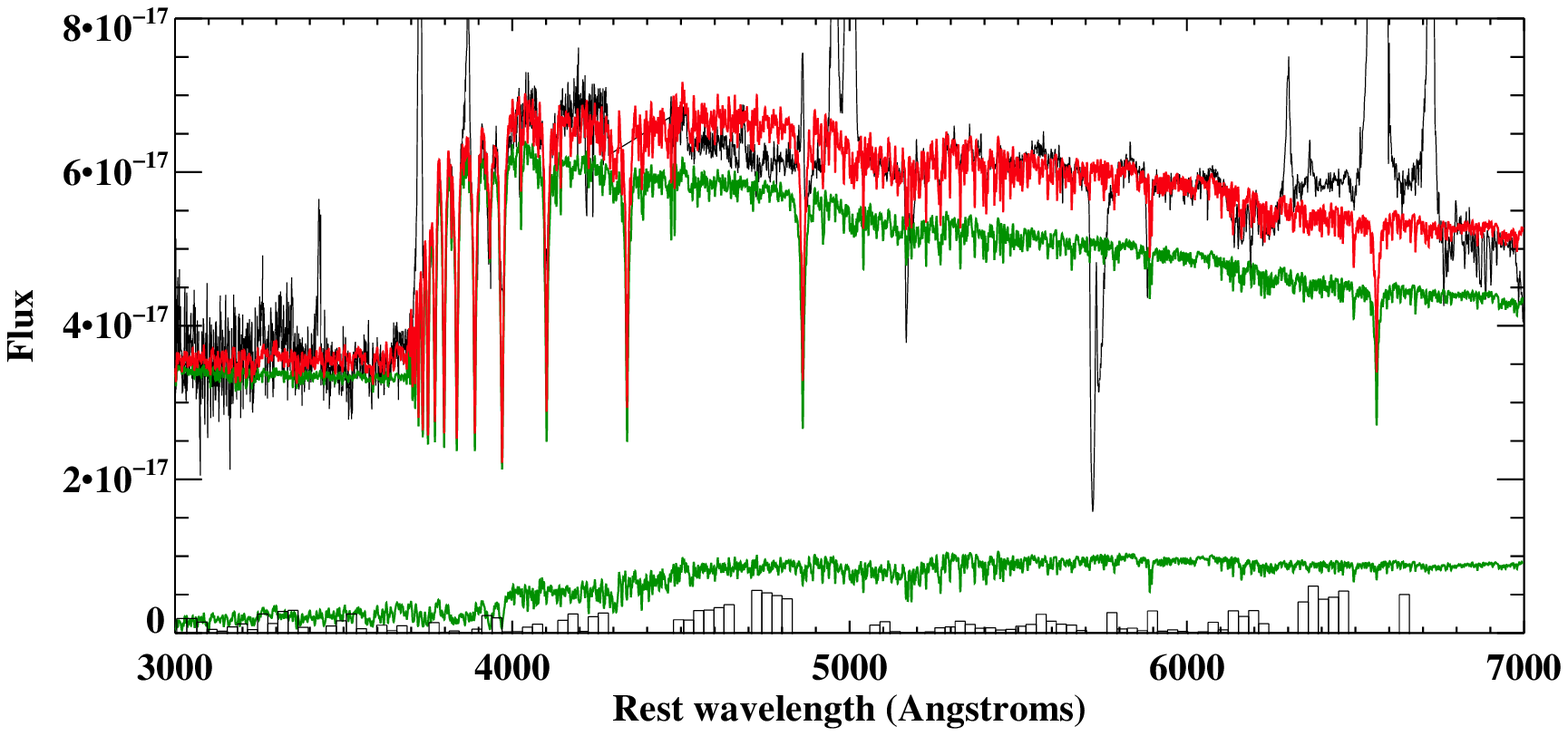}
\label{q1337-8np-b-bf}}

\subfloat{
\includegraphics[scale = 0.45, trim = 7mm 0mm 0mm 0mm]{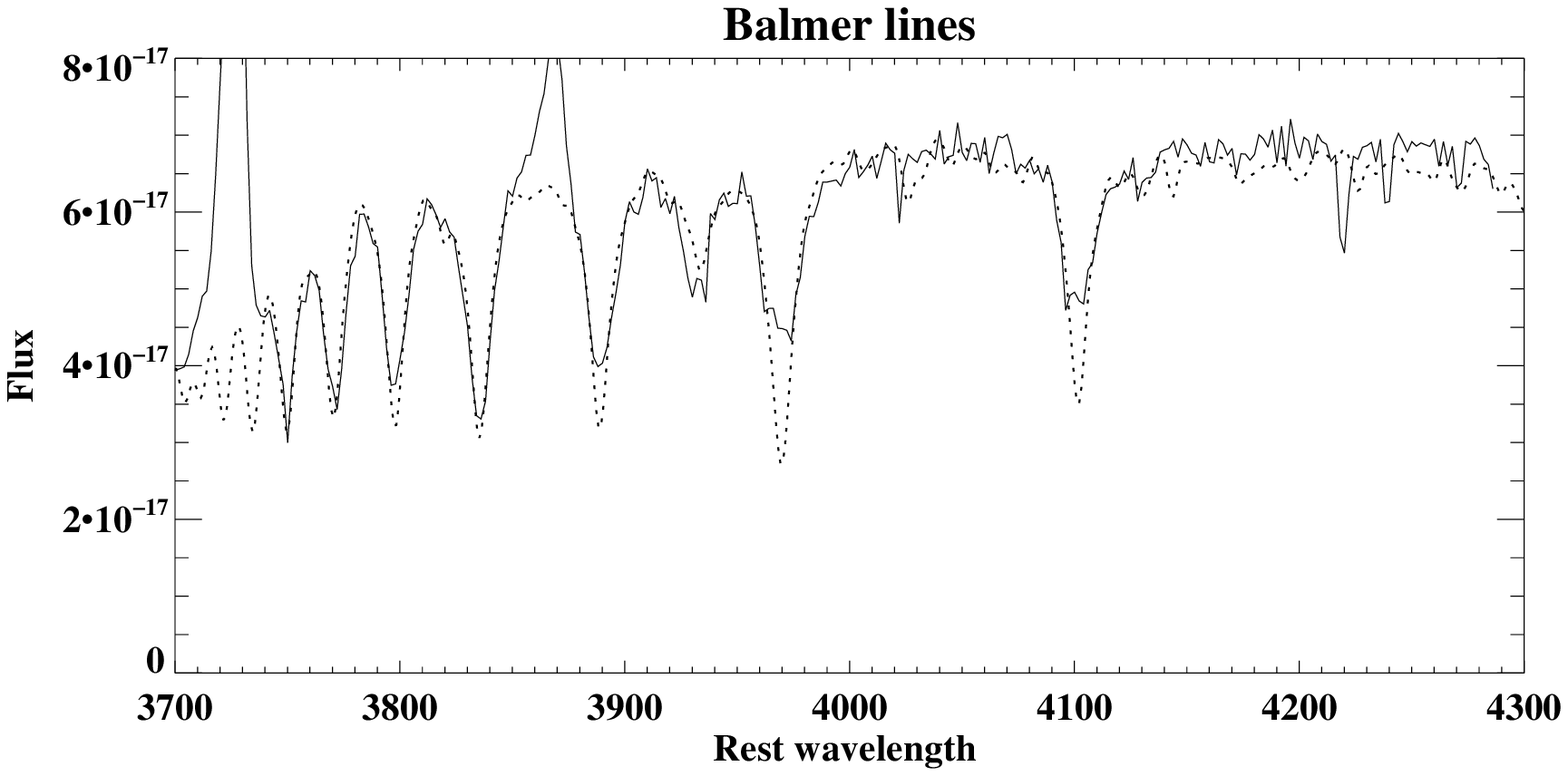}
\label{q1337-8np-b-bl}}

\caption{An example of an acceptable fit produced by \confit for J1337-01. The fit shown here was generated using combination 2 and includes a YSP with \tysp = 0.06 Gyr and \ebv = 0.4.}
\label{spec:q1337_confit}
\end{figure}

The initial modelling was carried out for J1337-01 using combination 2 which produced a best fitting model with a YSP \tysp = 0.04 and \ebv = 0.5. This stellar model was subtracted from the data and the \hb\ value flux measured in order to construct the nebular continuum. This nebular model was found to contribute $\sim 8\%$ of the total flux just below the Balmer edge. The internal reddening was measured using both the \hb/\ha\ and \hd/\hb\ ratios and were found to be consistent with Case B and therefore, no reddening correction was applied to the nebular before subtraction. All the modelling combinations outlined in section \ref{modelling} were applied and it can be seen in Tables \ref{results8} and \ref{results2} that all combinations that include a YSP are successful in producing acceptable results, some of which will allow for a 100\% contribution to the flux in the normalising bin by the YSP. Figure \ref{spec:q1337_confit} shows an example of an acceptable fit generated using combination 2 and including a YSP with \tysp = 0.06 Gyr and \ebv = 0.4.

\bsp	
\label{lastpage}
\end{document}